\documentclass[11pt,a4paper]{article}

\pdfoutput=1

\usepackage{jheppub}
\usepackage{graphicx}
\usepackage{dcolumn}
\usepackage{bm}
\usepackage{amssymb}
\usepackage{amsmath}
\usepackage{color}
\usepackage{slashed}
\usepackage{hyperref}
\usepackage{multirow,array}
\usepackage{subcaption}
\usepackage{enumitem}
\usepackage[normalem]{ulem}

\usepackage{physics}
\usepackage{listings}
\usepackage{cprotect}

\allowdisplaybreaks[1]

\makeatletter
\gdef\@fpheader{}
\makeatother

\preprint{OU-HET 1282, SAGA-HE-294}

\title{Radiative corrections to decays of the 125 GeV Higgs boson in the complex Higgs triplet model}

\author[a]{Masashi Aiko,}
\author[b]{Shinya Kanemura,}
\author[c]{Mariko Kikuchi,}
\author[d,e]{Kodai Sakurai,}
\author[b]{Sora Taniguchi}
\author[f]{and Kei Yagyu}

\affiliation[a]{National Institute of Technology, Miyakonojo College, Miyakonojo, Miyazaki 885-8567, Japan}
\affiliation[b]{Department of Physics, University of Osaka, Toyonaka, Osaka 560-0043, Japan}
\affiliation[c]{Department of Physics, Saga University, 1 Honjomachi, Saga 840-8502, Japan}
\affiliation[d]{Department of Physics, Tohoku University, Sendai, Miyagi 980-8578, Japan}
\affiliation[e]{National Institute of Technology, Tsuruoka College, Tsuruoka, Yamagata 997-0842, Japan}
\affiliation[f]{Department of Physics, Tokyo University of Science, 1-3, Kagurazaka, Shinjuku-ku, Tokyo 162-8601, Japan}

\emailAdd{m-aiko@miyakonojo.kosen-ac.jp}
\emailAdd{kanemu@het.phys.sci.osaka-u.ac.jp}
\emailAdd{mkikuchi@cc.saga-u.ac.jp}
\emailAdd{kodai.sakurai@tsuruoka-nct.ac.jp}
\emailAdd{taniguchi@het.phys.sci.osaka-u.ac.jp}
\emailAdd{yagyu@rs.tus.ac.jp}

\abstract{
The extension of the Higgs sector with an additional complex triplet field is often considered for generating the neutrino mass by the Type-II seesaw mechanism. 
Such an extension generally predicts $\rho\neq1$, where $\rho$ is the electroweak rho parameter at the tree level, so that the renormalization of the electroweak parameters is different from models like the standard model (SM) and two Higgs doublet models. 
In this paper, we present a full set of radiative corrections to decays of the 125 GeV Higgs boson ($h$) in this model. 
One-loop contributions of the extra Higgs bosons as well as SM fermions and gauge bosons to the decay rates of $h$ are calculated in the on-shell scheme. 
Gauge dependence appearing in the counter terms of mixing angles is eliminated by the pinch technique.
Higher-order QCD corrections are also implemented. 
We find that the decay rates can significantly deviate from the predictions in the SM and other extensions such as the two Higgs doublet models and the singlet model. 
For example, the decay rates of $h\to WW^\ast$ and $h\to ZZ^\ast$ can be a few percent larger than the SM value under current experimental and theoretical constraints. 
In this case, deviations in $h\to \gamma\gamma$ and Higgs self-coupling can reach about $-20\%$ and $100\%$, respectively.  
The pattern of the deviations is different from the other extended Higgs models. 
These characteristic predictions are expected to be detected at the High-Luminosity LHC or future Higgs factories.
}

\begin{document}

\maketitle

\section{Introduction} \label{sec: introduction}

The Standard Model (SM) has been confirmed as a successful theory explaining physics at the electroweak scale. 
However, there remain various mysteries that cannot be explained in the SM, such as neutrino oscillations, dark matter, the baryon asymmetry of the Universe, and inflation. 
Constructing new physics models to solve these unsolved problems is a goal in high-energy physics. 
While the Higgs boson with a mass of 125~GeV has been discovered, the structure of the Higgs sector has not been determined yet. 
The dynamics of electroweak symmetry breaking and 
the nature of the electroweak phase transition are also to be understood. 
Extensions of the Higgs sector of the SM are often introduced in new physics models, in which additional Higgs bosons are predicted. 
Although there are various types of extended Higgs sectors, many of them can be described by the simplest extensions with an additional scalar field at least approximately. 

One of the simplest extensions is the model with an additional isospin triplet field with $Y=1$, which we call the complex Higgs triplet model (CHTM). 
This model may be motivated by generating neutrino masses via the Type-II seesaw mechanism~\cite{Cheng:1980qt, Schechter:1980gr, Lazarides:1980nt, Mohapatra:1980yp, Magg:1980ut},
for which allowed parameter regions consistent with current neutrino data are discussed in Ref.~\cite{FileviezPerez:2008jbu}. 
The triplet field can also be introduced for radiative seesaw models~\cite{Kanemura:2012rj,Nomura:2016run,Guo:2016dzl,Nomura:2017emk}. 
The CHTM can also appear in the low-energy theory of the left-right symmetric model~\cite{Pati:1974yy,Mohapatra:1974hk,Mohapatra:1974gc,Senjanovic:1975rk}. 
Another motivation of this model is that the electroweak phase transition can be realized in this model, leading to characteristic predictions such as stochastic gravitational waves~\cite{Zhou:2022mlz,Jangid:2023lny} and deviations from the SM in the Higgs boson self-coupling~\cite{Aoki:2012yt,Aoki:2012jj}. 

The experimental value of the electroweak rho parameter ($\rho$) is close to unity, which strongly constrains extended Higgs models. 
While $\rho=1$ is guaranteed at tree level in the model with doublets and singlets, $\rho\neq1$ is generally predicted in the CHTM\footnote{In the Georgi-Machacek model~\cite{Georgi:1985nv, Chanowitz:1985ug}, in which the CHTM is included as a part of the particles content, $\rho = 1$ is predicted at tree level.}. In order to satisfy the data, the vacuum expectation value (VEV) of the triplet field ($v_{\Delta}$) has to be taken to be small. The fact that the CHTM predicts $\rho\neq1$ at tree level requires a different framework of renormalization for electroweak parameters from the models with $\rho=1$ at tree level~\cite{Blank:1997qa,Kanemura:2012rs,Aoki:2012jj}. 

The CHTM provides rich phenomenology in collider experiments because it contains multiple extra Higgs bosons such as two neutral Higgs bosons, singly charged Higgs bosons ($H^\pm$), and doubly charged Higgs bosons ($H^{\pm\pm}$). 
Direct searches for $H^{\pm \pm}$ are a definite way to distinguish the CHTM from other extensions like multi-doublet models at collider experiments. 
The decay pattern of $H^{\pm\pm}$ largely depends on the mass spectrum of additional Higgs bosons and the VEV of the triplet field. 
For example, the mass of $H^{\pm\pm}$ is strongly constrained from below by the leptonic decays ($H^{++}\to \ell^+\ell^{+}$) when $v_\Delta$ is smaller than ${\cal O}(1)$~MeV~\cite{Melfo:2011nx}. 
On the other hand, for larger values of $v_\Delta$, $H^{\pm\pm}$ mainly decays to $W^\pm W^\pm$. 
In Ref.~\cite{Kanemura:2013vxa}, constraints for the mass of $H^{\pm \pm}$ via diboson searches have been investigated. 
In the case where mass differences among the additional Higgs bosons are relatively large, cascade decays of additional Higgs bosons become important. Such a case has been studied in Ref.~\cite{Aoki:2011pz}~\footnote{From the recent CDF II results, large mass differences are strongly indicated in the CHTM~\cite{Kanemura:2022ahw}.}.
In the triplet extension models, $H^\pm$ can, in general, decay into $W^\pm Z$ with a large branching ratio~\cite{Asakawa:2005gv,Asakawa:2006gm,Godfrey:2010qb}, while such a decay process is suppressed in the model with a multi-doublet structure~\cite{Grifols:1980uq}.
See also Refs.~\cite{Huitu:1996su, Akeroyd:2005gt, Garayoa:2007fw, Kadastik:2007yd, Akeroyd:2007zv, delAguila:2008cj, Akeroyd:2009hb, Akeroyd:2010ip, Akeroyd:2010je, Akeroyd:2011zza, Chiang:2012dk, Akeroyd:2012nd, Chun:2012zu, delAguila:2013mia, Chun:2013vma, Kanemura:2014goa, Kanemura:2014ipa, Kang:2014lwn, Han:2015hba, Han:2015sca, Babu:2016rcr, Mitra:2016wpr, Cai:2017mow, Antusch:2018svb, Primulando:2019evb, Ashanujjaman:2022ofg, Han:2023vme, Ashanujjaman:2023tlj, Ducu:2024xxf} for related studies.

Apart from the direct searches for additional particles in extended Higgs sectors, measurements of the discovered Higgs boson with a mass of 125~GeV ($h$) have been intensively performed. 
In general, new physics effects deviate the couplings of $h$ from the SM predictions, which are strongly model-dependent. 
Therefore, one can test various new physics models by precision measurements of the Higgs boson observables by detecting the pattern of deviations. 
The current measurement accuracy for the Higgs boson couplings is expected to be improved at the High-Luminosity LHC (HL-LHC)~\cite{Cepeda:2019klc,CMS:2025hfp} and future Higgs factories such as the International Linear Collider (ILC)~\cite{Baer:2013cma,Fujii:2017vwa,Asai:2017pwp,Fujii:2019zll}, the Circular Electron-Positron Collider (CEPC)~\cite{CEPC-SPPCStudyGroup:2015csa,Ai:2025cpj}, $e^+e^-$ collisions of the Future Circular Collider (FCC-ee)~\cite{Gomez-Ceballos:2013zzn,FCC:2025lpp}.
For example, the $hWW$ coupling is measured with an accuracy of 1.6 (0.35)$\%$ at the HL-LHC (at the International linear collider) while the current measurement uncertainty at the LHC is $6\%$ for ATLAS~\cite{ATLAS:2022vkf} and $8\%$ for CMS experiments~\cite{CMS:2022dwd}.
In order to determine the shape of the Higgs sector by utilizing such future high-precision data, the theoretical predictions for $h$ observables also should be accurately evaluated in a variety of extended Higgs models, including higher-order corrections. 

There are many previous works discussing radiative corrections in the context of the simple extended Higgs models with $\rho=1$ at tree level. 
In two-Higgs doublet models (2HDMs), the influence of one-loop corrections to the Higgs boson couplings~\cite{Kanemura:2004mg,Kanemura:2014dja,Kanemura:2015mxa,Arhrib:2015hoa,Kanemura:2016sos}, the decays of $h$~\cite{Kanemura:2019kjg,Kanemura:2018yai,Altenkamp:2017ldc} and heavy Higgs bosons~\cite{Aiko:2021can,Kanemura:2022ldq,Aiko:2022gmz,Krause:2019qwe,Akeroyd:1998uw,Akeroyd:2000xa,Krause:2016oke,Krause:2016xku} are investigated. 
Similar studies have also been performed in the framework of the Higgs singlet model (HSM)~\cite{Bojarski:2015kra,Kanemura:2015fra,Kanemura:2016lkz,Altenkamp:2018bcs,Egle:2023pbm}.  
Recently, two-loop corrections to $h$ observables are also discussed in Refs.~\cite{Braathen:2019pxr,Braathen:2019zoh,Braathen:2020vwo,Sturm:2022hak,Aiko:2023nqj,Degrassi:2023eii,Degrassi:2024qsf,Degrassi:2025pqt}. 
As a public tool to perform the numerical evaluations of higher-order corrections in the above extended Higgs models, the authors of this paper have developed the \texttt{H-COUP} program~\cite{Kanemura:2017gbi,Kanemura:2019slf,Aiko:2023xui}.
Apart from this program, other public tools such as {\tt 2HDECAY}~\cite{Krause:2018wmo}, {\tt ewN2HDECAY}~\cite{Engeln:2018mbg}, {\tt EWsHDECAY}~\cite{Egle:2023pbm}, {\tt profecy4f}~\cite{Denner:2019fcr}, and {\tt FlexibleDecay}~\cite{Athron:2021kve} are also available.
{\tt 2HDECAY}, {\tt ewN2HDECAY}, and {\tt EWsHDECAY} calculate the full NLO corrections to the two-body decays of all Higgs bosons in a given extended Higgs model. 
{\tt Prophecy4f} computes full NLO corrections for specific decay modes, namely the four-body decays of CP-even Higgs bosons into four-fermion final states, in the HSM and 2HDMs.
{\tt FlexibleDecay} incorporates various BSM models and computes higher-order corrections to the Higgs boson decays without BSM loop effects.
The difference between these programs and {\tt H-COUP} is the following.  
{\tt H-COUP} evaluates the full one-loop corrections to all major decay modes of $h$ and additional Higgs bosons within a common renormalization scheme for various extended Higgs models, 
thereby enabling a systematic comparison of predictions among different models.

The importance of precision calculations is also applied to the models with triplet scalar extensions.
In Ref.~\cite{Blank:1997qa}, a renormalization scheme was proposed to evaluate higher order corrections to the electroweak precision parameters in the model with a real triplet scalar field.
The renormalization scheme was extended to the CHTM in Ref.~\cite{Kanemura:2012rs}. 
In Refs.~\cite{Aoki:2012yt,Aoki:2012jj},  
the authors proposed an on-shell renormalization scheme for the Higgs sector in the CHTM, and clarified the importance of one-loop corrections of Higgs boson couplings to weak gauge bosons and Higgs self-couplings.
However, radiative corrections to the Yukawa interactions were not studied in Refs.~\cite{Aoki:2012yt,Aoki:2012jj}.
Furthermore, the scalar mixing angles were not determined in a gauge independent way~\cite{Yamada:2001px, Freitas:2002um, Espinosa:2002cd}.

In this paper, we present a full set of radiative corrections to decays of the 125 GeV Higgs boson ($h$) in this model. 
One-loop contributions of the extra Higgs bosons as well as SM fermions and gauge bosons to the decay rates of $h$ are calculated in the on-shell scheme. 
We first remove the gauge dependence of the scalar mixing angle by using the pinch technique~\cite{Papavassiliou:1994pr,Papavassiliou:1989zd,Cornwall:1981zr,Binosi:2009qm,Degrassi:1992ue,Cornwall:1989gv}~\footnote{Other gauge invariant renormalization schemes for extended Higgs sectors are discussed in e.g., Refs.~\cite{Denner:2018opp,Dittmaier:2022ivi}.
Focusing on two-body scattering processes, we demonstrate how the appropriate pinch terms can be extracted.}
We then investigate in detail the decays of the Higgs boson, including NLO EW corrections and higher-order QCD corrections. Furthermore, we analyze the deviations from the SM predictions appearing in the Higgs boson decays. After taking into account theoretical and experimental constraints, we show that a characteristic pattern of deviations can be obtained, which allows us to distinguish the CHTM from other extended Higgs models.
Compared to earlier works on the renormalization of the CHTM, this paper has the following new points.
(1) We construct, for the first time in the CHTM, a gauge-invariant renormalization scheme by removing the gauge dependence from the on-shell renormalization of the scalar mixing angle.
(2) Analytic formulae for the renormalized $hf\bar{f}$ vertex function are provided.
(3) In the CHTM, the full one-loop corrections to the Higgs boson decay rates are calculated for the first time, and this includes not only the two-body decays of the Higgs boson but also the three-body decays.

This paper is organized as follows.
In Sec.~\ref{sec: model}, we describe the Lagrangian of CHTM and discuss features of this model. 
After presenting how we perform renormalization in Sec.~\ref{sec: Renormalization}, we devote ourselves to discussions about the gauge dependence on the scalar mixing angles and the procedure to remove it in Sec.~\ref{sec:gauge invariant}.
We then describe the analytical formulae of decay rates of $h$ in Sec.~\ref{sec: decay of h125}, and discuss numerical results for the deviations from the SM in $h$ decays in Sec.~\ref{sec: Deviations from the SM in the Higgs boson decays}.
Conclusions are given in Sec.~\ref{sec: Conclusions}. 

%%%%%%%%%%%%%%%%%%%%%%%%%%%%%%%%%%%%%%%%%%%%%%%%%%%%%%%%%%%%%%%%%%
\section{The complex Higgs triplet model}
\label{sec: model}
%%%%%%%%%%%%%%%%%%%%%%%%%%%%%%%%%%%%%%%%%%%%%%%%%%%%%%%%%%%%%%%%%%

In this section, we define the Lagrangian of the CHTM and discuss theoretical constraints as well as the decoupling limit.
We follow the notation of Refs.~\cite{Aoki:2012yt,Aoki:2012jj}.
The CHTM consists of a scalar doublet field $\Phi$ with hypercharge $Y=1/2$ and a complex scalar triplet field $\Delta$ with $Y=1$.
After electroweak symmetry breaking, the scalar fields are parameterized as
\begin{align}
\label{eq:Higgs field}
\Phi = \mqty(\phi^{+} \\ \frac{1}{\sqrt{2}}(v_{\phi}+\phi+i\chi))\qc
\Delta =
\mqty(\frac{1}{\sqrt{2}}\Delta^{+} & \Delta^{++}\\
\Delta^{0} & -\frac{1}{\sqrt{2}}\Delta^{+})\qq{with}
\Delta^{0}=\frac{1}{\sqrt{2}}(v_{\Delta}+\delta+i\eta),
\end{align}
where $v_{\phi}$ and $v_{\Delta}$ denote the VEVs, which satisfy $v^{2}=v_{\phi}^{2}+2v_{\Delta}^{2}\simeq 246~\mathrm{GeV}$.

%=================================================================
\subsection{The Higgs potential}
%=================================================================

The most general Higgs potential constructed from $\Phi$ and $\Delta$ is given by
\begin{align}
V(\Phi, \Delta)
&=
m^{2}\Phi^{\dagger}\Phi+M^{2}\Tr(\Delta^{\dagger}\Delta)
+\qty[\mu\Phi^{T}i\sigma_{2}\Delta^{\dagger}\Phi+\mathrm{h.c.}]
\notag \\ &\quad
+\lambda_{1}\qty(\Phi^{\dagger}\Phi)^{2}+\lambda_{2}\qty[\Tr(\Delta^{\dagger}\Delta)]^{2}+\lambda_{3}\Tr\qty[(\Delta^{\dagger}\Delta)^{2}]
\notag \\ &\quad
+\lambda_{4}\qty(\Phi^{\dagger}\Phi)\Tr(\Delta^{\dagger}\Delta)+\lambda_{5}\Phi^{\dagger}\Delta\Delta^{\dagger}\Phi,
\label{eq: Higgs_potential}
\end{align}
where $m^{2}$, $M^{2}$ and $\lambda_{1\text{--}5}$ are real parameters, whereas $\mu$ is in general complex. Since the relative phase between $\Phi$ and $\Delta$ only affects the $\mu$ term in the Higgs potential, the phase of $\mu$ can be removed by an appropriate field redefinition. Thus, all parameters in the Higgs potential can be taken as real without loss of generality~\cite{Dey:2008jm}. In the following, we work in the basis where $\mu$ is real.

From the stationary conditions of the Higgs potential, we obtain
\begin{align}
m^{2}
&=
-\lambda_{1}v_{\phi}^{2}-\frac{1}{2}(\lambda_{4}+\lambda_{5})v_{\Delta}^{2}+\sqrt{2}\mu v_{\Delta},
\label{eq: stationary conditions 1} \\
M^{2}
&=
-(\lambda_{2}+\lambda_{3})v_{\Delta}^{2}-\frac{1}{2}(\lambda_{4}+\lambda_{5})v_{\phi}^{2}+\frac{\mu v_{\phi}^{2}}{\sqrt{2}v_{\Delta}}.
\label{eq: stationary conditions 2}
\end{align}
The mass matrices for the singly charged states ($\mathcal{M}_{\pm}^2$), the CP-odd states ($\mathcal{M}_{\rm odd}^2$), and the CP-even states ($\mathcal{M}_{\rm even}^2$) are obtained as
\begin{align}
\mathcal{M}_{\pm}^{2}
&=
\qty(\frac{\mu v_\phi^2}{\sqrt{2}v_\Delta}-\frac{\lambda_{5}}{4}v_{\phi}^{2})\mqty(
2v_{\Delta}^{2}/v_{\phi}^{2} &
-\sqrt{2}v_{\Delta}/v_{\phi} \\
-\sqrt{2}v_{\Delta}/v_{\phi} &
1
),\\
\mathcal{M}_{\mathrm{odd}}^{2}
&=
\frac{\mu v_\phi^2}{\sqrt{2}v_\Delta}\mqty(
4v_{\Delta}^{2}/v_{\phi}^{2} &
-2v_{\Delta}/v_{\phi} \\
-2v_{\Delta}/v_{\phi} &
1
)\qc
\mathcal{M}_{\mathrm{even}}^{2}
=
\mqty(
\mathcal{M}_{11}^{2} &
\mathcal{M}_{12}^{2} \\
\mathcal{M}_{21}^{2} &
\mathcal{M}_{22}^{2}
),
\label{eq: mass_mat_CPeven}
\end{align}
where the components of $\mathcal{M}_{\mathrm{even}}^{2}$ are given by
\begin{align}
\mathcal{M}_{11}^{2}
&=
2\lambda_{1}v_{\phi}^{2}\qc
\mathcal{M}_{22}^{2}
=
\frac{\mu v_\phi^2}{\sqrt{2}v_\Delta}
+2(\lambda_{2}+\lambda_{3})v_{\Delta}^{2},
\notag \\
\mathcal{M}_{12}^{2}
&=
\mathcal{M}_{21}^{2}
=
(\lambda_{4}+\lambda_{5})v_{\phi}v_{\Delta}-\sqrt{2}\mu v_{\phi}.
\end{align}
We define the mass eigenstates as
\begin{align}
\mqty(\phi^{\pm} \\ \Delta^{\pm})
=
R(\beta)\mqty(G^{\pm} \\ H^{\pm})\qc
\mqty(\chi \\ \eta)
=
R(\beta')
\mqty(G^{0} \\ A)\qc
\mqty(\phi \\ \delta)
=
R(\alpha)\mqty(h \\ H),
\end{align}
where $G^{\pm}$ and $G^{0}$ denote the Nambu--Goldstone (NG) bosons.
The rotation matrix is given by
\begin{align}
R(\theta) = \mqty(c_{\theta} & -s_{\theta} \\ s_{\theta} & c_{\theta}),
\end{align}
with the shorthand notation $c_{\theta}=\cos{\theta}$ and $s_{\theta}=\sin{\theta}$.
In this paper, we identify $h$ with the SM-like Higgs boson with a mass of 125~GeV.
The mixing angles are given by
\begin{align}
\tan{\beta} &= \frac{\sqrt{2}v_{\Delta}}{v_{\phi}}\qc
\tan{\beta'} = \frac{2v_{\Delta}}{v_{\phi}}\qc
\tan{2\alpha}
=
\frac{2\mathcal{M}_{12}^{2}}{\mathcal{M}_{11}^{2}-\mathcal{M}_{22}^{2}}.
\label{eq: definition of mixing angles}
\end{align}
The mixing angles of the charged states ($\beta$) and the CP-odd states ($\beta'$) differ because $\Phi$ and $\Delta$ belong to different representations of $\mathrm{SU}(2)_L$.
Nevertheless, they are not independent: $\tan{\beta}=\tan{\beta'}/\sqrt{2}$.
The masses of the physical Higgs bosons are obtained as
\begin{alignat}{2}
m_{H^{\pm\pm}}^{2}
&=
\frac{\mu v_\phi^2}{\sqrt{2}v_\Delta}-\lambda_{3}v_{\Delta}^{2}-\frac{\lambda_{5}}{2}v_{\phi}^{2},& &
\\
m_{H^{\pm}}^{2} &= \qty(\frac{\mu v_\phi^2}{\sqrt{2}v_\Delta}-\frac{\lambda_{5}}{4}v_{\phi}^{2})\qty(1+\frac{2v_{\Delta}^{2}}{v_{\phi}^{2}}),&
m_{A}^{2} &= \frac{\mu v_\phi^2}{\sqrt{2}v_\Delta}\qty(1+\frac{
4v_{\Delta}^{2}}{v_{\phi}^{2}}),
\label{eq: mass_A}
\\
m_{h}^{2} &= \mathcal{M}_{11}^{2}c_{\alpha}^{2}+\mathcal{M}_{22}^{2}s_{\alpha}^{2}+2\mathcal{M}_{12}^{2}s_{\alpha}c_{\alpha}\qc&
% \\
m_{H}^{2} &= \mathcal{M}_{11}^{2}s_{\alpha}^{2}+\mathcal{M}_{22}^{2}c_{\alpha}^{2}-2\mathcal{M}_{12}^{2}s_{\alpha}c_{\alpha}.
\end{alignat}
We note that the CP-odd Higgs boson becomes massless in the $\mu\to0$ limit with nonzero $v_{\Delta}$.
This is because the Higgs potential respects a global $\mathrm{U}(1)$ symmetry when $\mu=0$, in which case the CP-odd Higgs boson becomes an additional NG boson.

The original eight parameters in the Higgs potential can be expressed in terms of the masses of the Higgs bosons and the mixing angles.
The explicit relations are provided, for instance, in Ref.~\cite{Aoki:2012jj}.
We choose the following parameters as inputs:
\begin{align}
    v,\quad m_{h}^{2},\quad m_{H}^{2},\quad m_{A}^{2},\quad m_{H^{\pm}}^{2},\quad m_{H^{\pm\pm}}^{2},\quad \alpha,\quad \beta'. 
\end{align}
It is convenient to determine the value of $\alpha$ in terms of $\lambda_{4}$.
They are related by
\begin{align}
    \sin{(2\alpha)}
    =
    \frac{2v_{\phi}v_{\Delta}}{m_{H}^{2}-m_{h}^{2}}
    \left[
    \frac{4m_{H^{\pm}}^{2}}{v_{\phi}^{2}+2v_{\Delta}^{2}}
    -\frac{2m_{A}^{2}}{v_{\phi}^{2}+4v_{\Delta}^{2}}-\lambda_{4}
    \right].
    \label{eq:sin2alpha}
\end{align}
We note that the following condition must be satisfied since $\mathcal{M}_{\mathrm{even}}^{2}$ is a real symmetric matrix.
\begin{align}
    (m_{H}^{2}-m_{h}^{2})^{2} \geq 4\left(\mathcal{M}_{12}^{2}\right)^{2}.
\end{align}
This is equivalent to requiring $\abs{\sin{(2\alpha)}} \leq 1$.

%=================================================================
\subsection{The kinetic term and Yukawa interaction}
%=================================================================

The kinetic terms are given by 
\begin{align}
\mathcal{L}_{\mathrm{Kin}}
=
(D_{\mu}\Phi)^{\dagger}(D^{\mu}\Phi)+\Tr[(D_{\mu}\Delta)^{\dagger}D^{\mu}\Delta],
\end{align}
where the covariant derivatives are defined as
\begin{align}
D_{\mu}\Phi
&=
\qty(\partial_{\mu}-ig\tau^{a}W_{\mu}^{a}-ig'YB_{\mu})\Phi,\quad%\\
D_{\mu}\Delta
=
\partial_{\mu}\Delta-ig\qty[\tau^{a}W_{\mu}^{a},\Delta]-ig'YB_{\mu}\Delta,
\end{align}
with the commutator $[a,b]=ab-ba$. Here, $\tau^{a}$ denotes the $\mathrm{SU}(2)_{L}$ generators defined as $\tau^{a}=\sigma^{a}/2$ with the Pauli matrices $\sigma^{a}$.

The masses of the weak gauge bosons are obtained as
\begin{align}\label{eq:m_V}
m_{W}^{2} = \frac{g^{2}}{4}(v_{\phi}^{2}+2v_{\Delta}^{2})\qc
m_{Z}^{2} = \frac{g_{Z}^{2}}{4}(v_{\phi}^{2}+4v_{\Delta}^{2}),
\end{align}
with $g_{Z}=g/c_{W}$ and $\tan{\theta_{W}} = g'/g$.
At tree level, the electroweak $\rho$ parameter is given by
\begin{align}
\rho_{0} = \frac{m_W^2}{m_Z^2c_W^2}
=\frac{v_\phi^2+2v_\Delta^2}{v_\phi^2+4v_\Delta^2}.
\label{eq: rho_tree}
\end{align}
Thus, unlike in the SM, $\rho_{0}$ deviates from unity in the CHTM. The global fit result for the electroweak $\rho$ parameter, $\rho_{\mathrm{exp}} = 1.00031\pm 0.00019$~\cite{ParticleDataGroup:2024cfk}, implies the bound $v_{\Delta}\lesssim 1.45~\mathrm{GeV}$ from Eq.~\eqref{eq: rho_tree}. However, we note that the tree-level contribution is of the same order as loop corrections, i.e., $v_{\Delta}/v \simeq (16\pi^{2})^{-1}$.
Therefore, one-loop effects would modify the tree-level bound on $v_{\Delta}$.
The one-loop analysis of electroweak precision observables is discussed in Sec.~\ref{sec: Deviations from the SM in the Higgs boson decays}.

The Yukawa sector of the CHTM consists of the $\Phi$ and $\Delta$ parts: $\mathcal{L}^{Y}=\mathcal{L}^{Y}_{\Phi}+\mathcal{L}^{Y}_{\Delta}$.
The $\Phi$ part is identical to the Yukawa sector in the SM, while the $\Delta$ part is given by
\begin{align}
{\cal L}^Y_{\Delta}=h_{ij}\left(\overline{L^c_L}\right)_ii\tau_2\Delta \left(L_L\right)_j +{\rm h.c.},
\end{align}
where $h_{ij}$ is a $3\times 3$ complex symmetric matrix. 
The $\Delta$ part generates neutrino masses via the Type-II seesaw mechanism~\cite{Cheng:1980qt, Schechter:1980gr, Lazarides:1980nt, Mohapatra:1980yp, Magg:1980ut}.
In the following analysis of the SM-like Higgs boson decays, we neglect contributions from the neutrino Yukawa couplings $h_{ij}$. These are sufficiently small to reproduce the observed neutrino oscillation data in the parameter region of interest in this paper.

From $\mathcal{L}_{\Phi}^{Y}$, the Yukawa interactions in the mass basis are obtained as
\begin{align}
\mathcal{L}^{Y}_{\Phi}
&\supset
-\sum_{f=u, d, e}\frac{m_{f}}{v}\qty[\overline{f}f(\zeta_{h}h+\zeta_{H}H)-2iI_{f}\overline{f}\gamma_{5}f(\zeta_{G^{0}}G^{0}+\zeta_{A}A)]
\notag \\ &\quad
+\qty[\qty(\frac{\sqrt{2}m_{u}}{v}V_{ud}\overline{u}P_{L}d-\frac{\sqrt{2}m_{d}}{v}V_{ud}\overline{u}P_{R}d)(\zeta_{G^{\pm}}G^{+}+\zeta_{H^{\pm}}H^{+})+\mathrm{h.c.}]
\notag \\ &\quad
-\qty[\frac{\sqrt{2}m_{e}}{v}\overline{\nu}P_{R}e(\zeta_{G^{\pm}}G^{+}+\zeta_{H^{\pm}}H^{+})+\mathrm{h.c.}],
\end{align}
where $I_f$ denotes the weak isospin of a fermion $f$, and $V_{ud}$ the CKM matrix. In the following analysis, we neglect the effects of quark mixing because they are small, and take the CKM matrix to be unity.
The coupling modifiers $\zeta_{X}$ are given by
\begin{alignat}{3}
\zeta_{h} &= c_{\alpha}/c_{\beta}\qc&
\zeta_{G^{0}} &= c_{\beta'}/c_{\beta}\qc&
\zeta_{G^{\pm}} &= 1,
\notag \\
\zeta_{H} &= -s_{\alpha}/c_{\beta}\qc&
\zeta_{A} &= -s_{\beta'}/c_{\beta}\qc&
\zeta_{H^{\pm}} &= -s_{\beta}/c_{\beta}.
\label{eq: zeta_phi}
\end{alignat}

%=================================================================
\subsection{Theoretical constraints}
\label{subsec: th_const}
%=================================================================

The potential parameters are required to satisfy theoretical constraints from vacuum stability and tree-level unitarity.
Vacuum stability demands that the Higgs potential be bounded from below at large field values, thereby ensuring the stability of the electroweak vacuum. 
The necessary and sufficient conditions for vacuum stability are given~\cite{Arhrib:2011uy}
\begin{align}
&\lambda_{1}>0\qc
\lambda_{2}+\text{MIN}\qty[\lambda_{3}, \frac{\lambda_{3}}{2}]>0, \notag \\
\lambda_{4}+\text{MIN}[0,\lambda_{5}]&+2\text{MIN}\qty[\sqrt{\lambda_{1}(\lambda_{2}+\lambda_{3})}, \sqrt{\lambda_{1}\qty(\lambda_{2}+\frac{\lambda_{3}}{2})}]>0.
\label{eq: vacuum stability}
\end{align}

Tree-level unitarity requires that the eigenvalues of the two-to-two scalar scattering matrix do not exceed $\zeta_{\mathrm{LQT}}=1$~\cite{Lee:1977yc, Lee:1977eg} or $1/2$~\cite{Luscher:1988gc, Marciano:1989ns}, depending on the choice of unitarity criterion.
In the CHTM, there are twelve independent eigenvalues~\cite{Arhrib:2011uy}\footnote{The similar analysis was performed in the context of the Georgi-Machacek model~\cite{Aoki:2007ah}.}.
\begin{align}
y_{1} &= 2\lambda_{1}\qc
y_{2} = 2(\lambda_{2}+\lambda_{3})\qc
y_{3} = 2\lambda_{2},
\notag \\
y_{4, \pm} &=
\lambda_{1}+\lambda_{2}+2\lambda_{3}
\pm\sqrt{(\lambda_{1}-\lambda_{2}-2\lambda_{3})^{2}+\lambda_{5}^{2}},
\notag \\
y_{5, \pm} &= 3\lambda_{1}+4\lambda_{2}+3\lambda_{3}
\pm\sqrt{(3\lambda_{1}-4\lambda_{2}-3\lambda_{3})^{2}+\frac{3}{2}(2\lambda_{4}+\lambda_{5})^{2}},
\notag \\
y_{6} &= \lambda_{4}\qc
y_{7} = \lambda_{4}+\lambda_{5}\qc
y_{8} = \frac{1}{2}(2\lambda_{4}+3\lambda_{5}),
\notag \\
y_{9} &= \frac{1}{2}(2\lambda_{4}-\lambda_{5})\qc
y_{10} = 2\lambda_{2}-\lambda_{3}.
\end{align}
For $a_{i}=y_{i}/(16\pi)$, the tree-level unitarity bound is given by $\abs{a_{i}} \leq \zeta_{\mathrm{LQT}}$.
In this paper, we take $\zeta_{\mathrm{LQT}}=1$.
In addition to tree-level unitarity, we impose the perturbativity constraint $\abs{\lambda_{i}}\leq 4\pi$.

%=================================================================
\subsection{Decoupling limit}
\label{subsec: decoupling_limit}
%=================================================================

From Eq.~\eqref{eq: rho_tree}, it follows that $v_{\Delta}$ must vanish in the decoupling limit of the triplet field.
This requirement results from the stationary condition in Eq.~\eqref{eq: stationary conditions 2}, which yields
\begin{align}
    v_{\Delta} 
    =
    \frac{\mu v_{\phi}^{2}}{\sqrt{2}\left[M^{2}+(\lambda_{2}+\lambda_{3})v_{\Delta}^{2}+(\lambda_{4}+\lambda_{5})v_{\phi}^{2}/2\right]}.
    \label{eq: triplet vev}
\end{align}
Then, in the limit $M^{2}\to\infty$, decoupling is realized with $v_{\Delta}\sim \mu v_{\phi}^{2}/(\sqrt{2}M^{2})\to 0$.
Alternatively, $v_{\Delta}\to 0$ can be realized by taking $\mu \to 0$, even if $M^{2}$ is of the order of the electroweak scale. In this case, Eq.~\eqref{eq: triplet vev} shows that $v_{\Delta}\sim \mu$.
If one takes the limit $\mu \to 0$ while keeping nonzero $v_{\Delta}$, an additional NG boson ($A$) emerges as mentioned earlier in this section.

From Eq.~\eqref{eq: zeta_phi} and Table~\ref{tab:couplings} in Appendix~\ref{app: coupling modifiers}, one finds that the coupling modifiers of $hf\bar{f},\, hZZ$ and $hWW$ become unity when $v_{\Delta} = 0$.
Thus, the SM-like Higgs boson couplings also approach their SM values in the $v_{\Delta}\to 0$ limit. We note that $\sin{(2\alpha)}$ is proportional to $v_{\Delta}$ (see Eq.~\eqref{eq:sin2alpha}). Hence, as long as the $\lambda_{i}$ couplings remain perturbative, $v_{\Delta}\to 0$ implies $\alpha\to 0$.

%%%%%%%%%%%%%%%%%%%%%%%%%%%%%%%%%%%%%%%%%%%%%%%%%%%%%%%%%%%%%%%%%%
\section{Renormalization and electroweak precision observables}
\label{sec: Renormalization}
%%%%%%%%%%%%%%%%%%%%%%%%%%%%%%%%%%%%%%%%%%%%%%%%%%%%%%%%%%%%%%%%%%

In this section, we describe the renormalization of the CHTM and the $G_{F}$ input scheme adopted in our calculation. We also discuss the electroweak precision observables, which are employed to constrain the CHTM parameters in numerical analysis.

%=================================================================
\subsection{Renormalization of the complex Higgs triplet model}
%=================================================================

In models with $\rho_{0}=1$ such as the SM, the electroweak sector is characterized by three independent input parameters, conventionally chosen to be $\alpha_{\mathrm{em}}, m_{W}$, and $m_{Z}$~\cite{Bohm:1986rj, Hollik:1988ii}.
In contrast, in models with $\rho_{0}\neq 1$ including the CHTM, an additional input parameter is required to specify the electroweak sector.

Depending on the choice of the additional input parameter, several renormalization schemes can be defined.
In Ref.~\cite{Blank:1997qa}, the effective weak mixing angle of the electron, $s_{e}^{2}$, is adopted as the additional input in the real Higgs triplet model.
This renormalization scheme was extended to the CHTM in Ref.~\cite{Kanemura:2012rs}.
It was pointed out that the existence of the decoupling limit is non-trivial in this scheme~\cite{Chen:2005jx,Chen:2006pb,Chankowski:2006hs,Kanemura:2012rs}.
The reason is that $v_{\Delta}$ becomes an output parameter, whose value is fixed at some nonzero value when one inputs $s_{e}^{2}\neq (s_{e}^{2})_{\mathrm{SM}}$, where $(s_{e}^{2})_{\mathrm{SM}}$ denotes the SM prediction.
As a result, the limit $v_{\Delta}\to 0$, which is required to realize the decoupling of the triplet field, cannot be freely taken.
To make the decoupling behavior manifest, a renormalization scheme based on the $\overline{\mathrm{MS}}$-renormalized triplet VEV was proposed in Ref.~\cite{Chankowski:2006hs} (see also Ref.~\cite{Chen:2008jg}).

In this paper, we mainly follow renormalization scheme II of Ref.~\cite{Aoki:2012jj}, in which the mixing angle $\beta'$ is taken as an additional input parameter and is renormalized by imposing an on-shell condition on the $G^{0}\text{--}A$ mixing (see also Ref.~\cite{Kanemura:2004mg}).
In this scheme, all counterterms are fixed by the on-shell conditions, and the decoupling limit can be realized by taking $\beta'\to 0$.
The conventional input parameters, $\alpha_{\mathrm{em}}, m_{W}$, and $m_{Z}$, are also used to specify the electroweak sector and are renormalized according to the on-shell conditions~\cite{Bohm:1986rj, Hollik:1988ii}.

Our calculation differs from that in Ref.~\cite{Aoki:2012jj} in two aspects.
The first concerns the treatment of gauge dependence.
Mixing counterterms depend on the gauge choice when these are renormalized using off-diagonal components of the two-point function~\cite{Krause:2016oke}.
To eliminate the gauge dependence, we employ the pinch technique~\cite{Papavassiliou:1994pr,Papavassiliou:1989zd,Cornwall:1981zr,Binosi:2009qm,Degrassi:1992ue,Cornwall:1989gv} for the first time to the CHTM~\footnote{The similar analysis has been done in the Georgi-Machacek model~\cite{Chiang:2017vvo, Chiang:2018xpl}.}, which will be discussed in Sec.~\ref{sec:gauge invariant}~\footnote{One may use physical processes to renormalize the mixing angles in a gauge-invariant way, and applications for extended Higgs models are discussed in, e.g., Refs.~\cite{Krause:2016oke,Denner:2018opp,Dittmaier:2022ivi,Kanemura:2024ium}.}.

The second concerns the tadpole renormalization.
There are two schemes for tadpole renormalization: the standard tadpole scheme~\cite{Denner:1991kt} and the alternative tadpole scheme~\cite{Fleischer:1980ub}.
For convenience in the subsequent analysis, we adopt the alternative tadpole scheme, whereas the standard tadpole scheme is employed in Ref.~\cite{Aoki:2012jj}.
In the alternative tadpole scheme, tadpole-inserted diagrams are included in all self-energies.
\begin{align}
    \Pi_{ij}(p^{2}) = \Pi_{ij}^{\mathrm{1PI}}(p^{2})+\Pi_{ij}^{\mathrm{Tad}}. \label{eq:all_self_energy}
\end{align}
Then, the counterterms are determined with $\Pi_{ij}(p^{2})$. We also include tadpole inserted diagrams in all vertex contributions.
Although the expressions for the counterterms depend on the tadpole scheme, there is no scheme difference in the renormalized quantities in the CHTM\footnote{Such a cancellation is non-trivial when $\overline{\mathrm{MS}}$ renormalized counterterms are involved, as in the two Higgs doublet model and the Higgs singlet model~\cite{Kanemura:2017wtm}.}.

We here comment on the renormalization of the Yukawa Lagrangian.
Since the doublet part, ${\cal L}^{Y}_{\Phi}$, is identical to that in the SM, its renormalization is performed in the same manner as in the SM~\cite{Bohm:1986rj, Hollik:1988ii, Denner:1991kt}.
On the other hand, the triplet part, ${\cal L}^Y_{\Delta}$, is irrelevant for the decays of the SM-like Higgs boson and is therefore neglected.
It would become relevant when studying loop corrections to the decays of the additional Higgs bosons, but this is beyond the scope of this paper.

In the following, we discuss the input scheme for one-loop calculations and electroweak precision observables in the CHTM. We denote the renormalized two-point function of the gauge bosons by $\widehat{\Pi}_{VV}(p^{2})$, whose explicit expressions are given in Ref.~\cite{Aoki:2012jj}.

%=================================================================
\subsection{\texorpdfstring{The $G_F$}{} input scheme}
\label{subsec: G_f input scheme}
%=================================================================

In the OS scheme, the CHTM Lagrangian is written in terms of $\alpha_{\mathrm{em}}$, $m_{W}^{2}$, $m_{Z}^{2}$, and $\beta'$. 
However, we use the Fermi constant $G_F$ as a numerical input parameter instead of $m_W$, because $G_F$ is measured very precisely.
For this replacement, we employ the following relation.
\begin{align}
G_F
= \frac{1}{\sqrt{2}v^{2}(1-\Delta r)}
\qq{with}
v^{2} = \frac{m_{W}^{2}}{\pi\alpha_{\mathrm{em}}}\left(1-\frac{m_{W}^{2}}{\rho_{0}m_{Z}^{2}}\right).
\label{eq:G_F}
\end{align}
The $\Delta r$ parameter is determined by the muon decay process and is given by~\cite{Sirlin:1980nh, Hollik:1988ii}
\begin{align}
\Delta r
&=
\frac{\Re\widehat{\Pi}_{WW}(0)}{m_{W}^{2}}+\delta_{VB},
\label{eq: delta_r}
\end{align}
where $\widehat{\Pi}_{WW}(p^{2})$ is the renormalized two-point function of the $W$ boson, and it depends not only on the counterterms for $m_{W}^{2}, m_{Z}^{2}$ and $\alpha_{\mathrm{em}}$ but also on that for $\beta'$.
Its expression is given in Ref.~\cite{Aoki:2012jj}.
The $\delta_{VB}$ correction represents vertex and box contributions.
At the one-loop level, it is expressed as~\cite{Blank:1997qa}
\begin{align}
\delta_{VB}
&=
\frac{\alpha_{\rm em}}{4\pi s_{W}^{2}}\qty[
6+\frac{10(1-s_{W}^{2})-3(R/c_{W}^{2})(1-2s_{W}^{2})}{2(1-R)}\ln{R}]
\qq{with} R=\frac{m_{W}^{2}}{m_{Z}^{2}}.
\end{align}
The right-hand side of Eq.~\eqref{eq:G_F} is a function of $\alpha_{\mathrm{em}}, m_{W}, m_{Z}$ and $\beta'$\footnote{It also depends on the other inputs such as $m_{h}, m_{t}, m_{H^{\pm\pm}}$ and so on.}.
Thus, the value of $m_{W}$ that reproduces the observed value of $G_{F}$ for given $\alpha_{\mathrm{em}}, m_{Z}$ and $\beta'$ can be determined by solving Eq.~\eqref{eq:G_F} iteratively~\cite{Hollik:1988ii,Blank:1997qa}.

Instead of iteratively solving Eq.~\eqref{eq:G_F}, we adopt an alternative approach by introducing barred quantities, which are determined from the tree-level relations~\cite{Kanemura:2019kjg}.
The tree-level $W$ boson mass, $\overline{m}_{W}$, is defined as
\begin{align}
    \overline{m}_{W}^{2}
    =
    \rho_{0}m_{Z}^{2}\overline{c}_{W}^{2}
    \qq{with}
    \overline{c}_{W}^{2} = 1-\overline{s}_{W}^{2},
\end{align}
where $\overline{s}_{W}$ is given as a function of $\alpha_{\mathrm{em}}, G_{F}, m_{Z}$ and $\beta'$.
\begin{align}
    \overline{s}_{W}^{2}
    &=
    \frac{1}{2}\left(1-\sqrt{1-\frac{4A_{0}}{\rho_{0}m_{Z}^{2}}}\right)
    \qq{with}
    A_{0}=\frac{\pi\alpha_{\mathrm{em}}}{\sqrt{2}G_{F}}.
\end{align}
At the one-loop level, $\Delta r$ can be evaluated using $\overline{m}_{W}$ instead of $m_{W}$, since $\Delta r$ is a one-loop quantity and the difference would only appear at the two-loop level.
From Eq.~\eqref{eq:G_F}, the weak mixing angle and the $W$ boson mass are written as
\begin{align}
    s_{W, G_{F}}^{2}
    =
    \frac{1}{2}\qty[
    1-\sqrt{1-\frac{4A_{0}}{\rho_{0}m_{Z}^{2}(1-\Delta r)}}]\qc
    m_{W, G_{F}}^{2}
    =
    \rho_{0}m_{Z}^{2}c_{W, G_{F}}^{2},
    \label{eq: swsq and mwsq}
\end{align}
with $c_{W, G_{F}}^{2}=1-s_{W, G_{F}}^{2}$.
They coincide with $s_{W}^{2}$ and $m_{W}^{2}$, which are obtained by iteratively solving Eq.~\eqref{eq:G_F}, at the one-loop level.
In subsequent analysis, we use $\overline{m}_{W}$ and $\overline{s}_{W}$ for loop calculations similar to $\Delta r$.
Since $\overline{m}_{W}$ is not the on-shell mass and is not the Lagrangian parameter in the oh-shell scheme, we use $m_{W, G_{F}}$ instead of $\overline{m}_{W}$ for tree-level calculations. In addition, we use $m_{W, G_{F}}$ for a kinematical factor involving processes with external $W$ bosons, such as $h\to WW^{*}$.

Finally, we discuss the treatment of $\beta'$ and $\lambda_{4}$. Since $\beta'$ is not convenient as an input for numerical analysis, we introduce $\overline{v}_{\Delta}$ defined by
\begin{align}
    \overline{v}_{\phi}^{2}+2\overline{v}_{\Delta}^2 = \overline{v}^{2}
    \qq{with}
    \overline{v}^{2} = \frac{1}{\sqrt{2}G_{F}},
\end{align}
and take $\tan{\beta'} = 2\overline{v}_{\Delta}/\overline{v}_{\phi}$ according to the tree-level relation in Eq.~\eqref{eq: definition of mixing angles}. From Eq.~\eqref{eq:G_F}, $v_{\Delta}$ can be expressed as
\begin{align}
    v_{\Delta, G_{F}}^{2}
    =
    \frac{s_{\beta'}^{2}}{2(1+c_{\beta'}^{2})}
    \frac{1}{\sqrt{2}G_{F}(1-\Delta r)}.
\end{align}
Thus, similar to $s_{W, G_{F}}^{2}$ and $m_{W, G_{F}}^{2}$, $v_{\Delta, G_{F}}$ can be calculated for given $\alpha_{\mathrm{em}}, G_{F}, m_{Z}$ and $\beta'$.
Since $\beta'=0$ corresponds to $v_{\Delta, G_{F}}=0$, the decoupling limit can be realized by taking the renormalized mixing angle $\beta'$ to be zero in our scheme.
The mixing angle $\alpha$, one of the input parameters, relates to $\overline{v}_{\phi}$, $\overline{v}_{\Delta}$ and the tree-level coupling $\overline{\lambda}_{4}$ in Eq.~\eqref{eq:sin2alpha}.
At the one-loop level, $\lambda_{4}$ receives a $\Delta r$ shift:
\begin{align}
    \lambda_{4, G_{F}} = \overline{\lambda}_{4}(1-\Delta r).
\end{align}
Similarly, $\lambda_{1,2,3}$ and $\lambda_{5}$ also receive a $\Delta r$ shift.
For the theoretical constraints, we use the tree-level couplings $\overline{\lambda}_i$, which are equivalent to $\lambda_{i,G_F}$ at the tree level.

%=================================================================
\subsection{Electroweak precision observables}
\label{subsec:EWPOs}
%=================================================================

The electroweak precision observables, such as the effective mixing angle and $Z$-boson decay rates, can be expressed in terms of the effective $Zf\bar{f}$ couplings.
They are given by~\cite{Blank:1997qa}
\begin{align}
g_{V}^{f}
&=
2\qty(\frac{\rho_{0}(1-\Delta r)}{1+\widehat{\Pi}'_{Z}(m_{Z}^{2})})^{1/2}
\notag \\ &\quad\times
\qty[
\frac{I_{f}}{2}-Q_{f}s_{W}^{2}\left(
1+\frac{c_{W}}{s_{W}}\frac{\widehat{\Pi}_{Z\gamma}(m_{Z}^{2})}{m_{Z}^{2}+\widehat{\Pi}_{\gamma\gamma}(m_{Z}^{2})}
\right)
+\Gamma^{V,\mathrm{loop}}_{Z f\bar{f}}(0,0,m_{Z}^{2})], \\
g_{A}^{f}
&=
2\qty(\frac{\rho_{0}(1-\Delta r)}{1+\widehat{\Pi}'_{Z}(m_{Z}^{2})})^{1/2}
\qty[\frac{I_{f}}{2}+\Gamma^{A, \mathrm{loop}}_{Z f\bar{f}}(0,0,m_{Z}^{2})].
\end{align}
where $Q_f$ denotes the electric charge of a fermion.
\begin{align}
    \widehat{\Pi}'_{Z}(p^{2}) = \Re\eval{\dv{\widehat{\Pi}_{Z}(p^{2})}{p^{2}}}_{p^{2}=m_{Z}^{2}}
    \qq{with}
    \widehat{\Pi}_{Z}(p^{2}) = \widehat{\Pi}_{ZZ}(p^{2})-\frac{\qty[\widehat{\Pi}_{Z\gamma}(p^{2})]^{2}}{p^{2}+\widehat{\Pi}_{\gamma\gamma}(p^{2})}.
\end{align}
The expressions for the loop-corrected vertices $\Gamma^{V/A, \mathrm{loop}}_{Z f\bar{f}}(0,0,p^{2})$ are the same as those in the SM in the massless limit of the external fermions.
Explicit formulae can be found, for instance, in Ref.~\cite{Kanemura:2019kjg}.

The effective weak mixing angle at the $Z$ pole, $s_{f}^{2}$, is defined as 
\begin{align}
s_{f}^{2}
&=
\frac{1}{4|Q_f|}\left(
1 - \frac{\Re(g_V^f)}{\Re(g_A^f)}
\right).
\end{align}
The leptonic decay rate of the $Z$ boson is given by
\begin{align}
\Gamma(Z \to \ell \bar{\ell}) &= \frac{\sqrt{2}m_Z^{3}G_F}{12\pi}
\left[(g_V^\ell)^2 + (g_A^\ell)^2\right]
\left(1+Q_{\ell}^{2}\frac{3\alpha_{\mathrm{em}}}{4\pi}\right).
\end{align}
We also introduce the loop-corrected $\rho$ parameter
\begin{align}
\rho = \rho_{0}+\Delta \rho,    
\end{align}
where $\Delta \rho$ is given by~\cite{Aoki:2012jj}
\begin{align}
\Delta \rho_{}  &= \text{Re}\left[\frac{\Pi_{ZZ}(0)}{m_Z^2}-\frac{\Pi_{WW}(0)}{m_W^2} + \frac{2s_W}{c_W}\frac{\Pi_{Z\gamma }(0)}{m_Z^2}-\frac{s_{2\beta'}}{1 + c_{\beta'}^{2}}\delta \beta'\right],
\label{del_rho}
\end{align}
with the mixing counterterm $\delta \beta'$ given in Eq.~\eqref{eq:del_beta}.
The analytical expressions for the transverse part of the gauge-boson two-point functions $\Pi_{VV'}(p^{2})$ $(V,V'=\gamma,Z,W)$ are given in Ref.~\cite{Aoki:2012jj}.
We use these quantities to impose the experimental constraints. See the discussion in Sec.~\ref{sec: Deviations from the SM in the Higgs boson decays}.

%%%%%%%%%%%%%%%%%%%%%%%%%%%%%%%%%%%%%%%%%%%%%%%%%%%%%%%%%%%%%%%%%%
\section{Gauge invariant scalar two-point functions}
\label{sec:gauge invariant}
%%%%%%%%%%%%%%%%%%%%%%%%%%%%%%%%%%%%%%%%%%%%%%%%%%%%%%%%%%%%%%%%%%

In this section, we discuss a prescription for removing the gauge dependence from the two-point functions of the scalar bosons.
The unrenormalized two-point functions have the gauge dependence through 1PI diagrams involving gauge bosons, NG bosons, and Faddeev-Popov ghost fields. 
Thanks to the Nielsen identity~\cite{Nielsen:1975fs}
in the alternative tadpole scheme, the gauge-dependent part of the two-point function defined in Eq.~(\ref{eq:all_self_energy}) for a single scalar field $\Pi_{\phi\phi}^{}$ is proportional to $(p^2 - m_\phi^2)$. 
Therefore, by imposing the on-shell condition for the renormalized two-point function, the gauge dependence vanishes.
As a consequence, the mass counterterm becomes gauge-independent.
%%%%
In contrast, for the off-diagonal self-energy $\Pi_{\phi\phi'}^\textrm{}$ ($\phi\neq\phi'$), the gauge dependence is not removed even under the on-shell condition. 
Because of this, counterterms of the mixing angles such as $\delta \alpha$, acquire the gauge dependence.

We employ the pinch technique~\cite{Papavassiliou:1994pr,Papavassiliou:1989zd,Cornwall:1981zr,Binosi:2009qm,Degrassi:1992ue,Cornwall:1989gv}, in which the gauge dependence in the two-point functions is canceled by adding the pinch terms extracted from the $f\bar{f} \to f\bar{f}$ scattering process.
Namely, we extract the gauge dependent parts of the vertex corrections, the box corrections and the wave-function renormalization of the external fermions in the $f\bar{f}\to f\bar{f}$ process, and add them into the self-energy corrections such that the gauge dependence of the latter is canceled. 
We follow the procedure for extracting these pinch terms as described in Ref.~\cite{Kanemura:2017wtm}, where the pinch terms of two-point functions in the 2HDMs and the HSM are discussed in detail.
In this paper, the two-point functions of the CP-even and CP-odd scalar fields are used to determine the counterterms in the scalar sector, while those of the charged scalar fields are not used for the renormalization. 
We thus focus on the discussion of the pinch terms for the CP-even and CP-odd scalar sectors. 

%%%%%
The pinch terms $\Pi_{ij}^\textrm{PT}(p^2)$ derived in this section are included in the self-energies as follows, 
\begin{align} 
\widetilde{\Pi}_{ij}(p^2) = \Pi_{ij}^\textrm{1PI}(p^2)+\Pi_{ij}^\textrm{Tad} + 
\Pi_{ij}^\textrm{PT}(p^2). \label{eq:Pi_GI}
\end{align}
which is gauge independent. 
As a result, the pinch terms are also added in counterterms of the mixing angles as follows,
\begin{align} 
%%%%%
\delta \alpha &=
-\frac{1}{2(m_H^2-m_h^2)}\left(
\widetilde{\Pi}_{Hh}(m_H^2)+\widetilde{\Pi}_{Hh}(m_h^2)
\right),\label{eq:del_al}\\
%%%%
\delta\beta' &=
-\frac{1}{2m_A^2}\left(
\widetilde{\Pi}_{AG^0}(m_A^2)+\widetilde{\Pi}_{AG^0}(0)
\right). \label{eq:del_beta}
\end{align}
Since $\beta$ is related to $\beta'$ as given in Eq.~(\ref{eq: definition of mixing angles}), $\delta\beta$ can be written as $\delta\beta =(1+s_{\beta}^2)\,\delta\beta'/\sqrt{2}$.  
We note that these pinched counterterms 
$\delta\alpha$, $\delta\beta$ and $\delta\beta'$ are applied to the counterterms generated via the parameter shifts but not to those generated via the field shifts for off-diagonal components (see Eq.~(93) of Ref.~\cite{Aoki:2012jj}).

%%%%%
\begin{figure}
    \centering
    \includegraphics[width=0.5\linewidth]{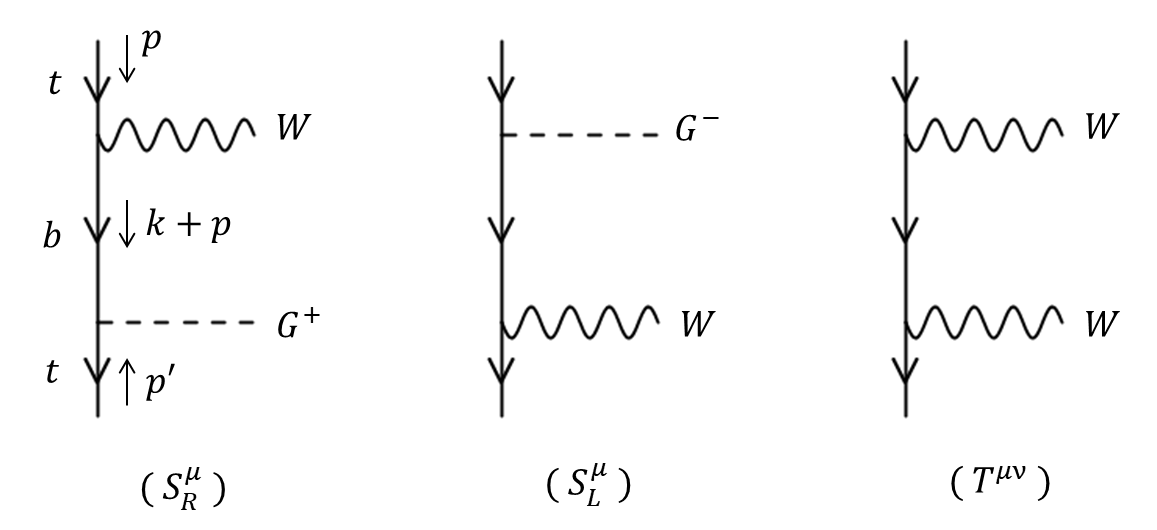}
    \caption{Sub-diagrams relevant to the pinch technique.}
    \label{fig:placeholder}
\end{figure}
We here explain several notations used in Sec.~\ref{sec:gauge invariant}. 
The amplitude discussed in this section is referred to as the ``pinched amplitude'', which is obtained by pinching out the internal fermion lines. 
%The procedure for the pinching operation for self energies of the scalar fields is given in Ref.~\cite{Papavassiliou:1997pb}. 
Focusing on the expressions involving the W-boson in the SM, 
we briefly explain the procedure of the pinching operation of the amplitude. 
The scattering diagrams relevant for the extraction of the pinch terms, which will be discussed in detail in the following subsections, are given by the vertex and box corrections with diagram topologies as shown in Fig.~\ref{fig:placeholder}~\footnote{For drawing the Feynman diagrams, we have used \texttt{Feynman diagram maker}~\cite{AidanSean:Feynman} and \texttt{TikZ-FeynHand}~\cite{Ellis:2016jkw, Dohse:2018vqo}.}. 
Each of these contributions can be expressed as~\cite{Papavassiliou:1997pb}
\begin{align}
S_R^\mu &=\frac{g}{v}\overline{v}(p')(m_t P_L-m_b P_R)
\frac{1}{\slashed{k}+\slashed{p}-m_b}\gamma^\mu P_L u(p), \\
S_L^\mu &=\frac{g}{v}\overline{v}(p')
\gamma^\mu P_L \frac{1}{\slashed{k}+\slashed{p}-m_b}
(m_t P_R-m_b P_L)u(p), \\
T^{\mu\nu} &=-
\frac{g^2}{2}\overline{v}(p')
\gamma^\mu P_L \frac{1}{\slashed{k}+\slashed{p}-m_b}
\gamma^\nu P_L  u(p). 
\end{align}
By acting with the loop momentum $k_\mu$ ($k_\mu k_\nu$) on these expressions, they can be rewritten as 
\begin{align}
k_\mu S_L^\mu &=\frac{g}{v}\overline{v}(p')m_t P_L u(p)+\cdots , \label{eq:kSL}\\
k_\mu S_R^\mu &=\frac{g}{v}\overline{v}(p')m_t P_R u(p)+\cdots , \label{eq:kSR}\\
k_\mu k_\nu T^{\mu\nu} &=
\frac{g^2}{2v}\overline{v}(p')m_t P_R u(p) - \frac{g^2}{2v}\overline{v}(p')\slashed{k} P_L u(p)+\cdots.  \label{eq:kkT} 
\end{align}
In Eq.~(\ref{eq:kSL})-(\ref{eq:kkT}), only the terms that the internal fermion propagators are pinched out are explicitly shown, 
while the remaining terms are denoted by ellipses. 
The first terms give the contributions that cancel the gauge dependence of the self-energies of the scalar fields. 
The second term of Eq.~(\ref{eq:kkT}) contributes to canceling those of the self-energies of the vector fields.  
Therefore, in this section, the amplitudes contain only the contributions from the first terms of Eq.~(\ref{eq:kSL})-(\ref{eq:kkT}). 
%%%%%

%%%%%%
In addition, we introduce the following expression calculated by the $R_\xi$ gauge
\begin{align}
\mathcal{O}_{\xi_V} = \mathcal{O}|_{\xi_V=1} + \mathcal{O}|_\textrm{G.D.}, 
\end{align}
where the first term on the RHS represents the quantity calculated in the ’t Hooft–Feynman gauge, and the second term depends on the gauge parameter $\xi_V$ ($V=W, Z, \gamma$). 
We also introduce the shorthand notation for the Passarino–Veltman functions~\cite{Passarino:1978jh}, defined as
\begin{align}
C_0(p^2;X, Y)&\equiv \frac{1}{m_X^2-m_Y^2}[B_0(p^2;X,X)-B_0(p^2;Y,Y)], \\
C_0(p^2;X, Y, Z)&\equiv \frac{1}{m_X^2-m_Y^2}[B_0(p^2;X,Z)-B_0(p^2;Y,Z)],
\end{align}
which are convenient in the following calculations. 

%=================================================================
\subsection{CP-even sector}
%=================================================================

\begin{figure}
    \centering
    \includegraphics[width=1\linewidth]{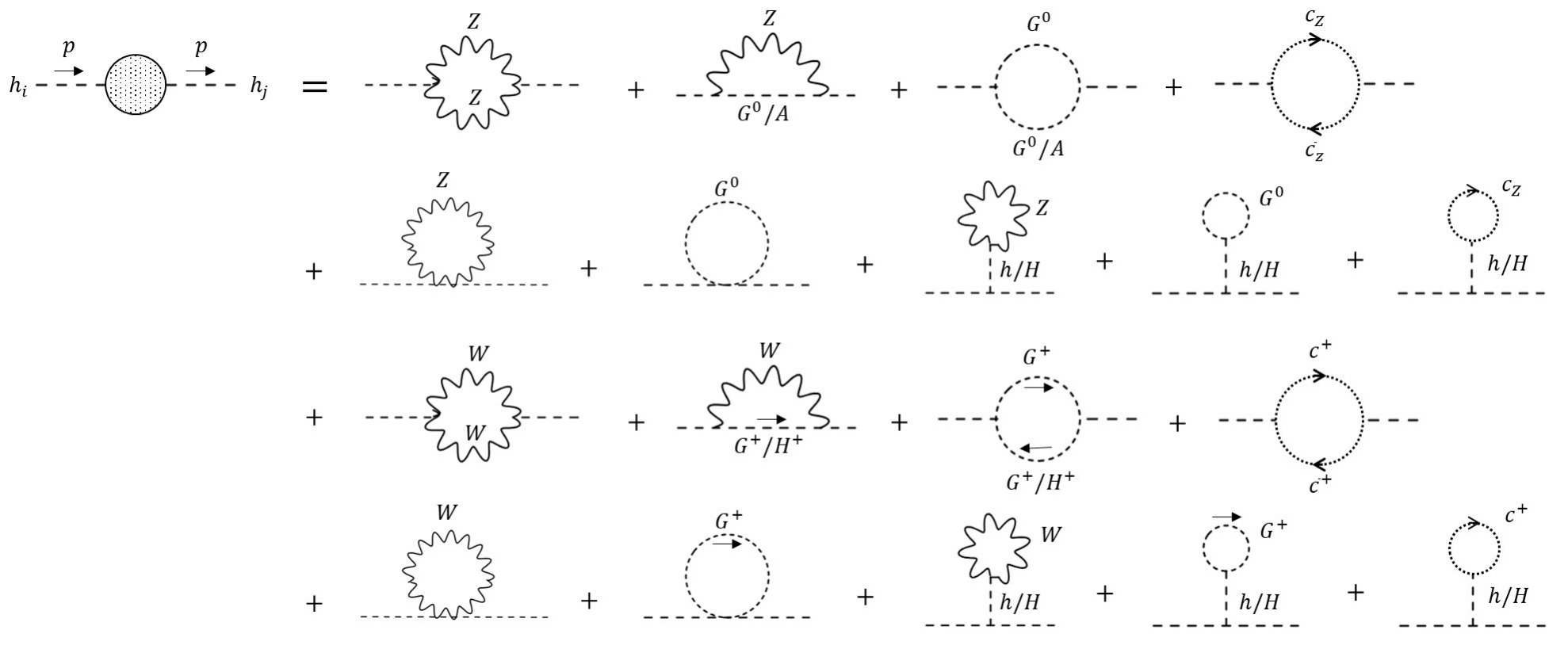}
    \caption{Diagrams dependent on the gauge parameters $\xi_V^{}$ for the self-energy of the CP-even Higgs bosons.}
    \label{fig:CPeven_self}
\end{figure}
\begin{figure}
    \centering
    \includegraphics[width=1\linewidth]{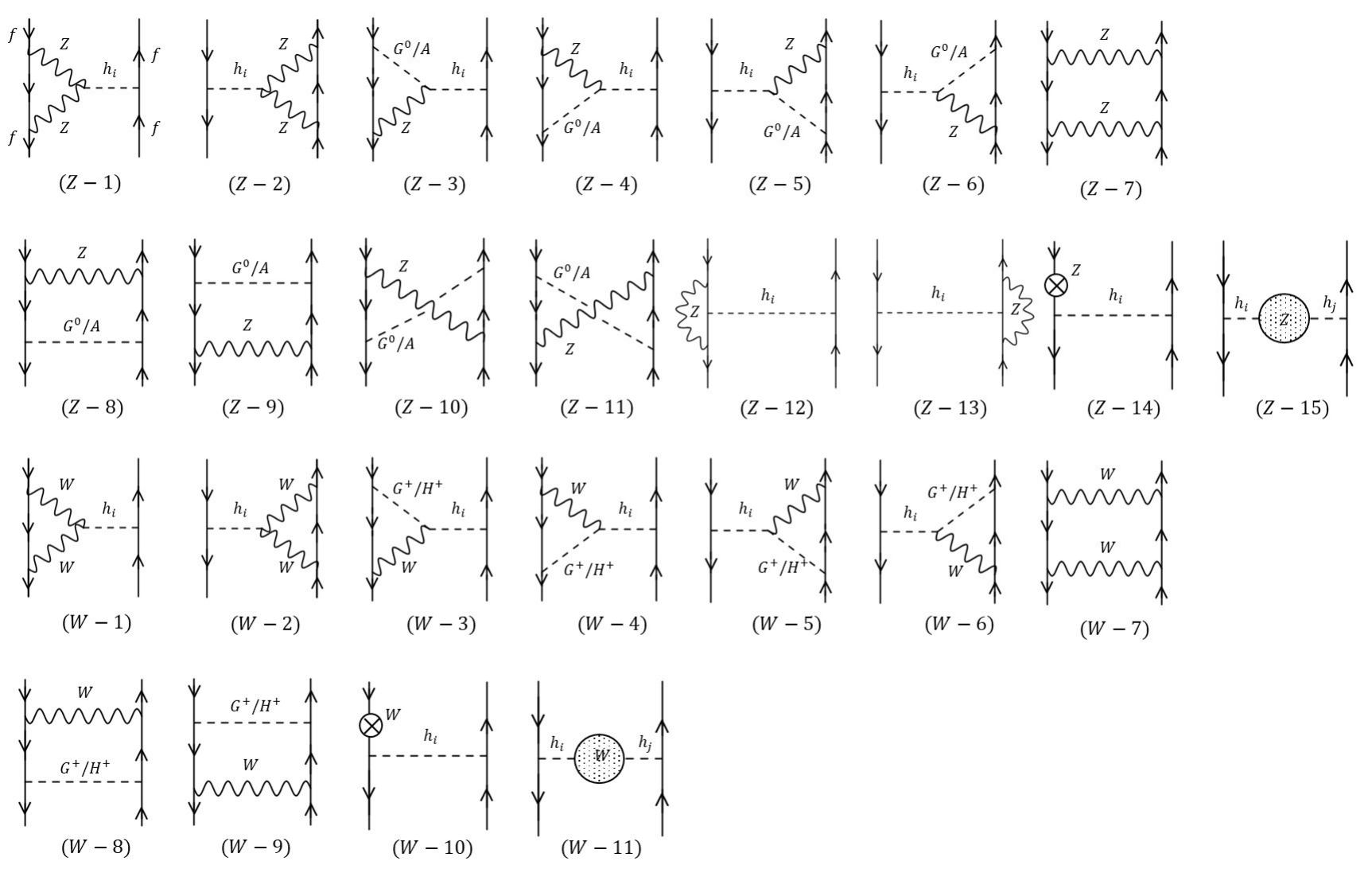}    
 \caption{Diagrams giving pinch terms for two-point functions of the CP-even Higgs bosons in the $f\bar{f}\to f\bar{f}$ scattering.}
    \label{fig:CPeven_WZ}
\end{figure}

%%%%%%
We first discuss the calculation of the pinch terms for two-point functions of the CP-even scalar fields $h_{i}$ ($h_1=h,\ h_2=H$). 
Fig.~\ref{fig:CPeven_self} presents the diagrams for the gauge-dependent two-point functions of the CP-even scalar fields. 
To extract the pinch terms for the CP-even scalar boson two-point functions, we consider the $f\bar{f}\to f\bar{f}$ scattering processes shown in Fig.~\ref{fig:CPeven_WZ}, which contain the gauge dependence and contributions proportional to $(m_f/v)^2$.   
We introduce the reduced amplitude $\overline{\mathcal{M}}$, defined as 
\begin{align}
\mathcal{M} =\overline{\mathcal{M}} \left(\frac{m_f}{v}\right)^2(\overline{f} f)\times(\overline{f}f). 
\end{align} 
This amplitude is defined as that with pinching the internal fermion lines.

The self-energy diagrams shown in Fig.~\ref{fig:CPeven_self} also contribute to the scattering amplitude, and correspond to diagrams (Z-15) and (W-11) in Fig.~\ref{fig:CPeven_WZ}. 
Since these self-energy-like contributions have an overall factor of $(m_f/v)^2$, 
in order to extract the pinch terms responsible for canceling their gauge dependence, 
it is sufficient to focus only on the contributions shown in Fig.~\ref{fig:CPeven_WZ}. 

%%%%%
%%%%%%

First, we confirm the cancellation of the gauge dependence of the diagrams shown in Fig.~\ref{fig:CPeven_WZ}. 
The gauge-dependent parts of the diagrams (Z-15) and (W-11) can be expressed as
\begin{align}
&\overline{\mathcal{M}}_{Z\text{-}15}^{hh}\big|_\textrm{G.D.} +\overline{\mathcal{M}}_{W\text{-}11}^{hh}\big|_\textrm{G.D.} =\ 
\frac{g_Z^2}{128\pi^2} \frac{\zeta_{h}^2}{p^2 - m_h^2} (1 - \xi_Z) 
 \big[
  -2 c_{hhZZ}^{}  B_0(0; Z, G^0) \notag \\
 &+\ c_{G^0 hZ}^2  (p^2 + m_h^2)  C_0(p^2; Z, G^0) 
 + 2 c_{AhZ}^2 (p^2-2 m_A^2 + m_h^2)  C_0(p^2; Z, G^0, A)
\big]\notag \\
&+\frac{g^2}{64\pi^2}\frac{\zeta_h^2}{p^2-m_h^2}(1-\xi_W)\bigg[
 -2c_{hhWW}^{}B_0(0; W, G^\pm)\notag\\
 &+2c_{H^\pm hW^\mp}^2(p^2+m_h^2 -2m_{H^\pm}^2)C_0(p^2; W, G^\pm, H^\pm)%\notag\\
 +c_{G^\pm hW^\mp}^2(p^2 +m_h^2)C_0(p^2; W, G^\pm)
\bigg], \label{eq:54}\\
%%%%%
&\overline{\mathcal{M}}_{Z\text{-}15}^{HH}\big|_\textrm{G.D.}+\overline{\mathcal{M}}_{W\text{-}11}^{HH}\big|_\textrm{G.D.} =\
 \frac{g_Z^2}{128\pi^2} \frac{\zeta_h^2}{p^2 - m_H^2}(1 - \xi_Z) 
  \big[
  -2 c_{HHZZ}^{}  B_0(0; Z, G^0) \notag \\
  &+ c_{G^0 HZ}^2 (p^2 + m_H^2)  C_0(p^2; Z, G^0)
  - 2 c_{AHZ}^2 (2 m_A^2 - m_H^2 - p^2)  C_0(p^2; Z, G^0, A)
\big]\notag\\
&+\frac{g^2}{64\pi^2}\frac{\zeta_H^2}{p^2-m_H^2}(1-\xi_W)\bigg[
 -2c_{HHWW}^{}B_0(0; W, G^\pm)\notag\\
 &+2c_{H^\pm H W^\mp}^2(p^2+m_H^2 -2m_{H^\pm}^2)C_0(p^2; W, G^\pm, H^\pm)
 +c_{G^\pm H W^\mp}^2(p^2 +m_H^2)C_0(p^2; W, G^\pm)
\bigg],\label{eq:55}\\
%%%%%
&\overline{\mathcal{M}}_{Z\text{-}15}^{hH}\big|_\textrm{G.D.} +\overline{\mathcal{M}}_{W\text{-}11}^{hH}\big|_\textrm{G.D.}=\
 \frac{g_Z^2}{256\pi^2} \frac{\, \zeta_H \zeta_h}{(p^2 - m_h^2)(p^2 - m_H^2)} (1 - \xi_Z) \notag \\
& \times \big[
  - c_{hHZZ}^{} (2p^2 - m_h^2 - m_H^2)  B_0(0; Z, G^0) 
 + 2 c_{G^0 hZ}^{} c_{G^0 HZ}^{} (p^4 - m_h^2 m_H^2)  C_0(p^2; Z, G^0) \notag \\
& \quad + 4\, c_{AhZ}^{} c_{AHZ}^{} \{p^4  + m_A^2 (m_h^2 + m_H^2 - 2p^2)- m_h^2 m_H^2 \}  C_0(p^2; Z, G^0, A)
\big]\notag\\
&+ \frac{g^2}{128\pi^2}\frac{\zeta_h\zeta_H}{(p^2-m_h^2)(p^2-m_H^2)}(1-\xi_W)\bigg[
 -c_{hHWW}^{}(2p^2-m_h^2-m_H^2)B_0(0; W, G^\pm)\notag\\
 &+4c_{H^\pm hW^\mp}^{} c_{H^\mp HW^\pm}^{}\{p^2(p^2-2m_{H^\pm}^2)
 -m_h^2m_H^2+m_h^2m_{H^\pm}^2+m_H^2m_{H^\pm}^2\}C_0(p^2; W, G^\pm, H^\pm) \notag\\
 &+2c_{G^\pm hW^\mp}^{} c_{G^\mp HW^\pm}^{}(p^4 -m_h^2 m_H^2)C_0(p^2; W, G^\pm)
\bigg], \label{eq:56}
\end{align}
where the mixing factors for the Yukawa couplings, $\zeta_X^{}$, are defined in Eq.~(\ref{eq: zeta_phi}), while those for the gauge couplings, $c_{XYZ(V)}$, are listed in Table~\ref{tab:couplings} in Appendix~\ref{app: coupling modifiers}.

The reduced amplitudes of the other diagrams shown in Fig.~\ref{fig:CPeven_WZ} are expressed as
\begin{align}
%%%%% Z-contributions %%%%%%%%
\sum_{i=1,2}\overline{\mathcal{M}}_{Z\text{-}i}\big|_\textrm{G.D.} &= 
\frac{c_{\beta'}^{}}{c_{\beta}^{}}\left(\zeta_h^{} c_{hZZ}^{} \overline{\mathcal{M}}_{\textrm{ver1}}[\textstyle{\frac{g_Z^{}}{\sqrt{2}}},Z,h]\big|_\textrm{G.D.}
+\zeta_H^{} c_{HZZ}^{}\overline{\mathcal{M}}_{\textrm{ver1}}[\textstyle{\frac{g_Z^{}}{\sqrt{2}}},Z,H]\big|_\textrm{G.D.}\right),\label{eq:57}\\
%%%%%%%%%
\sum_{i=3\text{-}6}\overline{\mathcal{M}}_{Z\text{-}i}\big|_\textrm{G.D.} &=  
\zeta_{G^0}^{} \zeta_h^{}c_{G^0 hZ}^{} \overline{\mathcal{M}}_\text{ver2}[\textstyle{\frac{g_Z^{}}{\sqrt{2}}},Z,h]\big|_\textrm{G.D.} 
+\zeta_{G^0}^{} \zeta_H^{} c_{G^0 HZ}^{} \overline{\mathcal{M}}_\text{ver2}[\textstyle{\frac{g_Z^{}}{\sqrt{2}}},Z,H]\big|_\textrm{G.D.} \notag\\
&+  \zeta_A^{} \zeta_h^{}c_{AhZ}^{} \overline{\mathcal{M}}_\text{ver3}[\textstyle{\frac{g_Z^{}}{\sqrt{2}}},Z,h]\big|_\textrm{G.D.}
+\zeta_A^{}\zeta_H^{} c_{AHZ}^{} \overline{\mathcal{M}}_\text{ver3}[\textstyle{\frac{g_Z^{}}{\sqrt{2}}},Z,H]\big|_\textrm{G.D.}, \\
%%%%%%%%%%
\overline{\mathcal{M}}_{Z\text{-}7}\big|_\textrm{G.D.} &=  
\frac{c_{\beta'}^{2}}{c_{\beta}^{2}}
\overline{\mathcal{M}}_\text{box1}[\textstyle{\frac{g_Z^{}}{\sqrt{2}}},Z]\big|_\textrm{G.D.}, \\
%%%%%%%%%
\sum_{i=8\text{-}11}\overline{\mathcal{M}}_{Z\text{-}i}\big|_\textrm{G.D.} &=  
\zeta_{G^0}^2\overline{\mathcal{M}}_\text{box2}[\textstyle{\frac{g_Z^{}}{\sqrt{2}}},Z,G^0]\big|_\textrm{G.D.}
%%%%%%%
+\zeta_A^2 \overline{\mathcal{M}}_\text{box2}[\textstyle{\frac{g_Z^{}}{\sqrt{2}}},Z,A]\big|_\textrm{G.D.},\\
%%%%%%%
\sum_{i=12\text{-}14}\overline{\mathcal{M}}_{Z\text{-}i}\big|_\textrm{G.D.} &=  
 \zeta_h^2 \overline{\mathcal{M}}_\text{ver+$\delta$}[\textstyle{\frac{g_Z^{}}{\sqrt{2}}},Z,h]\big|_\textrm{G.D.} 
+ \zeta_H^2 \overline{\mathcal{M}}_\text{ver+$\delta$}[\textstyle{\frac{g_Z^{}}{\sqrt{2}}},Z,H]\big|_\textrm{G.D.}, \label{eq:61}\\
%%%%% W-contributions %%%%%%%%
\sum_{i=1,2}\overline{\mathcal{M}}_{W\text{-}i}\big|_\textrm{G.D.} &= 
\zeta_h^{} c_{hWW}^{} \overline{\mathcal{M}}_{\textrm{ver1}}[g,W,h]\big|_\textrm{G.D.}
+\zeta_H^{} c_{HWW}^{}\overline{\mathcal{M}}_{\textrm{ver1}}[g,W,H]\big|_\textrm{G.D.},\\
%%%%%%%%%
\sum_{i=3\text{-}6}\overline{\mathcal{M}}_{W\text{-}i}\big|_\textrm{G.D.} &=  
\zeta_{G^\pm}^{}\zeta_h^{} c_{G^\pm hW^\mp}^{} \overline{\mathcal{M}}_\text{ver2}[g,W,h]\big|_\textrm{G.D.} 
+\zeta_{G^\pm}^{} \zeta_H^{}c_{G^\pm HW^\mp}^{}  \overline{\mathcal{M}}_\text{ver2}[g,W,H]\big|_\textrm{G.D.} \notag\\
&+  \zeta_{H^\pm}^{}\zeta_h^{}c_{H^\pm hW^\mp}^{} \overline{\mathcal{M}}_\text{ver3}[g,W,h]\big|_\textrm{G.D.}
+\zeta_{H^\pm}^{}\zeta_H^{}c_{H^\pm HW^\mp}^{} \overline{\mathcal{M}}_\text{ver3}[g,W,H]\big|_\textrm{G.D.}, \\
%%%%%%%%%%
\overline{\mathcal{M}}_{W\text{-}7}\big|_\textrm{G.D.} &=  
\overline{\mathcal{M}}_\text{box1}[g,W]\big|_\textrm{G.D.}, \\
%%%%%%%%%
\sum_{i=8,9}\overline{\mathcal{M}}_{W\text{-}i}\big|_\textrm{G.D.} &=  
\zeta_{G^\pm}^2\overline{\mathcal{M}}_\text{box2}[g,W,G^\pm]\big|_\textrm{G.D.}
%%%%%%%
+\zeta_{H^\pm}^2 \overline{\mathcal{M}}_\text{box2}[g,W,H^\pm]\big|_\textrm{G.D.},\\
%%%%%%%
\overline{\mathcal{M}}_{W\text{-}10}\big|_\textrm{G.D.} &=  
 \zeta_h^2 \overline{\mathcal{M}}_\text{ver+$\delta$}[g,W,h]\big|_\textrm{G.D.} 
+ \zeta_H^2 \overline{\mathcal{M}}_\text{ver+$\delta$}[g,W,H]\big|_\textrm{G.D.}, \label{eq:66}
\end{align}
where 
\begin{align}
\overline{\mathcal{M}}_\text{ver1}[g_i,V,\phi]\big|_\textrm{G.D.}
&= 
\frac{g_i^2}{16\pi^2} 
\frac{1}{p^2 - m_\phi^2} \bigg[
 -\left(1 + \frac{p^2}{2 m_V^2} \right) B_0(p^2; V, V) \notag\\
 &
 + \left(1 - \xi_V + \frac{p^2}{m_V^2} \right) B_0(p^2; V, G_V)
 + \left( \xi_V - \frac{p^2}{2 m_V^2} \right) B_0(p^2; G_V, G_V)
\bigg],\\
%%%%%
\overline{\mathcal{M}}_\text{ver2}[g_i,V,\phi]\big|_\textrm{G.D.}
&= \frac{g_i^2}{16\pi^2} 
\frac{1}{p^2 - m_\phi^2} 
\bigg[
B_0(p^2; V, V)
 - \left(1 - \xi_V + \frac{p^2}{m_V^2} \right) B_0(p^2; V, G_V) \notag \\
&
 - \left( \xi_V - \frac{p^2}{m_V^2} \right) B_0(p^2; G_V, G_V)
 + (1 - \xi_V) B_0(0; V, G_V)
\bigg],\\
%%%%%
\overline{\mathcal{M}}_\text{ver3}[g_i,V,\phi]\big|_\textrm{G.D.}
&= 
\frac{g_i^2}{16 \pi^2} 
\frac{1}{p^2 - m_\phi^2}(1 - \xi_V)
\bigg[
B_0(0; V, G_V)
 - (p^2 - m_\phi^2) C_0(p^2; V, G_V, \phi)
\bigg],  \\
%%%%%%
\overline{\mathcal{M}}_\text{box1}[g_i,V]\big|_\textrm{G.D.}
&= 
\frac{g_i^2}{64\pi^2} 
\frac{1}{m_V^2} 
\bigg[
B_0(p^2; V, V)
 - 2 B_0(p^2; V, G_V)
 + B_0(p^2; G_V, G_V)
\bigg], \\
%%%%%%
\overline{\mathcal{M}}_\text{box2}[g_i,V,\phi]\big|_\textrm{G.D.}
&= 
\frac{g_i^2}{32\pi^2} 
\frac{1}{m_V^2} 
\bigg[
B_0(p^2; V, \phi)
 - B_0(p^2; G_V, \phi)
\bigg], \\
%%%%%%%
\overline{\mathcal{M}}_\text{ver+$\delta$}[g_i,V,\phi]\bigg|_\textrm{G.D.}
&= 
- \frac{g_{i}^2 }{32 \pi^2}\frac{1}{p^2 - m_\phi^2}
(1 - \xi_V)\,
B_0(0; V, G_V), 
\end{align}
where $G_V$ denotes the NG boson associated with the vector boson $V$.  
%%%
% It can be confirmed analytically that the sum of all the contributions $\overline{\mathcal{M}}_{Z\text{-}1}\big|_\textrm{G.D.}$ ($\overline{\mathcal{M}}_{W\text{-}1}\big|_\textrm{G.D.}$)
% to $\overline{\mathcal{M}}_{Z\text{-}15}\big|_\textrm{G.D.}$ 
% ($\overline{\mathcal{M}}_{W\text{-}11}\big|_\textrm{G.D.}$) completely cancel each other. 
% In particular, in the 't Hooft-Feynman gauge, the gauge-dependent parts $\overline{\mathcal{M}}\big|_\text{G.D.}$ survive only in diagrams (Z-3) to (Z-6), (Z-15), (W-3) to (W-6) and (W-11). 
% Therefore, in the 't Hooft-Feynman gauge, the pinch terms with $\xi_V$ of the two-point functions for CP-even scalar bosons can be extracted from diagrams (Z-3) to (Z-6) and (W-3) to (W-6). 
% Since the pinch terms extracted from the reduced amplitude have the dimension of a scattering amplitude, it is necessary to multiply them by the factor $(p^2-m_{\phi}^2)(p^2-m_{\phi'}^2)/(\zeta_{\phi} \zeta_{\phi'})$, 
% in order to obtain quantities with the dimension of a self-energy. 
% As a consequence, the pinch terms for the self-energies in the ’t Hooft–Feynman gauge are given by 
It can be confirmed analytically that the sum of all the contributions $\overline{\mathcal{M}}_{Z\text{-}1}\big|_\textrm{G.D.}$ ($\overline{\mathcal{M}}_{W\text{-}1}\big|_\textrm{G.D.}$)
to $\overline{\mathcal{M}}_{Z\text{-}15}\big|_\textrm{G.D.}$ 
($\overline{\mathcal{M}}_{W\text{-}11}\big|_\textrm{G.D.}$) completely cancel each other.

Next, we move on to the extraction of the pinch terms for the self-energies from the amplitude. 
The pinch terms can be extracted from the reduced amplitudes except for the self-energy-like diagrams, i.e. (Z\text{-}15) and (W\textrm{-}11). %\blue{[Remind that (Z\text{-}15) and (W\textrm{-}11) are SE diagrams.]}
In the ’t Hooft–Feynman gauge, the pinch terms originating only from diagrams (Z-3)–(Z-6) and (W-3)–(W-6) survive. %\blue{[How to extract PT from G.D.?]}
The contributions in the reduced amplitude corresponding to the pinch terms of the scalar-field self-energies are expressed as 
\begin{align}
\sum_{i=3-6}\overline{\mathcal{M}}_{Z\text{-}i} %&=
%-\frac{g_{Z}^2}{32 \pi^2}\sum_{\varphi = G^0, A}\bigg\{
%\left(\frac{\zeta_{\varphi}^{}\zeta_{h}^{}c_{\varphi hZ}^{}}%{p^2 - m_h^2} +\frac{\zeta_{\varphi}^{}\zeta_{H}^{}c_{\varphi HZ}^{}}{p^2 - m_H^2}\right)%\notag\\
%\bigg[B_0(p^2;Z,\varphi)\notag\\
%&-(1-\xi_Z)B_0(0;Z,G^0)
%+\frac{1}{m_Z^2}\left(p^2-%m_{\varphi}^2\right)\left(B_0(p^2;Z,\varphi)-%B_0(p^2;G^0,\varphi)\right)
%\bigg]\bigg\}, \notag\\
%\xrightarrow{\xi_Z=1} 
&=-\frac{g_{Z}^2}{32 \pi^2}\bigg\{\left(\frac{\zeta_{G^0}^{}\zeta_{h}^{}c_{G^0 hZ}^{}}{p^2 - m_h^2} +\frac{\zeta_{G^0}^{}\zeta_{H}^{}c_{G^0 HZ}^{}}{p^2 - m_H^2}\right)
B_0(p^2;Z,Z)\notag\\
&\quad\quad\quad +\left(\frac{\zeta_{A}^{}\zeta_{h}^{}c_{A hZ}^{}}{p^2 - m_h^2} +\frac{\zeta_{A}^{}\zeta_{H}^{}c_{A HZ}^{}}{p^2 - m_H^2}\right)
B_0(p^2;Z,A)\bigg\},\\
%%%%%%%%
\sum_{i=3-6}\overline{\mathcal{M}}_{W\text{-}i} %&=
%-\frac{g^2}{16 \pi^2}\sum_{\varphi = G^\pm, H^\pm}\bigg\{
%\left(\frac{\zeta_{\varphi}^{}\zeta_{h}^{}c_{\varphi h %W^{\mp}}^{}}{p^2 - m_h^2} %+\frac{\zeta_{\varphi}^{}\zeta_{H}^{}c_{\varphi HW^{\mp}}^{}}%{p^2 - m_H^2}\right)%\notag\\
%\bigg[B_0(p^2;W,\varphi)\notag\\
%&-(1-\xi_W)B_0(0;W,G^\pm)
%+\frac{1}{m_W^2}\left(p^2-%m_{\varphi}^2\right)\left(B_0(p^2;W,\varphi)-%B_0(p^2;G^\pm,\varphi)\right)
%\bigg]\bigg\}, \notag\\
%\xrightarrow{\xi_W=1} 
&=-\frac{g^2}{16 \pi^2}\bigg\{\left(\frac{\zeta_{G^\pm}^{}\zeta_{h}^{}c_{G^\pm hW^{\mp}}^{}}{p^2 - m_h^2} +\frac{\zeta_{G^\pm}^{}\zeta_{H}^{}c_{G^\pm HW^{\mp}}^{}}{p^2 - m_H^2}\right)
B_0(p^2;W,W)\notag\\
&\quad\quad\quad +\left(\frac{\zeta_{H^\pm}^{}\zeta_{h}^{}c_{H^\pm hW^{\mp}}^{}}{p^2 - m_h^2} +\frac{\zeta_{H^\pm}^{}\zeta_{H}^{}c_{H^\pm HW^{\mp}}^{}}{p^2 - m_H^2}\right)
B_0(p^2;W,H^\pm)\bigg\}.  
\end{align}
%\blue{[Are the first and third lines needed?]}
The amplitude can be decomposed into pinch contributions to the $hh$, $hH$ and $HH$ self-energies as 
\begin{align}
\sum_{i=3-6}\left(\overline{\mathcal{M}}_{Z\text{-}i} + \overline{\mathcal{M}}_{W\text{-}i} \right)&=
\frac{\zeta_h^2}{(p^2-m_h^2)^2}\Pi_{hh}^\textrm{PT}(p^2)
+\frac{\zeta_H^2}{(p^2-m_H^2)^2}\Pi_{HH}^\textrm{PT}(p^2) \notag\\
&+\frac{2\zeta_h\zeta_H}{(p^2-m_h^2)(p^2-m_H^2)}\Pi_{hH}^\textrm{PT}(p^2). \label{eq:decompose}
\end{align}
These pinch contributions $\Pi_{\phi\phi'}^\textrm{PT}$ are combined with the original self-energy diagrams shown in Fig.~\ref{fig:CPeven_self}.
As a consequence, the pinch terms for the self-energies in the ’t Hooft–Feynman gauge are given by
 \begin{align}
\Pi_{hh}^\mathrm{PT}(p^2) 
&= -\frac{g^2}{16\pi^2} (p^2 - m_h^2) 
\left[
  c_{G^\pm hW^\mp}^2 B_0(p^2; W, W) 
+ c_{H^\pm hW^\mp}^2 B_0(p^2; W, H^\pm)
\right] \notag\\
&\quad -\frac{g_Z^2}{32\pi^2} (p^2 - m_h^2) 
\left[
  c_{G^0 hZ}^2 B_0(p^2; Z, Z) 
+ c_{AhZ}^2 B_0(p^2; Z, A)
\right], \\[1ex]
%%%%%%
\Pi_{HH}^\mathrm{PT}(p^2) 
&= -\frac{g^2}{16\pi^2} (p^2 - m_H^2) 
\left[
  c_{G^\pm HW^\mp}^2 B_0(p^2; W, W) 
+ c_{H^\pm HW^\mp}^2 B_0(p^2; W, H^\pm)
\right] \notag\\
&\quad -\frac{g_Z^2}{32\pi^2} (p^2 - m_H^2) 
\left[
  c_{G^0 HZ}^2 B_0(p^2; Z, Z) 
+ c_{AHZ}^2 B_0(p^2; Z, A)
\right], \\[1ex]
%%%%%%
\Pi_{hH}^\mathrm{PT}(p^2) 
&= -\frac{g^2}{32\pi^2} \left( 2p^2 - m_h^2 - m_H^2 \right) 
[
  c_{G^\pm hW^\mp}^{} c_{G^\mp HW^\pm}^{} \, B_0(p^2; W, W) \notag\\
&\hspace{4.5cm}
+ c_{H^\pm hW^\mp}^{} c_{H^\mp HW^\pm}^{} \, B_0(p^2; W, H^\pm)] \notag\\
&\quad -\frac{g_Z^2}{64\pi^2} \left( 2p^2 - m_h^2 - m_H^2 \right) 
\left[
  c_{G^0 hZ}^{} c_{G^0 HZ}^{} \, B_0(p^2; Z, Z) 
+ c_{AhZ}^{} c_{AHZ}^{} \, B_0(p^2; Z, A)
\right]. 
 \end{align}
In performing the decomposition shown in Eq.~(\ref{eq:decompose}),
%In order to distribute the contributions of Eqs.~(\ref{eq:57})-(\ref{eq:66}) into the contributions of Eqs.~(\ref{eq:54}), (\ref{eq:55}) and (\ref{eq:56}), 
we use the following relations among the scaling factors:
\begin{align}
\zeta_h^{} &= \zeta_A^{}\,c_{AhZ}^{} + \zeta_{G^0}^{}\,c_{G^0 hZ}^{}
= \zeta_{G^\pm}^{}\,c_{G^\pm hW^\mp}^{} + \zeta_{H^\pm}^{}\,c_{H^\pm hW^\mp}^{},\label{eq:sum rule1} \\
\zeta_H^{} &= \zeta_A^{}\,c_{AHZ}^{} + \zeta_{G^0}^{}\,c_{G^0 HZ}^{} 
= \zeta_{G^\pm}^{}\,c_{G^\pm HW^\mp}^{} + \zeta_{H^\pm}^{}\,c_{H^\pm HW^\mp}^{}, \label{eq:sum rule2} \\
\zeta_{G^0}^{} &= \zeta_h^{}\,c_{G^0 hZ}^{} + \zeta_H^{}\,c_{G^0 HZ}^{}
= \zeta_{G^\pm}^{}\,c_{G^\pm G^0W^\mp}^{} + \zeta_{H^\pm}^{}\,c_{H^\pm G^0W^\mp}^{}, \label{eq:sum rule3} \\
\zeta_A^{} &= \zeta_h^{}\,c_{AhZ}^{} + \zeta_H^{}\,c_{AHZ}^{} 
= \zeta_{G^\pm}^{}\,c_{G^\pm AW^\mp}^{} + \zeta_{H^\pm}^{}\,c_{H^\pm AW^\mp}^{}, \label{eq:sum rule4} \\
\zeta_{G^\pm}^{} &= \zeta_h^{}\,c_{G^\pm hW^\mp}^{} + \zeta_H^{}\,c_{G^\pm HW^\mp}^{}
=\zeta_{G^0}^{}\,c_{G^\pm G^0 W^\mp}^{}, \label{eq:sum rule5} 
\\
\zeta_{H^\pm}^{} &= \zeta_h^{}\,c_{H^\pm hW^\mp}^{} + \zeta_H^{}\,c_{H^\pm HW^\mp}^{}
=\zeta_A^{}\,c_{H^\pm AW^\mp}^{} + \zeta_{G^0}^{}\,c_{H^\pm G^0 W^\mp}^{}. \label{eq:sum rule6} 
\end{align}

%=================================================================
\subsection{CP-odd sector}
%=================================================================

\begin{figure}
    \centering
    \includegraphics[width=1.00\linewidth]{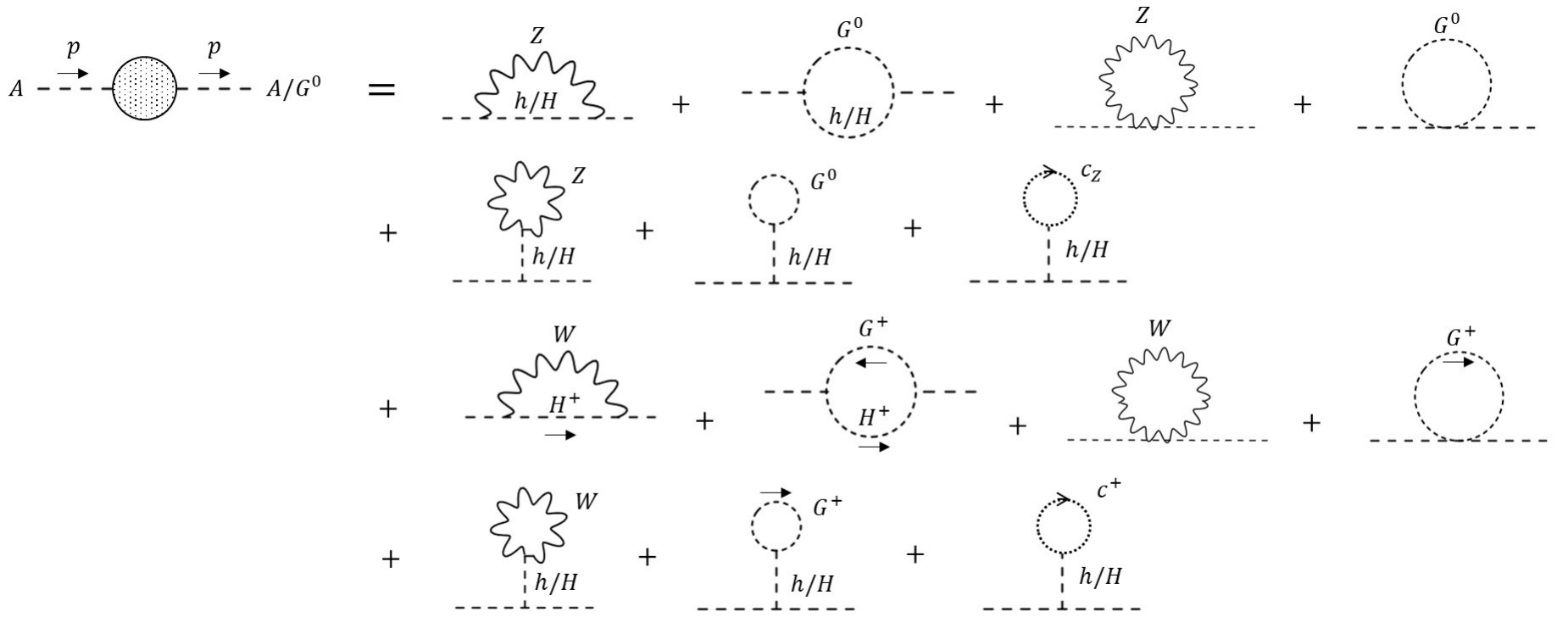}
    \caption{Diagrams dependent on the gauge parameters $\xi_V^{}$ for two-point functions of the CP-even Higgs bosons in the $f\bar{f}\to f\bar{f}$ scattering.} 
    \label{fig:self_odd}
\end{figure}

\begin{figure}
    \centering
    \includegraphics[width=1\linewidth]{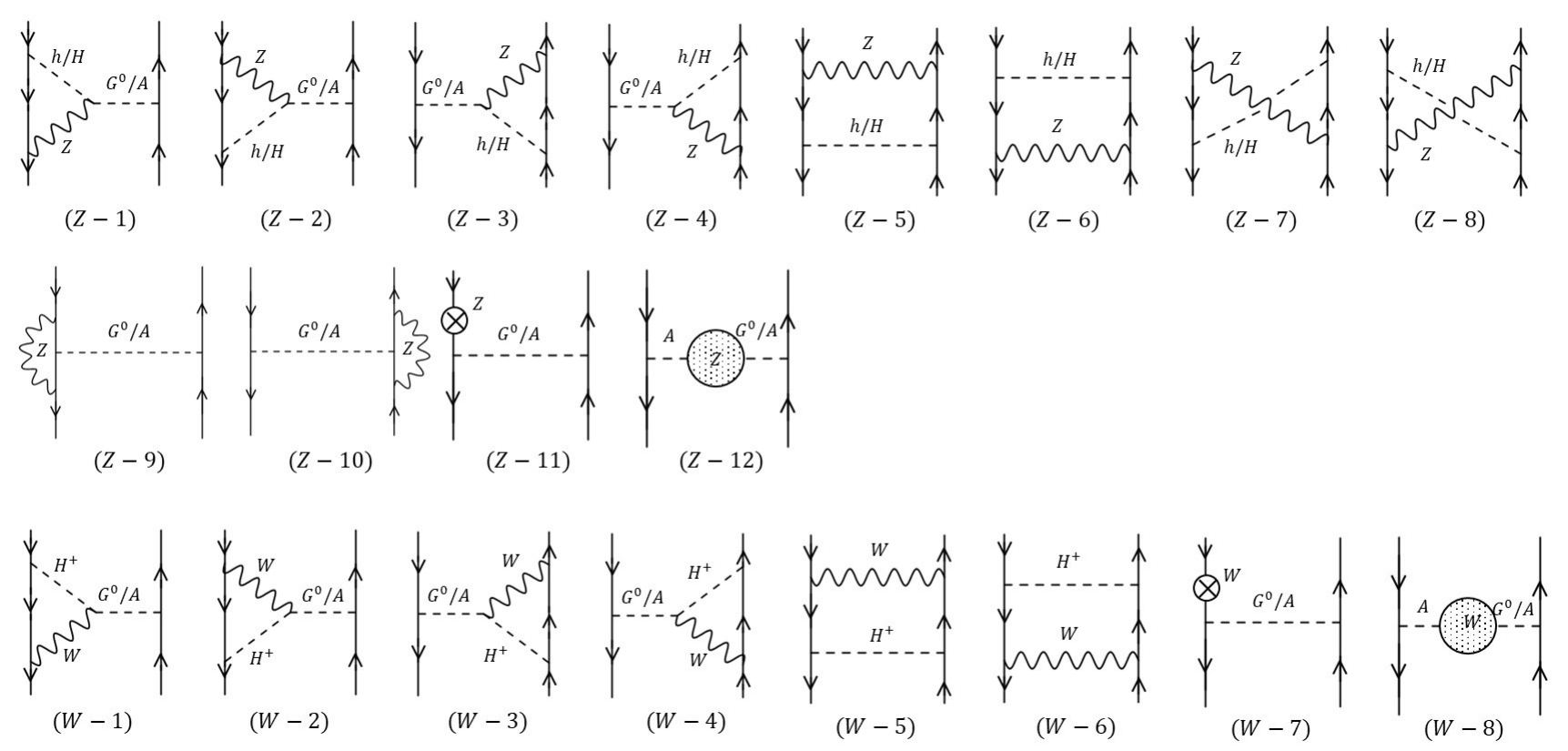}
    \caption{Diagrams giving pinch terms for two-point functions of the CP-even Higgs bosons in the $f\bar{f}\to f\bar{f}$ scattering.}
    \label{fig:CPodd_WZ}
\end{figure}

In this subsection, we derive the pinch terms for the two-point functions of the CP-odd scalar fields, which eliminate the gauge dependence arising from the diagrams shown in Fig.~\ref{fig:self_odd}.
The $f\bar{f}\to f\bar{f}$ scattering amplitude that eliminates the gauge dependence of the two-point functions of CP-odd scalar fields has the following structure:
\begin{align}
\mathcal{M} =-\overline{\mathcal{M}} \left(\frac{m_f}{v}\right)^2(\overline{f}\gamma_5 f)\times(\overline{f}\gamma_5 f). 
\end{align}
Diagrams having the structure are shown in Fig.~\ref{fig:CPodd_WZ}. 
The contributing diagrams for the $AA$ two-point function are the same as those in the 2HDMs. 
However, for the $AG^0$ two-point function, the situation differs from the 2HDMs in the sense that there are contributions from the ($W, H^\pm$), ($G^\pm, H^\pm$) and ($W$)-loop diagrams. 
Although those diagrams generate an additional $\xi_W$-dependent part, this contribution is canceled by the diagrams (W-1) to (W-4) in Fig.~\ref{fig:CPodd_WZ}, in which the $G^0$ propagates.
Because of these contributions, the $AG^0$ two-point function also acquires a pinch term that contains the $\xi_W^{}$ part. 
As a result, the gauge-dependent contributions of diagrams (Z-12) and (W-8) are derived as
 \begin{align}
& \overline{\mathcal{M}}^{AA}_{W\textrm{-}8}\big|_\textrm{G.D.} 
 +\overline{\mathcal{M}}^{AA}_{Z\textrm{-}12}\big|_\textrm{G.D.} =\frac{g^2}{16\pi^2}
\frac{\zeta_A^2 \zeta_{G^0}^2}{p^2-m_A^2}(1-\xi_W)\notag\\
&\times\bigg[
 -B_0(0;W,G^\pm)
 +(p^2+m_A^2 -2m_{H^\pm}^2)C_0(p^2;W,G^\pm,H^\pm)
\bigg] \notag\\
&+\frac{g_Z^2}{64 \pi^2 }\frac{\zeta_A^2}{p^2 - m_A^2} (1 - \xi_Z) \bigg[
   -(c_{AHZ}^2 + c_{AhZ}^2) \, B_0(0; Z, G^0) \notag\\
  & + c_{AHZ}^2 (p^2 + m_A^2 - 2 m_H^2) \, C_0(p^2; Z, G^0, H) \notag\\
  & + c_{AhZ}^2 (p^2 + m_A^2 - 2 m_h^2) \, C_0(p^2; Z, G^0, h)
\bigg], \\
%%%%
&\overline{\mathcal{M}}^{AG^0}_{W\textrm{-}8}\big|_\textrm{G.D.}
+\overline{\mathcal{M}}^{AG^0}_{Z\textrm{-}12}\big|_\textrm{G.D.}=\frac{g^2}{32\pi^2}
\frac{\zeta_A^{}\zeta_{G^0}^{}c_{H^\pm AW^\mp}^{}c_{H^\pm G^0W^\mp}^{}}{(p^2-m_A^2)(p^2-m_{G^0}^2)}(1-\xi_W)\notag\\
&\times\bigg[-(2p^2-m_A^2)B_0(0;W,G^\pm)
 +2\{p^2(p^2-2m_{H^\pm}^2)+m_A^2 m_{H^\pm}^2\}C_0(p^2;W,G^\pm,H^\pm) \bigg]\notag\\
&+\frac{g_Z^2}{128 \pi^2} \frac{2 \, \zeta_A \, \zeta_{G^0}}{(p^2 - m_A^2)(p^2 - m_{G^0}^2)} 
(1 - \xi_Z)\notag\\ &\times \bigg[
  -(c_{AhZ}^{} \, c_{G^0 hZ}^{} + c_{AHZ}^{} \, c_{G^0 HZ}^{}) (2p^2 - m_A^2) \, B_0(0; Z, G^0) \notag\\
  & + 2 c_{AhZ}^{} \, c_{G^0 hZ}^{} \{m_A^2 m_h^2 + p^2( p^2-2m_h^2 )\} \, C_0(p^2; Z, G^0, h) \notag\\\
  & + 2 c_{AHZ}^{} \, c_{G^0 HZ}^{} \{m_A^2 m_H^2 + p^2( p^2-2m_H^2 )\} \, C_0(p^2; Z, G^0, H)
\bigg]. 
 \end{align}
Contributions from other diagrams are calculated as
\begin{align}
\sum_{i=1\text{-}11}\overline{\mathcal{M}}_{Z\text{-}i}\big|_\textrm{G.D.} 
&=\frac{g_Z^2}{64 \pi^2} (1 - \xi_Z) \Bigg[
   \frac{\zeta_A^{}}{p^2 - m_A^2} \Big\{ 
      \zeta_A^{} \, B_0(0; Z, G^0) \notag\\
      &- 2 c_{AHZ}^{}\,\zeta_H^{}\,(p^2-m_H^2) C_0(p^2; Z, G^0, H) 
 - 2 (p^2 - m_h^2) \, c_{AhZ}^{}\,\zeta_h^{}\, C_0(p^2; Z, G^0, h) \Big\} \notag\\
&  + \frac{\zeta_{G^0}^{}}{p^2 - m_{G^0}^2} \Big\{
      \zeta_{G^0}^{} \, B_0(0; Z, G^0) 
      - 2 c_{G^0 HZ}^{}\,\zeta_H^{}\,(p^2 -m_H^2 ) C_0(p^2; Z, G^0, H) \notag\\
& \hspace{2cm}- 2 (p^2-m_h^2) \, c_{G^0 hZ}^{}\,\zeta_h^{}\, C_0(p^2; Z, G^0, h) \Big\} \notag\\
&  + \zeta_h^2 \, C_0(p^2; Z, G^0, h) 
        + \zeta_H^2 \, C_0(p^2; Z, G^0, H)
\Bigg] ,\label{eq:M_odd_Z}\\
%%%
\sum_{i=1\text{-}7}\overline{\mathcal{M}}_{W\text{-}i}\big|_\textrm{G.D.}&=
\frac{g^2 (1 - \xi_W^{})}{32 \pi^2} \Bigg[ 
   \frac{\zeta_A^{}\,\zeta_{H^\pm}\,c_{H^\pm AW^\mp}^{}}{p^2 - m_A^2}
   \Big\{ B_0(0; W, G^\pm) - 2 (p^2 - m_{H^+}^2) C_0(p^2; W, G^\pm, H^\pm) \Big\} \notag\\
&  +\frac{\zeta_{G^0}}{p^2 - m_{G^0}^2} 
   \Big\{ -\big(\zeta_{G^0} - 2\zeta_{H^\pm}^{} c_{H^\pm G^0W^\mp}^{}\big) B_0(0; W, G^\pm) \notag\\
   & \hspace{2cm}- 2 (p^2 - m_{H^+}^2) c_{H^\pm G^0W^\mp}^{}\,\zeta_{H^\pm}^{}\, C_0(p^2; W, G^\pm, H^\pm) \Big\} \notag\\
  &+ \zeta_{H^\pm}^2 \, C_0(p^2; W, G^\pm, H^\pm) 
\Bigg]. \label{eq:M_odd_W}
\end{align}
Part of contributions of Eqs.~(\ref{eq:M_odd_Z}) and (\ref{eq:M_odd_W}) 
cancels those of (Z-12) and (W-8).
However, they also contain contributions that cancel the gauge-dependent part of the $AZ^\mu$ mixing. 
In order to extract only the pinch term of the gauge-dependent parts from the $AG^0$ mixing diagram, we use the following relation implied by the Ward–Takahashi identity,
\begin{align}
\Lambda_{G^0}=\frac{p^2}{p^2-m_{G^0}^2}\Lambda_{G^0}^{}-im_Z^{}\Lambda_Z^\mu(\Delta_Z)_{\mu\nu}^{}p^\nu,  
\end{align}
where the $\bar{f}fG^0 $ and $\bar{f}fZ^\mu$ vertex functions and the Z-boson propagator are expressed by 
\begin{align}
\Lambda_{G^0} = -\frac{m_f}{v}\,\bar{f}\gamma_5 f, 
\quad
\Lambda_Z^\mu = i g_Z\, \bar{f}\gamma^\mu (v_f - a_f \gamma_5) f, \notag\\
(\Delta_Z)_{\mu\nu}
= \frac{1}{p^2 - m_Z^2}
\left(
  -g_{\mu\nu}
  + \frac{(1 - \xi_Z) p_\mu p_\nu}{p^2 - \xi_Z m_Z^2}
\right),
\end{align}
where the vector and axial-vector couplings are denoted by $v_f = I_f/\,2 -s_W^2 Q_f$ and $a_f=I_f/\,2$, respectively.
The product of the $\bar{f}fG^0$ vertex function and the propagators in the diagrams shown in Fig.~\ref{fig:CPodd_WZ}, except for (Z-12) and (W-8), can be rewritten as follows:
\begin{align}
\left(\frac{1}{m^2}+\frac{1}{p^2-m_{G^0}^2}+\frac{1}{p^2-m_A^2}\right)\Lambda_{G^0}^{}&=
\Bigg(\frac{p^2(p^2-m_A^2)}{m^2(p^2-m_{G^0}^2)(p^2-m_A^2)}
+\frac{p^2-m_A^2}{(p^2-m_{G^0}^2)(p^2-m_A^2)}\notag\\
&+\frac{p^2}{(p^2-m_{G^0}^2)(p^2-m_A^2)}\Bigg) \Lambda_{G^0}^{}+\cdots.
\end{align}
The term enclosed in parentheses can be extracted as the pinch term of the $AG^0$ mixing. 
 
As in the CP-even sector, in the 't Hooft-Feynman gauge, the pinch terms of the $AA$ and $AG^0$ two-point functions can be extracted from the diagrams (Z-1) to (Z-4) and (W-1) to (W-4) as
\begin{align}
\sum_{i=1-4}\left(\overline{\mathcal{M}}_{Z\text{-}i} + \overline{\mathcal{M}}_{W\text{-}i} \right)&=
\frac{\zeta_A^2}{(p^2-m_A^2)^2}\Pi_{AA}^\textrm{PT}(p^2)
+\frac{\zeta_{G^0}^2}{(p^2-m_{G^0}^2)^2}\Pi_{G^0G^0}^\textrm{PT}(p^2) \notag\\
&+\frac{2\zeta_A\zeta_{G^0}^{}}{(p^2-m_A^2)(p^2-m_{G^0}^2)}\Pi_{AG^0}^\textrm{PT}(p^2). \label{eq:decompose2}
\end{align}
where the LHS is calculated as
\begin{align}
\sum_{i=1-4}\left(\overline{\mathcal{M}}_{Z\text{-}i}
+\overline{\mathcal{M}}_{W\text{-}i}\right)&=
-\frac{g^2}{16 \pi^2}\left(\frac{\zeta_{A}^{}\zeta_{H^\pm}^{}c_{H^\pm A W^{\mp}}^{}}{p^2 - m_A^2} +\frac{\zeta_{G^0}^{}\zeta_{H^\pm}^{}c_{H^\pm G^0W^{\mp}}^{}}{p^2 - m_{G^0}^2}\right)
B_0(p^2;W,H^\pm)\notag\\
&-\frac{g_{Z}^2}{32 \pi^2}\sum_{\phi=h,H}\left(\frac{\zeta_{A}^{}\zeta_{\phi}^{}c_{G^0 \phi Z}^{}}{p^2 - m_A^2} +\frac{\zeta_{G^0}^{}\zeta_{\phi}^{}c_{G^0 \phi Z}^{}}{p^2 - m_{G^0}^2}\right)
B_0(p^2;Z,\phi). 
%%%%%%%%
\end{align}
By using the relations given in Eqs.~(\ref{eq:sum rule1}), (\ref{eq:sum rule2}), (\ref{eq:sum rule5}), and (\ref{eq:sum rule6}), the explicit forms of the pinch terms for the $AA$ and $AG^0$ self-energies can be derived as
\begin{align}
\Pi_{AA}^\mathrm{PT}(p^2) 
&= -\frac{g^2}{16\pi^2} \, c_{H^\pm A W^\mp}^2 \, (p^2 - m_A^2) \, B_0(p^2; W, H^\pm) \notag\\
&\quad -\frac{g_Z^2}{32\pi^2} (p^2 - m_A^2) 
\left[
  c_{AhZ}^2 B_0(p^2; Z, h) 
+ c_{AHZ}^2 B_0(p^2; Z, H)
\right], \label{eq:PT_AA} \\[1ex]
%%%%%%
\Pi_{A G^0}^\mathrm{PT}(p^2) 
&= -\frac{g^2}{32\pi^2} \, c_{H^\pm G^0  W^\mp}^{} \, c_{H^\pm A W^\mp}^{} \, (2p^2 - m_A^2) \, B_0(p^2; W, H^\pm) \notag\\
&\quad -\frac{g_Z^2}{64\pi^2} (2p^2 - m_A^2) 
\left[
  c_{hG^0 Z}^{} c_{AhZ}^{} \, B_0(p^2; Z, h) 
+ c_{HG^0 Z}^{} c_{AHZ}^{} \, B_0(p^2; Z, H)
\right]. \label{eq:PT_AG0}
\end{align}
Although the above pinch term formulae take the same form as those in the 2HDMs, the $H^\pm G^0 W$ coupling does not appear in multi-doublet structures. 
As a result, the first term in Eq.~(\ref{eq:PT_AG0}) is absent in the 2HDMs. 

%%%%%%%%%%%%%%%%%%%%%%%%%%%%%%%%%%%%%%%%%%%%%%%%%%%%%%%%%%%%%%%%%%
\section{Decays of the SM-like Higgs boson}
\label{sec: decay of h125}
%%%%%%%%%%%%%%%%%%%%%%%%%%%%%%%%%%%%%%%%%%%%%%%%%%%%%%%%%%%%%%%%%%

In this section, we discuss the decay of the SM-like Higgs boson including higher-order corrections.
In Sec.~\ref{subsec: renormalized vertex function}, we define the renormalized $hf\bar{f},\ hVV$ and $h\mathcal{V}\mathcal{V}'$ vertex functions, where $V=Z, W^{\pm}$ and $(\mathcal{V}, \mathcal{V'}) = (g,g), (\gamma, \gamma)$, and $(Z, \gamma)$. The renormalized $hhh$ vertex is given in Ref.~\cite{Aoki:2012jj}.
In Sec.~\ref{subsec: decay rates of h}, we present the decay rates in terms of the renormalized vertex functions.

%=================================================================
\subsection{Renormalized vertex functions}
\label{subsec: renormalized vertex function}
%=================================================================

%-----------------------------------------------------------------
\subsubsection{\texorpdfstring{$hf\bar{f}$}{} vertex}
%-----------------------------------------------------------------

%-----------------------------------------
\begin{figure}[t]
 \begin{minipage}{0.45\hsize}
 \centering
 \includegraphics[scale=0.45]{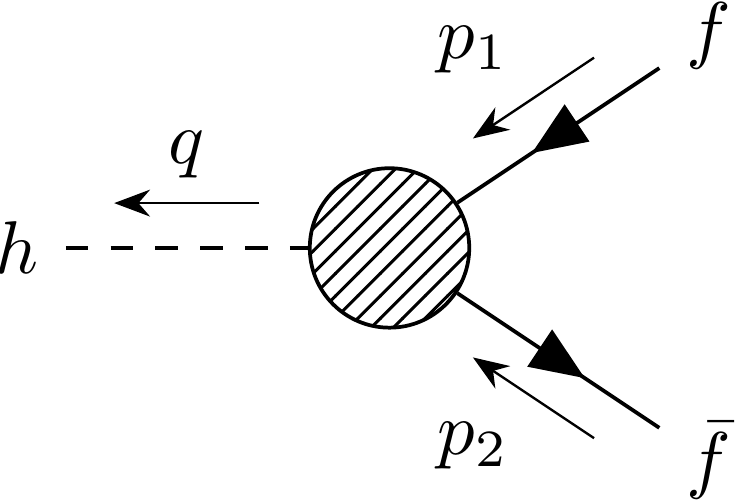}
 \end{minipage}
 \begin{minipage}{0.45\hsize}
 \centering
 \includegraphics[scale=0.45]{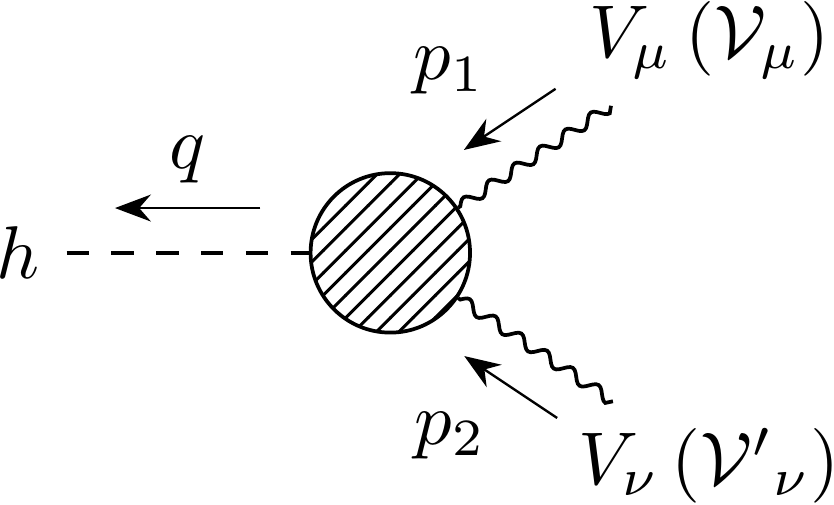} 
 \end{minipage}
  \caption{Momentum assignment for the $hf\bar{f}$ vertex (left) and the $hVV\, (h\mathcal{V}\mathcal{V'})$ vertex (right).}
 \label{fig: momentum assignment}
\end{figure}
%-----------------------------------------

The renormalized $hf\bar{f}$ vertex can be decomposed as~\cite{Kanemura:2015mxa}
\begin{align}
\widehat{\Gamma}_{hf\bar{f}}(p_{1}, p_{2}, q)&=
\widehat{\Gamma}_{hf\bar{f}}^{S}+\gamma_{5}\widehat{\Gamma}_{hf\bar{f}}^{P}
+\slashed{p}_{1}\widehat{\Gamma}_{hf\bar{f}}^{V_{1}}
+\slashed{p}_{2}\widehat{\Gamma}_{hf\bar{f}}^{V_{2}} \notag\\
&\quad
+\slashed{p}_{1}\gamma_{5}\widehat{\Gamma}_{hf\bar{f}}^{A_{1}}
+\slashed{p}_{2}\gamma_{5}\widehat{\Gamma}_{hf\bar{f}}^{A_{2}}
+\slashed{p}_{1}\slashed{p}_{2}\widehat{\Gamma}_{hf\bar{f}}^{T}
+\slashed{p}_{1}\slashed{p}_{2}\gamma_{5}\widehat{\Gamma}_{hf\bar{f}}^{PT}, \label{eq:hff-form}
\end{align}
where $p_{1}^{\mu}\, (p_{2}^{\mu})$ is the incoming four-momentum of the fermion (anti-fermion), and $q^{\mu}$ is the outgoing four-momentum of the SM-like Higgs boson (see the left panel of Fig.~\ref{fig: momentum assignment}).
The following relations hold for the on-shell fermions:
\begin{align}
\widehat{\Gamma}_{h f\bar{f}}^{P} &= \widehat{\Gamma}_{h f\bar{f}}^{PT} = 0, \quad 
\widehat{\Gamma}_{h f\bar{f}}^{V_1} = -\widehat{\Gamma}_{h f\bar{f}}^{V_2},\quad 
\widehat{\Gamma}_{h f\bar{f}}^{A_1} = -\widehat{\Gamma}_{h f\bar{f}}^{A_2}
\qq{with}
p_{1}^{2}=p_{2}^{2}=m_{f}^{2}.
\end{align}
The renormalized form factors consist of the tree-level and one-loop parts as
\begin{align}
\widehat{\Gamma}^X_{hf\bar{f}}={\Gamma}^{X,\mathrm{tree}}_{hf\bar{f}}+{\Gamma}^{X,\mathrm{loop}}_{hf\bar{f}}\qc (X = S, P, V_{1}, V_{2}, A_{1}, A_{2}, T, PT),
\end{align}
where the tree-level parts are given by
\begin{align}
\Gamma^{S,{\rm tree}}_{hf\bar{f}}=-\frac{m_{f}}{v}\zeta_{h}\qc
\Gamma^{X,{\rm tree}}_{hf\bar{f}}=0 \qq{for} X\neq S. 
\end{align}
The one-loop parts are decomposed into the 1PI diagrams and the counterterms.
\begin{align}
\Gamma^{X,{\rm loop}}_{hf\bar{f}} = \Gamma^{X,{\rm 1PI}}_{hf\bar{f}}+\delta \Gamma^{X}_{hf\bar{f}}.
\end{align}
The analytic expressions for $\Gamma^{X,{\rm 1PI}}_{hf\bar{f}}$ are presented in Appendix.~\ref{app: hff_1PI}, while $\delta \Gamma^{X}_{hf\bar{f}}$ are given by
\begin{align}
\delta \Gamma^{S}_{hf\bar{f}}
&=
\Gamma^{S, \mathrm{tree}}_{hf\bar{f}}\qty(
\frac{\delta m_{f}}{m_{f}}-\frac{\delta v}{v}+\tan{\beta}\delta\beta
+\delta Z_{V}^{f}+\frac{\delta Z_{h}}{2})
+\Gamma^{S, \mathrm{tree}}_{Hf\bar{f}}\qty(\delta\alpha+\frac{\delta Z_{Hh}}{2}),
\end{align}
where $\Gamma^{S, \mathrm{tree}}_{Hf\bar{f}}=-m_{f}\zeta_{H}/v$, and $\delta \Gamma^{X}_{hf\bar{f}} = 0$ for $X\neq S$.
The expressions for the counterterms are given in Ref.~\cite{Aoki:2012jj}, while we include the tadpole contributions as in Eq.~\eqref{eq:all_self_energy} and use gauge-independent mixing counterterms given in Eqs.~\eqref{eq:del_al} and \eqref{eq:del_beta}.

%-----------------------------------------------------------------------------
\subsubsection{\texorpdfstring{$hVV$}{} vertex}
\label{sec: hVV_vetex}
%-----------------------------------------------------------------------------

The renormalized $hVV$ vertex ($V=Z, W^{\pm}$) can be decomposed as~\cite{Kanemura:2004mg,Kanemura:2015mxa}
\begin{align}
\widehat{\Gamma}_{hVV}^{\mu\nu}(p_{1}, p_{2}, q) &=
g^{\mu\nu}\widehat{\Gamma}_{hVV}^{1}
+\frac{p_1^\nu p_2^\mu}{m_V^2}\widehat{\Gamma}_{hVV}^{2}
+i\epsilon^{\mu\nu\rho\sigma}\frac{p_{1\rho} p_{2\sigma}}{m_V^2}\widehat{\Gamma}_{hVV}^{3},
\end{align}
where $p_{1}^{\mu}$ and $p_{2}^{\mu}$ are the incoming four-momenta of the weak bosons, and $q^{\mu}$ is the outgoing four-momentum of the SM-like Higgs boson (see the right panel of Fig.~\ref{fig: momentum assignment}).
When we neglect the effects of the CP violation, we have $\widehat{\Gamma}_{hVV}^{3}(p_{1}^{2}, p_{2}^{2}, q^{2}) = 0$.
The renormalized form factors are composed of the tree-level and one-loop-level parts as
\begin{align}
\widehat{\Gamma}^{i}_{hVV}={\Gamma}^{i,{\rm tree}}_{hVV}+{\Gamma}^{i,{\rm loop}}_{hVV}\qc (i=1,2,3).
\end{align}
The tree-level couplings are given by
\begin{align}
\Gamma^{1, \mathrm{tree}}_{hVV}= c_{V}g_{hVV}\qc
\Gamma^{2, \mathrm{tree}}_{hVV}= \Gamma^{3, \mathrm{tree}}_{hVV}=0,
\end{align}
where $c_{Z\, (W)} = 2\, (1)$, and $g_{hVV}$ is given by
\begin{align}
    g_{hZZ} = \frac{1}{2}c_{hZZ}g_{Z}m_{Z}\qc
    g_{hWW} = c_{hWW}g m_{W}.
\end{align}
The one-loop parts are decomposed into the 1PI diagrams, tadpole diagrams, and the counterterms.
\begin{align}
{\Gamma}^{i, \mathrm{loop}}_{hVV} = \Gamma^{i, \mathrm{1PI}}_{hVV}+\Gamma^{i, \mathrm{Tad}}_{hVV}+\delta \Gamma^{i}_{hVV}.
\end{align}
The analytic expressions for $\Gamma^{i, \mathrm{1PI}}_{hVV}$ and $\delta \Gamma^{i}_{hVV}$ are presented in Ref.~\cite{Aoki:2012jj}.
In the alternative tadpole scheme, the tadpole contributions appear and are given by
\begin{align}
\Gamma^{1, \mathrm{Tad}}_{hVV}
=
c_{V}\qty(2g_{hhVV}\frac{\Gamma^{\rm 1PI}_{h}}{m_{h}^{2}}+g_{hHVV}\frac{\Gamma^{\rm 1PI}_{H}}{m_{H}^{2}})\qc
\Gamma^{2, \mathrm{Tad}}_{hVV} = \Gamma^{3, \mathrm{Tad}}_{hVV} = 0,
\end{align}
where $g_{hhVV}$ and $g_{hHVV}$ are given by
\begin{alignat}{2}
    g_{hhZZ} &= \frac{1}{8}c_{hhZZ}g_{Z}^{2}\qc&
    g_{hhWW} &= \frac{1}{4}c_{hhWW}g^{2}, \\
    g_{hHZZ} &= \frac{1}{8}c_{hHZZ}g_{Z}^{2}\qc&
    g_{hHWW} &= \frac{1}{4}c_{hHWW}g^{2}.
\end{alignat}
The analytic expressions for the 1PI tadpoles, $\Gamma^{\rm 1PI}_{h}$ and $\Gamma^{\rm 1PI}_{H}$, are presented in Ref.~\cite{Aoki:2012jj}.

%------------------------------------------------------------
\subsubsection{\texorpdfstring{$h\mathcal{V}\mathcal{V'}$}{} vertex}
\label{sec: hVV'_vetex}
%------------------------------------------------------------

The loop-induced $h\mathcal{V}\mathcal{V'}$ vertex with $(\mathcal{V}, \mathcal{V'}) = (g,g), (\gamma, \gamma), (Z, \gamma)$ can be decomposed as~\cite{Kanemura:2019kjg}
\begin{align}
\widehat{\Gamma}_{h\mathcal{V}\mathcal{V'}}^{\mu\nu}(p_{1}, p_{2}, q)
&=
g^{\mu\nu}\widehat{\Gamma}_{h\mathcal{V}\mathcal{V'}}^{1}
+{p_{1}^{\nu}p_{2}^{\mu}}\widehat{\Gamma}_{h\mathcal{V}\mathcal{V'}}^{2}
+i\epsilon^{\mu\nu\rho\sigma}p_{1\rho}p_{2\sigma}\widehat{\Gamma}_{h\mathcal{V}\mathcal{V'}}^{3},
\end{align}
where $p_{1}^{\mu}$ and $p_{2}^{\mu}$ are the incoming four-momenta of $\mathcal{V}$ and $\mathcal{V'}$, and $q^{\mu}$ is the outgoing four-momentum of the SM-like Higgs boson (see the right panel of Fig.~\ref{fig: momentum assignment}).
When we neglect the effects of the CP violation, $\widehat{\Gamma}_{h\mathcal{V}\mathcal{V'}}^{3} = 0$.
For the on-shell photon and gluon with a four-momentum $p_{2, \nu}$ ($\mathcal{V'}=\gamma\ \text{or}\ g$), the Ward-Takahashi identity holds, i.e. $p_{2, \nu}\widehat{\Gamma}_{h\mathcal{V}\mathcal{V'}}^{\mu\nu} = 0$, and we obtain
\begin{align}
\widehat{\Gamma}_{h\mathcal{V}\mathcal{V'}}^{1}
=
-p_{1}\cdot p_{2}\widehat{\Gamma}_{h\mathcal{V}\mathcal{V'}}^{2}.
\end{align}
The renormalized form factors are composed of the 1PI diagrams and the counterterms.
\begin{align}
\widehat{\Gamma}^{i}_{h\mathcal{V}\mathcal{V'}} = \Gamma^{i, \mathrm{1PI}}_{h\mathcal{V}\mathcal{V'}}+\delta \Gamma^{i}_{h\mathcal{V}\mathcal{V'}}\qc (i=1,2,3).
\end{align}
The analytic expressions for $\Gamma^{i, \mathrm{1PI}}_{h\mathcal{V}\mathcal{V'}}$ are presented in Appendix~\ref{app: hVV'_1PI}. For $(\mathcal{V}, \mathcal{V'}) = (g,g), (\gamma, \gamma)$, we have no counterterm at the one-loop level, i.e. $\delta \Gamma^{i}_{hgg} = \delta \Gamma^{i}_{h\gamma\gamma} = 0\ (i=1,2,3)$.
On the other hand, $\delta \Gamma^{i}_{hZ\gamma}$ are obtained as
\begin{align}
\delta \Gamma^{1}_{hZ\gamma}
=
g_{hZZ}\qty(\delta Z_{Z\gamma}-\frac{\delta s_{W}^{2}}{c_{W}s_{W}})\qc
\delta \Gamma^{2}_{hZ\gamma}=\delta \Gamma^{3}_{hZ\gamma}=0,
\end{align}
where the expressions for the counterterms are given in Ref.~\cite{Aoki:2012jj}.

%============================================================
\subsection{Decay rates of the SM-like Higgs boson}
\label{subsec: decay rates of h}
%============================================================

%------------------------------------------------------------
\subsubsection{\texorpdfstring{$h\to f\bar{f}$}{}}
%------------------------------------------------------------

The decay rate for $h\to f\bar{f}$ $(f \neq t)$ is given by
\begin{align}
\Gamma(h\to f\bar{f}) =
\Gamma_{\mathrm{LO}}(h\to f\bar{f})(1-\Delta r)
+\Gamma^{\mathrm{EW}}(h\to f\bar{f})
+\Gamma^{\mathrm{QCD}}(h\to f\bar{f}),
\label{eq: Gamma_hff_NLO}
\end{align} 
where the LO decay rate is given by
\begin{align}
\Gamma_{\mathrm{LO}}(h\to f\bar{f})
=
N_{c}^{f}\frac{m_{h}}{8\pi}\abs{\Gamma_{hf\bar{f}}^{S,\text{tree}}}^2\lambda^{3/2}\qty(\frac{m_{f}^{2}}{m_{h}^{2}}, \frac{m_{f}^{2}}{m_{h}^{2}}),
\label{eq:hff-lo}
\end{align}
with $N_{c}^{f} = 3\ (1)$ for $f$ being quarks (leptons).
The kinematical factor is defined by
\begin{align}
\lambda(x,y) = (1-x-y)^2-4xy. \label{eq: lambda_kinematical}
\end{align}
The correction $\Delta r$ appears when one replaces the electroweak VEV into the Fermi constant in $\Gamma_{\mathrm{LO}}(h\to f\bar{f})$ as discussed in Sec.~\ref{subsec: G_f input scheme}. 

The electroweak correction is given by~\cite{Kanemura:2019kjg}
\begin{align}
\Gamma^{\mathrm{EW}}(h\to f\bar{f})
&=
N_{c}^{f}\frac{m_{h}}{8\pi}\lambda^{3/2}\qty(\frac{m_{f}^{2}}{m_{h}^{2}}, \frac{m_{f}^{2}}{m_{h}^{2}})
\notag \\ &\quad
\times 2\Re
\qty{\Gamma_{hf\bar{f}}^{S,{\rm tree}}\qty[
\Gamma_{hf\bar{f}}^{S,{\rm loop}}
+2m_f\Gamma^{V_1,{\rm loop}}_{hf\bar{f}} 
+m_h^2\left(1-\frac{m_f^2}{m_h^2}\right)\Gamma_{hf\bar{f}}^{T,{\rm loop}}]^{*}}
\notag \\ &\quad
+\Gamma(h\to f\bar{f}\gamma),
\end{align}
where $\Gamma(h\to f\bar{f}\gamma)$ is the decay rate of the real-photon emission process and is obtained as
\begin{align}
\Gamma(h\to f\bar{f}\gamma)
=
N_{c}^{f}\frac{2\alpha_{\mathrm{em}}Q_{f}^{2}}{16\pi^{2}m_{h}}\abs{\Gamma_{hf\bar{f}}^{S,{\rm tree}}}^{2}\qty[
\Omega_{11}^{LL}+\Omega_{11}^{LR}
+\Omega_{22}^{LL}+\Omega_{22}^{LR}
+\Omega_{12}^{LL}+\Omega_{12}^{LR}],
\end{align}
where $\Omega_{ij}^{XY}$ $(X,Y = L,R;\, i,j=1,2)$ are presented in Ref.~\cite{Goodsell:2017pdq}.
We regulate infrared divergences in 1PI diagrams by introducing the photon mass. The photon mass dependence is canceled by the soft-photon emission part of $\Gamma(h\to f\bar{f}\gamma)$. We also add the hard-photon emission part so that $\Gamma^{\mathrm{EW}}(h\to f\bar{f})$ is independent of the detector cutoff.
For decays into quarks, we include the QCD correction $\Gamma^{\mathrm{QCD}}(h\to f\bar{f})$ up to NNLO in the $\overline{\rm MS}$ scheme~\cite{Braaten:1980yq, Gorishnii:1991zr, Chetyrkin:1995pd, Larin:1995sq}.
For their formulae, for instance, see Refs.~\cite{Kanemura:2019kjg, Aiko:2020ksl}.

%-----------------------------------------------------------------
\subsubsection{\texorpdfstring{$h \to ZZ^* \to Zf\bar{f}$}{}}
\label{sec:hzz}
%-----------------------------------------------------------------

We assign the momenta of the external particles as
\begin{align}
    h(p_{h}) \to Z(k_{Z})+f(k_{f})+\bar{f}(k_{\bar{f}}),
\end{align}
where $p_{i}\, (k_{i})$ is the incoming (outgoing) momentum.
In the following, we neglect the masses of final-state fermions.
Then, the Mandelstam variables are given by $s = (k_{\bar{f}}+k_{f})^{2},\, t = (k_{Z}+k_{\bar{f}})^{2}$, and $ u = (k_{Z}+k_{f})^{2}$ and satisfy $s+t+u=m_{h}^{2}+m_{Z}^{2}$.
Instead of $t$, we use the polar angle $\theta$ of $f$ in the rest frame of $Z^{*}$, and the $t$ parameter is written as
\begin{align}
t = \frac{1}{2}\qty[m_{h}^{2}+m_{Z}^{2}-s+m_{h}^{2}\,\lambda^{1/2}\qty(\frac{m_{Z}^{2}}{m_{h}^{2}}, \frac{s}{m_{h}^{2}})\cos{\theta}].
\label{eq: t param}
\end{align}

The decay rate for $h\to Z Z^{*}$ can be written as
\begin{align}
    \Gamma(h\to ZZ^{*}) = \sum_{f\neq t}\Gamma(h\to Z f\bar{f}),
\end{align}
where the partial decay rate for $h\to Z f\bar{f}$ is given by
\begin{align}
 \Gamma(h\to Z f\bar{f})
 =
 \Gamma_{\mathrm{LO}}(h\to Z f\bar{f})\qty(1-2\Delta r)
 +\Gamma^{\mathrm{EW}}(h\to Z f\bar{f})
 +\Gamma^{\mathrm{QCD}}(h\to Z f\bar{f}).
 \label{eq: Gamma_hZff_NLO}
\end{align}
The LO decay rate is given by
\begin{align}
\Gamma_{\mathrm{LO}}(h\to Z f\bar{f})
&=
N_{c}^{f}\qty(\frac{c_{hZZ}}{1+s_{\beta}^{2}})^{2}
\frac{G_{F}^{2}m_{h}m_{Z}^{4}}{24\pi^{3}}(v_{f}^{2}+a_{f}^{2})F\qty(\epsilon_{Z}),
\end{align}
where the kinematical function $F(x)$ is defined as~\cite{Keung:1984hn}
\begin{align}
 F(x)
 &=
 \frac{3(1-8x^{2}+20x^{4})}{\sqrt{4x^{2}-1}}\arccos{\qty(\frac{3x^{2}-1}{2x^{3}})}
 \notag \\ &\quad
 -(1-x^{2})\qty(\frac{47}{2}x^{2}-\frac{13}{2}+\frac{1}{x^{2}}-3(1-6x^{2}+x^{4})\ln{x})
\end{align}
with $\epsilon_{Z}=m_{Z}/m_{h}$.
We use $s_{W, G_{F}}$ for $v_{f} = I_{f}/2-Q_{f}s_{W}^{2}$ in the LO decay rate.

The electroweak correction is obtained as~\cite{Kanemura:2019kjg}
\begin{align}
 \Gamma^{\mathrm{EW}}(h\to Z f\bar{f})
 &=
 \int_{0}^{(m_{h}-m_{Z})^{2}}\dd s \frac{\dd \Gamma_{hZZ}}{\dd s}
 +
 \int_{0}^{(m_{h}-m_{Z})^{2}}\dd s \int_{-1}^{1}\dd \cos{\theta} \frac{\dd \Gamma_{hZZ}}{\dd s \dd \cos{\theta}}
 \notag \\ &\quad
 +\Gamma^{\mathrm{QED}}(h\to Z f\bar{f}),
\end{align}
where we have separated the weak correction and the QED correction. The weak correction is divided into two parts. The first term is obtained as
\begin{align}
\frac{\dd \Gamma_{hZZ}}{\dd s}
&=
N_{c}^{f}\frac{m_{h}^{3}}{96\pi^{3}}\frac{g_{Z}^{2}\Gamma_{hZZ}^{1, \mathrm{tree}}}{4m_{Z}^{2}}(v_{f}^{2}+a_{f}^{2})\frac{c_{a}^{\mathrm{kin}}}{(s-m_{Z}^{2})^{2}}
\Re\Bigg\{
%--- rdPiZZ ---
-\Gamma_{hZZ}^{1, \mathrm{tree}}\frac{\widehat{\Pi}_{ZZ}^{\prime}(m_{Z}^{2})}{2}
\notag \\ &\quad
%--- hZZ ---
+\left[
\Gamma_{hZZ}^{1, \mathrm{loop}}
+\frac{c_{b}^{\mathrm{kin}}}{c_{a}^{\mathrm{kin}}}\frac{\Gamma_{hZZ}^{2, \mathrm{loop}}}{m_{Z}^{2}}
\right](m_{Z}^{2}, s, m_{h}^{2})
%--- rPiZZ ---
-\Gamma_{hZZ}^{1, \mathrm{tree}}\frac{\widehat{\Pi}_{ZZ}(s)}{s-m_{Z}^{2}}
\notag \\ &\quad
%--- hZgam and rPiZG ---
+\frac{v_{f}Q_{f}c_{W}s_{W}}{v_{f}^{2}+a_{f}^{2}}\qty(
\frac{s-m_{Z}^{2}}{s}
\qty[
\widehat{\Gamma}_{hZ\gamma}^{1}
+\frac{c_{b}^{\mathrm{kin}}}{c_{a}^{\mathrm{kin}}}\widehat{\Gamma}_{hZ\gamma}^{2}
](m_{Z}^{2}, s, m_{h}^{2})
-\Gamma_{hZZ}^{1, \mathrm{tree}}\frac{\widehat{\Pi}_{Z\gamma}(s)}{s} 
)
\notag\\ &\quad
%--- Zff ---
+\Gamma_{hZZ}^{1, \mathrm{tree}}\frac{\qty[v_{f}\Gamma^{V, \mathrm{loop}}_{Zff} + a_{f}\Gamma_{Zff}^{A, \mathrm{loop}}](0,0,s)}{v_{f}^{2}+a_{f}^{2}}
\Bigg\},
\end{align}
where the term $\widehat{\Pi}_{ZZ}^{\prime}(m_{Z}^{2})=\dd\widehat{\Pi}_{ZZ}(p^{2})/\dd p^{2}|_{p^{2}=m_{Z}^{2}}$ is included since the residue of the renormalized $Z$-boson propagator is not unity~\cite{Bohm:1986rj, Hollik:1988ii}.
The kinematical factors are given by 
\begin{align}
c_{a}^{\mathrm{kin}} &= \sqrt{x_{1}^{2}-4x_{Z}}\qty[(x_{1}-6x_{Z})^{2}+8x_{Z}(1-3x_{Z})], \label{eq: c_a_kin} \\
c_{b}^{\mathrm{kin}} &= \frac{m_{h}^{2}}{2}(x_{1}^{2}-4x_{Z})^{3/2}(x_{1}-2x_{Z}), \label{eq: c_b_kin}
\end{align}
with $x_{Z}=m_{Z}^{2}/m_{h}^{2},\ x_{s}=s/m_{h}^{2}$ and $x_{1} = 1+x_{Z}-x_{s}$.
The second term involves the $hf\bar{f}$ vertex correction and the box contributions and is obtained as
\begin{align}
 \frac{\dd \Gamma_{hZZ}}{\dd s \dd \cos{\theta}}
 &=
 N_{c}^{f}\frac{1}{512\pi^{3}m_{h}}\sqrt{x_{1}^{2}-4x_{Z}}\frac{g_{Z}\Gamma_{hZZ}^{1, \mathrm{tree}}}{s-m_{Z}^{2}}\Re\qty[c_{1}^{\mathrm{kin}}{T}_{hf\bar{f}}^{Z}+{B}_{Z}],
\end{align}
with $c_{1}^{\mathrm{kin}} = 4s+2(tu-m_{h}^{2}m_{Z}^{2})/m_{Z}^{2}$.
The $hf\bar{f}$ vertex correction is expressed as
\begin{align}
 T_{hf\bar{f}}^{Z}
 &=
 \frac{2g_{Z}}{16\pi^{2}}\Bigg\{
 4g_{Z}^{3}m_{Z}c_{hZZ}(v_{f}^{4}+6v_{f}^{2}a_{f}^{2}+a_{f}^{4})
 \notag \\ &\qquad
 \times\Bigg[
 \qty(\frac{C_{0}+C_{11}}{-2})(t,0,m_{h}^{2}; Z,0,Z)
 +\qty(\frac{C_{12}}{2})(0,u,m_{h}^{2}; Z,0,Z)\Bigg]
 \notag \\ &\quad
 +g^{3}m_{W}c_{hWW}(v_{f}+a_{f})^{2}
 \notag \\ &\qquad
 \times\Bigg[
 \qty(\frac{C_{0}+C_{11}}{-2})(t,0,m_{h}^{2}; W,0,W)
 +\qty(\frac{C_{12}}{2})(0,u,m_{h}^{2}; W,0,W)\Bigg]
 \Bigg\}.
\end{align}
The box contribution can be written as
\begin{align}
B_{Z}
&=
\frac{1}{16\pi^{2}}\qty[
4(v_{f}+a_{f})\sum_{i\neq 6}C^{BZi}B^{BZi}
+8(v_{f}^{4}+6v_{f}^{2}a_{f}^{2}+a_{f}^{4})C^{BZ6}B^{BZ6}
],
\end{align}
where the coefficients $C^{BZi}$ and the box functions $B^{BZi}$ are presented in Appendix.~\ref{app: hToZZstar box diagrams}.

For the implementation of QED and QCD corrections, we follow Ref.~\cite{Kanemura:2019kjg}.
In the massless limit of the external fermions, they only appear in the $Zf\bar{f}$ vertex. Then, QED and QCD corrections are given by~\cite{Kniehl:1993ay}
\begin{align}
\Gamma^{\mathrm{QED}}(h\to Z f\bar{f})
&=
\qty(Q_{f}^{2}\frac{3\alpha_{\text{em}}}{4\pi})
\Gamma_{\mathrm{LO}}(h\to Z f\bar{f}), \\
\Gamma^{\mathrm{QCD}}(h\to Z f\bar{f})
&=
\qty(\frac{\alpha_{s}}{\pi})
\Gamma_{\mathrm{LO}}(h\to Z f\bar{f})
\qq{for} f=q.
\label{eq: hZZstar_qcd_qed}
\end{align}

%-----------------------------------------------------------------
\subsubsection{\texorpdfstring{$h\to WW^{*}\to Wf'\bar{f}$}{}}
\label{subsubsec: decay_rate_hToWWstar}
%-----------------------------------------------------------------

We assign the momenta of the external particles as
\begin{align}
    h(p_{h}) \to W^{-}(k_{W})+f'(k_{f'})+\bar{f}(k_{\bar{f}}),
\end{align}
where $p_{i}\, (k_{i})$ is the incoming (outgoing) momentum, and $f'\ (f)$ denotes a fermion with $I_{f}=+1/2\ (-1/2)$.
Similarly to the $h\to ZZ^{*}$ decay, we neglect the masses of the final-state fermions.
The Mandelstam variables are given by $s = (k_{\bar{f}}+k_{f'})^{2},\, t = (k_{W}+k_{\bar{f}})^{2}$, and $ u = (k_{W}+k_{f'})^{2}$ and satisfy $s+t+u=m_{h}^{2}+m_{W}^{2}$.
Instead of $t$, we use the polar angle $\theta$ of $f'$ in the rest frame of $W^{+*}$. The $t$ parameter is obtained by replacing $m_{Z}\to m_{W}$ in Eq.~\eqref{eq: t param}.

The decay rate for $h\to W W^{*}$ can be written as
\begin{align}
    \Gamma(h\to WW^{*}) = 2\sum_{f, f'}\Gamma(h\to W^{-} f'\bar{f}).
\end{align}
The factor of two accounts for the charge-conjugate process $h\to W^{+} f\bar{f'}$.
The partial decay rate for $h\to W^{-} f'\bar{f}$ is given by
\begin{align}
 \Gamma(h\to W^{-} f'\bar{f})
 &=
 \Gamma_{\mathrm{LO}}(h\to W^{-} f'\bar{f})\qty(1-2\Delta r)
 +\Gamma^{\mathrm{EW}}(h\to W^{-} f'\bar{f})
 \notag \\ &\quad 
 +\Gamma^{\mathrm{QCD}}(h\to W^{-} f'\bar{f}).
 \label{eq: Gamma_hWff'_NLO}
\end{align}
In the following, we neglect the effects of quark mixing and assume that the CKM matrix is the identity matrix.
Then, the LO decay rate is given by
\begin{align}
\Gamma_{\mathrm{LO}}(h\to W^{-} f' \bar{f})
&=
N_{c}^{f}c_{hWW}^{2}
\frac{G_{F}^{2}m_{h}m_{W}^{4}}{96\pi^{3}}F\qty(\epsilon_{W}),
\label{eq:h to WW* at LO}
\end{align}
with $\epsilon_{W}=m_{W}/m_{h}$.
We use $m_{W, G_{F}}$ for $m_{W}$ in the LO decay rate instead of $\overline{m}_{W}$ (See Sec.~\ref{subsec: G_f input scheme}).
We note that $\Delta r$ shifts the $W$-mass with ${\cal O}(100)~{\rm MeV}$~\cite{Aoki:2012jj}, and there are artificially large corrections if one uses $\overline{m}_{W}$ in Eq.~\eqref{eq:h to WW* at LO}.

The electroweak correction is given by~\cite{Kanemura:2019kjg}
\begin{align}
\Gamma^{\mathrm{EW}}(h\to W^{-} f'\bar{f})
&=
R_{f'\bar{f}}\Bigg[
\int_{0}^{(m_{h}-m_{W})^{2}}\dd s
\frac{\dd \Gamma_{hWW}}{\dd s}
+
\int_{0}^{(m_{h}-m_{W})^{2}}\dd s \int_{-1}^{1}\dd \cos{\theta}
\frac{\dd \Gamma_{hWW}}{\dd s \dd \cos{\theta}}
\notag \\ &\quad
+\Gamma({h\to W^{-} f' \bar{f} \gamma})
\Bigg],
\label{eq:NLOEW for h to WW*}
\end{align}
where we have multiplied $R_{f'\bar{f}}=\Gamma_{\mathrm{LO}}(h\to W^{-}f'\bar{f})|_{m_{W}=m_{W, G_{F}}}/\Gamma_{\mathrm{LO}}(h\to W^{-}f'\bar{f})|_{m_{W}=\overline{m}_{W}}$ to replace $\overline{m}_{W}$ in the kinematical factor with $m_{W, G_{F}}$.
We note that, different from $h \to Zf\bar{f}$, the weak and QED corrections cannot be separated for $h\to W^{-} f'\bar{f}$ since virtual photons appear together with virtual $W$ bosons in vertex corrections.
Therefore, we add the real-photon emission $\Gamma({h\to W^{-}f' \bar{f} \gamma})$ and cancel IR divergences~\cite{Kanemura:2019kjg}.
The first term on the right-hand side of Eq.~\eqref{eq:NLOEW for h to WW*} is obtained as
\begin{align}
\frac{\dd \Gamma_{hWW}}{\dd s}
&=
N_{c}^{f}\frac{m_{h}^{3}}{384\pi^{3}}\frac{g^{2}\Gamma_{hWW}^{1, \mathrm{tree}}}{4m_{W}^{2}}\frac{c_{a}^{\mathrm{kin}}}{(s-m_{W}^{2})^{2}}
\Re\Bigg\{
%--- rdPiWW ---
-\Gamma_{hWW}^{1, \mathrm{tree}}\frac{\widehat{\Pi}_{WW}^{\prime}(m_{W}^{2})}{2}
\notag \\  &\quad
%--- hWW ---
+\left[\Gamma_{hWW}^{1, \mathrm{loop}}
+\frac{c_{b}^{\mathrm{kin}}}{c_{a}^{\mathrm{kin}}}\frac{\Gamma_{hWW}^{2, \mathrm{loop}}}{m_{W}^{2}}\right](m_{W}^{2}, s, m_{h}^{2})
%--- rPiWW ---
-\Gamma_{hWW}^{1, \mathrm{tree}}\frac{\widehat{\Pi}_{WW}(s)}{s-m_{W}^{2}}
\notag \\  &\quad
%--- Wff ---
+\Gamma_{hWW}^{1, \mathrm{tree}}\qty[\Gamma^{V, \mathrm{loop}}_{Wff}+\Gamma^{A, \mathrm{loop}}_{Wff}](0,0,s) 
\Bigg\},
\label{eq:del_ew_w}
\end{align}
where the term $\widehat{\Pi}_{WW}^{\prime}(m_{W}^{2})=\dd\widehat{\Pi}_{WW}(p^{2})/\dd p^{2}|_{p^{2}=m_{W}^{2}}$ is included since the residue of the renormalized $W$-boson propagator is not unity~\cite{Bohm:1986rj, Hollik:1988ii}.
The expressions for the loop-corrected $Wf\bar{f'}$ vertices, $\Gamma^{V/A, \mathrm{loop}}_{Wff}(0,0,p^{2})$, are the same as those in the SM in the massless limit of the external fermions, and their explicit formulae can be found, for instance, in Ref.~\cite{Kanemura:2019kjg}.
The second term is given by
\begin{align}
\frac{\dd \Gamma_{hWW}}{\dd s \dd \cos{\theta}}
=
N_{c}^{f}\frac{1}{512\pi^{3}m_{h}}\sqrt{x_{1}^{2}-4x_{W}}\frac{g_{W}^{}\Gamma_{hWW}^{1, \mathrm{tree}}}{s-m_{W}^{2}}\Re\qty[c_{1}^{\mathrm{kin}}T_{hf\bar{f}}^{W}+B_{W}],
\end{align}
where $g_{W}^{}=g/\sqrt{2}$.
The kinematical factors $c_{a}^{\mathrm{kin}}$, $c_{b}^{\mathrm{kin}}$, $x_{1}$, and $c_{1}^{\mathrm{kin}}$ are obtained from those for $h\to ZZ^{*}$ by the replacements $m_{Z} \to m_{W}$ and $x_{Z} \to x_{W}$.
The $hf\bar{f}$ contribution is expressed as
\begin{align}
T_{hf\bar{f}}^{W}
&=
\frac{g_{W}^{}}{16\pi^{2}}\Bigg\{
4g_{Z}^{3}m_{Z}c_{hZZ}\Bigg[
(v_{f'}+a_{f'})^{2}\qty(\frac{C_{0}+C_{11}}{-2})(t,0,m_{h}^{2}; Z,0,Z)
\notag \\ &\qquad
+(v_{f}+a_{f})^{2}\qty(\frac{C_{12}}{2})(0,u,m_{h}^{2}; Z,0,Z)\Bigg]
\notag \\ &
+2g^{3}m_{W}c_{hWW}\Bigg[
\qty(\frac{C_{0}+C_{11}}{-2})(t,0,m_{h}^{2}; W,0,W)
+\qty(\frac{C_{12}}{2})(0,u,m_{h}^{2}; W,0,W)\Bigg]\Bigg\}.
\end{align}
The box contribution can be written as
\begin{align}
B_{W}
&=
\frac{4}{16\pi^{2}}\sum C^{BWi}B^{BWi},
\end{align}
where the coefficients $C^{BWi}$ and the box functions $B^{BWi}$ are given in Appendix.~\ref{app: hToZZstar box diagrams}.

In the massless limit of the external fermions, the QCD correction only appears in the $Wf\bar{f}'$ vertex and is given by~\cite{Kniehl:1993ay}
\begin{align}
\Gamma^{\mathrm{QCD}}(h\to W^{-} f'\bar{f})
=
\qty(\frac{\alpha_{s}}{\pi})
\Gamma_{\mathrm{LO}}(h\to W^{-} f'\bar{f})
\qq{for} f=q.
\label{eq: hWWstar_qcd}
\end{align}

%-----------------------------------------------------------------
\subsubsection{\texorpdfstring{$h \to gg,\, \gamma\gamma,\, Z\gamma$}{}}
%-----------------------------------------------------------------

The LO decay rates for the loop-induced processes $h \to gg,\, \gamma\gamma$ and $Z\gamma$ can be expressed as~\cite{Kanemura:2019kjg}
\begin{align}
\Gamma_{\mathrm{LO}}(h \to\mathcal{V} \mathcal{V'})
=
\frac{1}{8\pi(1+\delta_{\mathcal{V}\mathcal{V'}}) m_{h}}\abs{\widehat{\Gamma}_{h\mathcal{V}\mathcal{V'}}^{1}(m_{\mathcal{V}}^{2}, m_{\mathcal{V'}}^{2}, m_{h}^{2})}^{2}\lambda^{1/2}\qty(\frac{m_{\mathcal{V}}^{2}}{m_{h}^{2}}, \frac{m_{\mathcal{V'}}^{2}}{m_{h}^{2}}).
\end{align}
The quark-loop contributions receive QCD corrections. For $h\to gg$ and $\gamma\gamma$, we include them up to NNLO following Refs.~\cite{Dawson:1993qf, Spira:1995rr, Chetyrkin:1997iv}.
For $h\to Z\gamma$, we replace a quark mass in the Yukawa coupling with a running mass.
For their explicit formulae, see Ref.~\cite{Aiko:2020ksl} for instance.

The LO decay rate for $h\to \gamma\gamma$ can be simplified as follows~\cite{Djouadi:2005gi,Djouadi:2005gj,Chun:2012jw,BhupalDev:2013xol,Aoki:2012jj}.
\begin{align}
\Gamma_{\mathrm{LO}}(h \to\gamma \gamma)
&=
\frac{\sqrt{2}G_{F}\alpha_{\mathrm{em}}^{2}m_{h}^{3}}{256\pi^{3}}
% \notag \\ &\quad \times
\Bigg|
\zeta_{h}\sum_{f}N_{c}^{f}Q_{f}^{2}I_{F}^{h}(f)
+c_{hWW}I_{V}^{h}(W)
\notag \\ &\quad
-\frac{\lambda_{H^{+}H^{-}h}}{v}I_{S}^{h}(H^{\pm})
-\frac{4\lambda_{H^{++}H^{--}h}}{v}I_{S}^{h}(H^{\pm\pm})
\Bigg|^{2},
\label{eq: decay_rate_hgamgam}
\end{align}
where the loop functions are defined as~\cite{Kanemura:2015mxa}
\begin{align}
I_{F}^{h}(f)
&=
-\frac{8m_{f}^{2}}{m_{h}^{2}}
\qty[1+\qty(2m_{f}^{2}-\frac{m_{h}^{2}}{2})C_0(0, 0, h; f, f, f)],
\label{eq: Floop_hgamgam} \\
I_{V}^{h}(W)
&=
\frac{2m_{W}^{2}}{m_{h}^{2}}
\qty[6+\frac{m_{h}^{2}}{m_{W}^{2}}+(12m_{W}^{2}-6m_{h}^{2})C_0(0, 0, h; W, W, W)],
\label{eq: Wloop_hgamgam}\\
I_{S}^{h}(H^{\pm})
&=
\frac{2v^{2}}{m_{h}^{2}}
\qty[1+2m_{H^{\pm}}^{2} C_0(0, 0, h; H^{\pm}, H^{\pm}, H^{\pm})].
\label{eq: Sloop_hgamgam}
\end{align}
The $H^{\pm\pm}$ contribution can be obtained by replacing $H^{\pm}$ with $H^{\pm\pm}$.

Similarly to $h\to \gamma\gamma$, we write the decay rate for $h\to Z\gamma$ as~\cite{Djouadi:2005gi,Djouadi:2005gj,Arbabifar:2012bd,BhupalDev:2013xol,Chen:2013dh}
\begin{align}
\Gamma_{\mathrm{LO}}(h \to Z\gamma)
&=
\frac{\sqrt{2}G_{F}\alpha_{\mathrm{em}}^{2}m_{h}^{3}}{128\pi^{3}}
\qty(1-\frac{m_{Z}^{2}}{m_{h}^{2}})^{3}
% \notag \\ &\quad \times
\abs{
\zeta_{h}\sum_{f}N_{c}^{f}Q_{f}v_{f}J_{F}^{h}(f)
+J_{B}^{h}
}^{2},
\end{align}
where the fermionic loop function is given by~\cite{Kanemura:2015mxa}
\begin{align}
J_{F}^{h}(f)
&=
-\frac{4m_{f}^{2}}{s_{W}c_{W}(m_{h}^{2}-m_{Z}^{2})}
\bigg[2+(4m_{f}^{2}-m_{h}^{2}+m_{Z}^{2})C_{0}(Z, 0, h; f, f, f)\notag\\
&\quad
+\frac{2m_{Z}^{2}}{m_{h}^{2}-m_{Z}^{2}}\qty[B_{0}(h; f, f)-B_{0}(Z; f, f)]\bigg],
\label{eq: Floop_hZgam}
\end{align}
and we have defined the bosonic loop function as follows.
\begin{align}
J_{B}^{h}
=
\frac{16\pi^{2}v}{e^{2}(m_{h}^{2}-m_{Z}^{2})}
\widehat{\Gamma}_{hZ\gamma}^{1}(m_{Z}^2, 0, m_{h}^{2})_{B}.
\label{eq: J_B^h}
\end{align}
When $\beta'=0$, $J_{B}^{h}$ is consistent with that in Ref.~\cite{Chen:2013dh}.

%=================================================================
\subsection{Theoretical behaviors of radiative corrections}
%=================================================================

We here study the behavior of one-loop contributions, especially focusing on the decoupling of the additional Higgs bosons.
To study the decoupling behavior, we assume that all the additional Higgs bosons are degenerate in their mass: $m_{\Phi}\equiv m_{A}=m_{H}=m_{H^{\pm}}=m_{H^{\pm\pm}}$.
We take $\mu$ as an input parameter instead of $\overline{v}_{\Delta}$ and take $\overline{v}_{\Delta} = \mu \overline{v}^{2}/(\sqrt{2}m_{A}^{2})$ according to the decoupling relation (see the discussion of the decoupling limit in Sec.~\ref{subsec: th_const}).
When $m_{A}\gg \mu$, $\overline{v}_{\Delta}$ is negligible and the Higgs quartic couplings are obtained as
\begin{align}
    \lambda_{1} = \frac{m_{h}^{2}}{2v^{2}}\qc
    \lambda_{2} = -\frac{v^{2}}{2(m_{\Phi}^{2}-m_{h}^{2})}\left(\lambda_{4}^{2}-\frac{4m_{\Phi}^{2}}{v^{2}}\lambda_{4}+\frac{4m_{\Phi}^{2}m_{h}^{2}}{v^{4}}\right)\qc
    \lambda_{3} = \lambda_{5} = 0.
    \label{eq: lambdas small vtri limit}
\end{align}

%-----------------------------
\begin{figure}[t]
 \centering
 \includegraphics[scale=0.5]{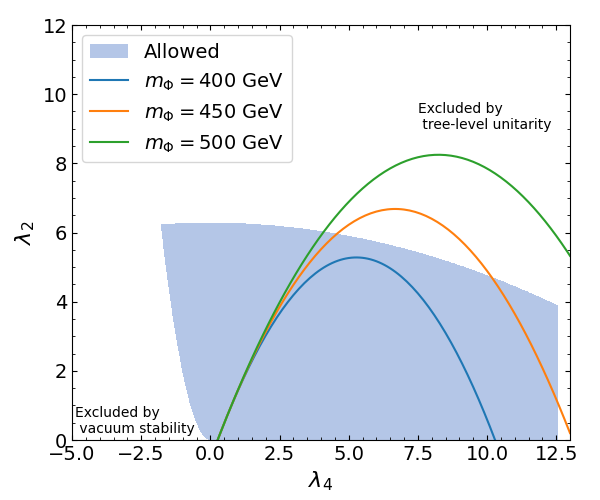}
 \caption{Allowed region in the $\lambda_{4}\text{--}\lambda_{2}$ plane for the decoupling scenario under the vacuum stability and the tree-level unitarity conditions as well as the perturbativity criterion.}
 \label{fig: decoupling_scenario}
\end{figure}
%-----------------------------

In Fig.~\ref{fig: decoupling_scenario}, we show the allowed region in the
$\lambda_{4}$--$\lambda_{2}$ plane with $\lambda_{1}=m_{h}^{2}/(2v^{2})$ and $\lambda_{3}=\lambda_{5}=0$ under the vacuum stability and the tree-level unitarity conditions as well as the perturbativity criterion.
The blue, orange and green curves correspond to $\lambda_{2}$ given in Eq.~\eqref{eq: lambdas small vtri limit} for $m_{\Phi}=400,\, 450$, and $500~\mathrm{GeV}$, respectively.
From the vacuum stability conditions in Eq.~\eqref{eq: vacuum stability}, $\lambda_{2}$ must be positive for $\lambda_{3}=0$.
In addition, for $m_{\Phi}>m_{h}$, $\lambda_{2}$ is a convex function of $\lambda_{4}$.
For $m_{\Phi}\gg m_{h}$, the solutions of $\lambda_{2}=0$ are given by
\begin{align}
\lambda_{4}^{-} = \frac{m_{h}^{2}}{v^{2}}\qc
\lambda_{4}^{+} = \frac{4m_{\Phi}^{2}}{v^{2}}\qty(1-\frac{m_{h}^{2}}{4m_{\Phi}^{2}}).
\end{align}
Therefore, since $\lambda_{2}>0$ and $\lambda_{4}^{\pm}>0$, $\lambda_{4}$ must be positive in the decoupling limit.

The maximal value of $\lambda_{4}$ depends on $m_{\Phi}$, and there are three distinct cases.
The first case corresponds to small $m_{\Phi}$, which is represented by the blue line with $m_{\Phi}=400$~GeV in Fig.~\ref{fig: decoupling_scenario}. In this case, the whole parabola lies in the allowed region, and $\lambda_{4}$ takes its maximal value at $\lambda_{4}^{\mathrm{max}} \simeq \lambda_{4}^{+}$ with $\lambda_{2}\simeq 0$, while $\lambda_{2}$ becomes maximal at the top of the parabola with $\lambda_{2}^{\mathrm{max}}=2m_{\Phi}^{2}/v^{2}$.
The second case corresponds to an intermediate value of $m_{\Phi}$, e.g. $m_{\Phi}=450$~GeV, where a part of the parabola is outside the allowed region as shown by the orange line, and the intermediate range of $\lambda_{4}$ is not allowed.
In this case, $\lambda_{4}^{\mathrm{max}} = 4\pi < \lambda_{4}^{+}$ due to the perturbativity criterion. We note that $\lambda_{4}\lesssim 20$ is allowed only by perturbative unitarity with $\zeta_{\mathrm{LQT}}=1$.
The third case corresponds to large $m_{\Phi}$. In this case,
as shown by the green line with $m_{\Phi}=500$~GeV, there is no solution on the right-hand side of the parabola, and $\lambda_{4}$ and $\lambda_{2}$ take their maximal values at the same time.

%-----------------------------
\begin{figure}[t]
 \centering
 \includegraphics[scale=0.55]{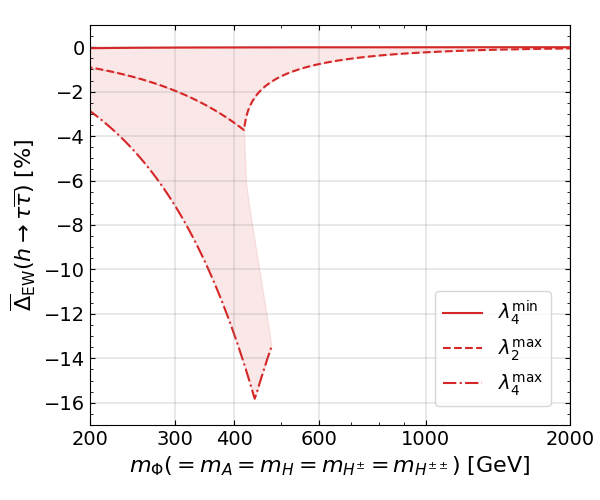}
 \caption{Decoupling behavior in the NLO EW corrections to the $h\to \tau\bar{\tau}$ decay.}
 \label{fig: barDelta_EW}
\end{figure}
%-----------------------------

We define $\overline{\Delta}_{\mathrm{EW}}(h\to XY)$ as~\cite{Kanemura:2019kjg}
\begin{align}
    \overline{\Delta}_{\mathrm{EW}}(h\to XY)
    =
     \left. \frac{\Gamma(h\to XY)}{\Gamma_{\mathrm{LO}}(h\to XY)} \right|_{\mathrm{CHTM}}
    -\left. \frac{\Gamma(h\to XY)}{\Gamma_{\mathrm{LO}}(h\to XY)} \right|_{\mathrm{SM}},
\end{align}
where $\Gamma(h\to XY)$ and $\Gamma_{\mathrm{LO}}(h\to XY)$ are given in Sec.~\ref{subsec: decay rates of h}.
We note that $\overline{\Delta}_{\mathrm{EW}}(h\to XY)$ represents scalar mixing and additional scalar loop contributions. Therefore, it vanishes in the decoupling limit.

In Fig.~\ref{fig: barDelta_EW}, we show the allowed range of $\overline{\Delta}_{\mathrm{EW}}(h\to\tau\overline{\tau})$ as a function of $m_{\Phi}$.
We take $\mu=1~\mathrm{GeV}$ and scan $\lambda_{4}$ under the vacuum stability, tree-level unitarity and perturbativity constraints.
The solid line corresponds to the prediction with $\lambda_{4}^{\mathrm{min}}\sim 0$.
In this case, all Higgs quartic couplings take small values, so that $\overline{\Delta}_{\mathrm{EW}}(h\to\tau\overline{\tau})$ is almost zero.
The dashed and dot-dashed lines correspond to the predictions with $\lambda_{2}^{\mathrm{max}}$ and $\lambda_{4}^{\mathrm{max}}$, respectively.
For $m_{\Phi} \lesssim 450~\mathrm{GeV}$, $\lambda_{2}$ becomes maximal at the top of the parabola in Fig.~\ref{fig: decoupling_scenario}, $\lambda_{2}^{\mathrm{max}} = 2m_{\Phi}^{2}/v^{2}$, while $\lambda_{4}^{\mathrm{max}}\simeq \lambda_{4}^{+}$ with $\lambda_{2}\simeq 0$.
Since both increase as $m_{\Phi}$ becomes larger, the magnitude of $\overline{\Delta}_{\mathrm{EW}}(h\to\tau\overline{\tau})$ also increases.
For $450~\mathrm{GeV} \lesssim m_{\Phi} \lesssim 470~\mathrm{GeV}$, there exist two distinct ranges of $\overline{\Delta}_{\mathrm{EW}}(h\to\tau\overline{\tau})$.
This corresponds to the second case discussed in Fig.~\ref{fig: decoupling_scenario}, where a part of the parabola lies outside the allowed region.
For example, for $m_{\Phi}=450~\mathrm{GeV}$, $-2\%\lesssim \overline{\Delta}_{\mathrm{EW}}(h\to\tau\overline{\tau}) \le 0\%$ and $-15\%\lesssim \overline{\Delta}_{\mathrm{EW}}(h\to\tau\overline{\tau}) \lesssim -11.5\%$ are allowed.
For $m_{\Phi} \gtrsim 470~\mathrm{GeV}$, $\overline{\Delta}_{\mathrm{EW}}(h\to\tau\overline{\tau})$ exhibits decoupling behavior as $m_{\Phi}$ increases.

The results for the other decay modes are almost identical except for $h\to WW^{*}$.
Since $\overline{\Delta}_{\mathrm{EW}}(h\to WW^{*})$ is sensitive to corrections to $m_{W}$ as discussed in Sec.~\ref{subsubsec: decay_rate_hToWWstar}, it exhibits nontrivial behavior for small $m_{\Phi}$, but shows decoupling behavior as $m_{\Phi}$ increases.

We briefly discuss the case with $\overline{v}_{\Delta}\neq 0$.
For $\overline{v}_{\Delta}\ll \overline{v}$, we have $\mu \simeq \sqrt{2}m_{\Phi}^{2}\overline{v}_{\Delta}/\overline{v}^{2}$, and $\mu$ becomes large as $m_{\Phi}$ increases. As a result, the Higgs quartic couplings become large, and the new physics contributions do not decouple. Since the size of the Higgs quartic couplings is constrained by the theoretical requirements, there is an upper bound on $m_{\Phi}$ for $\overline{v}_{\Delta}\neq 0$.
For example, $m_{\Phi}$ should be smaller than 7.5~TeV for $\overline{v}_{\Delta}=10~\mathrm{GeV}$.

%%%%%%%%%%%%%%%%%%%%%%%%%%%%%%%%%%%%%%%%%%%%%%%%%%%%%%%%%%%%%%%%%%
\section{Deviations from the SM predictions in the Higgs boson decays}
\label{sec: Deviations from the SM in the Higgs boson decays}
%%%%%%%%%%%%%%%%%%%%%%%%%%%%%%%%%%%%%%%%%%%%%%%%%%%%%%%%%%%%%%%%%%

%=================================================================
\subsection{Constraints on the parameter space}
\label{subsec: Constraints}
%=================================================================

From the tree-level analysis of $\rho_{0}$ in Sec.~\ref{sec: model}, we find that $v_{\Delta} \ll v$.
Then, the following mass relations are approximately satisfied.
\begin{align}\label{eq:mass spectrum}
    m_{A}^{2} = m_{H}^{2}\qc
    m_{H^{\pm}}^{2}-m_{A}^{2} = m_{H^{\pm\pm}}^{2}-m_{H^{\pm}}^{2}.
\end{align}
There are three scenarios depending on the value of $\xi=m_{H^{\pm}}^{2}-m_{A}^{2}$.
The first is the degenerate scenario with $\xi=0$, where all of the additional Higgs bosons are degenerate in their mass.
The second is the lightest $H^{\pm\pm}$ scenario (LS) with $\xi<0$ and $m_{H^{\pm\pm}} < m_{H^{\pm}}<m_{A}$.
The third is the heaviest $H^{\pm\pm}$ scenario (HS) with $\xi>0$ and $m_{A}<m_{H^{\pm}}<m_{H^{\pm\pm}}$.

Constraints from direct searches for the additional Higgs bosons have been studied in Ref.~\cite{Ashanujjaman:2021txz}.
In the degenerate scenario, $m_{H^{\pm\pm}} < 950~\mathrm{GeV}$ is excluded for $v_{\Delta} < \mathcal{O}(10^{-4})~\mathrm{GeV}$ by multi-lepton searches.
On the other hand, for $v_{\Delta} > \mathcal{O}(10^{-4})~\mathrm{GeV}$, multi-boson searches exclude $m_{H^{\pm\pm}} < 400~\mathrm{GeV}$.
In the HS, $m_{H^{\pm\pm}} \lesssim 900~\mathrm{GeV}$ is excluded for $v_{\Delta} < \mathcal{O}(10^{-6})~\mathrm{GeV}$ and $m_{H^{\pm\pm}}-m_{H^{\pm}}=10~\mathrm{GeV}$, similarly to the degenerate scenario.
By contrast, for $v_{\Delta} > \mathcal{O}(10^{-6})~\mathrm{GeV}$, the constraints weaken, and $m_{H^{\pm\pm}} \lesssim 400~\mathrm{GeV}$ is excluded for $v_{\Delta} > \mathcal{O}(10^{-1})~\mathrm{GeV}$ with $m_{H^{\pm\pm}}-m_{H^{\pm}}=10~\mathrm{GeV}$.
In the LS, $m_{H^{\pm\pm}} \lesssim 1000~\mathrm{GeV}$ is excluded for $v_{\Delta} < \mathcal{O}(10^{-4})~\mathrm{GeV}$ by multi-lepton searches, while $m_{H^{\pm\pm}} \lesssim 600~\mathrm{GeV}$ is excluded for $v_{\Delta} > \mathcal{O}(10^{-4})~\mathrm{GeV}$.

The CHTM parameters, especially $v_{\Delta}$, are constrained by the electroweak precision observables~\cite{Kanemura:2012rs,Aoki:2012jj,Kanemura:2022ahw}.
We calculate $m_{W, G_{F}},\, \rho,\, s_{e}^{2}$ and $\Gamma(Z\to \ell\overline{\ell})$ at the one-loop level and require that the differences between the CHTM and the SM predictions are within the experimental uncertainties.
The $2\sigma$ errors of these quantities are given by\footnote{We take the value of $|\rho-1|$ from a result corresponding to a three-parameter fit including the $T$ parameter.}~\cite{ParticleDataGroup:2024cfk}
\begin{align}
    |\Delta m_{W}| \leq 2.66~\mathrm{MeV}\qc
    |\rho-1| &\leq 1.8\times 10^{-3}\qc
    |\Delta s_{e}^{2}| \leq 6.6\times 10^{-4},
    \notag \\
    |\Delta \Gamma(Z\to \ell\overline{\ell})| &\leq 0.17~\mathrm{MeV}.
\end{align}

In addition, measurements of the signal strength of $h\to\gamma\gamma$ constrain the CHTM parameters. For $\mu_{\gamma\gamma}^{\mathrm{CHTM}} = [\sigma(gg\to h)\mathrm{Br}(h\to\gamma\gamma)]_{\mathrm{CHTM}}/[\sigma(gg\to h)\mathrm{Br}(h\to\gamma\gamma)]_{\mathrm{SM}}$, we define $\Delta \mu_{\gamma\gamma}$ as
\begin{align}
    \Delta \mu_{\gamma\gamma} = \mu_{\gamma\gamma}^{\mathrm{CHTM}}-1
    = \zeta_{h}^{2}\frac{\mathrm{Br}(h\to\gamma\gamma)_{\mathrm{CHTM}}}{\mathrm{Br}(h\to\gamma\gamma)_{\mathrm{SM}}}-1,
\end{align}
where we have used $\sigma(gg\to h)_{\mathrm{CHTM}}=\zeta_{h}^{2}\sigma(gg\to h)_{\mathrm{SM}}$ at LO.
The current bound is $\abs{\Delta\mu_{\gamma\gamma}}<0.18$ at $2\sigma$, which is taken in light of the experimental uncertainties, rather than the central values, reported by ATLAS~\cite{ATLAS:2022tnm} and CMS~\cite{CMS:2021kom}.

%-----------------------------
\begin{figure}[t]
 \centering
 \includegraphics[scale=0.4]{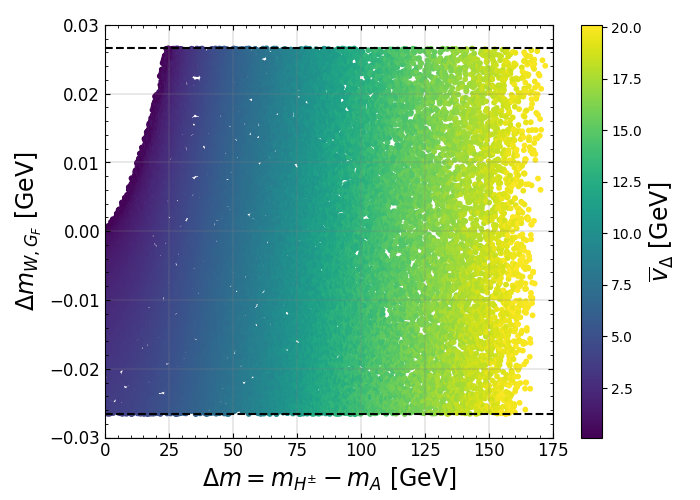}
 \includegraphics[scale=0.4]{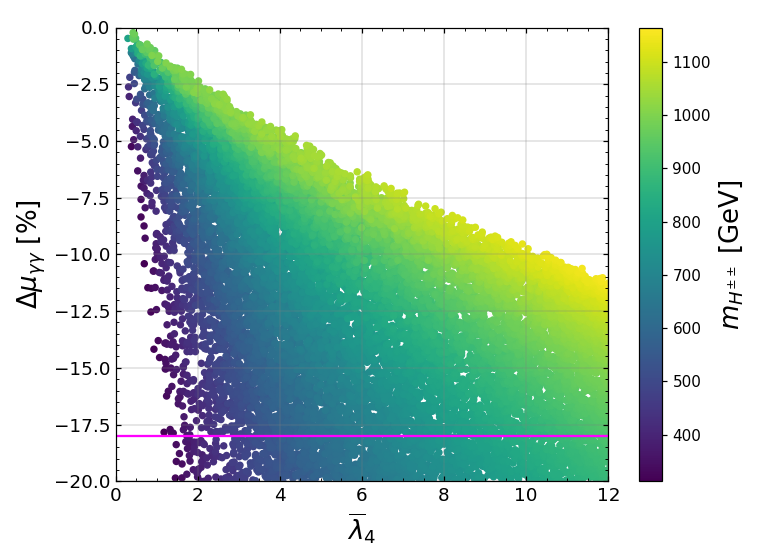} \\
 \includegraphics[scale=0.4]{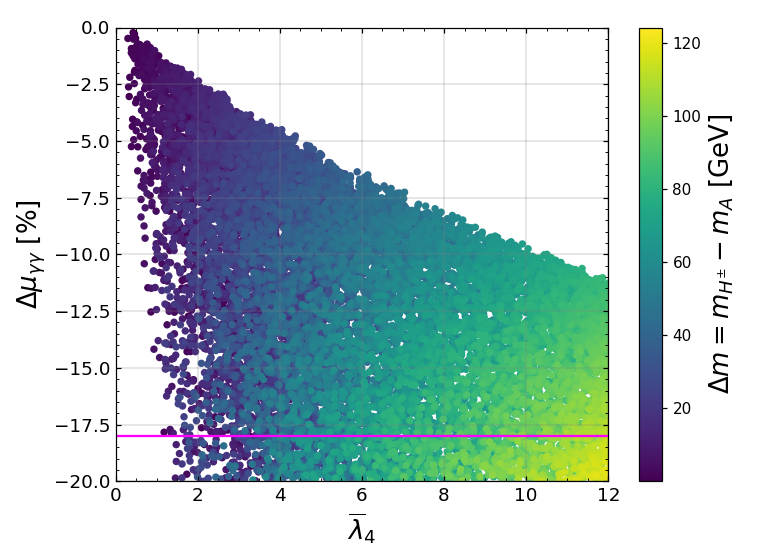}
 \includegraphics[scale=0.4]{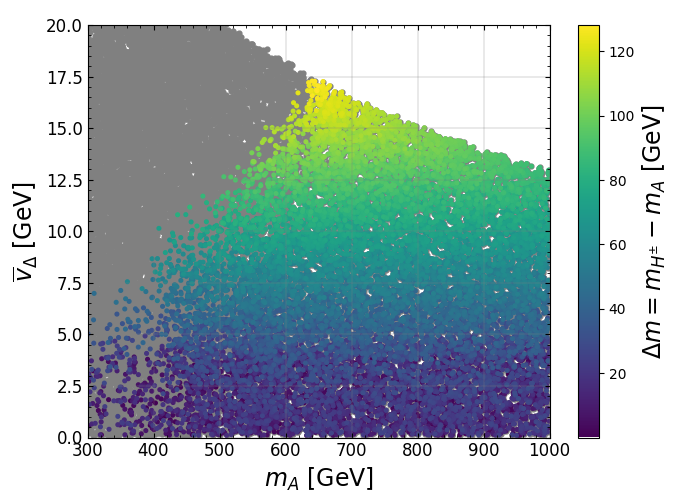}
 \caption{Predictions for $\Delta m_{W, G_{F}}$ (upper left) and $\Delta \mu_{\gamma\gamma}$ (upper right and lower left), together with the allowed parameter points (lower right) in the HS.}
 \label{fig: param_scan_mWGF_mugamgam}
\end{figure}
%-----------------------------

We first discuss the allowed parameter region in the HS.
In the upper left panel of Fig.~\ref{fig: param_scan_mWGF_mugamgam}, we show the prediction for $\Delta m_{W, G_{F}}$ as a function of $\Delta m = m_{H^{\pm}}-m_{A}$.
The color indicates the value of $\overline{v}_{\Delta}$.
We scan the ranges $300~\mathrm{GeV} \leq m_{A} \leq 1000~\mathrm{GeV}$ and $-5 \leq \overline{\lambda}_{4} \leq 4\pi$.
At the tree level, $m_{W, G_{F}}$ becomes smaller than the SM value as $\overline{v}_{\Delta}$ increases.
On the other hand, one-loop effects induced by the mass-squared difference $\xi$ increase the value of $m_{W}$. Thus, a larger $\Delta m$ is preferred for a larger $\overline{v}_{\Delta}$.

In the upper right (lower left) panel of Fig.~\ref{fig: param_scan_mWGF_mugamgam}, we show the prediction for $\Delta \mu_{\gamma\gamma}$ at LO as a function of $\overline{\lambda}_{4}$, where the color indicates the value of $m_{H^{\pm\pm}}\ (\Delta m)$ and the solid line represents the current bound.
In the HS, the $H^{\pm\pm}$ contribution (the fourth term in the right-hand side of Eq.~\eqref{eq: decay_rate_hgamgam}), which is proportional to $\lambda_{hH^{++}H^{--}}\simeq -\lambda_{4}v$, decreases $\mu_{\gamma\gamma}$ as $\overline{\lambda}_{4}$ increases. This loop effect is suppressed by $1/m_{H^{\pm\pm}}^{2}$, and thus a heavier $H^{\pm\pm}$ is preferred for larger $\overline{\lambda}_{4}$.
On the other hand, a large $\Delta m$ requires a sizable $\lambda_{4}$. This is because a large $\Delta m$ leads to a large negative value of $\lambda_{5}$ $(\simeq -4(m_{H^{\pm}}^{2}-m_{A}^{2})/v^{2}$).
From vacuum stability, we require $\lambda_{4}+\lambda_{5}\gtrsim 0$, so a larger $\overline{\lambda}_{4}$ is favored for more negative $\lambda_{5}$.

In the lower right panel of Fig.~\ref{fig: param_scan_mWGF_mugamgam}, we show the allowed parameter points in the HS under the constraints from the electroweak precision observables and $\Delta \mu_{\gamma\gamma}$.
The gray points are consistent with the electroweak precision observables but are excluded by $\Delta \mu_{\gamma\gamma}$, while the colored points satisfy both constraints.
If $m_{A} \lesssim 650~\mathrm{GeV}$, the points with large $\overline{v}_{\Delta}$ are excluded by $\Delta \mu_{\gamma\gamma}$.
As a result of the combined constraints, an upper bound on $\overline{v}_\Delta$ is obtained as $v_\Delta \lesssim 17.5~\mathrm{GeV}$ in the HS.

%-----------------------------
\begin{figure}[t]
 \centering
 \includegraphics[scale=0.4]{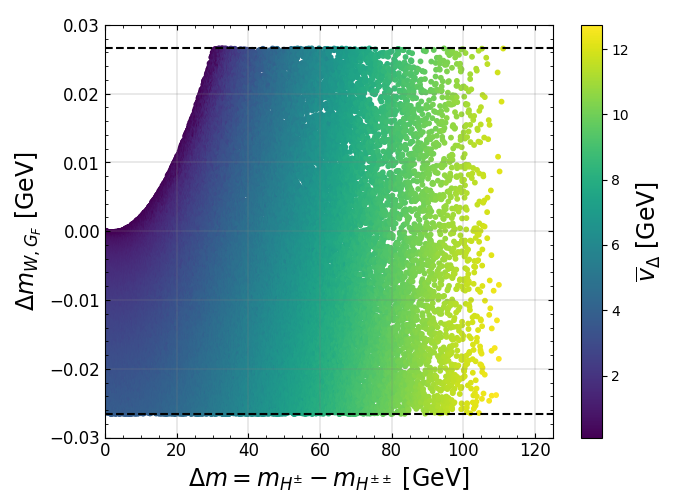}
 \includegraphics[scale=0.4]{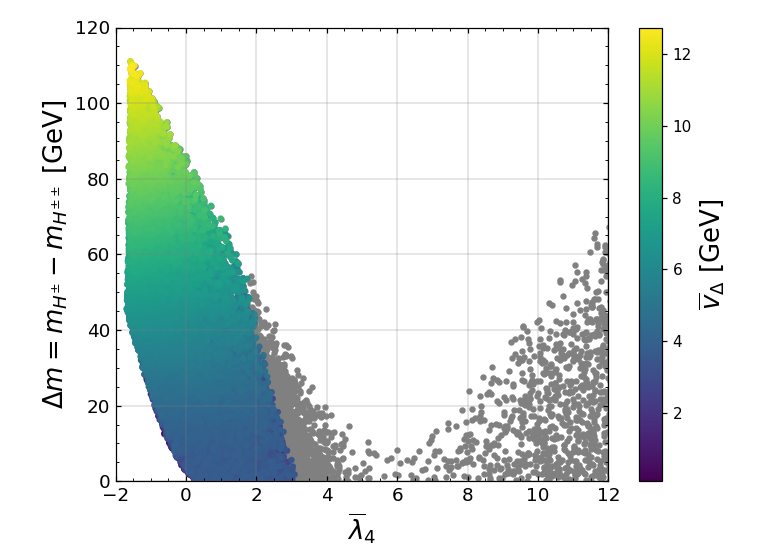}
 \includegraphics[scale=0.4]{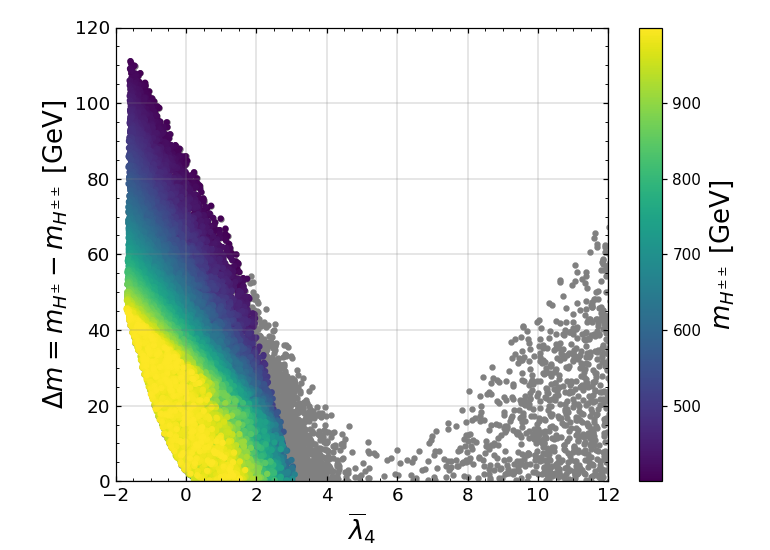}
 \includegraphics[scale=0.4]{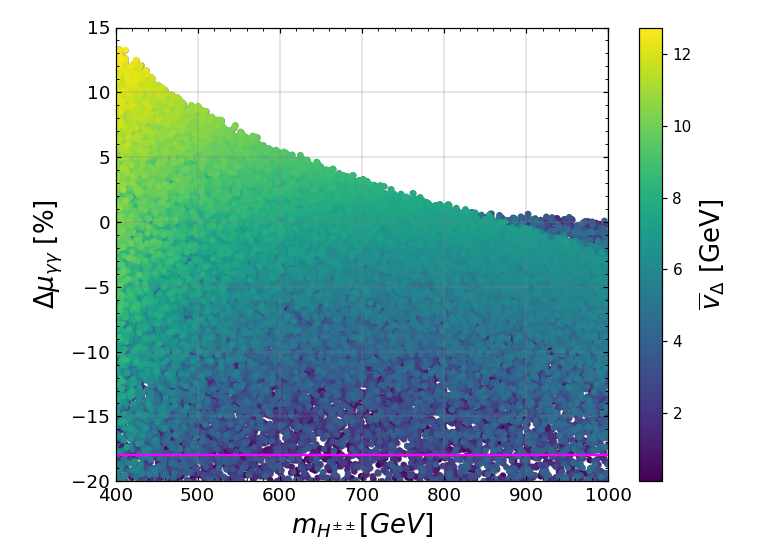}
 \caption{Predictions for $\Delta m_{W, G_{F}}$ (upper left) and $\Delta \mu_{\gamma\gamma}$ (upper right and lower left), together with the allowed parameter points (lower right) in the LS.}
 \label{fig: param_scan_mWGF_mugamgam_LS}
\end{figure}
%-----------------------------

Next, we discuss the allowed parameter region in the LS.
In the upper left panel of Fig.~\ref{fig: param_scan_mWGF_mugamgam_LS}, we show the prediction for $\Delta m_{W, G_{F}}$ as a function of $\Delta m = m_{H^{\pm}}-m_{H^{\pm\pm}}$.
The color indicates the value of $\overline{v}_{\Delta}$.
We scan the ranges $400~\mathrm{GeV} \leq m_{H^{\pm\pm}} \leq 1000~\mathrm{GeV}$ and $-5 \leq \overline{\lambda}_{4} \leq 4\pi$.
As in the HS, loop effects compensate for the tree-level reduction of $m_{W}$, and thus a larger $\Delta m$ is preferred for a larger $\overline{v}_{\Delta}$.

In the upper right (lower left) panel of Fig.~\ref{fig: param_scan_mWGF_mugamgam_LS}, we show the allowed parameter points in the LS in the $\overline{\lambda}_{4}\text{--}\Delta m$ plane, where the color indicates the value of $\overline{v}_{\Delta}$ ($m_{H^{\pm\pm}}$).
The gray points are excluded by the constraint from $\Delta \mu_{\gamma\gamma}$.
For $m_{H^{\pm\pm}} \lesssim 450~\mathrm{GeV}$, values of $\overline{\lambda}_{4}$ as large as $4\pi$ are allowed, whereas we find $\overline{\lambda}_{4}\lesssim 5$ for $m_{H^{\pm\pm}} \gtrsim 450~\mathrm{GeV}$.
These limits on $\overline{\lambda}_{4}$ are obtained from perturbativity and tree-level unitarity, in particular from $a_{5,+}$, and parameter points with large $\overline{\lambda}_{4}$ and light $m_{H^{\pm\pm}}$ satisfy the theoretical constraints and are consistent with the electroweak precision observables.
However, such points are excluded by the constraint from $\Delta \mu_{\gamma\gamma}$.
As a result of the combined constraints, an upper bound $\overline{\lambda}_{4}^{\mathrm{max}} \simeq 3$ is obtained for small $\overline{v}_{\Delta}$.

We now turn to the region with large $\overline{v}_{\Delta}$, where the constraint from $\Delta \mu_{\gamma\gamma}$ becomes less effective, and the allowed parameter region is instead determined by the electroweak precision observables, vacuum stability, and tree-level unitarity.
Since a larger $\Delta m$ is preferred for a larger $\overline{v}_{\Delta}$, $\lambda_{3}$ $(\simeq \lambda_{5}\simeq 4(m_{H^{\pm}}^{2}-m_{H^{\pm\pm}}^{2})/v^{2})$ becomes sizable and positive.
As a consequence, negative values of $\overline{\lambda}_{4}$ are allowed under the vacuum stability condition,
$\lambda_{4}+2\sqrt{\lambda_{1}(\lambda_{2}+\lambda_{3}/2)} > 0$.
As $\lambda_{3}$ increases, or equivalently as $\Delta m$ and $\overline{v}_{\Delta}$ increase, the upper bound on $\overline{\lambda}_{4}$ imposed by tree-level unitarity ($a_{5,+}$) becomes negative.
Since the lower bound on $\overline{\lambda}_{4}$ is determined by vacuum stability, this interplay leads to an upper bound on $\lambda_{3}$.
Consequently, the maximal allowed value of $\Delta m$, which depends on $m_{H^{\pm\pm}}$, is about $110~\mathrm{GeV}$ for $m_{H^{\pm\pm}} \simeq 400~\mathrm{GeV}$ in the LS.

In the lower right panel of Fig.~\ref{fig: param_scan_mWGF_mugamgam_LS}, we show the prediction for $\Delta \mu_{\gamma\gamma}$ at LO as a function of $m_{H^{\pm\pm}}$, where the color indicates the value of $\overline{v}_{\Delta}$ and the solid line represents the current bound.
Since the $H^{\pm\pm}$ contribution is proportional to $\lambda_{4}$, $\Delta \mu_{\gamma\gamma}$ is sensitive to the sign of $\lambda_{4}$, and a positive $\Delta \mu_{\gamma\gamma}$ is a clear signature of the LS with a large $\overline{v}_{\Delta}$.

%=================================================================
\subsection{Numerical results in the heaviest \texorpdfstring{$H^{\pm\pm}$}{} scenario} 
%=================================================================

In this section, we discuss deviations from the SM predictions in the Higgs boson decay rates and the branching ratios.
For convenience, we quantify the deviations by  
\begin{align}
\Delta R(h\to XX)=\frac{\Gamma(h\to XX)^{\rm NP}}{\Gamma(h\to XX)^{\rm SM}} -1,
\end{align}
where $\Gamma(h\to XX)^{{\rm NP (SM)}}$ is the decay rate for a new physics model (the SM). 
We also introduce
\begin{align}
\Delta \mu(h\to XX)=\frac{{\rm BR}(h\to XX)^{\rm NP}}{{\rm BR}(h\to XX)^{\rm SM}}-1,
\label{eq: Delta_mu_hToXX}
\end{align}
where ${\rm BR}(h\to XX)^{\rm NP {\rm (SM)}}$ is the branching ratio for a new physics model (SM).
The quantity $\Delta \mu$ is more useful than $\Delta R$ for directly comparing the model predictions with experimental results at the HL-LHC and future colliders.
On the other hand, $\Delta R$ is convenient for examining the theoretical behavior of deviations from the SM predictions.

\subsubsection{Correlations in the decay rates of \texorpdfstring{$h$}{}}

To demonstrate how one can distinguish CHTM from other extended Higgs models, we perform a scan analysis, calculating the deviations in the decay rates and the branching ratios with higher-order corrections.
We impose the theoretical constraints discussed in Sec.~\ref{sec: model} and the experimental constraints presented above (direct searches, electroweak precision observables, and measurements of $h\to \gamma\gamma$ at the LHC).
As discussed above, the upper limit of $\overline{v}_\Delta$ is $\overline{v}_\Delta \lesssim 17.5$~GeV in the HS. 
Thus, we fix $\overline{v}_\Delta=1, 10$ and $15$~GeV in the analysis of the HS.  
Other parameters are in the following ranges, 
\begin{align}
\mbox{ HS: }&
300~{\rm GeV}\le m_A\le 1000~{\rm GeV},\   0\le \overline{\lambda}_{4} \le 4\pi,\ 0\le\Delta m\le150~{\rm GeV},
\end{align}
where $\Delta m$ gives the mass difference among the heavy Higgs bosons; $\Delta m=m_{H^\pm}-m_A$ for the HS.~\footnote{In the degenerate scenario $m_{H^{\pm\pm}}=m_{H^{\pm}}=m_A=m_H$, the upper bound on $\overline{v}_\Delta$ becomes more stringent, reaching approximately $1\,\mathrm{GeV}$. Consequently, the predictions for the Higgs decay rates are close to those obtained in the HS with $\overline{v}_\Delta = 1\,\mathrm{GeV}$. Note that the deviations from the SM in the decay rate are relatively smaller than the HS with $\overline{v}_\Delta = 1\;{\rm GeV}$ since the lower bound of $m_{H^{\pm\pm}}$ is 400 {\rm GeV} in the degenerate case~\cite{Ashanujjaman:2021txz}.}

To discuss the distinguishability of the CHTM, we not only display the results for the CHTM, but also present the results of four types of 2HDMs, the inert doublet model (IDM), and the HSM. 
The predictions for the decays of $h$ for all of these models are calculated by {\tt H-COUP}~\cite{Kanemura:2017gbi,Kanemura:2019slf,Aiko:2023xui}. 
For these models, the input parameters, which are defined in Ref.~\cite{Aiko:2023xui}, are set as 
~\footnote{Although, in {\tt H-COUP 3.0}, $\lambda_{\Phi S}$ in the HSM is taken as an input parameter instead of $m_s$, we here scan $m_s$ with a relation written in Ref.~\cite{Kanemura:2015fra}.}
\begin{align}
\mbox{2HDM Type-I:~}&400~{\rm GeV}<m_{H}<1~{\rm TeV},~0.99<\sin(\beta-\alpha)<1\;,~ \notag\\
    &1.5<\tan\beta<10\;,~ 0<M<m_H+500~{\rm GeV}, \\
\mbox{2HDM Type-II:~}&800~{\rm GeV}<m_{H}<1{\rm TeV},~0.995<\sin(\beta-\alpha)<1\;,~ \notag\\
    & 1<\tan\beta<10\;,~ 0<M<m_H+500~{\rm GeV}, \\
\mbox{HSM:~}~&400~{\rm GeV}<m_H<1000~{\rm GeV}\;,~\lambda_S=0.1,\;~\mu_s=0\;,~ \notag \\
            &0.95<\cos\alpha<1\;, 0<m_s<m_H+500~{\rm GeV}, \\
\mbox{IDM:~}~&100~{\rm GeV}<m_{H^\pm}<1000~{\rm GeV}\;,m_H=60~{\rm GeV},m_A=m_{H^\pm}\notag \\
& 0<\lambda_2<4\pi,~\mu_{2}^{2} = 3581~{\rm GeV}^2. 
\end{align}
We assume the degenerate mass scenario for the 2HDMs, i.e.,  $m_A=m_{H^\pm}=m_{H}$. 
For Type-I, the lower bounds of $m_H$ and $\tan \beta$ are taken under the consideration of constraints from $b\to s\gamma$~\cite{Misiak:2017bgg, Misiak:2020vlo}. The same is true for the lower bounds of $m_H$ for Type-II 2HDM. 
For IDM, we consider the scenario where $H$ is dark matter with a mass of $60$~GeV. 
We note that, in this scenario, $\mu_2^2$ is fixed, so that loop corrections by additional Higgs bosons do not decouple in the decay rates of $h$ as can be seen in the following numerical results~\cite{Kanemura:2016sos, Aiko:2023nqj}.
In this scan region, particularly in the 2HDMs, the predicted Higgs decay rates can deviate significantly from the SM. Hence, we conservatively impose the current LHC constraints on $2\sigma$ allowed ranges of the Higgs coupling modifiers $\kappa^{}_{i}$ without their central values under the assumption that the branching ratio to beyond-the-Standard-Model (BSM) final states is zero~\cite{CMS:2025jwz}.
We apply the bounds on $\kappa^{}_\tau$, $\kappa^{}_b$, $\kappa^{}_\gamma$, $\kappa^{}_g$, $\kappa^{}_\lambda$. 
Including higher-order corrections, they are theoretically evaluated by $\kappa_{X}^2=\Gamma^{\rm 2HDM}(h\to XX)/\Gamma^{\rm SM}(h\to XX)$ ($X=\tau,b,g,\gamma$) and $\kappa^{}_\lambda=\lambda_{hhh}^{\rm 2HDM}/\lambda_{hhh}^{\rm SM}$~\footnote{We confirmed that for the CHTM, IDM and HSM there are no excluded points by the constraint of the Higgs coupling in the chosen scan region.}${}^{,\,}$\footnote{The one-loop corrections to $\lambda_{hhh}$ depend on the external incoming momenta $p_1$ and $p_2$ and the outgoing momentum $q=p_1+p_2$. We set them as $p_1^2=p_2^2=m_h^2$ and $q^2=(2m_h+10\;\mathrm{GeV})^2$.}.

%-----------------------------
\begin{figure}[t]
 \centering
 \includegraphics[scale=0.8]{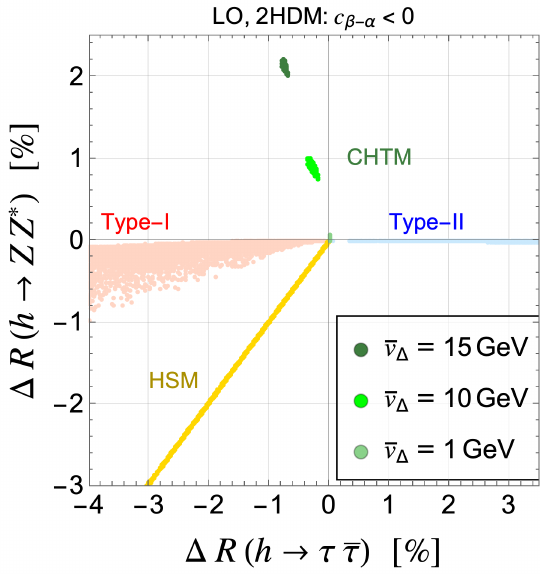}
  \includegraphics[scale=0.8]{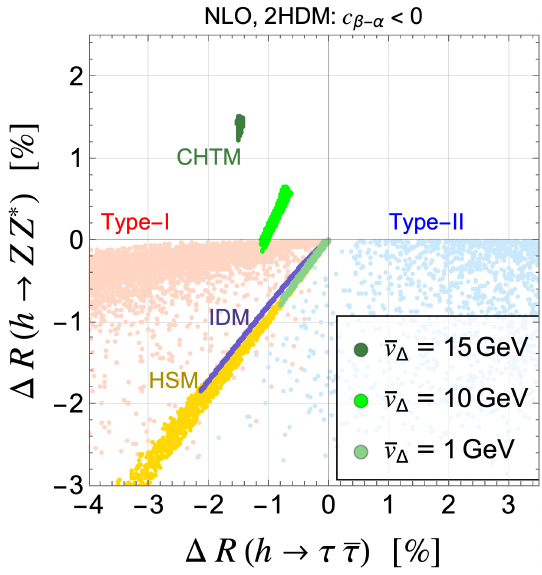} \\
   \includegraphics[scale=0.8]{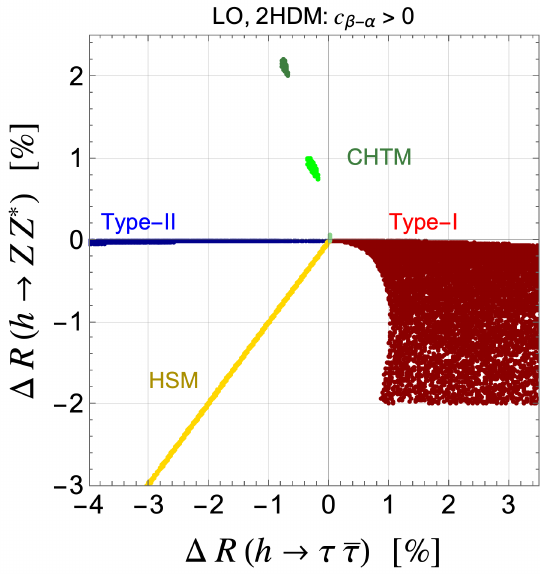}
  \includegraphics[scale=0.8]{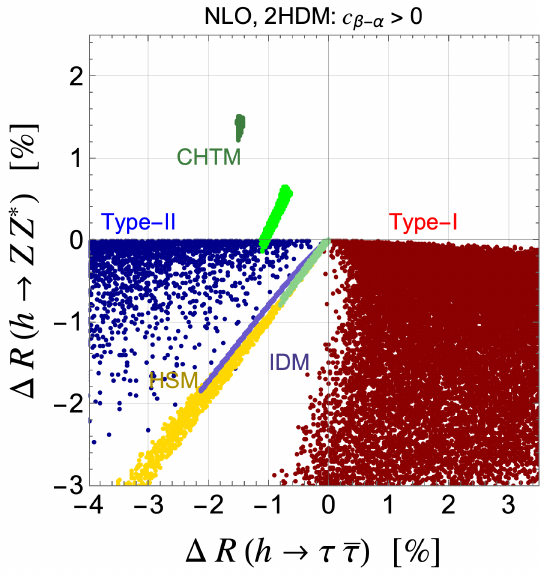}
 \caption{Correlations for $h\to ZZ^\ast$ and $h\to \tau\overline{\tau}$ in the HS. The left (right) panels show the results at the LO (NLO). Upper (lower) panels show 2HDM predictions with $c_{\beta-\alpha}<0$ ($c_{\beta-\alpha}>0$).}
 \label{fig:DR:ZZ:tautau}
\end{figure}
%-----------------------------

In this section, we present the numerical results for each model, distinguished by different colors in the figures. The results of the CHTM are shown in several shades of green, where each shade represents a different value of $\overline{v}_\Delta$, i.e., dark green, green, and light green correspond to the results for $\overline{v}_\Delta =$ 15, 10, and 1~GeV, respectively. The results for the Type-I (Type-II) 2HDM with $c_{\beta-\alpha}^{} < 0$ and those with $c_{\beta-\alpha}^{} > 0$ are shown by the light magenta (light blue) and dark red (dark blue) regions, respectively. 
In Fig.~\ref{fig:DR:ZZ:tautau}, the results for the two signs of $c_{\beta-\alpha}^{}$ are plotted in separate panels. 
In the other figures, however, the results are shown in the same panel without separation. 
We also plot the results of the HSM and the IDM using yellow and purplish blue points, respectively. This color scheme is used consistently in the following numerical figures.

While we will discuss numerical results with NLO EW and N(N)LO QCD corrections below, let us first discuss the behavior of the deviations in the LO decay rate.
At the tree-level, $\Delta R(h\to XX)$ is governed by the square of coupling modifiers, i.e., $\zeta_h^2$ for $h\to f\bar{f}$, $c_{hVV}^2$ for $h\to VV^\ast$. 
The shift of the gauge boson masses is also relevant for the latter process, i.e., Eqs.~\eqref{eq:m_V} and~\eqref{eq: swsq and mwsq}. 
The coupling modifiers are controlled by the scalar mixing parameters. 
Larger mixing leads to larger deviations from unity in them. 
Due to the larger $\overline{v}_\Delta$, decays into weak gauge bosons deviate in the positive direction as shown in the left panels of Fig.~\ref{fig:DR:ZZ:tautau}, which are characteristic predictions for the exotic Higgs sector~\cite{Grifols:1980uq,Kanemura:2013mc, Kanemura:2014bqa, Jurciukonis:2024bzx}. 
Thus, the CHTM can be separated from the simple extended Higgs models with $\rho_{\rm 0}=1$ by looking at the correlation between the $\Delta R(h\to ZZ^\ast)$ and $\Delta R(h\to \tau\overline{\tau})$. 

The NLO EW corrections significantly alter the quantitative results.
As a concrete example illustrating this point, we show the correlation between $\Delta R(h \to \tau\overline{\tau})$ and $\Delta R(h \to ZZ^\ast)$ obtained from the LO and NLO analyses in the left and right panels of Fig.~\ref{fig:DR:ZZ:tautau}, respectively.
As can be seen from a comparison of these figures, the theoretical predictions of the CHTM are shifted by almost $-1\%$ in the negative direction relative to the LO results for both $h \to ZZ^\ast$ and $h \to \tau\overline{\tau}$, independently of the value of $\overline{v}_\Delta$. 
This behavior originates from the loop corrections involving the scalar couplings $\lambda_{h\phi\phi}$ of the charged and doubly charged Higgs bosons (for instance, the counterterm $\delta Z_h$ contains a term proportional to $\lambda_{h\phi\phi}^2$). 
Assuming a small mixing angle $\alpha$, it scales as $\alpha \sim \overline{v}_\Delta$. 
Because of this, the three scenarios with different values of $\overline{v}_\Delta$ correspond to different possible values of $\alpha$. 
In addition, the mass splitting $\Delta m$ increases with $\overline{v}_\Delta$.
For example, after imposing the theoretical and experimental constraints discussed above, we find that for $\overline{v}_\Delta = 15~(1)\,\mathrm{GeV}$, one typically obtains $\sin\alpha \sim 0.1~(0)$ and $\Delta m_{\rm max} \sim 120~(30)\,\mathrm{GeV}$.
When $\sin\alpha \neq 0$, the contributions proportional to $\sin\alpha$ become sizable. 
(As shown in Appendix A of Ref.~\cite{Aoki:2012jj}, the coupling $\lambda_{h\phi\phi}$ can be expressed as $\lambda_{h\phi\phi} \sim A c_{\alpha} + B s_{\alpha}$, where $A$ and $B$ are coefficients with mass dimension one.) 
This enhances the loop contributions from both the doubly and singly charged Higgs bosons.
As a result, $\Delta R(h\to ZZ^\ast)$ and $\Delta R(h\to \tau\overline{\tau})$ deviate from their tree-level predictions by $\mathcal{O}(1)\%$.
In the case of $\overline{v}_\Delta = 1\,\mathrm{GeV}$, since $\sin\alpha$ takes a value close to zero, there exists a parameter region where the loop contributions from the additional Higgs bosons almost decouple, namely reproducing the LO results. 
A similar behavior can also be observed in the 2HDMs. Namely, the NLO predictions deviate from the tree-level results at the level of a few percent by loop corrections. While the LO predictions are well separated among the 2HDMs, HSM, and IDM, these predictions overlap with each other at the NLO level.

%-----------------------------
\begin{figure}[t]
 \centering
 \includegraphics[scale=0.8]{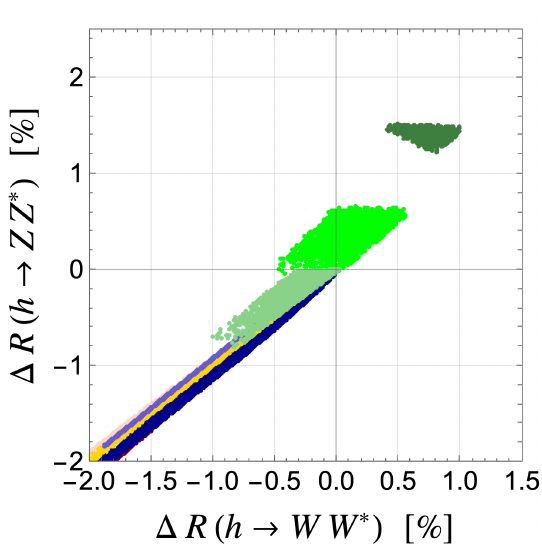}
 \caption{Correlations for the $h\to ZZ^\ast$ and $h\to WW^\ast$  in the HS.}
 \label{fig:DR:ZZ:WW}
\end{figure}
%-----------------------------

{We also show the correlation between $\Delta R(h \to ZZ^\ast)$ and $\Delta R(h \to WW^\ast)$ in Fig.~\ref{fig:DR:ZZ:WW}. 
A point worth emphasizing is that, in the CHTM, the deviation of $h \to ZZ^\ast$ tends to be relatively larger than that of $h \to WW^\ast$ in the case of $\Delta R(h \to ZZ^\ast) > 0$.
In contrast, $\Delta R(h \to ZZ^\ast)$ and $\Delta R(h \to WW^\ast)$ are approximately equal in other models.
This characteristic difference between the $h \to ZZ^\ast$ and $h \to WW^\ast$ decay modes in the CHTM originates from the difference in the coupling modifiers. In addition, it also comes from the fact that $h \to WW^\ast$  is affected by corrections of $\Delta r$ associated with the $W$-boson mass as discussed in Sec.~\ref{subsec: decay rates of h}.}

Let us discuss whether one can distinguish the CHTM from the other models\footnote{Discriminatation of the 2HDMs, HSM and IDM by the Higgs precision measurement is discussed in Refs.~\cite{Kanemura:2014bqa,Kanemura:2018yai,Kanemura:2019kjg,Aiko:2021nkb} in detail.}.
Although the inclusion of NLO corrections shifts the predictions in the negative direction regardless of the decay mode, the tendency that $h \to ZZ^\ast$ and $h \to WW^\ast$ can have positive deviations remains unchanged. 
Therefore, if such a positive deviation is observed in future collider experiments, one could identify the CHTM.
The result of $h\to ZZ^\ast$ for $\overline{v}_\Delta= 15\,{\rm GeV}$ is indeed comparable with the precision of the future Higgs factory, e.g. the ILC with $250\,{\rm GeV}$,  $0.58\%$, referring to the accuracy of $\kappa^{}_Z$~\cite{deBlas:2019rxi} (although $h\to \tau\overline{\tau}$ is challenging due to around the $4\%$ measurement accuracy).

%-----------------------------
\begin{figure}[t]
 \centering
 \includegraphics[scale=0.8]{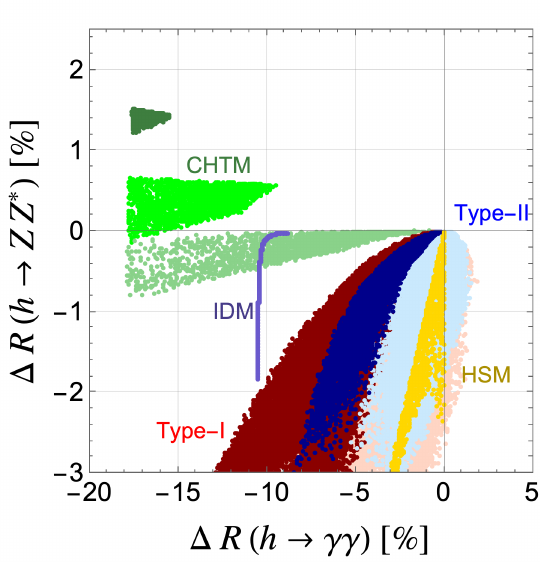}
  \includegraphics[scale=0.8]{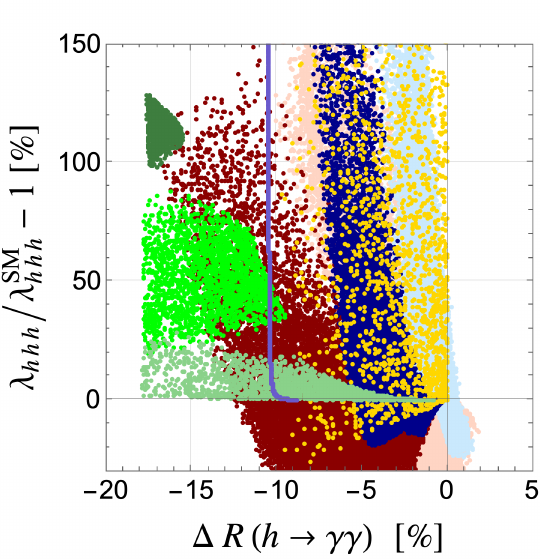}
 \caption{Correlations for the deviations in the decay rates and the Higgs self-coupling in the HS.}
 \label{fig:DR:gammagamma:hhh}
\end{figure}
%-----------------------------

When $h \to ZZ^\ast$ and $h \to WW^\ast$ have positive deviations, other observables also show characteristic predictions. 
In particular, both $h \to \gamma\gamma$ and the Higgs self-coupling receive significant corrections originating from the charged and doubly charged Higgs bosons through the $\lambda_{h\phi\phi}$ coupling. 
As shown in Fig.~\ref{fig:DR:gammagamma:hhh}, for $\overline{v}_\Delta = 15~\mathrm{GeV}$, $\Delta R(h \to \gamma\gamma)$ can be as large as $-18\%$, while the deviation of $\lambda_{hhh}$ from the SM prediction can reach about 130\%.\footnote{In the 2HDMs, we see that $\lambda_{hhh}^{}$ can deviate in a negative direction in certain regions of parameter space. 
The main sources of the negative shift arise from the tree-level contribution of order 
$c_{\beta-\alpha}^{2}$ and the one-loop correction of order $c_{\beta-\alpha}$.
These negative contributions appear in the case where $c_{\beta-\alpha}$ is not too small and $M\sim m_H^{}$~\cite{Kanemura:2004mg}.}
The projected sensitivity for the measurement of the triple Higgs coupling at the HL-LHC is below $30\%$ at the $68\%$ confidence level~\cite{CMS:2025hfp}, whereas that for the $h\gamma\gamma$ coupling is around 1.8\%~\cite{CMS:2025hfp}. 
Therefore, if the characteristic positive deviations in $h\to ZZ^\ast$ and $h\to WW^\ast$ predicted by the CHTM are realized, their indications should become visible already at the HL-LHC stage. 

\subsubsection{Correlations of the branching ratios of \texorpdfstring{$h$}{}}

We then discuss the deviations from the SM predictions in the branching ratio of $h$.
We can approximate $\Delta \mu(h\to XX)$ in Eq.~\eqref{eq: Delta_mu_hToXX} as
\begin{align} \label{eq:Dmu_ap}
\Delta \mu(h\to XX) \simeq \Delta R(h\to XX)-\Delta R^{\rm tot},
\end{align}
where $\Delta R^{\rm tot}$ denotes the deviation in the total decay width. It is defined by 
$\Delta R^{\rm tot}=\Gamma_{h}^{\rm tot, NP}/\Gamma_{h}^{\rm tot, SM}-1$. 
In the derivation of this equation, we have assumed that $\Delta R^{\rm tot}$, $\Delta R(h\to XX)\ll 1$.

%-----------------------------
\begin{figure}[t]
 \centering
 \includegraphics[scale=0.8]{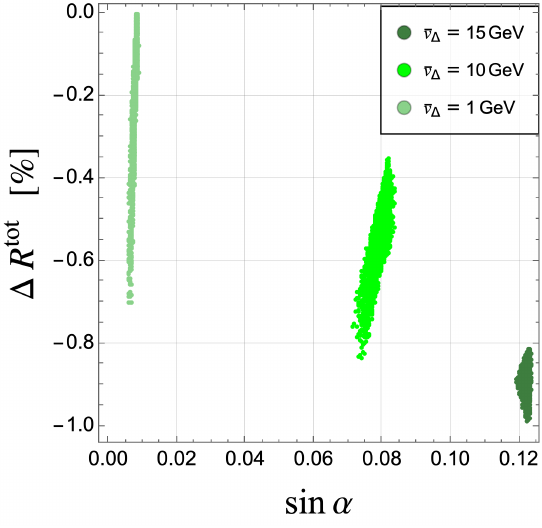}
 \caption{Deviations from the SM predictions in the total decay width defined of $h$ are shown in the CHTM by $\Delta R^{\rm tot}=\Gamma_{h}^{\rm tot, NP}/\Gamma_{h}^{\rm tot, SM}-1$ with $\Gamma^{\rm tot, NP(SM)}$ being the total decay with for new physics models (SM).}
\label{fig:dR_tot}
\end{figure}
%-----------------------------

In Fig.~\ref{fig:dR_tot}, to understand the behavior of $\Delta \mu(h\to XX)$, we display the deviations of the total decay width of $h$ as a function of $\sin\alpha$.  
Since the dominant contribution to the total width comes from $h \to b\bar{b}$, $\Delta R^{\rm tot}$ mainly reflects the effects of new physics on this decay mode. 
In $h\to b\bar{b}$, the loop contributions of the additional Higgs bosons induced by the $\lambda_{h\phi\phi}$ coupling also become significant, especially in the regime of large $\overline{v}_\Delta$. 
As a consequence, the deviation in the total width increases as the size of $\overline{v}_\Delta$ grows.

%-----------------------------
\begin{figure}[t]
 \centering
 \includegraphics[scale=0.8]{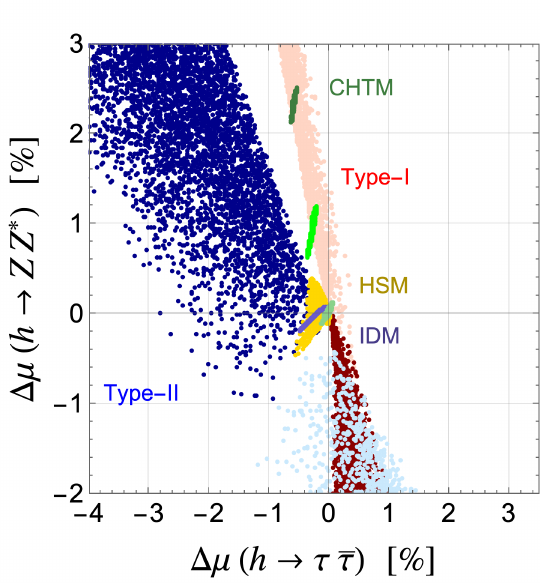}
 \includegraphics[scale=0.8]{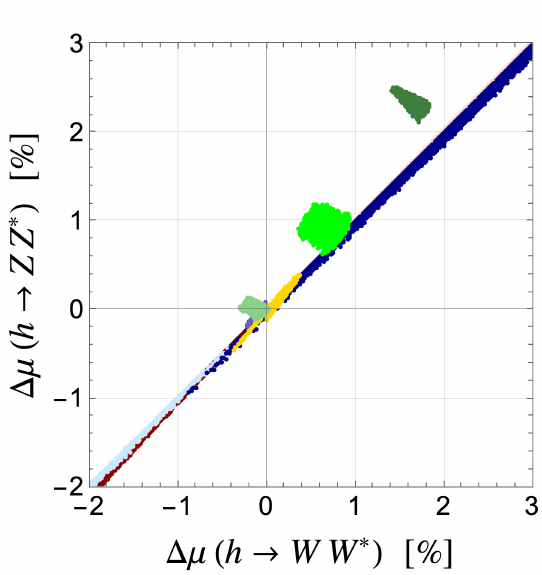}
 \includegraphics[scale=0.8]{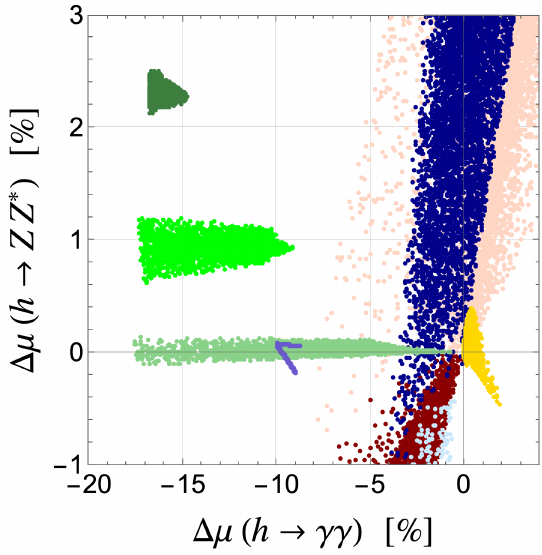}
 \caption{Correlations for the deviations in the branching ratios for the HS.}
 \label{fig:Dmu}
\end{figure}
%-----------------------------

The behavior of the deviations in the Higgs boson branching ratios from the SM predictions can be understood from Eq.~\eqref{eq:Dmu_ap}.
As an example, let us focus on the deviation in the branching ratio of $h \to ZZ^*$. 
As shown in Fig.~\ref{fig:Dmu}, $\Delta \mu(h\to ZZ^\ast)$ is approximately $0\%$ for $\overline{v}_\Delta = 0$ and about $2.4\%$ for $\overline{v}_\Delta = 15\ {\rm GeV}$. 
For $\overline{v}_\Delta = 0$, since $\Delta R(h\to ZZ^\ast) \simeq \Delta R^{\rm tot} \simeq [0, -0.8\%]$, these two deviations largely cancel each other, leading to a small deviation in $\Delta \mu (h\to ZZ^\ast)$.
On the other hand, for $\overline{v}_\Delta = 15~\mathrm{GeV}$, since $\Delta R(h \to ZZ^{*}) \simeq [1.2\%, 1.5\%]$ and $\Delta R^{\text{tot}} \simeq [-0.8\%, -1\%]$, the contributions of $\Delta R(h \to ZZ^\ast)$ and $\Delta R^{\text{tot}}$ add up to enhance $\Delta \mu({h \to ZZ^\ast})$.
A similar behavior can be understood for $h \to \tau\overline{\tau}$, $h \to WW^*$, and $h \to \gamma\gamma$ modes as well, based on Eq.~\eqref{eq:Dmu_ap}.
As a result, the deviation of $h\to ZZ^\ast$ reaches a maximum of $2.5\%$; however, this would be challenging to detect compared to the measurement accuracy currently estimated for future collider experiments, e.g., 6.7$\%$ at the ILC at 250 GeV~\cite{Fujii:2017vwa}.
When looking at the deviations in the branching ratios, the characteristic positive shift in the $h \to VV^\ast$ modes overlaps with the theoretical predictions of the Type-I 2HDM in the plane of $\Delta \mu (h\to \tau\overline{\tau})$ and $\Delta \mu (h\to ZZ^\ast)$. 
As can be seen in Ref.~\cite{Kanemura:2019kjg}, when $c_{\beta-\alpha} < 0$, the total width in the Type-I 2HDM tends to shift in the negative direction, resulting in a positive deviation in $\Delta \mu(h\to VV^\ast)$.
However, the discrimination between the CHTM and the Type-I 2HDM is possible in the $h \to \gamma\gamma$ mode. 
In the CHTM, $\Delta \mu(h\to \gamma\gamma)$ shows a negative deviation exceeding 10\%, which is not observed in the 2HDMs.
Therefore, similar to the discussion on the deviations in decay rates from the SM predictions, the $h \to \gamma\gamma$ mode plays a crucial role in identifying the CHTM. 

\subsection{Numerical results in the lightest \texorpdfstring{$H^{\pm\pm}$}{} scenario} 

We discuss the deviations in the decay rates and branching ratios in the LS ($m_{H^{\pm\pm}} < m_{H^{\pm}} < m_{A}$).
In the following analysis, we take $\overline{v}_{\Delta} =1$ and $10$ GeV. As discussed in Sec.~\ref{subsec: Constraints}, $\overline{v}_{\Delta} =15$~GeV is excluded by the theoretical and experimental constraints.
Other parameters are in the following ranges, 
\begin{align}
\mbox{ LS: }&
400~{\rm GeV}\le m_{H^{\pm\pm}}\le 1000~{\rm GeV},\   -5\le \overline{\lambda}_{4} \le 4\pi,\ 0\le\Delta m\le150~{\rm GeV},
\end{align}
where $\Delta m = m_{H^{\pm}}-m_{H^{\pm\pm}}$.

%-----------------------------
\begin{figure}[t]
 \centering
 \includegraphics[scale=0.8]{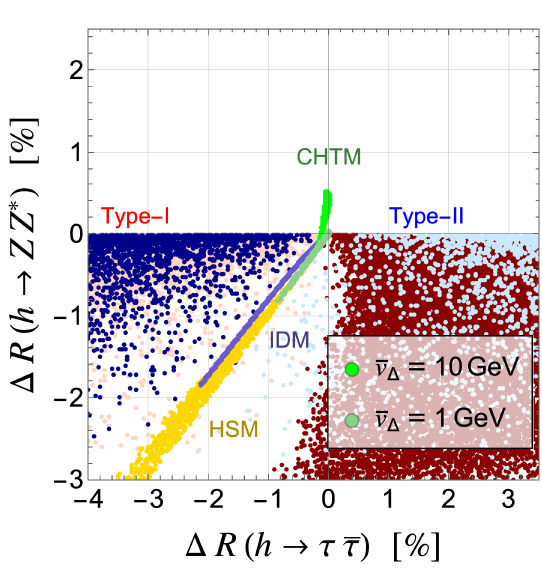}
 \includegraphics[scale=0.8]{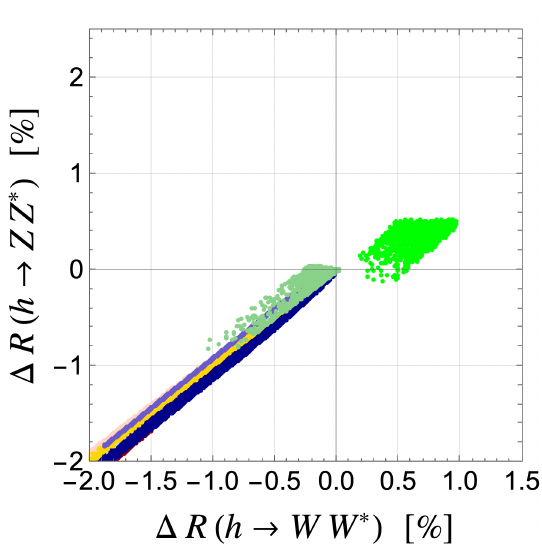}
 \caption{Correlations for the deviations in the decay rates in the LS. The color code is the same as that in the HS.}
 \label{fig:DR:ZZ:tautau_inv}
\end{figure}
%-----------------------------

%-----------------------------
\begin{figure}[t]
 \centering
 \includegraphics[scale=0.8]{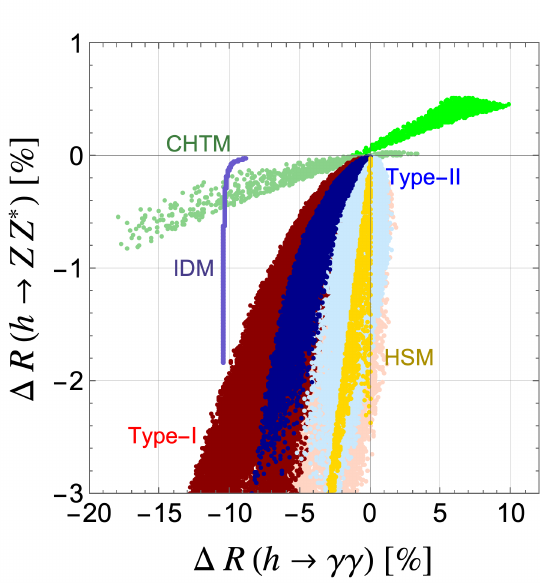}
  \includegraphics[scale=0.8]{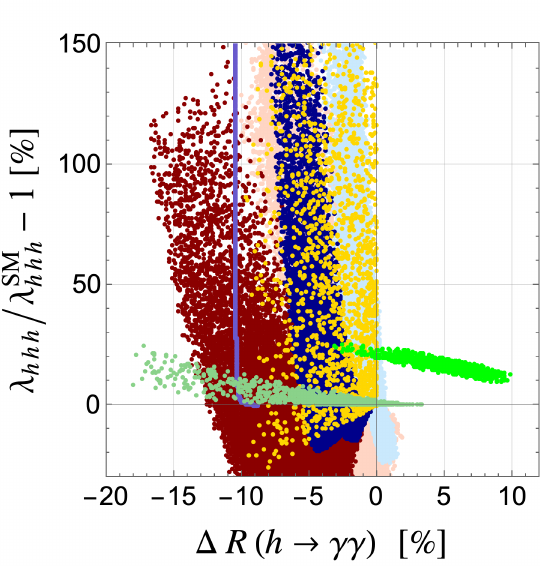}
 \caption{Correlations for the deviations in the decay rates in the LS.}
 \label{fig:DR:gammagamma:hhh_inv}
\end{figure}
%-----------------------------

In Fig.~\ref{fig:DR:ZZ:tautau_inv}, we show the correlation between $\Delta R(h\to\tau\overline{\tau})$ and $\Delta R(h\to ZZ^*)$ (left panel) and  $\Delta R(h\to WW^*)$ and 
$\Delta R(h\to ZZ^*)$ (right panel), where all the predictions except for the CHTM are the same as those shown in Fig.~\ref{fig:DR:ZZ:tautau} and Fig.~\ref{fig:DR:ZZ:WW}, respectively.
Since a smaller $\overline{\lambda}_{4}$ leads to a smaller $s_{\alpha}$, the deviations in the Higgs decays are suppressed compared to the HS.
The clear signal in the CHTM is a positive deviation in $h\to VV^*$ with large $\overline{v}_{\Delta}$, and the maximal values are $\Delta R(h\to ZZ^*)\simeq 0.5\%$ and $\Delta R(h\to WW^*)\simeq 1.0\%$.

In Fig.~\ref{fig:DR:gammagamma:hhh_inv}, we show the correlation between $\Delta R(h\to \gamma\gamma)$ and $\Delta R(h\to ZZ^*)$ (left panel) and $\Delta R(h\to \gamma\gamma)$ and $\lambda_{hhh}/\lambda_{hhh}^{\rm SM}-1$ similar to the plots shown in Fig.~\ref{fig:DR:gammagamma:hhh}. 
Unlike the HS, the deviation in the decay rate of $h \to \gamma\gamma$ tends to be predicted in the positive direction for large $\overline{v}_{\Delta}$ as discussed in Sec.~\ref{subsec: Constraints}.
Such a positive deviation in $h \to \gamma\gamma$ is rarely seen in the other models, so that this can be a smoking-gun signature in the LS.
Regarding the right panel, we see that the amount of the deviation in the $\lambda_{hhh}$ coupling is maximally given to be about 20\% due to the smallness of $\overline{\lambda}_{4}$ compared to the HS.

%-----------------------------
\begin{figure}[t]
 \centering
 \includegraphics[scale=0.8]{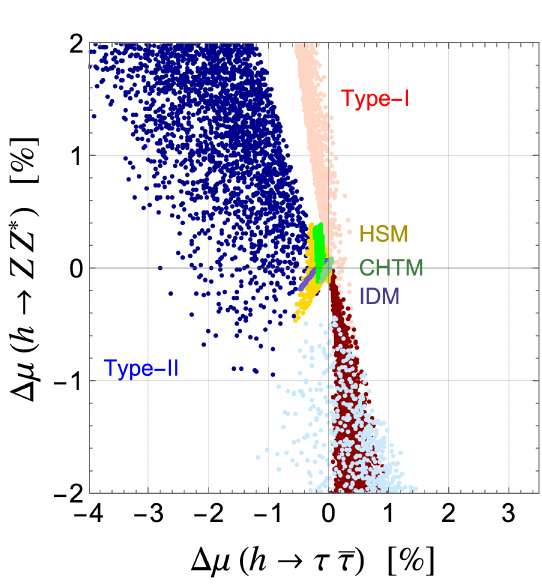}
 \includegraphics[scale=0.8]{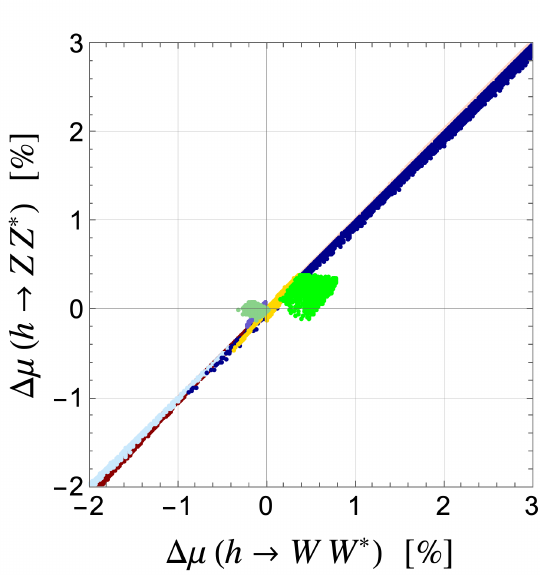}
 \includegraphics[scale=0.8]{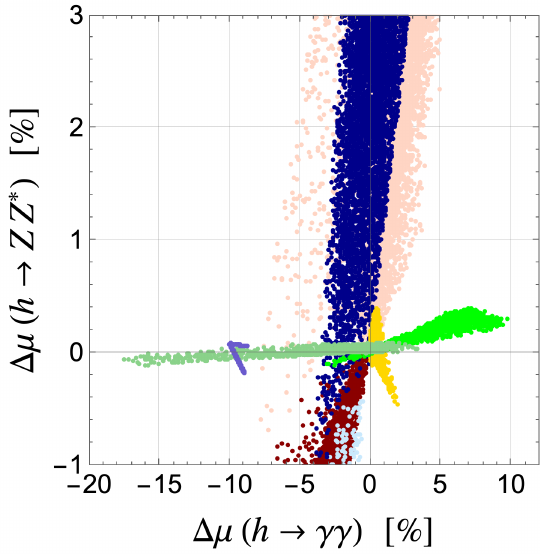}
 \caption{Correlations for the deviations in the branching ratios in the LS.}
 \label{fig:Dmu_inv}
\end{figure}
%-----------------------------

Finally, we show the correlation between the deviations in the branching ratios in Fig.~\ref{fig:Dmu_inv}. Similar to Fig.~\ref{fig:Dmu}, we display three correlations: $\Delta\mu(h \to \tau\overline{\tau})\text{--}\Delta\mu(h \to ZZ^*)$ (upper left), $\Delta\mu(h \to WW^*)\text{--}\Delta\mu(h \to ZZ^*)$ (upper right) and 
$\Delta\mu(h \to \gamma\gamma)\text{--}\Delta\mu(h \to ZZ^*)$ (lower). 
We see that the correlation $\Delta\mu(h \to \gamma\gamma)\text{--}\Delta\mu(h \to ZZ^*)$ shows a different pattern from that in the other models as we expect from the behavior of $\Delta R(h \to \gamma\gamma)$. Therefore, we can indirectly explore the LS by measuring the deviation in the branching ratio of $h \to \gamma\gamma$ and other modes such as $h \to ZZ^*$.

%%%%%%%%%%%%%%%%%%%%%%%%%%%%%%%%%%%%%%%%%%%%%%%%%%%%%%%%%%%%%%%%%%
\section{Conclusions} \label{sec: Conclusions}
%%%%%%%%%%%%%%%%%%%%%%%%%%%%%%%%%%%%%%%%%%%%%%%%%%%%%%%%%%%%%%%%%%

In this paper, we present a full set of radiative corrections to decays of the 125 GeV Higgs boson in the CHTM, which is one of the representative examples of an exotic Higgs sector.
We have analytically evaluated NLO EW corrections to the Higgs boson two-body decays ($h\to f\bar{f}$) and three-body decays ($h\to VV^\ast$).
For the loop-induced processes $h\to\gamma\gamma$, $Z\gamma$, and $gg$, we included the higher-order QCD corrections, while the EW corrections are at LO. 
To perform numerical evaluations of the 125 GeV Higgs boson decays, we have implemented these analytical results into the existing {\tt H-COUP v3}~\cite{Aiko:2023xui} with appropriate extensions. The program for computing the theoretical predictions of the CHTM will be released as the next version of {\tt H-COUP}.

To perform this calculation, we have employed a gauge-invariant on-shell renormalization scheme based on Ref.~\cite{Aiko:2023xui}. In the original work~\cite{Aoki:2012jj}, a residual gauge dependence arising from the scalar mixing angle remained. 
In this paper, we have eliminated this gauge dependence by applying the pinch technique. 
As discussed in detail in Sec.~\ref{sec:gauge invariant}, by focusing on the $f\bar{f}\to f\bar{f}$ scattering process, we have demonstrated that the relevant pinch terms can be extracted from the vertex, box, and external leg corrections. These pinch terms exactly cancel the gauge-dependent contributions in the Higgs boson self-energy diagrams.

Taking into account various theoretical and experimental constraints, we analyzed the correlations of the deviations in the Higgs boson decay rates and the branching ratios.
In the numerical analysis, we focused on the case with $v_{\Delta}=\mathcal{O}(1\text{--}10)$~GeV, where a characteristic prediction emerges as seen already at LO: the $h\to VV^\ast$ modes deviate from the SM in a positive direction in the case where $H^{\pm\pm}$ is the heaviest among the extra Higgs bosons. 
When NLO corrections are included, the deviations generally shift in the negative direction irrespective of the decay mode ($\sim -1\%$ for $h\to VV^\ast$ and $h\to f\bar{f}$, and $\sim -10\%$ for $h\to\gamma\gamma$).
We demonstrated that even after including NLO corrections, the characteristic predictions of the CHTM are retained.
In particular, at the level of decay rates, we have shown that the CHTM can be distinguished from other simple extended Higgs models, such as the 2HDMs, HSM and IDM, through this distinctive feature.
Moreover, when such a characteristic positive deviation appears in $h\to VV^\ast$, one expects corresponding deviations of $\sim-20\%$ in $h\to\gamma\gamma$ and  $\sim100\%$ in the $hhh$ coupling. Even in the case where $H^{\pm\pm}$ is lightest among the extra Higgs bosons, the CHTM can be identified by detecting $\mathcal{O}(10)\%$ positive deviations in $h\to \gamma\gamma$.
Future precision measurements of the Higgs boson at collider experiments will therefore provide crucial tests for exotic Higgs sectors such as the one investigated in this study.

Finally, although we mainly concentrated on the parameter region with $v_\Delta\sim {\cal O}(1\text{--}10)$ GeV where diboson decay is the dominant decay mode of $H^{\pm\pm}$, our calculation can be applied into wider parameter regions, including the case with a smaller $v_\Delta$  where $H^{\pm\pm}$ mainly decays into di-leptons.

\begin{acknowledgments}
MA was supported by the JSPS Grant-in-Aid for JSPS Fellows No. 22KJ3126.
SK was supported, in part, by Grants-in-Aid for Scientific Research(KAKENHI) Nos. 20H00160, 24KF0060, and 24KF0238.
KS was supported by the JSPS KAKENHI Grant Numbers 25K17379 and 21K20363.
ST was supported by JST SPRING, Grant Number JPMJSP2138.
\end{acknowledgments}

\appendix

%%%%%%%%%%%%%%%%%%%%%%%%%%%%%%%%%%%%%%%%%%%%%%%%%%%%%%%%%%%%%%%%%%
\section{Tree level scalar-vector couplings}
\label{app: coupling modifiers}
%%%%%%%%%%%%%%%%%%%%%%%%%%%%%%%%%%%%%%%%%%%%%%%%%%%%%%%%%%%%%%%%%%

In this appendix, we list the coupling modifiers for the Higgs–Higgs–gauge–gauge, Higgs–gauge–gauge, and Higgs–Higgs–gauge interactions, which are expressed as
\begin{align}
{\cal L}_{\rm kin} &\ni\;
\sum_{\substack{\phi\phi'=hh,\,hH,\,HH,\\ AA,\,G^0A,\,G^0G^0}}
\Big[
   \frac{g^2}{4}\,c_{\phi\phi'WW}^{}\,
   \phi\phi'\, W^{\pm\mu} W_\mu^\mp
 + \frac{g_Z^2}{8}\,c_{\phi\phi'ZZ}^{}\,
   \phi\phi'\, Z^{\mu} Z_\mu
\Big]
 \notag\\[1ex]
&+ \sum_{\phi=h,H}
\Big[
   g m_W\, c_{\phi WW}^{}\, \phi\, W^{\pm\mu} W_\mu^\mp
 + \frac{g_Z}{2} m_Z\, c_{\phi ZZ}^{}\, \phi\, Z^{\mu} Z_\mu \notag\\[1ex]
 &\pm \frac{g}{2}\, c_{H^{\pm}\phi W^{\mp}}^{}\,
   (p_1^{}-p_2^{})^\mu\, H^\pm \phi\, W_\mu^{\mp}
 \pm \frac{g}{2}\, c_{G^{\pm}\phi W^{\mp}}^{}\,
   (p_1^{}-p_2^{})^\mu\, G^\pm \phi\, W_\mu^{\mp}
 \notag\\[1ex]
&
 - i\frac{g_Z}{2}\, c_{G^0 \phi Z}^{}\,
   (p_1^{}-p_2^{})^\mu\, G^0 \phi\, Z_\mu
 - i\frac{g_Z}{2}\, c_{A \phi Z}^{}\,
   (p_1^{}-p_2^{})^\mu\, A \phi\, Z_\mu
\Big]\notag\\[1ex]
& - \sum_{\phi=G^0,A}
\Big[
i \frac{g}{2}\, c_{H^{\pm}\phi W^{\mp}}^{}\,
   (p_1^{}-p_2^{})^\mu\, H^\pm \phi\, W_\mu^{\mp}
 +i \frac{g}{2}\, c_{G^{\pm}\phi W^{\mp}}^{}\,
   (p_1^{}-p_2^{})^\mu\, G^\pm \phi\, W_\mu^{\mp}
 \Big],
\end{align} 
with $p_1$ and $p_2$ being the incoming momenta of $X$ and $Y$ in each $XYZ$-coupling, respectively.  
The explicit formula of the mixing factors $c_{XYZ(V)}^{}$ are given in Tab.~\ref{tab:couplings}.

\begin{table}[t]
\centering
\renewcommand{\arraystretch}{1.25}
\begin{tabular}{c|c||c|c}
\hline\hline
 Mixing factors & Expression & Mixing factors & Expression \\
\hline
$c_{hhWW}^{}$ & $\frac{1}{2}(3 - c_{2\alpha})$
& $c_{hhZZ}^{}$ & $\frac{1}{2}(5 - 3 c_{2\alpha})$ \\

$c_{hHWW}^{}$ & $s_{2\alpha}$
& $c_{hHZZ}^{}$ & $3 s_{2\alpha}$ \\

$c_{HHWW}^{}$ & $\frac{1}{2}(3 + c_{2\alpha})$
& $c_{HHZZ}^{}$ & $\frac{1}{2}(5 + 3 c_{2\alpha})$ \\

$c_{AAWW}^{}$ & $\frac{1}{2}(3 + c_{2\beta'})$
& $c_{AAZZ}^{}$ & $\frac{1}{2}(5 + 3 c_{2\beta'})$ \\

$c_{AG^0WW}^{}$ & $s_{2\beta'}$
& $c_{AG^0ZZ}^{}$ & $3 s_{2\beta'}$ \\

$c_{G^0G^0WW}^{}$ & $\frac{1}{2}(3 - c_{2\beta'}^{})$
& $c_{G^0G^0ZZ}^{}$ & $\frac{1}{2}(5 - 3 c_{2\beta'})$ \\

$c_{hWW}^{}$ & $c_\beta c_\alpha + \sqrt{2}\, s_\beta s_\alpha$
& $c_{hZZ}^{}$ & $c_{\beta'} c_\alpha + 2 s_{\beta'} s_\alpha$ \\

$c_{HWW}^{}$ & $-c_\beta s_\alpha + \sqrt{2}\, s_\beta c_\alpha$
& $c_{HZZ}^{}$ & $-c_{\beta'} s_\alpha + 2 s_{\beta'} c_\alpha$ \\

$c_{G^\pm G^0 W^\mp}^{}$ &  $c_{\beta'} c_\beta + \sqrt{2}\, s_{\beta'} s_\beta$
&$c_{G^\pm A W^\mp}^{}$ & 0 \\

$c_{H^\pm G^0 W^\mp}^{}$ & $-c_{\beta'}^{} s_{\beta}^{} + \sqrt{2}\, s_{\beta'}^{} c_{\beta}^{}$
&$c_{H^\pm A W^\mp}^{}$ & $s_{\beta'}^{} s_{\beta}^{} + \sqrt{2}\, c_{\beta'}^{} c_{\beta}^{}$\\

$c_{G^\pm h W^\mp}^{}$ & $c_\beta c_\alpha + \sqrt{2}\, s_\beta s_\alpha$
& $c_{G^0 h Z}^{}$ & $c_\alpha c_{\beta'} + 2 s_\alpha s_{\beta'}$ \\

$c_{H^\pm h W^\mp}^{}$ & $-s_\beta c_\alpha + \sqrt{2}\, c_\beta s_\alpha$
& $c_{G^0 H Z}^{}$ & $-c_{\beta'} s_\alpha + 2 c_\alpha s_{\beta'}$ \\

$c_{G^\pm H W^\mp}^{}$ & $-s_\alpha c_\beta + \sqrt{2}\, c_\alpha s_\beta$
& $c_{AhZ}^{}$ & $-c_\alpha s_{\beta'} + 2 s_\alpha c_{\beta'}$ \\

$c_{H^\pm H W^\mp}^{}$ & $s_\alpha s_\beta + \sqrt{2}\, c_\alpha c_\beta$
& $c_{AHZ}^{}$ & $2 c_\alpha c_{\beta'} + s_\alpha s_{\beta'}$ \\
\hline\hline
\end{tabular}
\caption{Mixing factors of Scalar--vector interaction.}
\label{tab:couplings}
\end{table}

%%%%%%%%%%%%%%%%%%%%%%%%%%%%%%%%%%%%%%%%%%%%%%%%%%%%%%%%%%%%%%%%%%
\section{1PI diagrams for \texorpdfstring{$hf\bar{f}$}{} vertex}
\label{app: hff_1PI}
%%%%%%%%%%%%%%%%%%%%%%%%%%%%%%%%%%%%%%%%%%%%%%%%%%%%%%%%%%%%%%%%%%

We present the form factors of the 1PI $hf\bar{f}$ vertex at the one-loop level in terms of the Passarino–Veltman functions~\cite{Passarino:1978jh}. We follow the convention of Ref.~\cite{Kanemura:2015mxa}. The momentum assignment is shown in the left panel of Fig.~\ref{fig: momentum assignment}.

\begin{align}
&\qty(\frac{m_{f}}{v})^{-1}16\pi^{2}\Gamma^{S, \mathrm{1PI}}_{hf\bar{f}}
\notag \\ &
=
%--- FVF ---
-4\zeta_{h}e^{2}Q_{f}^{2}C_{hff}^{FVF}(f,\gamma,f)
-4\zeta_{h}g_{Z}^{2}(v_{f}^{2}-a_{f}^{2})C_{hff}^{FVF}(f,Z,f)
\notag \\ &\quad
%--- FSF ---
+\frac{m_{f}^{2}}{v^{2}}\zeta_{h}\qty[
\zeta_{h}^{2}C_{hff}^{FSF}(f,h,f)+\zeta_{H}^{2}C_{hff}^{FSF}(f,H,f)
-\zeta_{G^{0}}^{2}C_{hff}^{FSF}(f,G^{0},f)-\zeta_{A}^{2}C_{hff}^{FSF}(f,A,f)
]
\notag \\ &\quad
-\frac{2m_{f'}^{2}}{v^{2}}\zeta_{h}\qty[
\zeta_{G^{\pm}}^{2}C_{hff}^{FSF}(f',G^{\pm},f')
+\zeta_{H^{\pm}}^{2}C_{hff}^{FSF}(f',H^{\pm},f')
]
\notag \\ &\quad
%--- VFV ---
-2g_{Z}^{4}v^{2}(v_{f}^{2}-a_{f}^{2})\sqrt{1+s_{\beta}^{2}}(c_{\beta'}c_{\alpha}+2s_{\beta'}s_{\alpha})C_0(Z,f,Z)
\notag \\ &\quad
%--- SFS ---
-\frac{m_{f}^{2}}{v}\Big\{
6\lambda_{hhh}\zeta_{h}^{2}C_{0}(h,f,h)
+2\lambda_{HHh}\zeta_{H}^{2}C_{0}(H,f,H)
+2\lambda_{Hhh}\zeta_{h}\zeta_{H}[C_{0}(h,f,H)+C_{0}(H,f,h)]
\notag \\ &\qquad
-2\lambda_{G^{0}G^{0}h}\zeta_{G^{0}}^{2}C_{0}(G^{0},f,G^{0})
-2\lambda_{AAh}\zeta_{A}^{2}C_{0}(A,f,A)
-\lambda_{AG^{0}h}\zeta_{G^{0}}\zeta_{A}
[C_{0}(A,f,G^{0})+C_{0}(G^{0},f,A)]
\Big\}
\notag \\ &\quad
+\frac{2m_{f'}^{2}}{v}\Big\{
\lambda_{G^{+}G^{-}h}\zeta_{G^{\pm}}^{2}C_{0}(G^{\pm},f',G^{\pm})
+\lambda_{H^{+}H^{-}h}\zeta_{H^{\pm}}^{2}C_{0}(H^{\pm},f',H^{\pm})
\notag\\ &\qquad
+\lambda_{H^{+}G^{-}h}\zeta_{G^{\pm}}\zeta_{H^{\pm}}
[C_{0}(G^{\pm},f',H^{\pm})+C_{0}(H^{\pm},f',G^{\pm})]
\Big\}
%=== W-boson loops ===
%--- VFS+SFV ---
\notag \\ &\quad
-\frac{g^{2}}{4}\zeta_{G^{\pm}}(c_{\beta}c_{\alpha}+\sqrt{2}s_{\beta}s_{\alpha})\qty[
C_{hff}^{VFS}(W,f',G^{\pm})+C_{hff}^{SFV}(G^{\pm},f',W)
]
\notag \\ &\quad
-\frac{g^{2}}{4}\zeta_{H^{\pm}}(-s_{\beta}c_{\alpha}+\sqrt{2}c_{\beta}s_{\alpha})\qty[
C_{hff}^{VFS}(W,f',H^{\pm})+C_{hff}^{SFV}(H^{\pm},f',W)
]
\notag \\ &\quad
%=== Z-boson loops ===
%--- VFS+SFV ---
-\frac{g_{Z}^{2}}{8}\zeta_{G^{0}}(c_{\beta'}c_{\alpha}+2s_{\beta'}s_{\alpha})\qty[
C_{hff}^{VFS}(Z,f,G^{0})+C_{hff}^{SFV}(G^{0},f,Z)
]
\notag \\ & \quad
-\frac{g_{Z}^{2}}{8}\zeta_{A}(-s_{\beta'}c_{\alpha}+2c_{\beta'}s_{\alpha})\qty[
C_{hff}^{VFS}(Z,f,A)+C_{hff}^{SFV}(A,f,Z)
], \\
&\qty(\frac{m_{f}}{v})^{-1}16\pi^{2}\Gamma^{P, \mathrm{1PI}}_{hf\bar{f}}
\notag \\ &
=
%--- VFS+SFV ---
\frac{g^{2}}{4}\zeta_{G^{\pm}}(c_{\beta}c_{\alpha}+\sqrt{2}s_{\beta}s_{\alpha})\qty[
C_{hff}^{VFS}(W,f',G^{\pm})-C_{hff}^{SFV}(G^{\pm},f',W)
]
\notag \\ &\quad
+\frac{g^{2}}{4}\zeta_{H^{\pm}}(-s_{\beta}c_{\alpha}+\sqrt{2}c_{\beta}s_{\alpha})\qty[
C_{hff}^{VFS}(W,f',H^{\pm})-C_{hff}^{SFV}(H^{\pm},f',W)
]
\notag \\ &\quad
+2g_{Z}^{2}v_{f}a_{f}\zeta_{G^{0}}(c_{\beta'}c_{\alpha}+2s_{\beta'}s_{\alpha})\qty[
C_{hff}^{VFS}(Z,f,G^{0})-C_{hff}^{SFV}(G^{0},f,Z)
]
\notag \\ &\quad
+2g_{Z}^{2}v_{f}a_{f}\zeta_{A}(-s_{\beta'}c_{\alpha}+2c_{\beta'}s_{\alpha})\qty[
C_{hff}^{VFS}(Z,f,A)-C_{hff}^{SFV}(A,f,Z)
], \\
&16\pi^{2}\Gamma^{V_{1}, \mathrm{1PI}}_{hf\bar{f}}
\notag \\
&=
%--- FVF ---
-\frac{2m_{f}^{2}}{v}\zeta_{h}\qty[
e^{2}Q_{f}^{2}(C_{0}+2C_{11})(f,\gamma,f)
+g_{Z}^{2}(v_{f}^{2}+a_{f}^{2})(C_{0}+2C_{11})(f,Z,f)
]
\notag \\ &\quad
-\frac{g^{2}}{2}\frac{m_{f'}^{2}}{v}\zeta_{h}(C_{0}+2C_{11})(f',W,f')
\notag \\ &\quad
%--- FSF ---
-\frac{m_{f}^{4}}{v^{3}}\zeta_{h}\Big[
\zeta_{h}^{2}(C_{0}+2C_{11})(f,h,f)+\zeta_{H}^{2}(C_{0}+2C_{11})(f,H,f)
\notag \\ &\qquad
+\zeta_{G^{0}}^{2}(C_{0}+2C_{11})(f,G^{0},f)+\zeta_{A}^{2}(C_{0}+2C_{11})(f,A,f)
\Big]
\notag \\ &\quad
-\frac{m_{f'}^{2}}{v^{3}}(m_{f}^{2}+m_{f'}^{2})\zeta_{h}\qty[
\zeta_{G^{\pm}}^{2}(C_{0}+2C_{11})(f',G^{\pm},f')
+\zeta_{H^{\pm}}^{2}(C_{0}+2C_{11})(f',H^{\pm},f')
]
\notag \\ &\quad
%--- VFV ---
+\sqrt{1+s_{\beta}^{2}}(c_{\beta'}c_{\alpha}+2s_{\beta'}s_{\alpha})
g_{Z}^{4}v(v_{f}^{2}+a_{f}^{2})(C_{0}+C_{11})(Z,f,Z)
\notag \\ &\quad
+(c_{\beta}c_{\alpha}+\sqrt{2}s_{\beta}s_{\alpha})
\frac{g^{4}}{4}v(C_{0}+C_{11})(W,f',W)
\notag \\ &\quad
%--- SFS ---
-\frac{m_{f}^{2}}{v^{2}}\Big\{
6\lambda_{hhh}\zeta_{h}^{2}(C_{0}+C_{11})(h,f,h)
+2\lambda_{HHh}\zeta_{H}^{2}(C_{0}+C_{11})(H,f,H)
\notag \\ &\qquad
+2\lambda_{Hhh}\zeta_{h}\zeta_{H}[(C_{0}+C_{11})(H,f,h)+(C_{0}+C_{11})(h,f,H)]
\notag \\ &\qquad
+2\lambda_{G^{0}G^{0}h}\zeta_{G^{0}}^{2}(C_{0}+C_{11})(G^{0},f,G^{0})
+2\lambda_{AAh}\zeta_{A}^{2}(C_{0}+C_{11})(A,f,A)
\notag \\ &\qquad
+\lambda_{AG^{0}h}\zeta_{G^{0}}\zeta_{A}
[(C_{0}+C_{11})(A,f,G^{0})+(C_{0}+C_{11})(G^{0},f,A)]
\Big\}
\notag \\ &\quad
-\frac{m_{f}^{2}+m_{f'}^{2}}{v^{2}}\Big\{
\lambda_{G^{+}G^{-}h}\zeta_{G^{\pm}}^{2}
(C_{0}+C_{11})(G^{\pm},f',G^{\pm})
+\lambda_{H^{+}H^{-}h}\zeta_{H^{\pm}}^{2}
(C_{0}+C_{11})(H^{\pm},f',H^{\pm})
\notag \\ &\qquad
+\lambda_{H^{+}G^{-}h}\zeta_{G^{\pm}}\zeta_{H^{\pm}}
[(C_{0}+C_{11})(G^{\pm},f',H^{\pm})+(C_{0}+C_{11})(H^{\pm},f',G^{\pm})]
\Big\}
\notag \\ &\quad
%--- VFS+SFV ---
+\frac{g^{2}}{4}\frac{m_{f'}^{2}}{v}\Big[
\zeta_{G^{\pm}}(c_{\beta}c_{\alpha}+\sqrt{2}s_{\beta}s_{\alpha})
\qty[(2C_{0}+C_{11})(W,f',G^{\pm})+(-C_{0}+C_{11})(G^{\pm},f',W)]
\notag \\ &\qquad
+\zeta_{H^{\pm}}(-s_{\beta}c_{\alpha}+\sqrt{2}c_{\beta}s_{\alpha})
\qty[(2C_{0}+C_{11})(W,f',H^{\pm})+(-C_{0}+C_{11})(H^{\pm},f',W)]
\Big]
\notag \\ &\quad
+\frac{g_{Z}^{2}}{8}\frac{m_{f}^{2}}{v}\Big[
\zeta_{G^{0}}(c_{\beta'}c_{\alpha}+2s_{\beta'}s_{\alpha})
\qty[(2C_{0}+C_{11})(Z,f,G^{0})+(-C_{0}+C_{11})(G^{0},f,Z)]
\notag \\ &\qquad
+\zeta_{A}(-s_{\beta'}c_{\alpha}+2c_{\beta'}s_{\alpha})
[(2C_{0}+C_{11})(Z,f,A)+(-C_{0}+C_{11})(A,f,Z)]
\Big], \\
&16\pi^{2}\Gamma^{V_{2}, \mathrm{1PI}}_{hf\bar{f}}
\notag \\ &
=
%--- FVF ---
-\frac{2m_{f}^{2}}{v}\zeta_{h}\qty[
e^{2}Q_{f}^{2}(C_{0}+2C_{12})(f,\gamma,f)
+g_{Z}^{2}(v_{f}^{2}+a_{f}^{2})(C_{0}+2C_{12})(f,Z,f)
]
\notag \\ &\quad
-\frac{g^{2}}{2}\frac{m_{f'}^{2}}{v}\zeta_{h}(C_{0}+2C_{12})(f',W,f')
\notag \\ &\quad
%--- FSF ---
-\frac{m_{f}^{4}}{v^{3}}\zeta_{h}\Big[
\zeta_{h}^{2}(C_{0}+2C_{12})(f,h,f)+\zeta_{H}^{2}(C_{0}+2C_{12})(f,H,f)
\notag \\ &\qquad
+\zeta_{G^{0}}^{2}(C_{0}+2C_{12})(f,G^{0},f)+\zeta_{A}^{2}(C_{0}+2C_{12})(f,A,f)
\Big]
\notag \\ &\quad
-\frac{m_{f'}^{2}}{v^{3}}(m_{f}^{2}+m_{f'}^{2})\zeta_{h}\qty[
\zeta_{G^{\pm}}^{2}(C_{0}+2C_{12})(f',G^{\pm},f')
+\zeta_{H^{\pm}}^{2}(C_{0}+2C_{12})(f',H^{\pm},f')
]
\notag \\ &\quad
%--- VFV ---
+\sqrt{1+s_{\beta}^{2}}(c_{\beta'}c_{\alpha}+2s_{\beta'}s_{\alpha})
g_{Z}^{4}v(v_{f}^{2}+a_{f}^{2})C_{12}(Z,f,Z)
\notag \\ &\quad
+(c_{\beta}c_{\alpha}+\sqrt{2}s_{\beta}s_{\alpha})
\frac{g^{4}}{4}vC_{12}(W,f',W)
\notag \\ &\quad
%--- SFS ---
-\frac{m_{f}^{2}}{v^{2}}\Big\{
6\lambda_{hhh}\zeta_{h}^{2}C_{12}(h,f,h)
+2\lambda_{HHh}\zeta_{H}^{2}C_{12}(H,f,H)
\notag \\ &\qquad
+2\lambda_{Hhh}\zeta_{h}\zeta_{H}
\qty[C_{12}(H,f,h)+C_{12}(h,f,H)]
\notag \\ &\qquad
+2\lambda_{G^{0}G^{0}h}\zeta_{G^{0}}^{2}C_{12}(G^{0},f,G^{0})
+2\lambda_{AAh}\zeta_{A}^{2}C_{12}(A,f,A)
\notag \\ &\qquad
+\lambda_{AG^{0}h}\zeta_{G^{0}}\zeta_{A}
\qty[C_{12}(A,f,G^{0})+C_{12}(G^{0},f,A)]
\Big\}
\notag \\ &\quad
-\frac{m_{f}^{2}+m_{f'}^{2}}{v^{2}}\Big\{
\lambda_{G^{+}G^{-}h}\zeta_{G^{\pm}}^{2}
C_{12}(G^{\pm},f',G^{\pm})
+\lambda_{H^{+}H^{-}h}\zeta_{H^{\pm}}^{2}
C_{12}(H^{\pm},f',H^{\pm})
\notag \\ &\qquad
+\lambda_{H^{+}G^{-}h}\zeta_{G^{\pm}}\zeta_{H^{\pm}}
\qty[C_{12}(G^{\pm},f',H^{\pm})+C_{12}(H^{\pm},f',G^{\pm})]
\Big\}
\notag \\ &\quad
%--- VFS+SFV ---
+\frac{g^{2}}{4}\frac{m_{f'}^{2}}{v}\Big[
\zeta_{G^{\pm}}(c_{\beta}c_{\alpha}+\sqrt{2}s_{\beta}s_{\alpha})
\qty[(2C_{0}+C_{12})(W,f',G^{\pm})+(-C_{0}+C_{12})(G^{\pm},f',W)]
\notag \\ &\qquad
+\zeta_{H^{\pm}}(-s_{\beta}c_{\alpha}+\sqrt{2}c_{\beta}s_{\alpha})
\qty[(2C_{0}+C_{12})(W,f',H^{\pm})+(-C_{0}+C_{12})(H^{\pm},f',W)]
\Big]
\notag \\ &\quad
+\frac{g_{Z}^{2}}{8}\frac{m_{f}^{2}}{v}\Big[
\zeta_{G^{0}}(c_{\beta'}c_{\alpha}+2s_{\beta'}s_{\alpha})
\qty[(2C_{0}+C_{12})(Z,f,G^{0})+(-C_{0}+C_{12})(G^{0},f,Z)]
\notag \\ &\qquad
+\zeta_{A}(-s_{\beta'}c_{\alpha}+2c_{\beta'}s_{\alpha})
\qty[(2C_{0}+C_{12})(Z,f,A)+(-C_{0}+C_{12})(A,f,Z)]
\Big], \\
&16\pi^{2}\Gamma^{A_{1}, \mathrm{1PI}}_{hf\bar{f}}
\notag \\ &
=
%--- FVF ---
4g_{Z}^{2}v_{f}a_{f}\frac{m_{f}^{2}}{v}\zeta_{h}(C_{0}+2C_{11})(f,Z,f)
+\frac{g^{2}}{2}\frac{m_{f'}^{2}}{v}\zeta_h(C_{0}+2C_{11})(f',W,f')
\notag \\ &\quad
%--- FSF ---
-\frac{m_{f'}^{2}}{v^{3}}(m_{f}^{2}-m_{f'}^{2})\zeta_{h}\qty[
\zeta_{G^{\pm}}^{2}(C_{0}+2C_{11})(f',G^{\pm},f')
+\zeta_{H^{\pm}}^{2}(C_{0}+2C_{11})(f',H^{\pm},f')
]
\notag\\ &\quad
%--- VFV ---
-\sqrt{1+s_{\beta}^{2}}(c_{\beta'}c_{\alpha}+2s_{\beta'}s_{\alpha})
2g_{Z}^{4}v_{f}a_{f}v(C_{0}+C_{11})(Z,f,Z)
\notag\\ &\quad
-(c_{\beta}c_{\alpha}+\sqrt{2}s_{\beta}s_{\alpha})
\frac{g^{4}}{4}v(C_{0}+C_{11})(W,f',W)
\notag\\ &\quad
%--- SFS ---
-\frac{m_{f}^{2}-m_{f'}^{2}}{v^2}\Big[
\zeta_{G^{\pm}}^{2}\lambda_{G^{+}G^{-}h}(C_{0}+C_{11})(G^{\pm},f',G^{\pm})
+\zeta_{H^{\pm}}^{2}\lambda_{H^{+}H^{-}h}(C_{0}+C_{11})(H^{\pm},f',H^{\pm})
\notag\\ &\quad
+\zeta_{G^{\pm}}\zeta_{H^{\pm}}\lambda_{H^{+}G^{-}h}
\qty[(C_{0}+C_{11})(G^{\pm},f',H^{\pm})+(C_{0}+C_{11})(H^{\pm},f',G^{\pm})]
\Big]
\notag\\ &\quad
%--- VFS+SFV ---
-\frac{g^{2}}{4}\frac{m_{f'}^2}{v}\Big[
\zeta_{G^{\pm}}(c_{\beta}c_{\alpha}+\sqrt{2}s_{\beta}s_{\alpha})
\qty[(2C_{0}+C_{11})(W,f',G^{\pm})+(-C_{0}+C_{11})(G^{\pm},f',W)]
\notag\\ &\qquad
+\zeta_{H^{\pm}}(-s_{\beta}c_{\alpha}+\sqrt{2}c_{\beta}s_{\alpha})
\qty[(2C_{0}+C_{11})(W,f',H^{\pm})+(-C_{0}+C_{11})(H^{\pm},f',W)]
\Big]
\notag\\ &\quad
-2g_{Z}^{2}v_{f}a_{f}\frac{m_{f}^{2}}{v}\Big[
\zeta_{G^{0}}(c_{\beta'}c_{\alpha}+2s_{\beta'}s_{\alpha})
\qty[(2C_{0}+C_{11})(Z,f,G^{0})+(-C_{0}+C_{11})(G^{0},f,Z)]
\notag\\ &\qquad
+\zeta_{A}(-s_{\beta'}c_{\alpha}+2c_{\beta'}s_{\alpha})
\qty[(2C_{0}+C_{11})(Z,f,A)+(-C_{0}+C_{11})(A,f,Z)]
\Big], \\
&16\pi^{2}\Gamma^{A_{2}, \mathrm{1PI}}_{hf\bar{f}}
\notag \\ &
=
%--- FVF ---
4g_{Z}^{2}v_{f}a_{f}\frac{m_{f}^{2}}{v}\zeta_{h}(C_{0}+2C_{12})(f,Z,f)
+\frac{g^{2}}{2}\frac{m_{f'}^{2}}{v}\zeta_h(C_{0}+2C_{12})(f',W,f')
\notag \\ &\quad
%--- FSF ---
-\frac{m_{f'}^{2}}{v^{3}}(m_{f}^{2}-m_{f'}^{2})\zeta_{h}\qty[
\zeta_{G^{\pm}}^{2}(C_{0}+2C_{12})(f',G^{\pm},f')
+\zeta_{H^{\pm}}^{2}(C_{0}+2C_{12})(f',H^{\pm},f')
]
\notag\\ &\quad
%--- VFV ---
-\sqrt{1+s_{\beta}^{2}}(c_{\beta'}c_{\alpha}+2s_{\beta'}s_{\alpha})
2g_{Z}^{4}v_{f}a_{f}vC_{12}(Z,f,Z)
\notag\\ &\quad
-(c_{\beta}c_{\alpha}+\sqrt{2}s_{\beta}s_{\alpha})
\frac{g^{4}}{4}vC_{12}(W,f',W)
\notag\\ &\quad
%--- SFS ---
-\frac{m_{f}^{2}-m_{f'}^{2}}{v^2}\Big[
\zeta_{G^{\pm}}^{2}\lambda_{G^{+}G^{-}h}C_{12}(G^{\pm},f',G^{\pm})
+\zeta_{H^{\pm}}^{2}\lambda_{H^{+}H^{-}h}C_{12}(H^{\pm},f',H^{\pm})
\notag\\ &\quad
+\zeta_{G^{\pm}}\zeta_{H^{\pm}}\lambda_{H^{+}G^{-}h}
\qty[C_{12}(G^{\pm},f',H^{\pm})+C_{12}(H^{\pm},f',G^{\pm})]
\Big]
\notag\\ &\quad
%--- VFS+SFV ---
-\frac{g^{2}}{4}\frac{m_{f'}^2}{v}\Big[
\zeta_{G^{\pm}}(c_{\beta}c_{\alpha}+\sqrt{2}s_{\beta}s_{\alpha})
\qty[(2C_{0}+C_{12})(W,f',G^{\pm})+(-C_{0}+C_{12})(G^{\pm},f',W)]
\notag\\ &\qquad
+\zeta_{H^{\pm}}(-s_{\beta}c_{\alpha}+\sqrt{2}c_{\beta}s_{\alpha})
\qty[(2C_{0}+C_{12})(W,f',H^{\pm})+(-C_{0}+C_{12})(H^{\pm},f',W)]
\Big]
\notag\\ &\quad
-2g_{Z}^{2}v_{f}a_{f}\frac{m_{f}^{2}}{v}\Big[
\zeta_{G^{0}}(c_{\beta'}c_{\alpha}+2s_{\beta'}s_{\alpha})
\qty[(2C_{0}+C_{12})(Z,f,G^{0})+(-C_{0}+C_{12})(G^{0},f,Z)]
\notag\\ &\qquad
+\zeta_{A}(-s_{\beta'}c_{\alpha}+2c_{\beta'}s_{\alpha})
\qty[(2C_{0}+C_{12})(Z,f,A)+(-C_{0}+C_{12})(A,f,Z)]
\Big], \\
&\qty(\frac{m_{f}}{v})^{-1}16\pi^{2}\Gamma^{T, \mathrm{1PI}}_{hf\bar{f}}
\notag \\ &
=
%--- FSF ---
-\frac{m_{f}^{2}}{v^{2}}\zeta_{h}\Big[
\zeta_{h}^{2}(C_{11}-C_{12})(f,h,f)+\zeta_{H}^{2}(C_{11}-C_{12})(f,H,f)
\notag\\ &\qquad
-\zeta_{G^{0}}^{2}(C_{11}-C_{12})(f,G^{0},f)-\zeta_{A}^{2}(C_{11}-C_{12})(f,A,f)
\Big]
\notag\\ &\quad
+\frac{2m_{f'}^{2}}{v^{2}}\zeta_{h}\Big[
\zeta_{G^{\pm}}^{2}(C_{11}-C_{12})(f',G^{\pm},f')
+\zeta_{H^{\pm}}^{2}(C_{11}-C_{12})(f',H^{\pm},f')
\Big]
\notag\\ &\quad
%--- VFS+SFV ---
-\frac{g^{2}}{4}\Big[
\zeta_{G^{\pm}}(c_{\beta}c_{\alpha}+\sqrt{2}s_{\beta}s_{\alpha})
[(2C_{0}+2C_{11}-C_{12})(W,f',G^{\pm})
+(C_{0}+C_{11}-2C_{12})(G^{\pm},f',W)]
\notag\\ &\qquad
+\zeta_{H^{\pm}}(-s_{\beta}c_{\alpha}+\sqrt{2}c_{\beta}s_{\alpha})
[(2C_{0}+2C_{11}-C_{12})(W,f',H^{\pm})
+(C_{0}+C_{11}-2C_{12})(H^{\pm},f',W)]
\Big]
\notag\\ &\quad
-\frac{g_{Z}^{2}}{8}\Big[
\zeta_{G^{0}}(c_{\beta'}c_{\alpha}+2s_{\beta'}s_{\alpha})
[(2C_{0}+2C_{11}-C_{12})(Z,f,G^{0})
+(C_{0}+C_{11}-2C_{12})(G^{0},f,Z)]
\notag\\ &\qquad
+\zeta_{A}(-s_{\beta'}c_{\alpha}+2c_{\beta'}s_{\alpha})
[(2C_{0}+2C_{11}-C_{12})(Z,f,A)
+(C_{0}+C_{11}-2C_{12})(A,f,Z)]
\Big], \\
&\qty(\frac{m_{f}}{v})^{-1}16\pi^{2}\Gamma^{PT, \mathrm{1PI}}_{hf\bar{f}}
\notag \\ &
=
%--- VFS+SFV ---
\frac{g^{2}}{4}\Big[
\zeta_{G^{\pm}}(c_{\beta}c_{\alpha}+\sqrt{2}s_{\beta}s_{\alpha})
[(2C_{0}+2C_{11}-C_{12})(W,f',G^{\pm})
-(C_{0}+C_{11}-2C_{12})(G^{\pm},f',W)]
\notag\\ &\qquad
+\zeta_{H^{\pm}}(-s_{\beta}c_{\alpha}+\sqrt{2}c_{\beta}s_{\alpha})
[(2C_{0}+2C_{11}-C_{12})(W,f',H^{\pm})
-(C_{0}+C_{11}-2C_{12})(H^{\pm},f',W)]
\Big]
\notag\\ &\quad
+2g_{Z}^{2}v_{f}a_{f}\Big[
\zeta_{G^{0}}(c_{\beta'}c_{\alpha}+2s_{\beta'}s_{\alpha})
[(2C_{0}+2C_{11}-C_{12})(Z,f,G^{0})
-(C_{0}+C_{11}-2C_{12})(G^{0},f,Z)]
\notag\\ &\qquad
+\zeta_{A}(-s_{\beta'}c_{\alpha}+2c_{\beta'}s_{\alpha})
[(2C_{0}+2C_{11}-C_{12})(Z,f,A)
-(C_{0}+C_{11}-2C_{12})(A,f,Z)]
\Big],
\end{align}
where the loop functions are defined as~\cite{Kanemura:2015mxa}
\begin{align}
&C_{hff}^{FVF}(X,Y,Z)
\notag\\
&=
[m_{f}^{2}C_{0}+p_{1}^{2}(C_{11}+C_{21})+p_{2}^{2}(C_{12}+C_{22})
+p_{1}\cdot p_{2}(2C_{23}-C_{0})+4C_{24}](X,Y,Z)-1,\\
&C_{hff}^{FSF}(X,Y,Z)
\notag \\
&=
[m_{f}^{2}C_{0}+p_{1}^{2}(C_{11}+C_{21})+p_{2}^{2}(C_{12}+C_{22})
+2p_{1}\cdot p_{2}(C_{11}+C_{23})+4C_{24}](X,Y,Z)-\frac{1}{2}, \\
&C_{hff}^{VFS}(X,Y,Z)
\notag\\
&=
[p_{1}^{2}(2C_{0}+3C_{11}+C_{21})+p_{2}^{2}(2C_{12}+C_{22})
+2p_{1}\cdot p_{2}(2C_{12}+C_{23})+4C_{24}](X,Y,Z)-\frac{1}{2}, \\
&C_{hff}^{SFV}(X,Y,Z)
\notag\\
&=
[p_{1}^{2}(-C_{0}+C_{21})+p_{2}^{2}(-C_{12}+C_{22})+2p_{1}\cdot p_{2}(C_{23}+C_{12}-C_{0}-C_{11})+4C_{24}](X,Y,Z)
\notag\\ &\quad
-\frac{1}{2}. 
\end{align}

%%%%%%%%%%%%%%%%%%%%%%%%%%%%%%%%%%%%%%%%%%%%%%%%%%%%%%%%%%%%%%%%%%
\section{1PI diagrams for \texorpdfstring{$h\mathcal{V}\mathcal{V'}$}{} vertex}
\label{app: hVV'_1PI}
%%%%%%%%%%%%%%%%%%%%%%%%%%%%%%%%%%%%%%%%%%%%%%%%%%%%%%%%%%%%%%%%%%

The form factors of the 1PI $hgg$ vertex are obtained as
\begin{align}
&16\pi^{2}\Gamma_{hg^{a}g^{b}}^{1, \mathrm{1PI}}(p_{1}^{2}, p_{2}^{2}, q^{2})
\notag \\
&=
-\zeta_{h}\sum_{q}\frac{4g_{s}^{2}m_{q}^{2}}{v}\qty[
8C_{24}(q, q, q)-2B_{0}(q^{2}; q, q)+(p_{1}^{2}+p_{2}^{2}-q^{2})C_{0}(q, q, q)
]\delta^{ab}, \\
&16\pi^{2}\Gamma_{hg^{a}g^{b}}^{2, \mathrm{1PI}}(p_{1}^{2}, p_{2}^{2}, q^{2})
=
-\zeta_{h}\sum_{q}\frac{8g_{s}^{2}m_{q}^{2}}{v}q^{2}\qty[
C_{0}+4C_{1223}](q, q, q)\delta^{ab},
\end{align}
where $C_{1223}=C_{12}+C_{23}$.

The $h\gamma\gamma$ and $hZ\gamma$ vertices are decomposed into fermionic and bosonic loop contributions.
The form factors of the 1PI $h\gamma\gamma$ vertex are obtained as
\begin{align}
&16\pi^{2}\Gamma_{h\gamma\gamma}^{1, \mathrm{1PI}}(p_{1}^{2}, p_{2}^{2}, q^{2})_{F}
\notag \\
&=
-\zeta_{h}\sum_{f}N_{c}^{f}\frac{4e^{2}Q_{f}^{2}m_{f}^{2}}{v}\qty[
8C_{24}(f, f, f)-2B_{0}(q^{2}; f, f)+(p_{1}^{2}+p_{2}^{2}-q^{2})C_{0}(f, f, f)
], \\
&16\pi^{2}\Gamma_{h\gamma\gamma}^{2, \mathrm{1PI}}(p_{1}^{2}, p_{2}^{2}, q^{2})_{F}
=
-\zeta_{h}\sum_{f}N_{c}^{f}\frac{8e^{2}Q_{f}^{2}m_{f}^{2}}{v}\qty[
C_{0}+4C_{1223}](f, f, f), \\
&16\pi^{2}\Gamma_{h\gamma\gamma}^{1, \mathrm{1PI}}(p_{1}^{2}, p_{2}^{2}, q^{2})_{B}
\notag \\
&=
\frac{2e^{2}m_{W}^{2}}{v}(c_{\beta}c_{\alpha}+\sqrt{2}s_{\beta}s_{\alpha})
\notag \\ &\quad \times
\Big\{
(5p_{1}^{2}+5p_{2}^{2}-7q^{2})C_{0}(W, W, W)
-6[B_{0}(q^{2}; W, W)-4C_{24}(W, W, W)]
\notag \\ &\quad
-\frac{m_{h}^{2}}{m_{W}^{2}}[B_{0}(q^{2}; W, W)-4C_{24}(W, W, W)]-m_{h}^{2}C_{0}(W, W, W)
\Big\}
\notag \\ &\quad
-8e^{2}\lambda_{hH^{++}H^{--}}[4C_{24}(H^{\pm\pm}, H^{\pm\pm}, H^{\pm\pm})-B_{0}(q^{2}; H^{\pm\pm}, H^{\pm\pm})]
\notag \\ &\quad
-2e^{2}\lambda_{hH^{+}H^{-}}[4C_{24}(H^{\pm}, H^{\pm}, H^{\pm})-B_{0}(q^{2}; H^{\pm}, H^{\pm})], \\
&16\pi^{2}\Gamma_{h\gamma\gamma}^{2, \mathrm{1PI}}(p_{1}^{2}, p_{2}^{2}, q^{2})_{B}
\notag \\
&=
\frac{8e^{2}m_{W}^{2}}{v}(c_{\beta}c_{\alpha}+\sqrt{2}s_{\beta}s_{\alpha})q^{2}\qty[
4C_{0}+6C_{1223}+\frac{m_{h}^{2}}{m_{W}^{2}}C_{1223}](W, W, W)
\notag \\ &\quad
-32e^{2}\lambda_{hH^{++}H^{--}}q^{2}C_{1223}(H^{\pm\pm}, H^{\pm\pm}, H^{\pm\pm})
-8e^{2}\lambda_{hH^{+}H^{-}}q^{2}C_{1223}(H^{\pm}, H^{\pm}, H^{\pm}).
\end{align}

The form factors of the 1PI $hZ\gamma$ vertex are obtained as
\begin{align}
&16\pi^{2}\Gamma_{hZ\gamma}^{1, \mathrm{1PI}}(p_{1}^{2}, p_{2}^{2}, q^{2})_{F}
\notag \\
&=
-\zeta_{h}\sum_{f}N_{c}^{f}v_{f}Q_{f}\frac{4eg_{Z}m_{f}^{2}}{v}\qty[
8C_{24}(f, f, f)-2B_{0}(q^{2}; f, f)+(p_{1}^{2}+p_{2}^{2}-q^{2})C_{0}(f, f, f)
], \\
&16\pi^{2}\Gamma_{hZ\gamma}^{2, \mathrm{1PI}}(p_{1}^{2}, p_{2}^{2}, q^{2})_{F}
=
-\zeta_{h}\sum_{f}N_{c}^{f}v_{f}Q_{f}\frac{8eg_{Z}m_{f}^{2}}{v}\qty[
C_{0}+4C_{1223}](f, f, f), \\
&16\pi^{2}\Gamma_{hZ\gamma}^{1, \mathrm{1PI}}(p_{1}^{2}, p_{2}^{2}, q^{2})_{B}
\notag \\
&=
%==============
%   SM-like part
%==============
%--- VVV ---
eg^{2}m_{W}c_{W}(c_{\alpha}c_{\beta}+\sqrt{2}s_{\alpha}s_{\beta})
\left[
2C^{VVV}_{hVV1}(W, W, W)
-6B_{0}(q^{2}; W, W)+4
-2C_{24}(c^{\pm}, c^{\pm}, c^{\pm})
\right]
\notag \\ &\quad
%--- SVV ---
+egg_{Z}m_{W}(s_{W}^{2}+s_{\beta}^{2})
(c_{\alpha}c_{\beta}+\sqrt{2}s_{\alpha}s_{\beta})
\notag \\ &\qquad
\times\left[
C^{SVV}_{hVV1}(G^{\pm}, W, W)
+2m_{W}^{2}C_{0}(W, G^{\pm}, W)
-2C_{24}(W, G^{\pm}, G^{\pm})
+B_{0}(p_{1}^{2}; G^{\pm}, W)
\right]
\notag \\ &\quad
%--- SSV ---
+egg_{Z}m_{W}(c_{2W}-s_{\beta}^{2})
(c_{\alpha}c_{\beta}+\sqrt{2}s_{\alpha}s_{\beta})
C_{24}(G^{\pm}, G^{\pm}, W)
\notag \\ &\quad
%--- VVS ---
-eg^{2}m_{W}c_{W}(c_{\alpha}c_{\beta}+\sqrt{2}s_{\alpha}s_{\beta})
C^{VVS}_{hVV1}(W, W, G^{\pm})
\notag \\ &\quad
%--- SVS ---
-2eg_{Z}m_{W}^{2}(s_{W}^{2}+s_{\beta}^{2})\lambda_{G^{+}G^{-}h}
C_{0}(G^{\pm}, W, G^{\pm})
\notag \\ &\quad
%--- SV2 ---
-egg_{Z}m_{W}\qty[-c_{\alpha}c_{\beta}s_{W}^{2}+\sqrt{2}s_{\alpha}s_{\beta}(c_{W}^{2}-2)]
B_{0}(p_{2}^{2}; G^{\pm}, W)
\notag \\ &\quad
%--- SSS ---
-4eg_{Z}(c_{2W}-s_{\beta}^{2})\lambda_{G^{+}G^{-}h}
\left[
C_{24}(G^{\pm}, G^{\pm}, G^{\pm})
-\frac{1}{4}B_{0}(q^{2}; G^{\pm}, G^{\pm})
\right]
\notag \\ &\quad
%==============
%   BSM part
%==============
%--- SVV ---
+egg_{Z}m_{W}s_{\beta}c_{\beta}
(-c_{\alpha}s_{\beta}+\sqrt{2}s_{\alpha}c_{\beta})
\left[
C^{SVV}_{hVV1}(H^{\pm}, W, W)
-2C_{24}(W, H^{\pm}, H^{\pm})
\right]
\notag \\ &\quad
%--- SVS ---
-2eg_{Z}m_{W}^{2}s_{\beta}c_{\beta}\lambda_{H^{+}G^{-}h}
C_{0}(H^{\pm}, W, G^{\pm})
\notag \\ &\quad
%--- SSV ---
-egg_{Z}m_{W}s_{\beta}c_{\beta}
(-c_{\alpha}s_{\beta}+\sqrt{2}s_{\alpha}c_{\beta})
\left[
C_{24}(H^{\pm}, G^{\pm}, W)
-B_{0}(p_{1}^{2}; H^{\pm}, W)
\right]
\notag \\ &\quad
%--- SSS+SS ---
-4eg_{Z}(c_{2W}-c_{\beta}^{2})\lambda_{H^{+}H^{-}h}
\left[
C_{24}(H^{\pm}, H^{\pm}, H^{\pm})-\frac{1}{4}B_{0}(q^{2}; H^{\pm}, H^{\pm})
\right]
\notag \\ &\quad
-16eg_{Z}c_{2W}\lambda_{H^{++}H^{--}h}
\left[
C_{24}(H^{\pm\pm}, H^{\pm\pm}, H^{\pm\pm})
-\frac{1}{4}B_{0}(q^{2}; H^{\pm\pm}, H^{\pm\pm})
\right]
\notag \\ &\quad
+4eg_{Z}s_{\beta}c_{\beta}\lambda_{H^{+}G^{-}h}
\left[
C_{24}(H^{\pm}, G^{\pm}, G^{\pm})
+C_{24}(G^{\pm}, H^{\pm}, H^{\pm})
-\frac{1}{2}B_{0}(q^{2}; H^{\pm}, G^{\pm})
\right], \\
&16\pi^{2}\Gamma_{hZ\gamma}^{2, \mathrm{1PI}}(p_{1}^{2}, p_{2}^{2}, q^{2})_{B}
\notag \\
&=
%==============
%   SM-like part
%==============
%--- VVV+ccc+VVS ---
eg^{2}m_{W}c_{W}(c_{\alpha}c_{\beta}+\sqrt{2}s_{\alpha}s_{\beta})
\left[
2C^{VVV}_{hVV2}(W, W, W)
-2C_{1223}(c^{\pm}, c^{\pm}, c^{\pm})
-C^{VVS}_{hVV2}(W, W, G^{\pm})
\right]
\notag \\ &\quad
%--- SVV+VSS ---
+egg_{Z}m_{W}(s_{W}^{2}+s_{\beta}^{2})
(c_{\alpha}c_{\beta}+\sqrt{2}s_{\alpha}s_{\beta})
\left[
C^{SVV}_{hVV2}(G^{\pm}, W, W)
-2C^{VSS}_{hVV2}(W, G^{\pm}, G^{\pm})
\right]
\notag \\ &\quad
%--- SSV ---
+egg_{Z}m_{W}(c_{2W}-s_{\beta}^{2})
(c_{\alpha}c_{\beta}+\sqrt{2}s_{\alpha}s_{\beta})
C^{SSV}_{hVV2}(G^{\pm}, G^{\pm}, W)
\notag \\ &\quad
%--- SSS ---
-4eg_{Z}(c_{2W}-s_{\beta}^{2})\lambda_{G^{+}G^{-}h}
C_{1223}(G^{\pm}, G^{\pm}, G^{\pm})
\notag \\ &\quad
%==============
%   BSM part
%==============
%--- SVV+SSV+VSS ---
+egg_{Z}m_{W}s_{\beta}c_{\beta}
(-c_{\alpha}s_{\beta}+\sqrt{2}s_{\alpha}c_{\beta})
\notag \\ &\qquad
\times\left[
C^{SVV}_{hVV2}(H^{\pm}, W, W)
-C^{SSV}_{hVV2}(H^{\pm}, G^{\pm}, W)
-2C^{VSS}_{hVV2}(W, H^{\pm}, H^{\pm})
\right]
\notag \\ &\quad
%--- SSS ---
-4eg_{Z}(c_{2W}-c_{\beta}^{2})\lambda_{H^{+}H^{-}h}
C_{1223}(H^{\pm}, H^{\pm}, H^{\pm})
\notag \\ &\quad
-16eg_{Z}c_{2W}\lambda_{H^{++}H^{--}h}
C_{1223}(H^{\pm\pm}, H^{\pm\pm}, H^{\pm\pm})
\notag \\ &\quad
+4eg_{Z}s_{\beta}c_{\beta}\lambda_{H^{+}G^{-}h}\left[
C_{1223}(H^{\pm}, G^{\pm}, G^{\pm})
+C_{1223}(G^{\pm}, H^{\pm}, H^{\pm})
\right],
\end{align}
where the loop functions are defined as~\cite{Kanemura:2015mxa}
\begin{align}
C_{hVV1}^{VVV}(X,Y,Z)
&=
\Big[18C_{24}+p_{1}^{2}(2C_{21}+3C_{11}+C_{0})+p_{2}^{2}(2C_{22}+C_{12})
\notag \\ &\quad
+p_{1}\cdot p_{2}(4C_{23}+3C_{12}+C_{11}-4C_{0})\Big](X,Y,Z)-3, \\
C_{hVV2}^{VVV}(X,Y,Z)&= \left(10C_{23}+9C_{12}+C_{11}+5C_0\right)(X,Y,Z), \\
%%%
C_{hVV1}^{SVV}(X,Y,Z)
&=
\Big[3C_{24}+p_1^2(C_{21}-C_0)+p_2^2(C_{22}-2C_{12}+C_0)
\notag \\ &\qquad
+2p_1\cdot p_2 (C_{23}-C_{11})\Big](X,Y,Z)-\frac{1}{2}, \\
C_{hVV2}^{SVV}(X,Y,Z)&= \left(4C_{11}-3C_{12}-C_{23}\right)(X,Y,Z), \\
%%%
C_{hVV1}^{VVS}(X,Y,Z)
&=
\Big[3C_{24}+p_1^2(C_{21}+4C_{11}+4C_0)+p_2^2(C_{22}+2C_{12})
\notag \\ &\qquad
+2p_1\cdot p_2 (C_{23}+2C_{12}+C_{11}+2C_0)\Big](X,Y,Z)-\frac{1}{2}, \\
C_{hVV2}^{VVS}(X,Y,Z)&= \left(2C_{11}-5C_{12}-2C_0-C_{23}\right)(X,Y,Z), \\
%%%
C_{hVV2}^{SSV}(X,Y,Z)&= (C_{23}-C_{12})(X,Y,Z), \\
C_{hVV2}^{VSS}(X,Y,Z)&= (C_{23}+C_{12}+2C_{11}+2C_0)(X,Y,Z).
\end{align}

%%%%%%%%%%%%%%%%%%%%%%%%%%%%%%%%%%%%%%%%%%%%%%%%%%%%%%%%%%%%%%%%%%
\section{Box diagrams for the \texorpdfstring{$h\to VV^{*}$}{}}
\label{app: Box diagrams}
%%%%%%%%%%%%%%%%%%%%%%%%%%%%%%%%%%%%%%%%%%%%%%%%%%%%%%%%%%%%%%%%%%

%=================================================================
\subsection{Box diagrams for \texorpdfstring{$h\to Zf\bar{f}$}{}}
\label{app: hToZZstar box diagrams}
%=================================================================

\begin{figure}[t]
\centering
\begin{minipage}{0.3\hsize}
\centering
\includegraphics[scale=0.25]{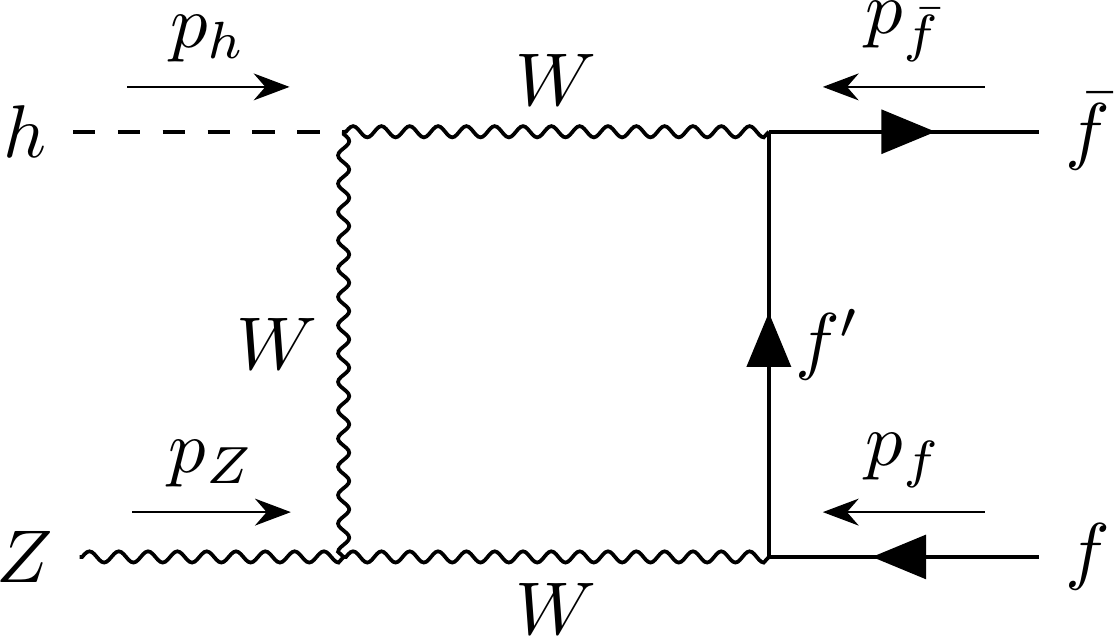}
\subcaption*{(1)}
\end{minipage}
\begin{minipage}{0.3\hsize}
\centering
\includegraphics[scale=0.25]{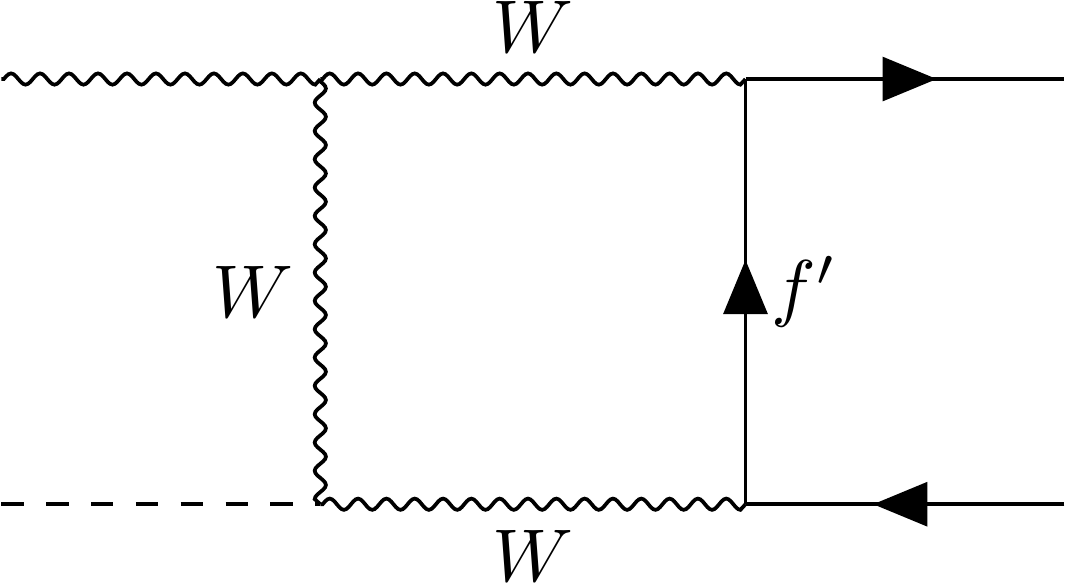}
\subcaption*{(2)}
\end{minipage}
\begin{minipage}{0.3\hsize}
\centering
\includegraphics[scale=0.25]{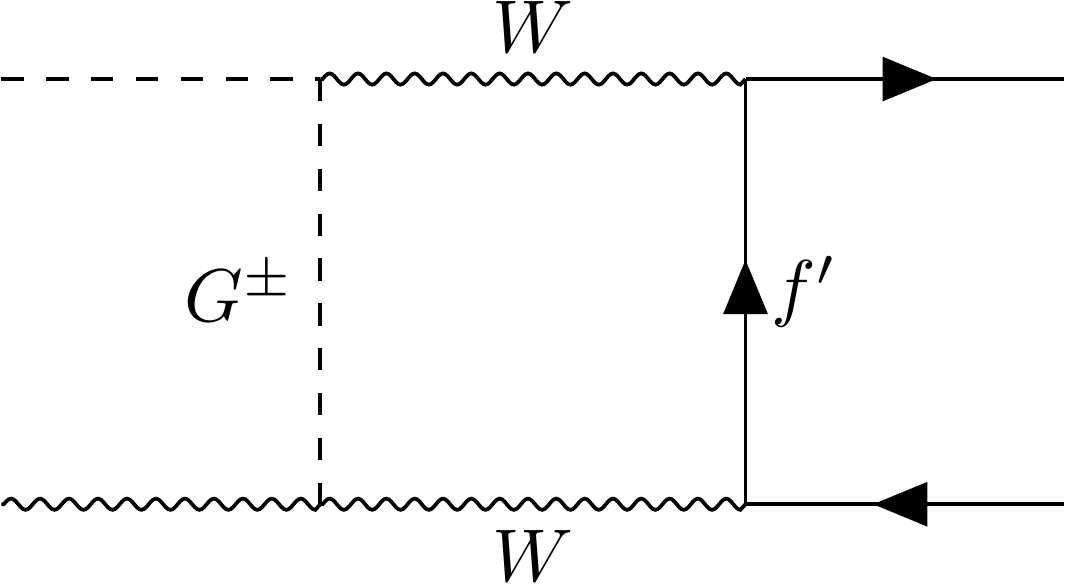}
\subcaption*{(3)}
\end{minipage} \\
%%%
\begin{minipage}{0.3\hsize}
\centering
\includegraphics[scale=0.25]{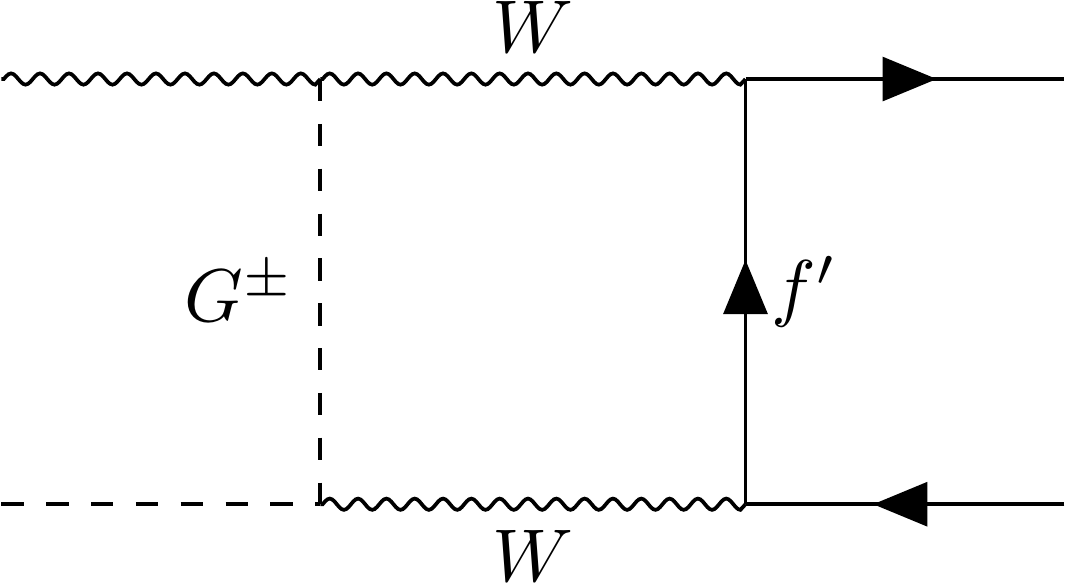}
\subcaption*{(4)}
\end{minipage}
\begin{minipage}{0.3\hsize}
\centering
\includegraphics[scale=0.25]{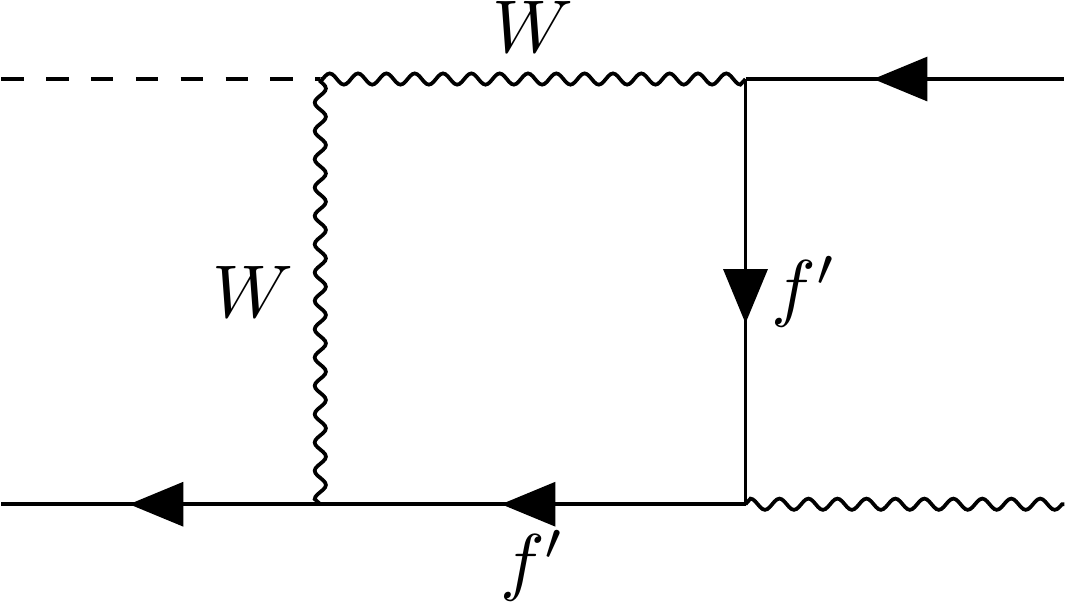}
\subcaption*{(5)}
\end{minipage}
\begin{minipage}{0.3\hsize}
\centering
\includegraphics[scale=0.25]{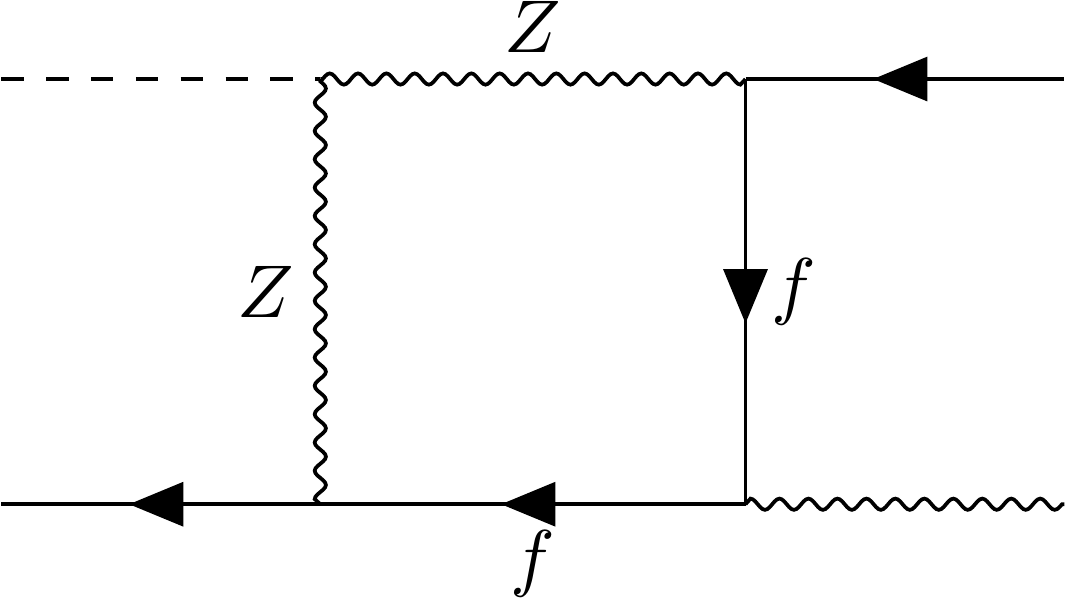}
\subcaption*{(6)}
\end{minipage} \\
%%%
\begin{minipage}{0.3\hsize}
\centering
\includegraphics[scale=0.25]{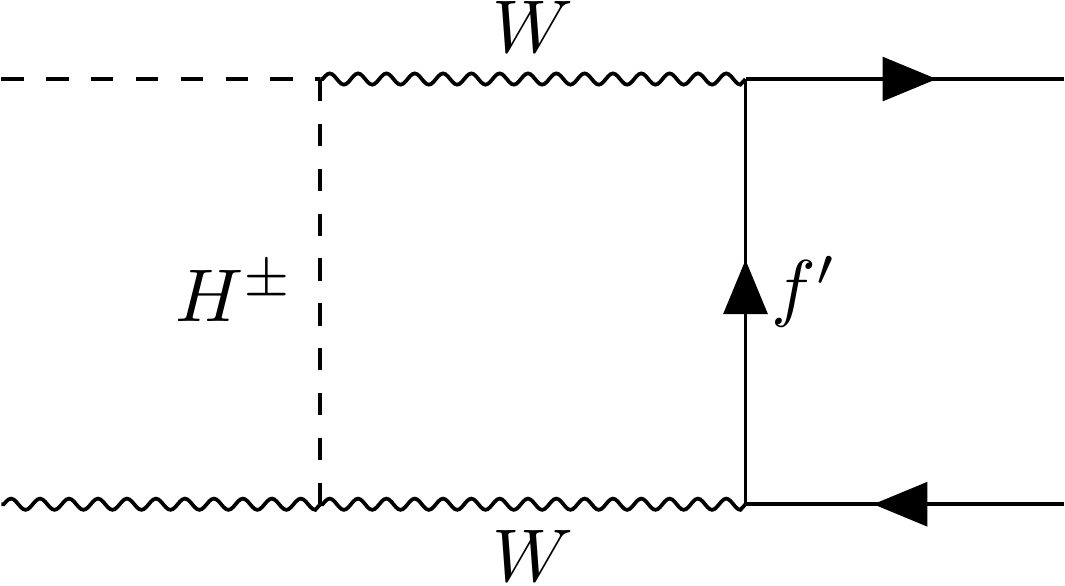}
\subcaption*{(7)}
\end{minipage}
\begin{minipage}{0.3\hsize}
\centering
\includegraphics[scale=0.25]{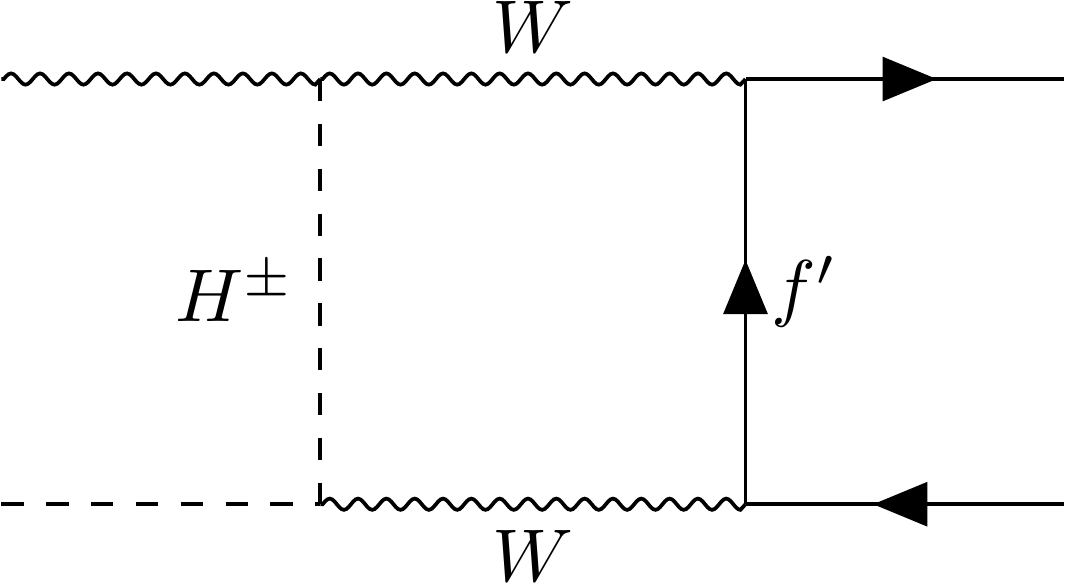}
\subcaption*{(8)}
\end{minipage}
\caption{Box diagrams for $h\to Zf\bar{f}$.}
\label{fig: Box_diagrams_hZZstar}
\end{figure}

We here give the analytic expressions for the box diagrams for $h\to Zf\bar{f}$.
For the calculation of box diagrams, we assign all momenta in the incoming way (see Diagram (1) in Fig.~\ref{fig: Box_diagrams_hZZstar}).
These incoming momenta are related to the outgoing momenta, such as $k_{Z} = -p_{Z},\ k_{\bar{f}} = -p_{f}$ and $k_{f} = -p_{\bar{f}}$.
Then, the Mandelstam variables are obtained as $s = (p_{f}+p_{\bar{f}})^{2},\, t = (p_{Z}+p_{f})^{2}$, and $u = (p_{Z}+p_{\bar{f}})^{2}$.
The LO amplitude is obtained as
\begin{align}
\mathcal{M}_{\mathrm{LO}}
=
C_{\mathrm{tree}}^{Z}
\overline{v}(p_{\bar{f}})\gamma^{\mu}(v_{f}-a_{f}\gamma_{5})u(p_{f})\epsilon_{\mu}^{*}(p_{Z}),
\end{align}
where $C_{\mathrm{tree}}^{Z}=g_{Z}\Gamma_{hZZ}^{1, \mathrm{tree}}/(s-m_{Z}^{2})$.

At the one-loop level, there are six SM-like diagrams and two diagrams with new particles as shown in Fig.~\ref{fig: Box_diagrams_hZZstar}.
The box amplitude can be parameterized as
\begin{align}
\mathcal{M}_{BZi} = \frac{1}{16\pi^{2}}\overline{v}(p_{\bar{f}})\mathcal{M}_{BZi}^{\mu}u(p_{f})\epsilon_{\mu}^{*}(p_{Z}),
\end{align}
where
\begin{align}
\mathcal{M}_{BZi}^{\mu}
&=
C^{BZi}\qty[\slashed{p}_{h}(F_{1,3}^{BZi}p_{f}^{\mu}+F_{1,4}^{BZi}p_{\bar{f}}^{\mu})+\gamma^{\mu}F_{\gamma}^{BZi}]P_{L}\qquad (i\neq6), \\
\mathcal{M}_{BZ6}^{\mu}
&=
C^{BZ6}\qty[\slashed{p}_{h}(F_{1,3}^{BZ6}p_{f}^{\mu}+F_{1,4}^{BZ6}p_{\bar{f}}^{\mu})+\gamma^{\mu}F_{\gamma}^{BZ6}](v_{f}-a_{f}\gamma_{5})^{3}.
\end{align}
Then, the NLO contribution is obtained as
\begin{align}
(16\pi^{2})2\Re\qty[\sum \mathcal{M}_{\mathrm{tree}}^{*}\mathcal{M}_{BZi}]
&=
4(v_{f}+a_{f})C_{\mathrm{tree}}^{Z}C^{BZi}B^{BZi}\qquad (i\neq6), \\
(16\pi^{2})2\Re\qty[\sum \mathcal{M}_{\mathrm{tree}}^{*}\mathcal{M}_{BZ6}]
&=
8(v_{f}^{4}+6v_{f}^{2}a_{f}^{2}+a_{f}^{4})C_{\mathrm{tree}}^{Z}C^{BZ6}B^{BZ6},
\end{align}
where
\begin{align}
B^{BZi} = sF_{\gamma}^{BZi}+\frac{tu-m_{Z}^{2}m_{h}^{2}}{4m_{Z}^{2}}\qty[2F_{\gamma}^{BZi}-(t-m_{Z}^{2})F_{1,3}^{BZi}-(u-m_{Z}^{2})F_{1,4}^{BZi}].
\end{align}
In the following, we use the shorthand notation for $D$ functions.
\begin{align}
&D_{i,\, ij}(p_{1}^{2}, p_{2}^{2}, p_{3}^{2}, p_{4}^{2}, p_{12}^{2}, p_{23}^{2}; X, Y, Z, W)
\notag \\
&=
D_{i,\, ij}(p_{1}^{2}, p_{2}^{2}, p_{3}^{2}, p_{123}^{2}, p_{12}^{2}, p_{23}^{2}; m_{X}^{2}, m_{Y}^{2}, m_{Z}^{2}, m_{W}^{2}),    
\end{align}
with $p_{123}^{2}=(p_{1}+p_{2}+p_{3})^{2},\, p_{12}^{2}=(p_{1}+p_{2})^{2},\, p_{23}^{2}=(p_{2}+p_{3})^{2}$.
The coefficients and the form factors of each diagram are obtained as follows.

\begin{itemize}[labelsep=-1.2cm]
\item[$\lbrack \text{Diagram (1)} \rbrack$]
\begin{align}
C^{BZ1}
&=
-2a_{f}g^{4}m_{W}c_{W}(c_{\beta}c_{\alpha}+\sqrt{2}s_{\beta}s_{\alpha}),
\notag \\
F^{BZ1}_{1,3}
&=
F_{p\cdot\gamma1}^{VFVV}(0, 0, m_{Z}^{2}, m_{h}^{2}, s, t; W, 0, W, W),
\notag \\
F^{BZ1}_{1,4}
&=
F_{p\cdot\gamma2}^{VFVV}(0, 0, m_{Z}^{2}, m_{h}^{2}, s, t; W, 0, W, W),
\notag \\
F^{BZ1}_{\gamma}
&=
F_{\gamma}^{VFVV}(0, 0, m_{Z}^{2}, m_{h}^{2}, s, t; W, 0, W, W),
\end{align}
\end{itemize}
where the loop functions are defined as
\begin{align}
F_{p\cdot\gamma1}^{VFVV}
&=
2(D_{13}-D_{12}+2D_{26}), \\
F_{p\cdot\gamma2}^{VFVV}
&=
4(D_{0}+D_{11}+D_{13}+D_{25}), \\
F_{\gamma}^{VFVV}
&=
-2C_{0}(p_{12}^{2}, p_{3}^{2}, p_{123}^{2}; m_{V_{1}}^{2}, m_{V_{3}}^{2}, m_{V_{4}}^{2})
-[4D_{27}+(2p_{12}^{2}+p_{23}^{2}-p_{123}^{2})(D_{0}+D_{11})
\notag \\ &\quad
+2(p_{23}^{2}-p_{3}^{2})D_{12}+p_{23}^{2}D_{13}],
\end{align}

The coefficient of Diagram (2) is the same as that of Diagram (1).
Its form factors are obtained by interchanging $F_{1,3}^{BZ1}$ and $F_{1,4}^{BZ1}$ and replacing $t \to u$.

\begin{itemize}[labelsep=-1.2cm]
\item[$\lbrack \text{Diagram (3)} \rbrack$]
\begin{align}
C^{BZ3}
&=
a_{f}g^{3}g_{Z}m_{W}(s_{W}^{2}+s_{\beta}^{2})(c_{\beta}c_{\alpha}+\sqrt{2}s_{\beta}s_{\alpha}),
\notag \\
F_{1,3}^{BZ3}
&=
F_{p\cdot\gamma}^{VFVS}(0, 0, m_{Z}^{2}, m_{h}^{2}, s, t; W, 0, W, G^{\pm}),
\notag \\
F_{1,4}^{BZ3} &= 0,
\notag \\
F_{\gamma}^{BZ3}
&=
F_{\gamma}^{VFVS}(0, 0, m_{Z}^{2}, m_{h}^{2}, s, t; W, 0, W, G^{\pm}),
\end{align}
\end{itemize}
where the loop functions are defined as
\begin{align}
F_{p\cdot\gamma}^{VFVS}
&=
-4(D_{12}-D_{13}), \\
F_{\gamma}^{VFVS}
&=
C_{0}(p_{12}^{2}, p_{3}^{2}, p_{123}^{2}; m_{V_{1}}^{2}, m_{V_{3}}^{2}, m_{S_{4}}^{2})
\notag \\ &\quad
+2\qty[-(p_{23}^{2}-p_{123}^{2})(D_{0}+D_{11})+p_{23}^{2}D_{13}].
\end{align}

The coefficient of Diagram (4) is the same as that of Diagram (3).
Its form factors are obtained by interchanging $F_{1,3}^{BZ3}$ and $F_{1,4}^{BZ3}$ and replacing $t \to u$.

\begin{itemize}[labelsep=-1.2cm]
\item[$\lbrack \text{Diagram (5)} \rbrack$]
\begin{align}
C^{BZ5}
&=
-(v_{f'}+a_{f'})g^{3}g_{Z}m_{W}(c_{\beta}c_{\alpha}+\sqrt{2}s_{\beta}s_{\alpha}),
\notag \\
F_{1,3}^{BZ5}
&=
F_{p\cdot\gamma1}^{VFFV}(0, m_{Z}^{2}, 0, m_{h}^{2}, t, u; W, 0, 0, W),
\notag \\
F_{1,4}^{BZ5}
&=
F_{p\cdot\gamma2}^{VFFV}(0, m_{Z}^{2}, 0, m_{h}^{2}, t, u; W, 0, 0, W),
\notag \\
F_{\gamma}^{BZ5}
&=
F_{\gamma}^{VFFV}(0, m_{Z}^{2}, 0, m_{h}^{2}, t, u; W, 0, 0, W).
\end{align}
\end{itemize}
where the loop functions are defined as
\begin{align}
F_{p\cdot\gamma1}^{VFFV}
&=
-2(D_{0}+D_{11}+D_{12}+D_{24}), \\
F_{p\cdot\gamma2}^{VFFV}
&=
-2D_{26}, \\
F_{\gamma}^{VFFV}
&=
-C_{0}(p_{12}^{2}, p_{3}^{2}, p_{123}^{2}; m_{V_{1}}^{2}, m_{F_{3}}^{2}, m_{V_{4}}^{2})
\notag \\ &\quad
-\qty[-2D_{27}+(p_{12}^{2}-p_{2}^{2})(D_{0}+D_{11})+p_{2}^{2}D_{12}].
\end{align}

The coefficient of Diagram (6) is obtained as
\begin{align}
C^{BZ6}
&=
-2g_{Z}^{4}m_{Z}(c_{\beta'}c_{\alpha}+2s_{\beta'}s_{\alpha}).
\end{align}
Its form factors are obtained from those of Diagram (5) by replacing $W \to Z$.

\begin{itemize}[labelsep=-1.2cm]
\item[$\lbrack \text{Diagram (7)} \rbrack$]
\begin{align}
C^{BZ7}
&=
a_{f}g^{3}g_{Z}m_{W}s_{\beta}c_{\beta}(-s_{\beta}c_{\alpha}+\sqrt{2}c_{\beta}s_{\alpha}),
\notag \\
F_{1,3}^{BZ7}
&=
F_{p\cdot\gamma}^{VFVS}(0, 0, m_{Z}^{2}, m_{h}^{2}, s, t; W, 0, W, H^{\pm})
\notag \\
F_{1,4}^{BZ7} &= 0, \notag\\
F_{\gamma}^{BZ7}
&=
F_{\gamma}^{VFVS}(0, 0, m_{Z}^{2}, m_{h}^{2}, s, t; W, 0, W, H^{\pm}).
\end{align}
\end{itemize}

The coefficient of Diagram (8) is the same as that of Diagram (7).
Its form factors are obtained by interchanging $F_{1,3}^{BZ7}$ and $F_{1,4}^{BZ7}$ and replacing $t \to u$.

%=================================================================
\subsection{Box diagrams for \texorpdfstring{$h\to W^{-}f'\bar{f}$}{}}
%=================================================================

\begin{figure}[t]
\centering
\begin{minipage}{0.225\hsize}
\centering
\includegraphics[scale=0.18]{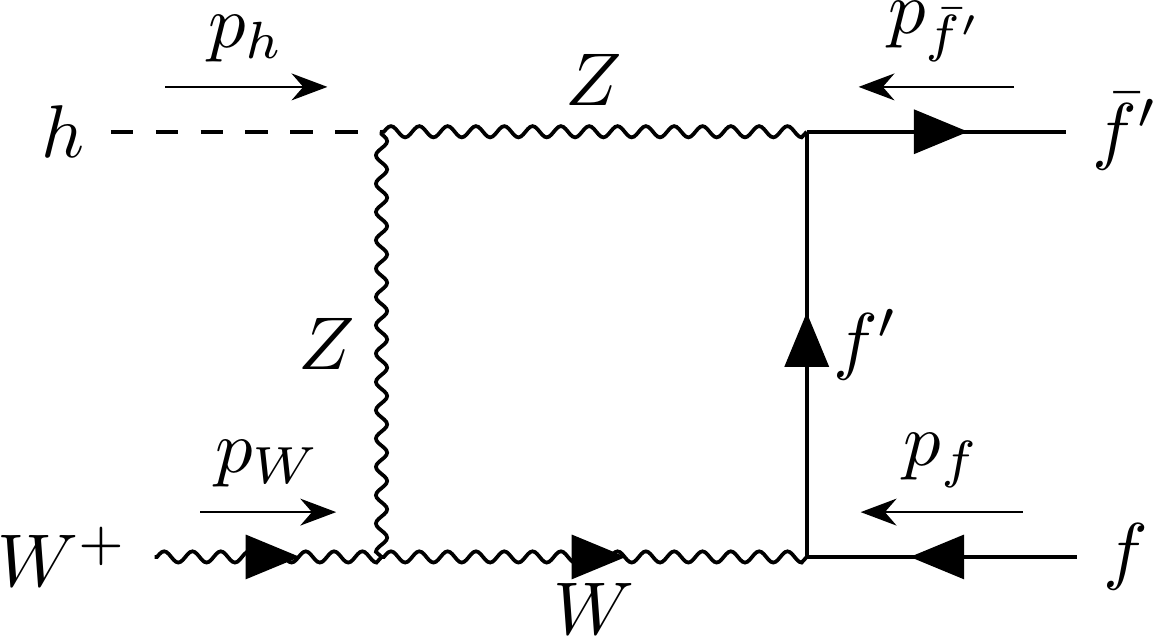}
\subcaption*{(1)}
\end{minipage}
\begin{minipage}{0.225\hsize}
\centering
\includegraphics[scale=0.18]{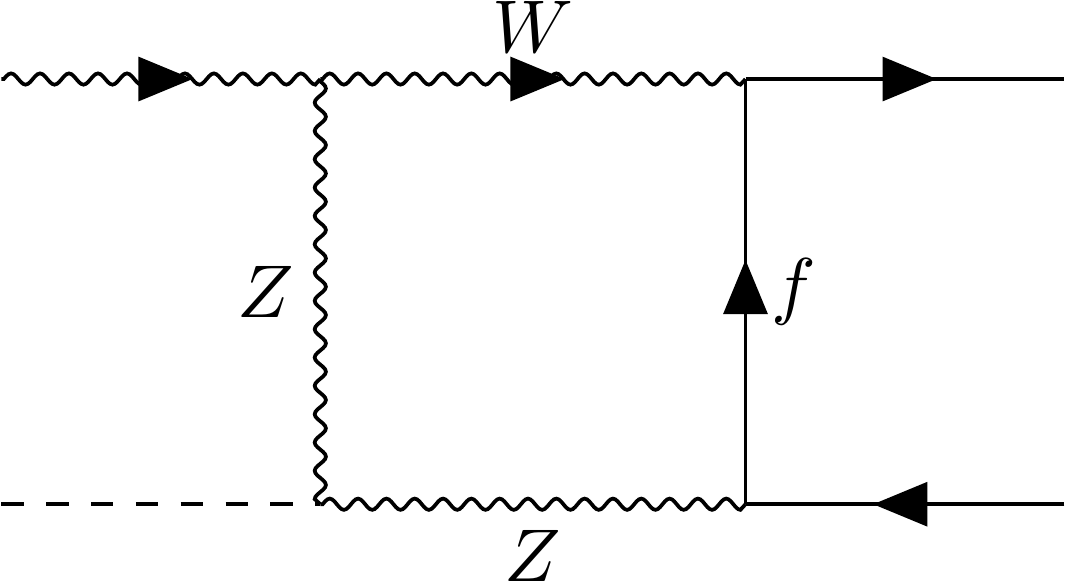}
\subcaption*{(2)}
\end{minipage}
\begin{minipage}{0.225\hsize}
\centering
\includegraphics[scale=0.18]{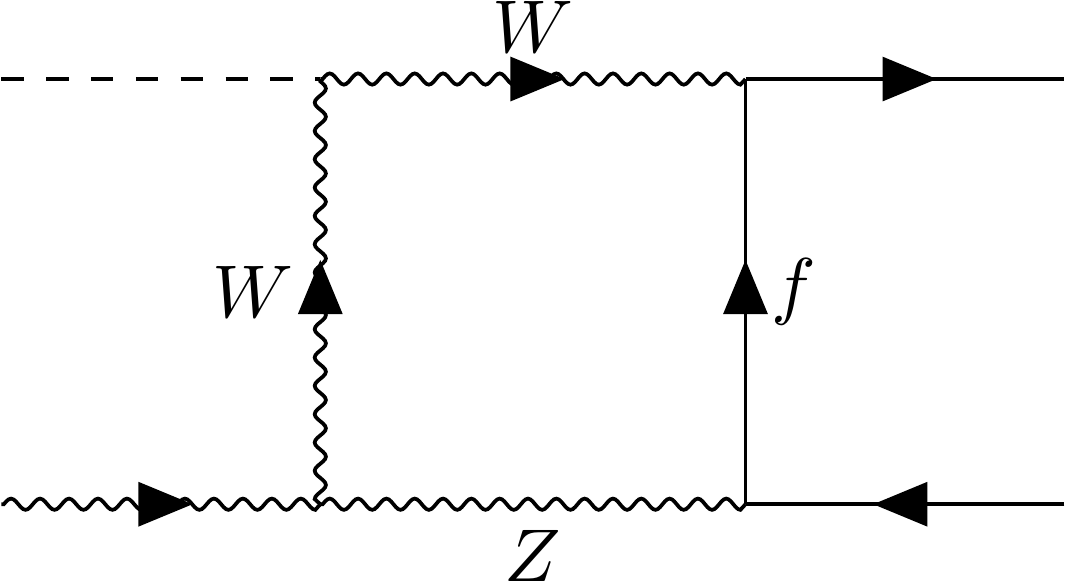}
\subcaption*{(3)}
\end{minipage}
\begin{minipage}{0.225\hsize}
\centering
\includegraphics[scale=0.18]{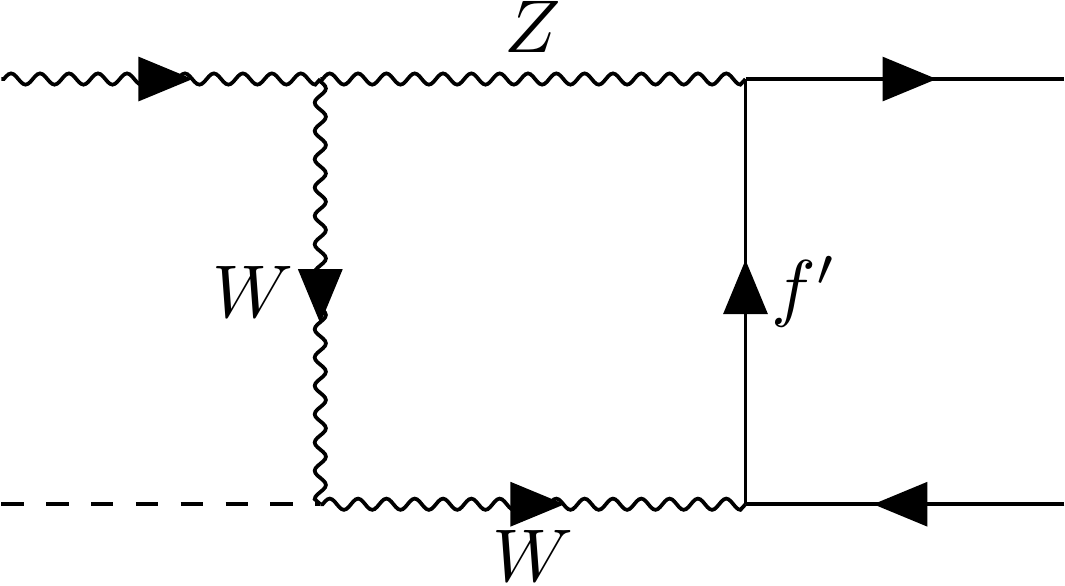}
\subcaption*{(4)}
\end{minipage} \\
%%%
\begin{minipage}{0.225\hsize}
\centering
\includegraphics[scale=0.18]{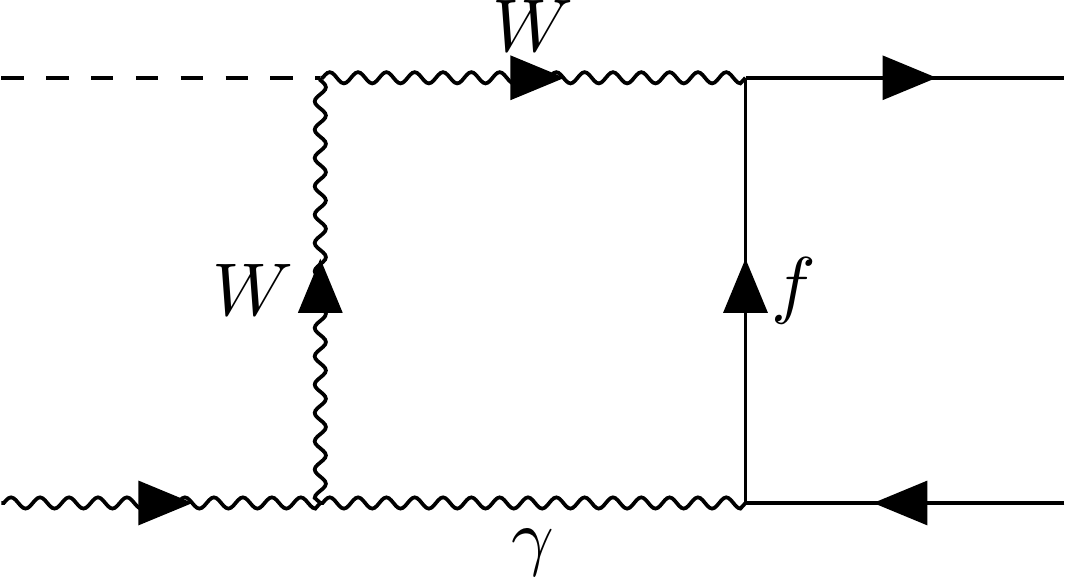}
\subcaption*{(5)}
\end{minipage}
\begin{minipage}{0.225\hsize}
\centering
\includegraphics[scale=0.18]{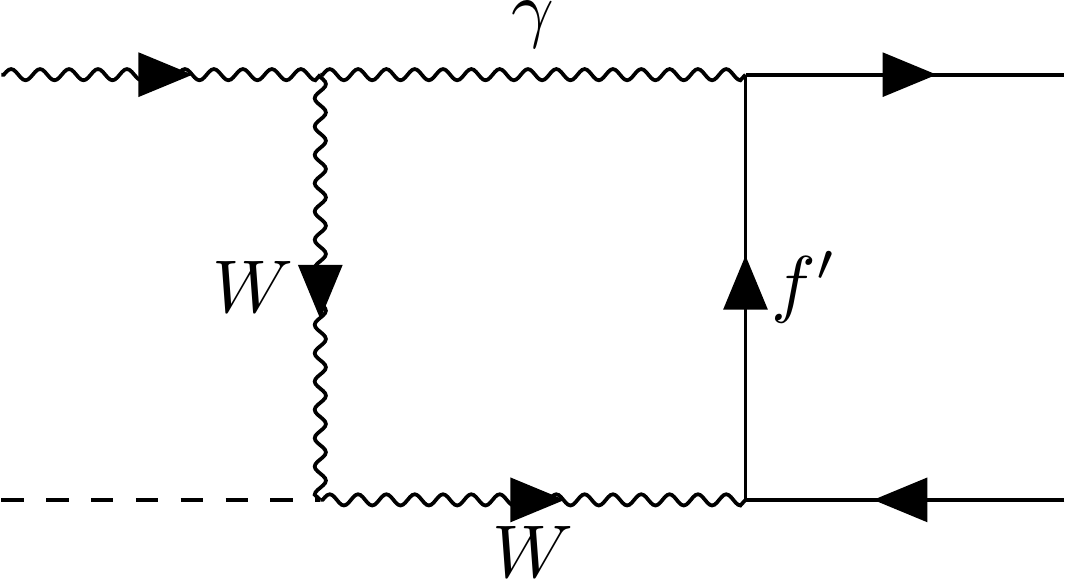}
\subcaption*{(6)}
\end{minipage}
\begin{minipage}{0.225\hsize}
\centering
\includegraphics[scale=0.18]{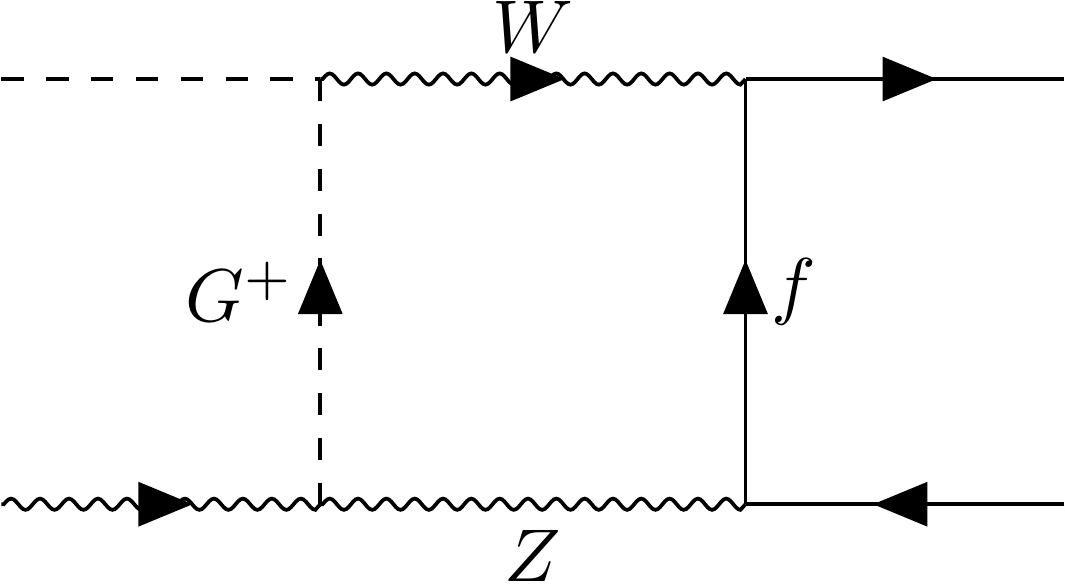}
\subcaption*{(7)}
\end{minipage}
\begin{minipage}{0.225\hsize}
\centering
\includegraphics[scale=0.18]{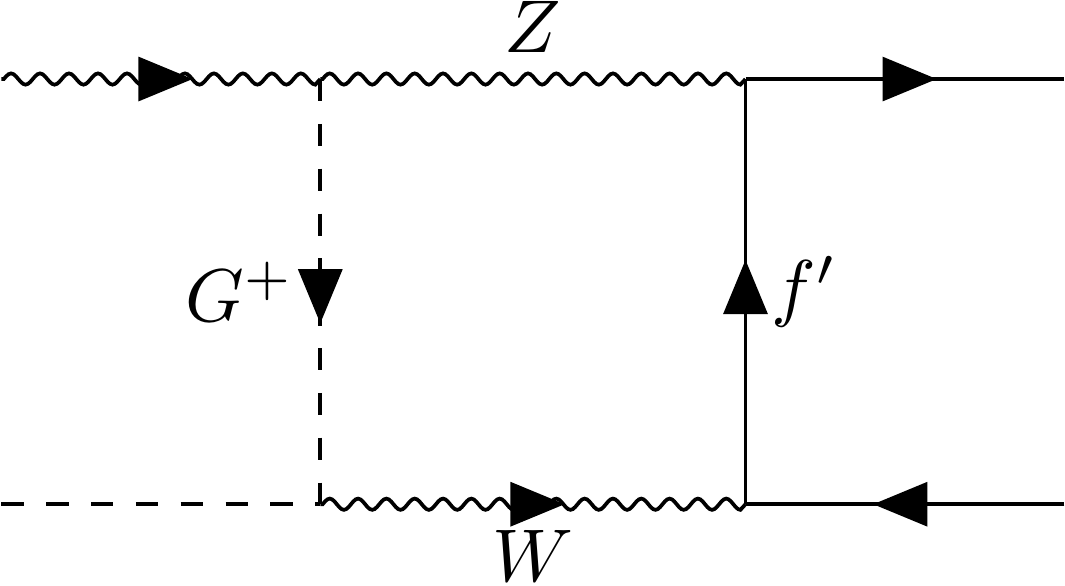}
\subcaption*{(8)}
\end{minipage} \\
%%%
\begin{minipage}{0.225\hsize}
\centering
\includegraphics[scale=0.18]{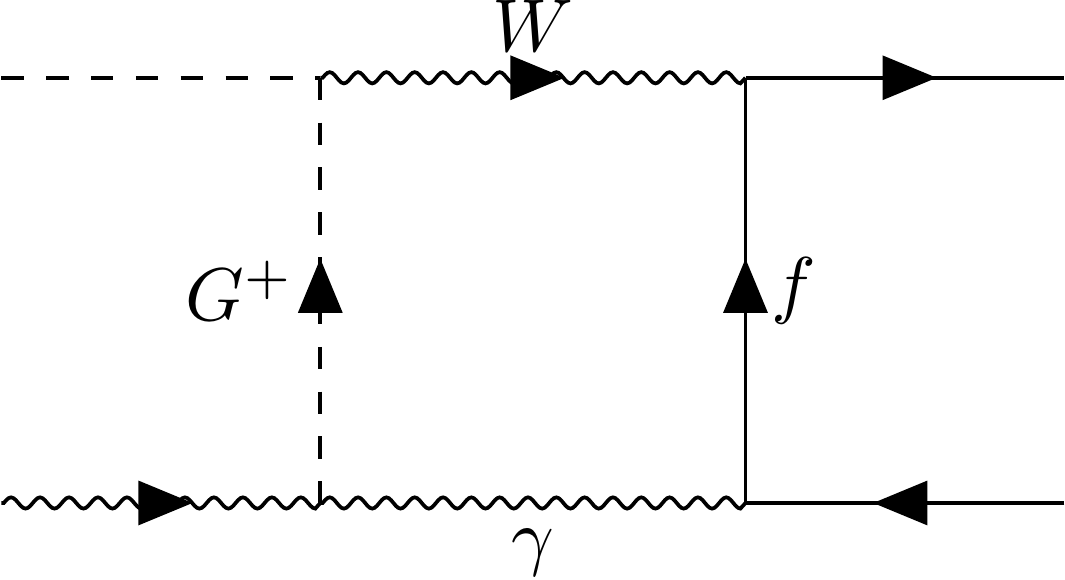}
\subcaption*{(9)}
\end{minipage}
\begin{minipage}{0.225\hsize}
\centering
\includegraphics[scale=0.18]{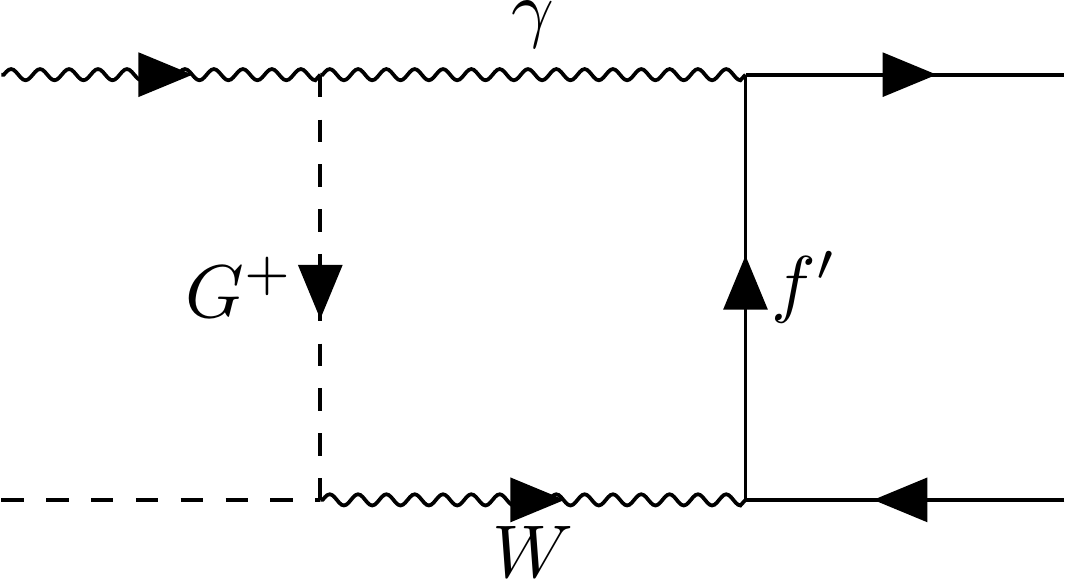}
\subcaption*{(10)}
\end{minipage}
\begin{minipage}{0.225\hsize}
\centering
\includegraphics[scale=0.18]{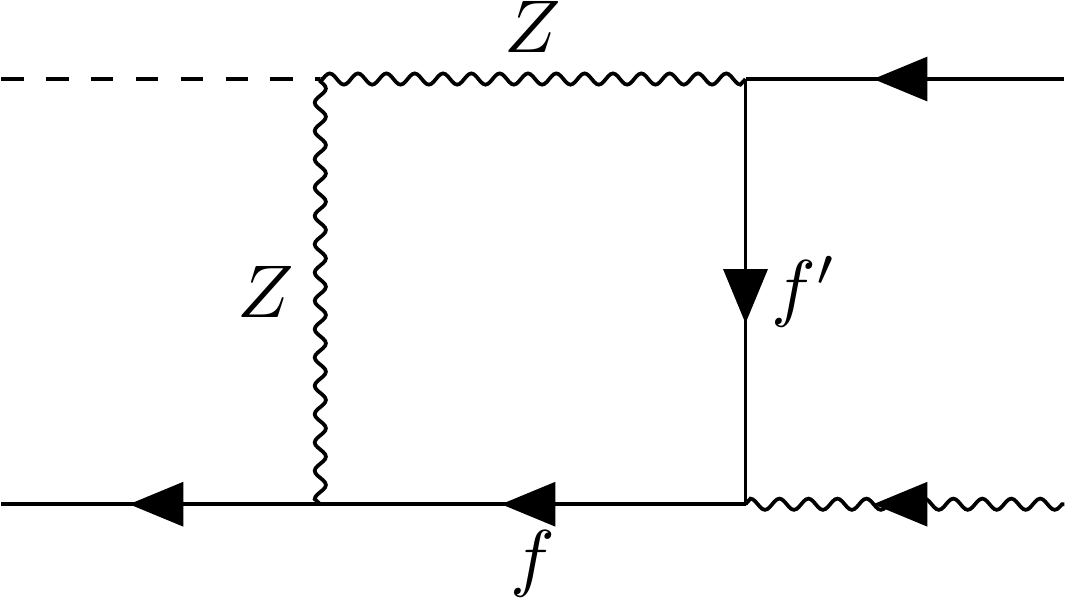}
\subcaption*{(11)}
\end{minipage} \\
%%%
\begin{minipage}{0.225\hsize}
\centering
\includegraphics[scale=0.18]{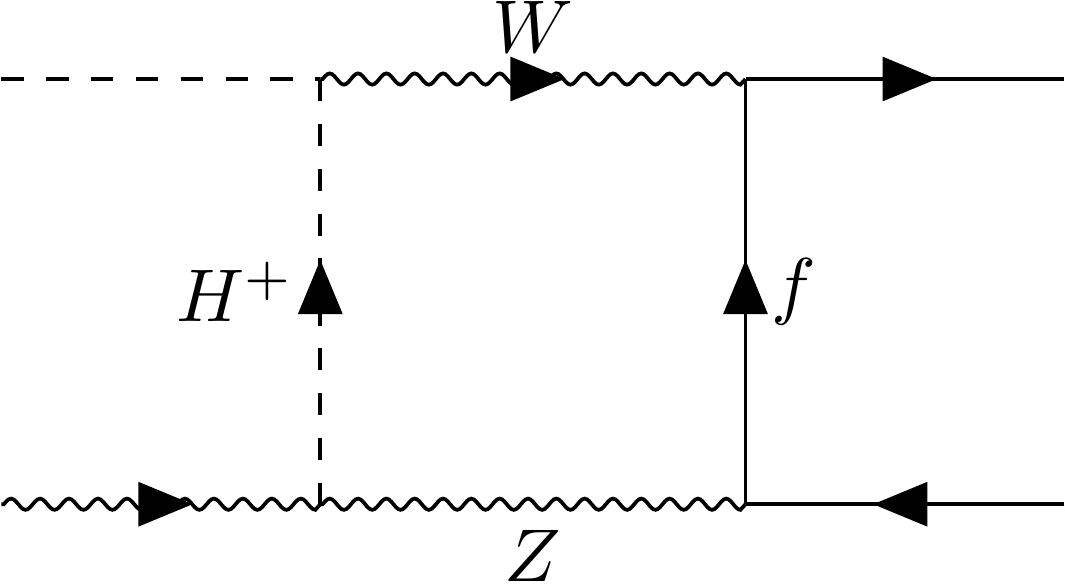}
\subcaption*{(12)}
\end{minipage}
\begin{minipage}{0.225\hsize}
\centering
\includegraphics[scale=0.18]{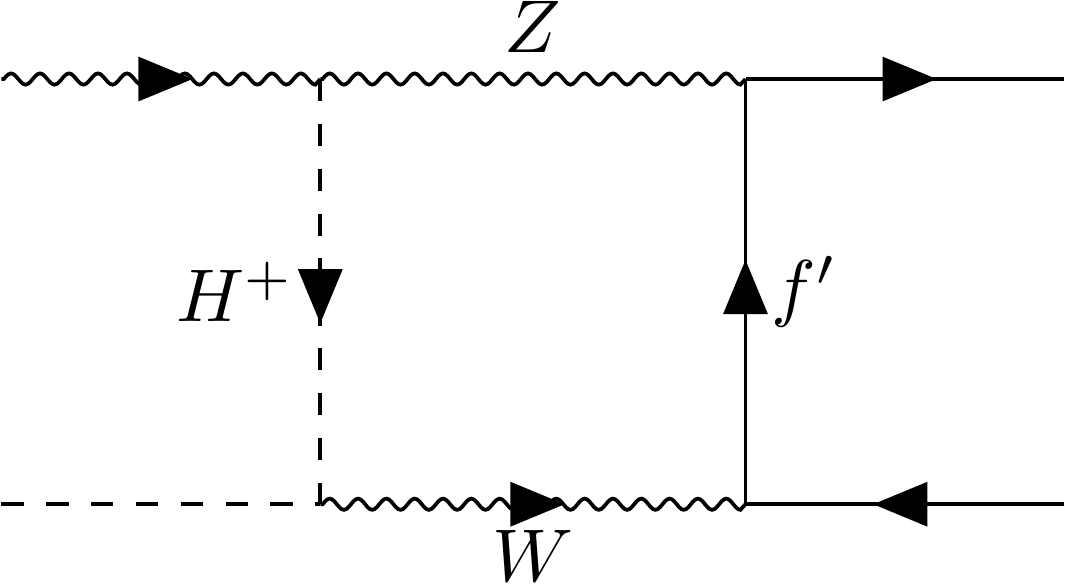}
\subcaption*{(13)}
\end{minipage}
\caption{Box diagrams for $h\to W^{-}f'\bar{f}$.}
\label{fig: Box_diagrams_hWffp}
\end{figure}

We here give the analytic expressions for the box diagrams for $h\to W^{-}f'\bar{f}$.
Similarly to $h\to Zf\bar{f}$, we assign all momenta in the incoming way (see Diagram (1) in Fig.~\ref{fig: Box_diagrams_hWffp}).
These incoming momenta are related to the out-going momenta, such as 
$k_{W} = -p_{W},\ k_{\bar{f}} = -p_{f}$ and $k_{f'} = -p_{\bar{f'}}$.
Then, the Mandelstam variables are obtained as $s = (p_{f}+p_{\bar{f'}})^{2},\, t = (p_{W}+p_{f})^{2}$, and $u = (p_{W}+p_{\bar{f'}})^{2}$.
The LO amplitude is obtained as
\begin{align}
\mathcal{M}_{\mathrm{tree}}
=
C_{\mathrm{tree}}^{W}
\overline{v}(p_{\bar{f'}})\gamma^{\mu}P_{L}u(p_{f})\epsilon_{\mu}^{*}(p_{W}),
\end{align}
where $C_{\mathrm{tree}}^{W} = g_{W}\Gamma_{hWW}^{1, \mathrm{tree}}/(s-m_{W}^{2})$.

At the one-loop level, there are eleven SM-like diagrams and two diagrams with new particles as shown in Fig.~\ref{fig: Box_diagrams_hWffp}.
The box amplitude can be parameterized as
\begin{align}
\mathcal{M}_{BWi} = \frac{1}{16\pi^{2}}\overline{v}(p_{\bar{f'}})\mathcal{M}_{BWi}^{\mu}u(p_{f})\epsilon_{\mu}^{*}(p_{W}),
\end{align}
where
\begin{align}
\mathcal{M}_{BWi}^{\mu}
&=
C^{BWi}\qty[\slashed{p}_{h}(F_{1,3}^{BWi}p_{f}^{\mu}+F_{1,4}^{BWi}p_{\bar{f'}}^{\mu})+\gamma^{\mu}F_{\gamma}^{BWi}]P_{L}.
\end{align}
Then, the NLO contribution is obtained as
\begin{align}
(16\pi^{2})2\Re\qty[\sum \mathcal{M}_{\mathrm{tree}}^{*}\mathcal{M}_{BWi}]
&=
4C_{\mathrm{tree}}^{W}C^{BWi}\Re\qty[B^{BWi}],
\end{align}
where
\begin{align}
B^{BWi} = sF_{\gamma}^{BWi}+\frac{tu-m_{W}^{2}m_{h}^{2}}{4m_{W}^{2}}\qty[2F_{\gamma}^{BWi}-(t-m_{W}^{2})F_{1,3}^{BWi}-(u-m_{W}^{2})F_{1,4}^{BWi}].
\end{align}
The coefficients and the form factors of each diagram are obtained as follows.
\begin{itemize}[labelsep=-1.2cm]
\item[$\lbrack \text{Diagram (1)} \rbrack$]
\begin{align}
C^{BW1}
&=
-4a_{f'}(v_{f'}+a_{f'})\frac{g^{2}g_{Z}^{2}}{\sqrt{2}}c_{W}m_{Z}c_{hZZ},
\notag \\
F_{1,3}^{BW1}
&=
F_{p\cdot\gamma1}^{VFVV}(m_{f'}^{2}, m_{f}^{2}, m_{W}^{2}, m_{h}^{2}, s, t; Z, f', W, Z),
\notag \\
F_{1,4}^{BW1}
&=
F_{p\cdot\gamma2}^{VFVV}(m_{f'}^{2}, m_{f}^{2}, m_{W}^{2}, m_{h}^{2}, s, t; Z, f', W, Z),
\notag \\
F_{\gamma}^{BW1}
&=
F_{\gamma}^{VFVV}(m_{f'}^{2}, m_{f}^{2}, m_{W}^{2}, m_{h}^{2}, s, t; Z, f', W, Z).
\end{align}
\end{itemize}

The coefficient of Diagram (2) is obtained by replacing $f'\to f$.
Its form factors are obtained by interchanging $F_{1,3}^{BW1}$ and $F_{1,4}^{BW1}$, exchanging $f'$ and $f$, and replacing $t \to u$.

\begin{itemize}[labelsep=-1.2cm]
\item[$\lbrack \text{Diagram (3)} \rbrack$]
\begin{align}
C^{BW3}
&=
-4a_{f}(v_{f}+a_{f})\frac{g^{3}g_{Z}}{\sqrt{2}}c_{W}m_{W}c_{hWW},
\notag \\
F_{1,3}^{BW3}
&=
F_{p\cdot\gamma1}^{VFVV}(m_{f'}^{2}, m_{f}^{2}, m_{W}^{2}, m_{h}^{2}, s, t; W, f, Z, W),
\notag \\
F_{1,4}^{BW3}
&=
F_{p\cdot\gamma2}^{VFVV}(m_{f'}^{2}, m_{f}^{2}, m_{W}^{2}, m_{h}^{2}, s, t; W, f, Z, W),
\notag \\
F_{\gamma}^{BW3}
&=
F_{\gamma}^{VFVV}(m_{f'}^{2}, m_{f}^{2}, m_{W}^{2}, m_{h}^{2}, s, t; W, f, Z, W).
\end{align}
\end{itemize}

The coefficient of Diagram (4) is obtained by replacing $f'\to f$.
Its form factors are obtained by interchanging $F_{1,3}^{BW3}$ and $F_{1,4}^{BW3}$, exchanging $f'$ and $f$, and replacing $t \to u$.

\begin{itemize}[labelsep=-1.2cm]
\item[$\lbrack \text{Diagram (5)} \rbrack$]
\begin{align}
C^{BW5}
&=
-4a_{f}eQ_{f}\frac{g^{3}}{\sqrt{2}}s_{W}m_{W}c_{hWW},
\notag \\
F_{1,3}^{BW5}
&=
F_{p\cdot\gamma1}^{VFVV}(m_{f'}^{2}, m_{f}^{2}, m_{W}^{2}, m_{h}^{2}, s, t; W, f, \gamma, W),
\notag \\
F_{1,4}^{BW5}
&=
F_{p\cdot\gamma2}^{VFVV}(m_{f'}^{2}, m_{f}^{2}, m_{W}^{2}, m_{h}^{2}, s, t; W, f, \gamma, W),
\notag \\
F_{\gamma}^{BW5}
&=
F_{\gamma}^{VFVV}(m_{f'}^{2}, m_{f}^{2}, m_{W}^{2}, m_{h}^{2}, s, t; W, f, \gamma, W).
\end{align}
\end{itemize}

The coefficient of Diagram (6) is obtained by replacing $f'\to f$.
Its form factors are obtained by interchanging $F_{1,3}^{BW5}$ and $F_{1,4}^{BW5}$, exchanging $f'$ and $f$, and replacing $t \to u$.

\begin{itemize}[labelsep=-1.2cm]
\item[$\lbrack \text{Diagram (7)} \rbrack$]
\begin{align}
C^{BW7}
&=
-2a_{f}(v_{f}+a_{f})\frac{g^{2}g_{Z}^{2}}{\sqrt{2}}(s_{W}^{2}+s_{\beta}^{2})m_{W}c_{hWW},
\notag \\
F_{1,3}^{BW7}
&=
F_{p\cdot\gamma}^{VFVS}(m_{f'}^{2}, m_{f}^{2}, m_{W}^{2}, m_{h}^{2}, s, t; W, f, Z, G^{\pm}),
\notag \\
F_{1,4}^{BW7} &= 0,
\notag \\
F_{\gamma}^{BW7}
&=
F_{\gamma}^{VFVS}(m_{f'}^{2}, m_{f}^{2}, m_{W}^{2}, m_{h}^{2}, s, t; W, f, Z, G^{\pm}).
\end{align}
\end{itemize}

The coefficient of Diagram (8) is obtained by replacing $f'\to f$.
Its form factors are obtained by interchanging $F_{1,3}^{BW7}$ and $F_{1,4}^{BW7}$, exchanging $f'$ and $f$, and replacing $t \to u$.

\begin{itemize}[labelsep=-1.2cm]
\item[$\lbrack \text{Diagram (9)} \rbrack$]
\begin{align}
C^{BW9}
&=
2a_{f}Q_{f}\frac{e^{2}g^{2}}{\sqrt{2}}m_{W}c_{hWW},
\notag \\
F_{1,3}^{BW9}
&=
F_{p\cdot\gamma}^{VFVS}(m_{f'}^{2}, m_{f}^{2}, m_{W}^{2}, m_{h}^{2}, s, t; W, f, \gamma, G^{\pm}),
\notag \\
F_{1,4}^{BW9} &= 0,
\notag \\
F_{\gamma}^{BW9}
&=
F_{\gamma}^{VFVS}(m_{f'}^{2}, m_{f}^{2}, m_{W}^{2}, m_{h}^{2}, s, t; W, f, \gamma, G^{\pm}).
\end{align}
\end{itemize}

The coefficient of Diagram (10) is obtained by replacing $f'\to f$.
Its form factors are obtained by interchanging $F_{1,3}^{BW9}$ and $F_{1,4}^{BW9}$, exchanging $f'$ and $f$, and replacing $t \to u$.

\begin{itemize}[labelsep=-1.2cm]
\item[$\lbrack \text{Diagram (11)} \rbrack$]
\begin{align}
C^{BW11}
&=
-2(v_{f}+a_{f})(v_{f'}+a_{f'})\frac{g_{Z}^{3}g}{\sqrt{2}}m_{Z}c_{hZZ},
\notag \\
F_{1,3}^{BW11}
&=
F_{p\cdot\gamma1}^{VFFV}(m_{f}^{2}, m_{W}^{2}, m_{f'}^{2}, m_{h}^{2}, t, u; Z, f, f', Z),
\notag \\
F_{1,4}^{BW11}
&=
F_{p\cdot\gamma2}^{VFFV}(m_{f}^{2}, m_{W}^{2}, m_{f'}^{2}, m_{h}^{2}, t, u; Z, f, f', Z),
\notag \\
F_{\gamma}^{BW11}
&=
F_{\gamma}^{VFFV}(m_{f}^{2}, m_{W}^{2}, m_{f'}^{2}, m_{h}^{2}, t, u; Z, f, f', Z).
\end{align}
\end{itemize}

\begin{itemize}[labelsep=-1.2cm]
\item[$\lbrack \text{Diagram (12)} \rbrack$]
\begin{align}
C^{BW12}
&=
-2a_{f}(v_{f}+a_{f})\frac{g^{2}g_{Z}^{2}}{\sqrt{2}}s_{\beta}c_{\beta}m_{W}c_{hH^{\pm}W^{\mp}},
\notag \\
F_{1,3}^{BW12}
&=
F_{p\cdot\gamma}^{VFVS}(m_{f'}^{2}, m_{f}^{2}, m_{W}^{2}, m_{h}^{2}, s, t; W, f, Z, H^{\pm}),
\notag \\
F_{1,4}^{BW12} &= 0,
\notag \\
F_{\gamma}^{BW12}
&=
F_{\gamma}^{VFVS}(m_{f'}^{2}, m_{f}^{2}, m_{W}^{2}, m_{h}^{2}, s, t; W, f, Z, H^{\pm}).
\end{align}
\end{itemize}

The coefficient of Diagram (13) is obtained by replacing $f'\to f$.
Its form factors are obtained by interchanging $F_{1,3}^{BW12}$ and $F_{1,4}^{BW12}$, exchanging $f'$ and $f$, and replacing $t \to u$.

\bibliographystyle{utphys28mod}
\bibliography{reference}

@article{Ducu:2024xxf,
    author = "Ducu, Otilia A. and Dumitriu, Ana E. and Jinaru, Adam and Kukla, Romain and Monnier, Emmanuel and Moultaka, Gilbert and Tudorache, Alexandra and Xu, Hanlin",
    title = "{Type-II Seesaw Higgs triplet productions and decays at the LHC}",
    eprint = "2410.14830",
    archivePrefix = "arXiv",
    primaryClass = "hep-ph",
    doi = "10.1007/JHEP06(2025)020",
    journal = "JHEP",
    volume = "06",
    pages = "020",
    year = "2025"
}

@article{Ashanujjaman:2023tlj,
    author = "Ashanujjaman, Saiyad and Maharathy, Siddharth P.",
    title = "{Probing compressed mass spectra in the type-II seesaw model at the LHC}",
    eprint = "2305.06889",
    archivePrefix = "arXiv",
    primaryClass = "hep-ph",
    doi = "10.1103/PhysRevD.107.115026",
    journal = "Phys. Rev. D",
    volume = "107",
    number = "11",
    pages = "115026",
    year = "2023"
}

@article{Han:2023vme,
    author = "Han, Chengcheng and Lei, Zhanhong and Liao, Weihao",
    title = "{Testing type II seesaw leptogenesis at the LHC*}",
    eprint = "2303.15709",
    archivePrefix = "arXiv",
    primaryClass = "hep-ph",
    doi = "10.1088/1674-1137/ace708",
    journal = "Chin. Phys. C",
    volume = "47",
    number = "9",
    pages = "093104",
    year = "2023"
}

@article{Ashanujjaman:2022ofg,
    author = "Ashanujjaman, Saiyad and Ghosh, Kirtiman and Sahu, Rameswar",
    title = "{Low-mass doubly charged Higgs bosons at the LHC}",
    eprint = "2211.00632",
    archivePrefix = "arXiv",
    primaryClass = "hep-ph",
    doi = "10.1103/PhysRevD.107.015018",
    journal = "Phys. Rev. D",
    volume = "107",
    number = "1",
    pages = "015018",
    year = "2023"
}

@article{Primulando:2019evb,
    author = "Primulando, R. and Julio, J. and Uttayarat, P.",
    title = "{Scalar phenomenology in type-II seesaw model}",
    eprint = "1903.02493",
    archivePrefix = "arXiv",
    primaryClass = "hep-ph",
    doi = "10.1007/JHEP08(2019)024",
    journal = "JHEP",
    volume = "08",
    pages = "024",
    year = "2019"
}

@article{Antusch:2018svb,
    author = "Antusch, Stefan and Fischer, Oliver and Hammad, A. and Scherb, Christiane",
    title = "{Low scale type II seesaw: Present constraints and prospects for displaced vertex searches}",
    eprint = "1811.03476",
    archivePrefix = "arXiv",
    primaryClass = "hep-ph",
    doi = "10.1007/JHEP02(2019)157",
    journal = "JHEP",
    volume = "02",
    pages = "157",
    year = "2019"
}

@article{Cai:2017mow,
    author = "Cai, Yi and Han, Tao and Li, Tong and Ruiz, Richard",
    title = "{Lepton Number Violation: Seesaw Models and Their Collider Tests}",
    eprint = "1711.02180",
    archivePrefix = "arXiv",
    primaryClass = "hep-ph",
    reportNumber = "PITT-PACC-1712, IPPP-17-74, COEPP-MN-17-17",
    doi = "10.3389/fphy.2018.00040",
    journal = "Front. in Phys.",
    volume = "6",
    pages = "40",
    year = "2018"
}

@article{Mitra:2016wpr,
    author = "Mitra, Manimala and Niyogi, Saurabh and Spannowsky, Michael",
    title = "{Type-II Seesaw Model and Multilepton Signatures at Hadron Colliders}",
    eprint = "1611.09594",
    archivePrefix = "arXiv",
    primaryClass = "hep-ph",
    doi = "10.1103/PhysRevD.95.035042",
    journal = "Phys. Rev. D",
    volume = "95",
    number = "3",
    pages = "035042",
    year = "2017"
}

@article{Babu:2016rcr,
    author = "Babu, K. S. and Jana, Sudip",
    title = "{Probing Doubly Charged Higgs Bosons at the LHC through Photon Initiated Processes}",
    eprint = "1612.09224",
    archivePrefix = "arXiv",
    primaryClass = "hep-ph",
    reportNumber = "OSU-HEP-16-11",
    doi = "10.1103/PhysRevD.95.055020",
    journal = "Phys. Rev. D",
    volume = "95",
    number = "5",
    pages = "055020",
    year = "2017"
}

@article{Han:2015sca,
    author = "Han, Zhi-Long and Ding, Ran and Liao, Yi",
    title = "{LHC phenomenology of the type II seesaw mechanism: Observability of neutral scalars in the nondegenerate case}",
    eprint = "1506.08996",
    archivePrefix = "arXiv",
    primaryClass = "hep-ph",
    doi = "10.1103/PhysRevD.92.033014",
    journal = "Phys. Rev. D",
    volume = "92",
    number = "3",
    pages = "033014",
    year = "2015"
}

@article{Han:2015hba,
    author = "Han, Zhi-Long and Ding, Ran and Liao, Yi",
    title = "{LHC Phenomenology of Type II Seesaw: Nondegenerate Case}",
    eprint = "1502.05242",
    archivePrefix = "arXiv",
    primaryClass = "hep-ph",
    doi = "10.1103/PhysRevD.91.093006",
    journal = "Phys. Rev. D",
    volume = "91",
    pages = "093006",
    year = "2015"
}

@article{Kang:2014lwn,
    author = "Kang, Zhaofeng and Li, Jinmian and Li, Tianjun and Liu, Yandong and Ning, Guo-Zhu",
    title = "{Light Doubly Charged Higgs Boson via the $WW^*$ Channel at LHC}",
    eprint = "1404.5207",
    archivePrefix = "arXiv",
    primaryClass = "hep-ph",
    doi = "10.1140/epjc/s10052-015-3774-1",
    journal = "Eur. Phys. J. C",
    volume = "75",
    number = "12",
    pages = "574",
    year = "2015"
}

@article{Kanemura:2014ipa,
    author = "Kanemura, Shinya and Kikuchi, Mariko and Yokoya, Hiroshi and Yagyu, Kei",
    title = "{LHC Run-I constraint on the mass of doubly charged Higgs bosons in the same-sign diboson decay scenario}",
    eprint = "1412.7603",
    archivePrefix = "arXiv",
    primaryClass = "hep-ph",
    reportNumber = "UT-HET-098",
    doi = "10.1093/ptep/ptv071",
    journal = "PTEP",
    volume = "2015",
    pages = "051B02",
    year = "2015"
}

@article{Kanemura:2014goa,
    author = "Kanemura, Shinya and Kikuchi, Mariko and Yagyu, Kei and Yokoya, Hiroshi",
    title = "{Bounds on the mass of doubly-charged Higgs bosons in the same-sign diboson decay scenario}",
    eprint = "1407.6547",
    archivePrefix = "arXiv",
    primaryClass = "hep-ph",
    reportNumber = "UT-HET-091",
    doi = "10.1103/PhysRevD.90.115018",
    journal = "Phys. Rev. D",
    volume = "90",
    number = "11",
    pages = "115018",
    year = "2014"
}

@article{Chun:2013vma,
    author = "Chun, Eung Jin and Sharma, Pankaj",
    title = "{Search for a doubly-charged boson in four lepton final states in type II seesaw}",
    eprint = "1309.6888",
    archivePrefix = "arXiv",
    primaryClass = "hep-ph",
    reportNumber = "KIAS-P13054",
    doi = "10.1016/j.physletb.2013.11.056",
    journal = "Phys. Lett. B",
    volume = "728",
    pages = "256--261",
    year = "2014"
}

@article{delAguila:2013mia,
    author = "del {\'A}guila, Francisco and Chala, Mikael",
    title = "{LHC bounds on Lepton Number Violation mediated by doubly and singly-charged scalars}",
    eprint = "1311.1510",
    archivePrefix = "arXiv",
    primaryClass = "hep-ph",
    doi = "10.1007/JHEP03(2014)027",
    journal = "JHEP",
    volume = "03",
    pages = "027",
    year = "2014"
}

@article{Chun:2012zu,
    author = "Chun, Eung Jin and Sharma, Pankaj",
    title = "{Same-Sign Tetra-Leptons from Type II Seesaw}",
    eprint = "1206.6278",
    archivePrefix = "arXiv",
    primaryClass = "hep-ph",
    reportNumber = "KIAS-P12035",
    doi = "10.1007/JHEP08(2012)162",
    journal = "JHEP",
    volume = "08",
    pages = "162",
    year = "2012"
}

@article{Akeroyd:2012nd,
    author = "Akeroyd, A. G. and Moretti, S. and Sugiyama, Hiroaki",
    title = "{Five-lepton and six-lepton signatures from production of neutral triplet scalars in the Higgs Triplet Model}",
    eprint = "1201.5047",
    archivePrefix = "arXiv",
    primaryClass = "hep-ph",
    reportNumber = "SHEP-11-32",
    doi = "10.1103/PhysRevD.85.055026",
    journal = "Phys. Rev. D",
    volume = "85",
    pages = "055026",
    year = "2012"
}

@article{Chiang:2012dk,
    author = "Chiang, Cheng-Wei and Nomura, Takaaki and Tsumura, Koji",
    title = "{Search for doubly charged Higgs bosons using the same-sign diboson mode at the LHC}",
    eprint = "1202.2014",
    archivePrefix = "arXiv",
    primaryClass = "hep-ph",
    doi = "10.1103/PhysRevD.85.095023",
    journal = "Phys. Rev. D",
    volume = "85",
    pages = "095023",
    year = "2012"
}

@article{Akeroyd:2011zza,
    author = "Akeroyd, A. G. and Sugiyama, Hiroaki",
    title = "{Production of doubly charged scalars from the decay of singly charged scalars in the Higgs Triplet Model}",
    eprint = "1105.2209",
    archivePrefix = "arXiv",
    primaryClass = "hep-ph",
    reportNumber = "SHEP-11-09",
    doi = "10.1103/PhysRevD.84.035010",
    journal = "Phys. Rev. D",
    volume = "84",
    pages = "035010",
    year = "2011"
}

@article{Akeroyd:2010je,
    author = "Akeroyd, A. G. and Chiang, Cheng-Wei",
    title = "{Phenomenology of Large Mixing for the CP-even Neutral Scalars of the Higgs Triplet Model}",
    eprint = "1003.3724",
    archivePrefix = "arXiv",
    primaryClass = "hep-ph",
    doi = "10.1103/PhysRevD.81.115007",
    journal = "Phys. Rev. D",
    volume = "81",
    pages = "115007",
    year = "2010"
}

@article{Akeroyd:2010ip,
    author = "Akeroyd, A. G. and Chiang, Cheng-Wei and Gaur, Naveen",
    title = "{Leptonic signatures of doubly charged Higgs boson production at the LHC}",
    eprint = "1009.2780",
    archivePrefix = "arXiv",
    primaryClass = "hep-ph",
    reportNumber = "SHEP-10-29",
    doi = "10.1007/JHEP11(2010)005",
    journal = "JHEP",
    volume = "11",
    pages = "005",
    year = "2010"
}

@article{Akeroyd:2009hb,
    author = "Akeroyd, A. G. and Chiang, Cheng-Wei",
    title = "{Doubly charged Higgs bosons and three-lepton signatures in the Higgs Triplet Model}",
    eprint = "0909.4419",
    archivePrefix = "arXiv",
    primaryClass = "hep-ph",
    doi = "10.1103/PhysRevD.80.113010",
    journal = "Phys. Rev. D",
    volume = "80",
    pages = "113010",
    year = "2009"
}

@article{delAguila:2008cj,
    author = "del Aguila, F. and Aguilar-Saavedra, J. A.",
    title = "{Distinguishing seesaw models at LHC with multi-lepton signals}",
    eprint = "0808.2468",
    archivePrefix = "arXiv",
    primaryClass = "hep-ph",
    doi = "10.1016/j.nuclphysb.2008.12.029",
    journal = "Nucl. Phys. B",
    volume = "813",
    pages = "22--90",
    year = "2009"
}

@article{Akeroyd:2007zv,
    author = "Akeroyd, A. G. and Aoki, Mayumi and Sugiyama, Hiroaki",
    title = "{Probing Majorana Phases and Neutrino Mass Spectrum in the Higgs Triplet Model at the CERN LHC}",
    eprint = "0712.4019",
    archivePrefix = "arXiv",
    primaryClass = "hep-ph",
    reportNumber = "SISSA-99-2007-EP",
    doi = "10.1103/PhysRevD.77.075010",
    journal = "Phys. Rev. D",
    volume = "77",
    pages = "075010",
    year = "2008"
}

@article{Kadastik:2007yd,
    author = "Kadastik, M. and Raidal, M. and Rebane, L.",
    title = "{Direct determination of neutrino mass parameters at future colliders}",
    eprint = "0712.3912",
    archivePrefix = "arXiv",
    primaryClass = "hep-ph",
    doi = "10.1103/PhysRevD.77.115023",
    journal = "Phys. Rev. D",
    volume = "77",
    pages = "115023",
    year = "2008"
}

@article{Garayoa:2007fw,
    author = "Garayoa, Julia and Schwetz, Thomas",
    title = "{Neutrino mass hierarchy and Majorana CP phases within the Higgs triplet model at the LHC}",
    eprint = "0712.1453",
    archivePrefix = "arXiv",
    primaryClass = "hep-ph",
    reportNumber = "CERN-PH-TH-2007-255, IFIC-07-75, FTUV-07-1210",
    doi = "10.1088/1126-6708/2008/03/009",
    journal = "JHEP",
    volume = "03",
    pages = "009",
    year = "2008"
}

@article{Akeroyd:2005gt,
    author = "Akeroyd, A. G. and Aoki, Mayumi",
    title = "{Single and pair production of doubly charged Higgs bosons at hadron colliders}",
    eprint = "hep-ph/0506176",
    archivePrefix = "arXiv",
    reportNumber = "KEK-TH-1023",
    doi = "10.1103/PhysRevD.72.035011",
    journal = "Phys. Rev. D",
    volume = "72",
    pages = "035011",
    year = "2005"
}

@article{Chiang:2017vvo,
    author = "Chiang, Cheng-Wei and Kuo, An-Li and Yagyu, Kei",
    title = "{Radiative corrections to Higgs couplings with weak gauge bosons in custodial multi-Higgs models}",
    eprint = "1707.04176",
    archivePrefix = "arXiv",
    primaryClass = "hep-ph",
    reportNumber = "NCTS-PH-1721",
    doi = "10.1016/j.physletb.2017.09.061",
    journal = "Phys. Lett. B",
    volume = "774",
    pages = "119--122",
    year = "2017"
}

@article{Arbabifar:2012bd,
    author = "Arbabifar, Fatemeh and Bahrami, Sahar and Frank, Mariana",
    title = "{Neutral Higgs Bosons in the Higgs Triplet Model with nontrivial mixing}",
    eprint = "1211.6797",
    archivePrefix = "arXiv",
    primaryClass = "hep-ph",
    reportNumber = "CUMQ-HEP-168",
    doi = "10.1103/PhysRevD.87.015020",
    journal = "Phys. Rev. D",
    volume = "87",
    number = "1",
    pages = "015020",
    year = "2013"
}

@article{BhupalDev:2013xol,
    author = "Bhupal Dev, P. S. and Ghosh, Dilip Kumar and Okada, Nobuchika and Saha, Ipsita",
    title = "{125 GeV Higgs Boson and the Type-II Seesaw Model}",
    eprint = "1301.3453",
    archivePrefix = "arXiv",
    primaryClass = "hep-ph",
    reportNumber = "MAN-HEP-2012-021",
    doi = "10.1007/JHEP03(2013)150",
    journal = "JHEP",
    volume = "03",
    pages = "150",
    year = "2013",
    note = "[Erratum: JHEP 05, 049 (2013)]"
}

@article{Chun:2012jw,
    author = "Chun, Eung Jin and Lee, Hyun Min and Sharma, Pankaj",
    title = "{Vacuum Stability, Perturbativity, EWPD and Higgs-to-diphoton rate in Type II Seesaw Models}",
    eprint = "1209.1303",
    archivePrefix = "arXiv",
    primaryClass = "hep-ph",
    reportNumber = "KIAS-P12055",
    doi = "10.1007/JHEP11(2012)106",
    journal = "JHEP",
    volume = "11",
    pages = "106",
    year = "2012"
}

@article{Djouadi:2005gj,
    author = "Djouadi, Abdelhak",
    title = "{The Anatomy of electro-weak symmetry breaking. II. The Higgs bosons in the minimal supersymmetric model}",
    eprint = "hep-ph/0503173",
    archivePrefix = "arXiv",
    reportNumber = "LPT-ORSAY-05-18",
    doi = "10.1016/j.physrep.2007.10.005",
    journal = "Phys. Rept.",
    volume = "459",
    pages = "1--241",
    year = "2008"
}

@article{Djouadi:2005gi,
    author = "Djouadi, Abdelhak",
    title = "{The Anatomy of electro-weak symmetry breaking. I: The Higgs boson in the standard model}",
    eprint = "hep-ph/0503172",
    archivePrefix = "arXiv",
    reportNumber = "LPT-ORSAY-05-17",
    doi = "10.1016/j.physrep.2007.10.004",
    journal = "Phys. Rept.",
    volume = "457",
    pages = "1--216",
    year = "2008"
}

@article{Chen:2008jg,
    author = "Chen, Mu-Chun and Dawson, Sally and Jackson, C. B.",
    title = "{Higgs Triplets, Decoupling, and Precision Measurements}",
    eprint = "0809.4185",
    archivePrefix = "arXiv",
    primaryClass = "hep-ph",
    doi = "10.1103/PhysRevD.78.093001",
    journal = "Phys. Rev. D",
    volume = "78",
    pages = "093001",
    year = "2008"
}

@article{Chen:2006pb,
    author = "Chen, Mu-Chun and Dawson, Sally and Krupovnickas, Tadas",
    title = "{Higgs triplets and limits from precision measurements}",
    eprint = "hep-ph/0604102",
    archivePrefix = "arXiv",
    reportNumber = "FERMILAB-PUB-06-065-T, BNL-06-2",
    doi = "10.1103/PhysRevD.74.035001",
    journal = "Phys. Rev. D",
    volume = "74",
    pages = "035001",
    year = "2006"
}

@article{Chen:2005jx,
    author = "Chen, Mu-Chun and Dawson, Sally and Krupovnickas, Tadas",
    title = "{Constraining new models with precision electroweak data}",
    eprint = "hep-ph/0504286",
    archivePrefix = "arXiv",
    reportNumber = "BNL-HET-05-15",
    doi = "10.1142/S0217751X0603388X",
    journal = "Int. J. Mod. Phys. A",
    volume = "21",
    pages = "4045--4070",
    year = "2006"
}

@article{Keung:1984hn,
    author = "Keung, Wai-Yee and Marciano, William J.",
    title = "{HIGGS SCALAR DECAYS: H ---{\ensuremath{>}} W+- X}",
    reportNumber = "BNL-34578",
    doi = "10.1103/PhysRevD.30.248",
    journal = "Phys. Rev. D",
    volume = "30",
    pages = "248",
    year = "1984"
}

@article{Chiang:2018xpl,
    author = "Chiang, Cheng-Wei and Kuo, An-Li and Yagyu, Kei",
    title = "{One-loop renormalized Higgs boson vertices in the Georgi-Machacek model}",
    eprint = "1804.02633",
    archivePrefix = "arXiv",
    primaryClass = "hep-ph",
    doi = "10.1103/PhysRevD.98.013008",
    journal = "Phys. Rev. D",
    volume = "98",
    number = "1",
    pages = "013008",
    year = "2018"
}

@article{Aoki:2007ah,
    author = "Aoki, Mayumi and Kanemura, Shinya",
    title = "{Unitarity bounds in the Higgs model including triplet fields with custodial symmetry}",
    eprint = "0712.4053",
    archivePrefix = "arXiv",
    primaryClass = "hep-ph",
    doi = "10.1103/PhysRevD.77.095009",
    journal = "Phys. Rev. D",
    volume = "77",
    number = "9",
    pages = "095009",
    year = "2008",
    note = "[Erratum: Phys.Rev.D 89, 059902 (2014)]"
}

@article{Aiko:2021nkb,
    author = "Aiko, Masashi and Kanemura, Shinya and Mawatari, Kentarou",
    title = "{Next-to-leading-order corrections to the Higgs strahlung process from electron{\textendash}positron collisions in extended Higgs models}",
    eprint = "2109.02884",
    archivePrefix = "arXiv",
    primaryClass = "hep-ph",
    reportNumber = "OU-HET-1099",
    doi = "10.1140/epjc/s10052-021-09764-8",
    journal = "Eur. Phys. J. C",
    volume = "81",
    number = "11",
    pages = "1000",
    year = "2021"
}

@article{Sirlin:1980nh,
    author = "Sirlin, A.",
    title = "{Radiative Corrections in the SU(2)-L x U(1) Theory: A Simple Renormalization Framework}",
    reportNumber = "PRINT-80-0267 (IAS,PRINCETON)",
    doi = "10.1103/PhysRevD.22.971",
    journal = "Phys. Rev. D",
    volume = "22",
    pages = "971--981",
    year = "1980"
}

@article{Kanemura:2024ium,
    author = "Kanemura, Shinya and Kikuchi, Mariko and Yagyu, Kei",
    title = "{New renormalization scheme in the two Higgs doublet models}",
    eprint = "2408.08033",
    archivePrefix = "arXiv",
    primaryClass = "hep-ph",
    reportNumber = "OU-HET-1239, NU-EHET 003",
    doi = "10.1016/j.physletb.2024.139050",
    journal = "Phys. Lett. B",
    volume = "858",
    pages = "139050",
    year = "2024"
}

@article{Kniehl:1993ay,
    author = "Kniehl, Bernd A.",
    title = "{Higgs phenomenology at one loop in the standard model}",
    reportNumber = "DESY-93-069",
    doi = "10.1016/0370-1573(94)90037-X",
    journal = "Phys. Rept.",
    volume = "240",
    pages = "211--300",
    year = "1994"
}

@article{Kanemura:2012rs,
    author = "Kanemura, Shinya and Yagyu, Kei",
    title = "{Radiative corrections to electroweak parameters in the Higgs triplet model and implication with the recent Higgs boson searches}",
    eprint = "1201.6287",
    archivePrefix = "arXiv",
    primaryClass = "hep-ph",
    reportNumber = "UT-HET-064",
    doi = "10.1103/PhysRevD.85.115009",
    journal = "Phys. Rev. D",
    volume = "85",
    pages = "115009",
    year = "2012"
}

@article{Chanowitz:1985ug,
    author = "Chanowitz, Michael S. and Golden, Mitchell",
    title = "{Higgs Boson Triplets With M ($W$) = M ($Z$) $\cos \theta \omega$}",
    reportNumber = "LBL-20269",
    doi = "10.1016/0370-2693(85)90700-2",
    journal = "Phys. Lett. B",
    volume = "165",
    pages = "105--108",
    year = "1985"
}

@article{Georgi:1985nv,
    author = "Georgi, Howard and Machacek, Marie",
    title = "{DOUBLY CHARGED HIGGS BOSONS}",
    reportNumber = "HUTP-85/A051",
    doi = "10.1016/0550-3213(85)90325-6",
    journal = "Nucl. Phys. B",
    volume = "262",
    pages = "463--477",
    year = "1985"
}

@misc{Dohse:2018vqo,
    author = "Dohse, Max",
    eprint = "1802.00689",
    archivePrefix = "arXiv",
    primaryClass = "cs.OH",
    month = "1",
}

@article{Ellis:2016jkw,
    author = "Ellis, Joshua",
    title = "{TikZ-Feynman: Feynman diagrams with TikZ}",
    eprint = "1601.05437",
    archivePrefix = "arXiv",
    primaryClass = "hep-ph",
    doi = "10.1016/j.cpc.2016.08.019",
    journal = "Comput. Phys. Commun.",
    volume = "210",
    pages = "103--123",
    year = "2017"
}

@article{Marciano:1989ns,
    author = "Marciano, William J. and Valencia, G. and Willenbrock, S.",
    title = "{Renormalization Group Improved Unitarity Bounds on the Higgs Boson and Top Quark Masses}",
    reportNumber = "BNL-42766",
    doi = "10.1103/PhysRevD.40.1725",
    journal = "Phys. Rev. D",
    volume = "40",
    pages = "1725",
    year = "1989"
}

@article{Luscher:1988gc,
    author = "Luscher, M. and Weisz, P.",
    title = "{Is There a Strong Interaction Sector in the Standard Lattice Higgs Model?}",
    reportNumber = "DESY-88-083",
    doi = "10.1016/0370-2693(88)91799-6",
    journal = "Phys. Lett. B",
    volume = "212",
    pages = "472--478",
    year = "1988"
}

@article{Lee:1977eg,
    author = "Lee, Benjamin W. and Quigg, C. and Thacker, H. B.",
    title = "{Weak Interactions at Very High-Energies: The Role of the Higgs Boson Mass}",
    reportNumber = "FERMILAB-PUB-77-030-T",
    doi = "10.1103/PhysRevD.16.1519",
    journal = "Phys. Rev. D",
    volume = "16",
    pages = "1519",
    year = "1977"
}

@article{Lee:1977yc,
    author = "Lee, Benjamin W. and Quigg, C. and Thacker, H. B.",
    title = "{The Strength of Weak Interactions at Very High-Energies and the Higgs Boson Mass}",
    reportNumber = "FERMILAB-PUB-77-022-T",
    doi = "10.1103/PhysRevLett.38.883",
    journal = "Phys. Rev. Lett.",
    volume = "38",
    pages = "883--885",
    year = "1977"
}

@article{Arhrib:2011uy,
    author = "Arhrib, A. and Benbrik, R. and Chabab, M. and Moultaka, G. and Peyranere, M. C. and Rahili, L. and Ramadan, J.",
    title = "{The Higgs Potential in the Type II Seesaw Model}",
    eprint = "1105.1925",
    archivePrefix = "arXiv",
    primaryClass = "hep-ph",
    doi = "10.1103/PhysRevD.84.095005",
    journal = "Phys. Rev. D",
    volume = "84",
    pages = "095005",
    year = "2011"
}

@article{Cheng:1980qt,
    author = "Cheng, T. P. and Li, Ling-Fong",
    title = "{Neutrino Masses, Mixings and Oscillations in SU(2) x U(1) Models of Electroweak Interactions}",
    reportNumber = "PRINT-80-0511 (CARNEGIE-MELLON), COO-3066-152",
    doi = "10.1103/PhysRevD.22.2860",
    journal = "Phys. Rev. D",
    volume = "22",
    pages = "2860",
    year = "1980"
}

@article{Schechter:1980gr,
    author = "Schechter, J. and Valle, J. W. F.",
    title = "{Neutrino Masses in SU(2) x U(1) Theories}",
    reportNumber = "SU-4217-167, COO-3533-167",
    doi = "10.1103/PhysRevD.22.2227",
    journal = "Phys. Rev. D",
    volume = "22",
    pages = "2227",
    year = "1980"
}

@article{Lazarides:1980nt,
    author = "Lazarides, George and Shafi, Q. and Wetterich, C.",
    title = "{Proton Lifetime and Fermion Masses in an SO(10) Model}",
    reportNumber = "FREIBURG-THEP-80-2",
    doi = "10.1016/0550-3213(81)90354-0",
    journal = "Nucl. Phys. B",
    volume = "181",
    pages = "287--300",
    year = "1981"
}

@article{Mohapatra:1980yp,
    author = "Mohapatra, Rabindra N. and Senjanovic, Goran",
    title = "{Neutrino Masses and Mixings in Gauge Models with Spontaneous Parity Violation}",
    reportNumber = "FERMILAB-PUB-80-061-THY, FERMILAB-PUB-80-061-T",
    doi = "10.1103/PhysRevD.23.165",
    journal = "Phys. Rev. D",
    volume = "23",
    pages = "165",
    year = "1981"
}

@article{Magg:1980ut,
    author = "Magg, M. and Wetterich, C.",
    title = "{Neutrino Mass Problem and Gauge Hierarchy}",
    reportNumber = "CERN-TH-2829",
    doi = "10.1016/0370-2693(80)90825-4",
    journal = "Phys. Lett. B",
    volume = "94",
    pages = "61--64",
    year = "1980"
}

@article{Kanemura:2012rj,
    author = "Kanemura, Shinya and Sugiyama, Hiroaki",
    title = "{Dark matter and a suppression mechanism for neutrino masses in the Higgs triplet model}",
    eprint = "1202.5231",
    archivePrefix = "arXiv",
    primaryClass = "hep-ph",
    reportNumber = "UT-HET-065",
    doi = "10.1103/PhysRevD.86.073006",
    journal = "Phys. Rev. D",
    volume = "86",
    pages = "073006",
    year = "2012"
}

@article{Nomura:2016run,
    author = "Nomura, Takaaki and Okada, Hiroshi and Orikasa, Yuta",
    title = "{Radiative neutrino mass in alternative left{\textendash}right model}",
    eprint = "1602.08302",
    archivePrefix = "arXiv",
    primaryClass = "hep-ph",
    reportNumber = "KIAS-P16020",
    doi = "10.1140/epjc/s10052-017-4657-4",
    journal = "Eur. Phys. J. C",
    volume = "77",
    number = "2",
    pages = "103",
    year = "2017"
}

@article{Guo:2016dzl,
    author = "Guo, Shu-Yuan and Han, Zhi-Long and Liao, Yi",
    title = "{Testing the type II radiative seesaw model: From dark matter detection to LHC signatures}",
    eprint = "1609.01018",
    archivePrefix = "arXiv",
    primaryClass = "hep-ph",
    doi = "10.1103/PhysRevD.94.115014",
    journal = "Phys. Rev. D",
    volume = "94",
    number = "11",
    pages = "115014",
    year = "2016"
}

@article{Nomura:2017emk,
    author = "Nomura, Takaaki and Okada, Hiroshi",
    title = "{Loop induced type-II seesaw model and GeV dark matter with $U(1)_{B-L}$ gauge symmetry}",
    eprint = "1704.08581",
    archivePrefix = "arXiv",
    primaryClass = "hep-ph",
    reportNumber = "KIAS-P17031",
    doi = "10.1016/j.physletb.2017.10.033",
    journal = "Phys. Lett. B",
    volume = "774",
    pages = "575--581",
    year = "2017"
}

@article{Pati:1974yy,
    author = "Pati, Jogesh C. and Salam, Abdus",
    title = "{Lepton Number as the Fourth Color}",
    reportNumber = "IC-74-7",
    doi = "10.1103/PhysRevD.10.275",
    journal = "Phys. Rev. D",
    volume = "10",
    pages = "275--289",
    year = "1974",
    note = "[Erratum: Phys.Rev.D 11, 703--703 (1975)]"
}

@article{Mohapatra:1974hk,
    author = "Mohapatra, Rabindra N. and Pati, Jogesh C.",
    title = "{Left-Right Gauge Symmetry and an Isoconjugate Model of CP Violation}",
    reportNumber = "MDDP-TR-74-085",
    doi = "10.1103/PhysRevD.11.566",
    journal = "Phys. Rev. D",
    volume = "11",
    pages = "566--571",
    year = "1975"
}

@article{Mohapatra:1974gc,
    author = "Mohapatra, R. N. and Pati, Jogesh C.",
    title = "{A Natural Left-Right Symmetry}",
    reportNumber = "CCNY-HEP-74-2",
    doi = "10.1103/PhysRevD.11.2558",
    journal = "Phys. Rev. D",
    volume = "11",
    pages = "2558",
    year = "1975"
}

@article{Senjanovic:1975rk,
    author = "Senjanovic, G. and Mohapatra, Rabindra N.",
    title = "{Exact Left-Right Symmetry and Spontaneous Violation of Parity}",
    reportNumber = "CCNY-HEP-75-5",
    doi = "10.1103/PhysRevD.12.1502",
    journal = "Phys. Rev. D",
    volume = "12",
    pages = "1502",
    year = "1975"
}

@article{Zhou:2022mlz,
    author = "Zhou, Ruiyu and Bian, Ligong and Du, Yong",
    title = "{Electroweak phase transition and gravitational waves in the type-II seesaw model}",
    eprint = "2203.01561",
    archivePrefix = "arXiv",
    primaryClass = "hep-ph",
    doi = "10.1007/JHEP08(2022)205",
    journal = "JHEP",
    volume = "08",
    pages = "205",
    year = "2022"
}

@article{Jangid:2023lny,
    author = "Jangid, Shilpa and Okada, Hiroshi",
    title = "{Electroweak phase transition with radiative symmetry breaking in a type-II seesaw model with an inert doublet}",
    eprint = "2310.12591",
    archivePrefix = "arXiv",
    primaryClass = "hep-ph",
    doi = "10.1103/PhysRevD.109.015001",
    journal = "Phys. Rev. D",
    volume = "109",
    number = "1",
    pages = "015001",
    year = "2024"
}

@article{Aoki:2012jj,
    author = "Aoki, Mayumi and Kanemura, Shinya and Kikuchi, Mariko and Yagyu, Kei",
    title = "{Radiative corrections to the Higgs boson couplings in the triplet model}",
    eprint = "1211.6029",
    archivePrefix = "arXiv",
    primaryClass = "hep-ph",
    doi = "10.1103/PhysRevD.87.015012",
    journal = "Phys. Rev. D",
    volume = "87",
    number = "1",
    pages = "015012",
    year = "2013"
}

@article{Blank:1997qa,
    author = "Blank, T. and Hollik, W.",
    title = "{Precision observables in SU(2) x U(1) models with an additional Higgs triplet}",
    eprint = "hep-ph/9703392",
    archivePrefix = "arXiv",
    reportNumber = "KA-TP-3-1997",
    doi = "10.1016/S0550-3213(97)00785-2",
    journal = "Nucl. Phys. B",
    volume = "514",
    pages = "113--134",
    year = "1998"
}

@article{Kanemura:2013vxa,
    author = "Kanemura, Shinya and Yagyu, Kei and Yokoya, Hiroshi",
    title = "{First constraint on the mass of doubly-charged Higgs bosons in the same-sign diboson decay scenario at the LHC}",
    eprint = "1305.2383",
    archivePrefix = "arXiv",
    primaryClass = "hep-ph",
    reportNumber = "UT-HET-080",
    doi = "10.1016/j.physletb.2013.08.054",
    journal = "Phys. Lett. B",
    volume = "726",
    pages = "316--319",
    year = "2013"
}

@article{Aoki:2011pz,
    author = "Aoki, Mayumi and Kanemura, Shinya and Yagyu, Kei",
    title = "{Testing the Higgs triplet model with the mass difference at the LHC}",
    eprint = "1110.4625",
    archivePrefix = "arXiv",
    primaryClass = "hep-ph",
    reportNumber = "KANAZAWA-11-17, UT-HET-059",
    doi = "10.1103/PhysRevD.85.055007",
    journal = "Phys. Rev. D",
    volume = "85",
    pages = "055007",
    year = "2012"
}

@article{Kanemura:2022ahw,
    author = "Kanemura, Shinya and Yagyu, Kei",
    title = "{Implication of the W boson mass anomaly at CDF II in the Higgs triplet model with a mass difference}",
    eprint = "2204.07511",
    archivePrefix = "arXiv",
    primaryClass = "hep-ph",
    reportNumber = "OU-HET-1141",
    doi = "10.1016/j.physletb.2022.137217",
    journal = "Phys. Lett. B",
    volume = "831",
    pages = "137217",
    year = "2022"
}

@article{Cepeda:2019klc,
    author = "Cepeda, M. and others",
    editor = "Dainese, Andrea and Mangano, Michelangelo and Meyer, Andreas B. and Nisati, Aleandro and Salam, Gavin and Vesterinen, Mika Anton",
    title = "{Report from Working Group 2}: {Higgs Physics at the HL-LHC and HE-LHC}",
    eprint = "1902.00134",
    archivePrefix = "arXiv",
    primaryClass = "hep-ph",
    reportNumber = "CERN-LPCC-2018-04",
    doi = "10.23731/CYRM-2019-007.221",
    journal = "CERN Yellow Rep. Monogr.",
    volume = "7",
    pages = "221--584",
    year = "2019"
}

@article{CMS:2025hfp,
    author = "{ATLAS and CMS Collaboration}", 
    title = "{Highlights of the HL-LHC physics projections by ATLAS and CMS}",
    eprint = "2504.00672",
    archivePrefix = "arXiv",
    primaryClass = "hep-ex",
    reportNumber = "{ATL-PHYS-PUB-2025-018; CMS-HIG-25-002}",
    journal = "arXiv preprint",
    month = "April",
    year = "2025"
}

@article{Baer:2013cma,
    editor = "Baer, Howard and others",
    collaboration = "ILC",
    title = "{The International Linear Collider Technical Design Report - Volume 2: Physics}",
    eprint = "1306.6352",
    archivePrefix = "arXiv",
    primaryClass = "hep-ph",
    reportNumber = "ILC-REPORT-2013-040, ANL-HEP-TR-13-20, BNL-100603-2013-IR, IRFU-13-59, CERN-ATS-2013-037, COCKCROFT-13-10, CLNS-13-2085, DESY-13-062, FERMILAB-TM-2554, IHEP-AC-ILC-2013-001, INFN-13-04-LNF, JAI-2013-001, JINR-E9-2013-35, JLAB-R-2013-01, KEK-REPORT-2013-1, KNU-CHEP-ILC-2013-1, LLNL-TR-635539, SLAC-R-1004, ILC-HIGRADE-REPORT-2013-003",
    journal = "arXiv preprint",
    month = "6",
    year = "2013"
}

@article{Fujii:2017vwa,
    author = "Fujii, Keisuke and others",
    title = "{Physics Case for the 250 GeV Stage of the International Linear Collider}",
    eprint = "1710.07621",
    archivePrefix = "arXiv",
    primaryClass = "hep-ex",
    reportNumber = "DESY-17-155, KEK-PREPRINT-2017-31, LAL-17-059, SLAC-PUB-17161",
    journal = "arXiv preprint",
    month = "10",
    year = "2017"
}

@article{Asai:2017pwp,
    author = "Asai, Shoji and Tanaka, Junichi and Ushiroda, Yutaka and Nakao, Mikihiko and Tian, Junping and Kanemura, Shinya and Matsumoto, Shigeki and Shirai, Satoshi and Endo, Motoi and Kakizaki, Mitsuru",
    title = "{Report by the Committee on the Scientific Case of the ILC Operating at 250 GeV as a Higgs Factory}",
    eprint = "1710.08639",
    archivePrefix = "arXiv",
    primaryClass = "hep-ex",
    journal = "arXiv preprint",
    month = "10",
    year = "2017"
}

@article{Fujii:2019zll,
    author = "Fujii, Keisuke and others",
    collaboration = "LCC Physics Working Group",
    title = "{Tests of the Standard Model at the International Linear Collider}",
    eprint = "1908.11299",
    archivePrefix = "arXiv",
    primaryClass = "hep-ex",
    reportNumber = "DESY 19-146, KEK Preprint 2019-22, SLAC-PUB-17467, DESY-19-146",
    journal = "arXiv preprint",
    month = "8",
    year = "2019"
}

@article{CEPC-SPPCStudyGroup:2015csa,
    author = "Ahmad, Muhammd and others",
    title = "{CEPC-SPPC Preliminary Conceptual Design Report. 1. Physics and Detector}",
    reportNumber = "IHEP-CEPC-DR-2015-01, IHEP-TH-2015-01, IHEP-EP-2015-01",
    journal = "arXiv preprint",
    month = "3",
    year = "2015"
}

@article{Ai:2025cpj,
    author = "Ai, Xiaocong and others",
    title = "{New Physics Search at the CEPC: a General Perspective}",
    eprint = "2505.24810",
    archivePrefix = "arXiv",
    primaryClass = "hep-ex",
    doi = "10.1088/1674-1137/ae1194",
    journal = "arXiv preprint",
    month = "5",
    year = "2025"
}

@article{Gomez-Ceballos:2013zzn,
    author = "Bicer, M. and others",
    editor = "Graf, Norman A. and Peskin, Michael E. and Rosner, Jonathan L.",
    collaboration = "TLEP Design Study Working Group",
    title = "{First Look at the Physics Case of TLEP}",
    eprint = "1308.6176",
    archivePrefix = "arXiv",
    primaryClass = "hep-ex",
    doi = "10.1007/JHEP01(2014)164",
    journal = "JHEP",
    volume = "01",
    pages = "164",
    year = "2014"
}

@article{FCC:2025lpp,
    author       = {M. Benedikt et al. (FCC Collaboration)},
    title        = {Future Circular Collider Feasibility Study Report: Volume 1, Physics, Experiments, Detectors},
    eprint       = {2505.00272},
    archivePrefix= {arXiv},
    primaryClass = {hep-ex},
    note         = {CERN-FCC-PHYS-2025-0002},
    doi          = {10.17181/CERN.9DKX.TDH9},
    journal = "arXiv preprint",
    month        = apr,
    year         = {2025}
}

@article{ATLAS:2022vkf,
    author = "Aad, Georges and others",
    collaboration = "ATLAS",
    title = "{A detailed map of Higgs boson interactions by the ATLAS experiment ten years after the discovery}",
    eprint = "2207.00092",
    archivePrefix = "arXiv",
    primaryClass = "hep-ex",
    reportNumber = "CERN-EP-2022-057",
    doi = "10.1038/s41586-022-04893-w",
    journal = "Nature",
    volume = "607",
    number = "7917",
    pages = "52--59",
    year = "2022",
    note = "[Erratum: Nature 612, E24 (2022)]"
}

@article{CMS:2022dwd,
    author = "Tumasyan, Armen and others",
    collaboration = "CMS",
    title = "{A portrait of the Higgs boson by the CMS experiment ten years after the discovery.}",
    eprint = "2207.00043",
    archivePrefix = "arXiv",
    primaryClass = "hep-ex",
    reportNumber = "CMS-HIG-22-001, CERN-EP-2022-039",
    doi = "10.1038/s41586-022-04892-x",
    journal = "Nature",
    volume = "607",
    number = "7917",
    pages = "60--68",
    year = "2022",
    note = "[Erratum: Nature 623, (2023)]"
}

@article{Kanemura:2004mg,
    author = "Kanemura, Shinya and Okada, Yasuhiro and Senaha, Eibun and Yuan, C. -P.",
    title = "{Higgs coupling constants as a probe of new physics}",
    eprint = "hep-ph/0408364",
    archivePrefix = "arXiv",
    doi = "10.1103/PhysRevD.70.115002",
    journal = "Phys. Rev. D",
    volume = "70",
    pages = "115002",
    year = "2004"
}

@article{Kanemura:2014dja,
    author = "Kanemura, Shinya and Kikuchi, Mariko and Yagyu, Kei",
    title = "{Radiative corrections to the Yukawa coupling constants in two Higgs doublet models}",
    eprint = "1401.0515",
    archivePrefix = "arXiv",
    primaryClass = "hep-ph",
    reportNumber = "UT-HET-086",
    doi = "10.1016/j.physletb.2014.02.022",
    journal = "Phys. Lett. B",
    volume = "731",
    pages = "27--35",
    year = "2014"
}

@article{Kanemura:2015mxa,
    author = "Kanemura, Shinya and Kikuchi, Mariko and Yagyu, Kei",
    title = "{Fingerprinting the extended Higgs sector using one-loop corrected Higgs boson couplings and future precision measurements}",
    eprint = "1502.07716",
    archivePrefix = "arXiv",
    primaryClass = "hep-ph",
    reportNumber = "UT-HET-099",
    doi = "10.1016/j.nuclphysb.2015.04.015",
    journal = "Nucl. Phys. B",
    volume = "896",
    pages = "80--137",
    year = "2015"
}

@article{Arhrib:2015hoa,
    author = "Arhrib, Abdesslam and Benbrik, Rachid and El Falaki, Jaouad and Jueid, Adil",
    title = "{Radiative corrections to the Triple Higgs Coupling in the Inert Higgs Doublet Model}",
    eprint = "1507.03630",
    archivePrefix = "arXiv",
    primaryClass = "hep-ph",
    doi = "10.1007/JHEP12(2015)007",
    journal = "JHEP",
    volume = "12",
    pages = "007",
    year = "2015"
}

@article{Kanemura:2016sos,
    author = "Kanemura, Shinya and Kikuchi, Mariko and Sakurai, Kodai",
    title = "{Testing the dark matter scenario in the inert doublet model by future precision measurements of the Higgs boson couplings}",
    eprint = "1605.08520",
    archivePrefix = "arXiv",
    primaryClass = "hep-ph",
    reportNumber = "UT-HET-114",
    doi = "10.1103/PhysRevD.94.115011",
    journal = "Phys. Rev. D",
    volume = "94",
    number = "11",
    pages = "115011",
    year = "2016"
}

@article{Kanemura:2019kjg,
    author = "Kanemura, Shinya and Kikuchi, Mariko and Mawatari, Kentarou and Sakurai, Kodai and Yagyu, Kei",
    title = "{Full next-to-leading-order calculations of Higgs boson decay rates in models with non-minimal scalar sectors}",
    eprint = "1906.10070",
    archivePrefix = "arXiv",
    primaryClass = "hep-ph",
    reportNumber = "OU-HET 1006, KA-TP-12-2019",
    doi = "10.1016/j.nuclphysb.2019.114791",
    journal = "Nucl. Phys. B",
    volume = "949",
    pages = "114791",
    year = "2019"
}

@article{Kanemura:2018yai,
    author = "Kanemura, Shinya and Kikuchi, Mariko and Mawatari, Kentarou and Sakurai, Kodai and Yagyu, Kei",
    title = "{Loop effects on the Higgs decay widths in extended Higgs models}",
    eprint = "1803.01456",
    archivePrefix = "arXiv",
    primaryClass = "hep-ph",
    reportNumber = "OU-HET 960, UT-HET 124, OU-HET-960, UT-HET-124",
    doi = "10.1016/j.physletb.2018.06.035",
    journal = "Phys. Lett. B",
    volume = "783",
    pages = "140--149",
    year = "2018"
}

@article{Altenkamp:2017ldc,
    author = "Altenkamp, Lukas and Dittmaier, Stefan and Rzehak, Heidi",
    title = "{Renormalization schemes for the Two-Higgs-Doublet Model and applications to h {\textrightarrow} WW/ZZ {\textrightarrow} 4 fermions}",
    eprint = "1704.02645",
    archivePrefix = "arXiv",
    primaryClass = "hep-ph",
    reportNumber = "FR-PHENO-2017-003, CP3-ORIGINS-2017-012",
    doi = "10.1007/JHEP09(2017)134",
    journal = "JHEP",
    volume = "09",
    pages = "134",
    year = "2017"
}

@article{Aiko:2021can,
    author = "Aiko, Masashi and Kanemura, Shinya and Sakurai, Kodai",
    title = "{Radiative corrections to decays of charged Higgs bosons in two Higgs doublet models}",
    eprint = "2108.11868",
    archivePrefix = "arXiv",
    primaryClass = "hep-ph",
    reportNumber = "OU-HET 1100, TU 1128",
    doi = "10.1016/j.nuclphysb.2021.115581",
    journal = "Nucl. Phys. B",
    volume = "973",
    pages = "115581",
    year = "2021"
}

@article{Kanemura:2022ldq,
    author = "Kanemura, Shinya and Kikuchi, Mariko and Yagyu, Kei",
    title = "{Next-to-leading order corrections to decays of the heavier CP-even Higgs boson in the two Higgs doublet model}",
    eprint = "2203.08337",
    archivePrefix = "arXiv",
    primaryClass = "hep-ph",
    reportNumber = "OU-HET 1113, NU-EHET 001",
    doi = "10.1016/j.nuclphysb.2022.115906",
    journal = "Nucl. Phys. B",
    volume = "983",
    pages = "115906",
    year = "2022"
}

@article{Aiko:2022gmz,
    author = "Aiko, Masashi and Kanemura, Shinya and Sakurai, Kodai",
    title = "{Radiative corrections to decay branching ratios of the CP-odd Higgs boson in two Higgs doublet models}",
    eprint = "2207.01032",
    archivePrefix = "arXiv",
    primaryClass = "hep-ph",
    reportNumber = "KEK-TH 2438, OU-HET 1150, TU 1162",
    doi = "10.1016/j.nuclphysb.2022.116047",
    journal = "Nucl. Phys. B",
    volume = "986",
    pages = "116047",
    year = "2023"
}

@article{Krause:2019qwe,
    author = {Krause, Marcel and M{\"u}hlleitner, Margarete},
    title = "{Impact of Electroweak Corrections on Neutral Higgs Boson Decays in Extended Higgs Sectors}",
    eprint = "1912.03948",
    archivePrefix = "arXiv",
    primaryClass = "hep-ph",
    doi = "10.1007/JHEP04(2020)083",
    journal = "JHEP",
    volume = "04",
    pages = "083",
    year = "2020"
}

@article{Akeroyd:1998uw,
    author = "Akeroyd, Andrew G. and Arhrib, Abdesslam and Naimi, El-Mokhtar",
    title = "{Yukawa coupling corrections to the decay $H^{+} \to W^{+}$ A0}",
    eprint = "hep-ph/9811431",
    archivePrefix = "arXiv",
    reportNumber = "IFIC-98-86A, FTUV-98-87, FISIST-14-98-CFIF, UFR-HEP-98-10",
    doi = "10.1007/s100529900266",
    journal = "Eur. Phys. J. C",
    volume = "12",
    pages = "451--460",
    year = "2000",
    note = "[Erratum: Eur.Phys.J.C 14, 371 (2000)]"
}

@article{Akeroyd:2000xa,
    author = "Akeroyd, A. G. and Arhrib, A. and Naimi, E.",
    title = "{Radiative corrections to the decay $H^{+} \to W^{+}$ A0}",
    eprint = "hep-ph/0002288",
    archivePrefix = "arXiv",
    reportNumber = "KEK-PREPRINT-99-182, UFR-HEP-2000-03",
    doi = "10.1007/s100520100653",
    journal = "Eur. Phys. J. C",
    volume = "20",
    pages = "51--62",
    year = "2001"
}

@article{Krause:2016oke,
    author = "Krause, Marcel and Lorenz, Robin and Muhlleitner, Margarete and Santos, Rui and Ziesche, Hanna",
    title = "{Gauge-independent Renormalization of the 2-Higgs-Doublet Model}",
    eprint = "1605.04853",
    archivePrefix = "arXiv",
    primaryClass = "hep-ph",
    doi = "10.1007/JHEP09(2016)143",
    journal = "JHEP",
    volume = "09",
    pages = "143",
    year = "2016"
}

@article{Krause:2016xku,
    author = "Krause, Marcel and Muhlleitner, Margarete and Santos, Rui and Ziesche, Hanna",
    title = "{Higgs-to-Higgs boson decays in a 2HDM at next-to-leading order}",
    eprint = "1609.04185",
    archivePrefix = "arXiv",
    primaryClass = "hep-ph",
    doi = "10.1103/PhysRevD.95.075019",
    journal = "Phys. Rev. D",
    volume = "95",
    number = "7",
    pages = "075019",
    year = "2017"
}

@article{Bojarski:2015kra,
    author = "Bojarski, F. and Chalons, G. and Lopez-Val, D. and Robens, T.",
    title = "{Heavy to light Higgs boson decays at NLO in the Singlet Extension of the Standard Model}",
    eprint = "1511.08120",
    archivePrefix = "arXiv",
    primaryClass = "hep-ph",
    doi = "10.1007/JHEP02(2016)147",
    journal = "JHEP",
    volume = "02",
    pages = "147",
    year = "2016"
}

@article{Kanemura:2015fra,
    author = "Kanemura, Shinya and Kikuchi, Mariko and Yagyu, Kei",
    title = "{Radiative corrections to the Higgs boson couplings in the model with an additional real singlet scalar field}",
    eprint = "1511.06211",
    archivePrefix = "arXiv",
    primaryClass = "hep-ph",
    reportNumber = "UT-HET-105",
    doi = "10.1016/j.nuclphysb.2016.04.005",
    journal = "Nucl. Phys. B",
    volume = "907",
    pages = "286--322",
    year = "2016"
}

@article{Kanemura:2016lkz,
    author = "Kanemura, Shinya and Kikuchi, Mariko and Yagyu, Kei",
    title = "{One-loop corrections to the Higgs self-couplings in the singlet extension}",
    eprint = "1608.01582",
    archivePrefix = "arXiv",
    primaryClass = "hep-ph",
    reportNumber = "UT-HET-116",
    doi = "10.1016/j.nuclphysb.2017.02.004",
    journal = "Nucl. Phys. B",
    volume = "917",
    pages = "154--177",
    year = "2017"
}

@article{Altenkamp:2018bcs,
    author = "Altenkamp, Lukas and Boggia, Michele and Dittmaier, Stefan",
    title = "{Precision calculations for $h \to WW/ZZ \to 4$ fermions in a Singlet Extension of the Standard Model with Prophecy4f}",
    eprint = "1801.07291",
    archivePrefix = "arXiv",
    primaryClass = "hep-ph",
    reportNumber = "FR-PHENO-2017-024",
    doi = "10.1007/JHEP04(2018)062",
    journal = "JHEP",
    volume = "04",
    pages = "062",
    year = "2018"
}

@article{Egle:2023pbm,
    author = {Egle, Felix and M{\"u}hlleitner, Margarete and Santos, Rui and Viana, Jo{\~a}o},
    title = "{Electroweak corrections to Higgs boson decays in a Complex Singlet extension of the SM and their phenomenological impact}",
    eprint = "2306.04127",
    archivePrefix = "arXiv",
    primaryClass = "hep-ph",
    doi = "10.1007/JHEP11(2023)116",
    journal = "JHEP",
    volume = "11",
    pages = "116",
    year = "2023"
}

@article{Krause:2018wmo,
    author = {Krause, Marcel and M{\"u}hlleitner, Margarete and Spira, Michael},
    title = "{2HDECAY {\textemdash}A program for the calculation of electroweak one-loop corrections to Higgs decays in the Two-Higgs-Doublet Model including state-of-the-art QCD corrections}",
    eprint = "1810.00768",
    archivePrefix = "arXiv",
    primaryClass = "hep-ph",
    doi = "10.1016/j.cpc.2019.08.003",
    journal = "Comput. Phys. Commun.",
    volume = "246",
    pages = "106852",
    year = "2020"
}

@article{Engeln:2018mbg,
    author = {Engeln, Isabell and M{\"u}hlleitner, Margarete and Wittbrodt, Jonas},
    title = "{N2HDECAY: Higgs Boson Decays in the Different Phases of the N2HDM}",
    eprint = "1805.00966",
    archivePrefix = "arXiv",
    primaryClass = "hep-ph",
    reportNumber = "DESY 18-064, KA-TP-09-2018, DESY-18-064",
    doi = "10.1016/j.cpc.2018.07.020",
    journal = "Comput. Phys. Commun.",
    volume = "234",
    pages = "256--262",
    year = "2019"
}

@article{Denner:2019fcr,
    author = {Denner, Ansgar and Dittmaier, Stefan and M{\"u}ck, Alexander},
    title = "{PROPHECY4F 3.0: A Monte Carlo program for Higgs-boson decays into four-fermion final states in and beyond the Standard Model}",
    eprint = "1912.02010",
    archivePrefix = "arXiv",
    primaryClass = "hep-ph",
    reportNumber = "FR-PHENO-2019-018, TTK-19-51",
    doi = "10.1016/j.cpc.2020.107336",
    journal = "Comput. Phys. Commun.",
    volume = "254",
    pages = "107336",
    year = "2020"
}

@article{Athron:2021kve,
    author = {Athron, Peter and B{\"u}chner, Adam and Harries, Dylan and Kotlarski, Wojciech and St{\"o}ckinger, Dominik and Voigt, Alexander},
    title = "{FlexibleDecay: An automated calculator of scalar decay widths}",
    eprint = "2106.05038",
    archivePrefix = "arXiv",
    primaryClass = "hep-ph",
    doi = "10.1016/j.cpc.2022.108584",
    journal = "Comput. Phys. Commun.",
    volume = "283",
    pages = "108584",
    year = "2023"
}

@article{Aoki:2012yt,
    author = "Aoki, Mayumi and Kanemura, Shinya and Kikuchi, Mariko and Yagyu, Kei",
    title = "{Renormalization of the Higgs Sector in the Triplet Model}",
    eprint = "1204.1951",
    archivePrefix = "arXiv",
    primaryClass = "hep-ph",
    reportNumber = "KANAZAWA-12-04, UT-HET-067",
    doi = "10.1016/j.physletb.2012.07.016",
    journal = "Phys. Lett. B",
    volume = "714",
    pages = "279--285",
    year = "2012"
}

@article{Yamada:2001px,
    author = "Yamada, Youichi",
    title = "{Gauge dependence of the on-shell renormalized mixing matrices}",
    eprint = "hep-ph/0103046",
    archivePrefix = "arXiv",
    reportNumber = "TU-612",
    doi = "10.1103/PhysRevD.64.036008",
    journal = "Phys. Rev. D",
    volume = "64",
    pages = "036008",
    year = "2001"
}

@article{Freitas:2002um,
    author = "Freitas, Ayres and Stockinger, Dominik",
    title = "{Gauge dependence and renormalization of tan beta in the MSSM}",
    eprint = "hep-ph/0205281",
    archivePrefix = "arXiv",
    reportNumber = "DESY-02-068",
    doi = "10.1103/PhysRevD.66.095014",
    journal = "Phys. Rev. D",
    volume = "66",
    pages = "095014",
    year = "2002"
}

@article{Espinosa:2002cd,
    author = "Espinosa, J. R. and Yamada, Y.",
    title = "{Scale independent and gauge independent mixing angles for scalar particles}",
    eprint = "hep-ph/0207351",
    archivePrefix = "arXiv",
    reportNumber = "IFT-UAM-CSIC-02-33, TU-665",
    doi = "10.1103/PhysRevD.67.036003",
    journal = "Phys. Rev. D",
    volume = "67",
    pages = "036003",
    year = "2003"
}

@article{Papavassiliou:1994pr,
    author = "Papavassiliou, Joannis",
    title = "{Gauge independent transverse and longitudinal self energies and vertices via the pinch technique}",
    eprint = "hep-ph/9406258",
    archivePrefix = "arXiv",
    reportNumber = "NYU-TH-94-05-02A, NYU-TH-94-05-02-",
    doi = "10.1103/PhysRevD.50.5958",
    journal = "Phys. Rev. D",
    volume = "50",
    pages = "5958--5970",
    year = "1994"
}

@article{Papavassiliou:1989zd,
    author = "Papavassiliou, Joannis",
    title = "{Gauge Invariant Proper Selfenergies and Vertices in Gauge Theories with Broken Symmetry}",
    reportNumber = "UCLA/89/TEP/36",
    doi = "10.1103/PhysRevD.41.3179",
    journal = "Phys. Rev. D",
    volume = "41",
    pages = "3179",
    year = "1990"
}

@article{Cornwall:1981zr,
    author = "Cornwall, John M.",
    title = "{Dynamical Mass Generation in Continuum QCD}",
    reportNumber = "UCLA-81-TEP-30",
    doi = "10.1103/PhysRevD.26.1453",
    journal = "Phys. Rev. D",
    volume = "26",
    pages = "1453",
    year = "1982"
}

@article{Binosi:2009qm,
    author = "Binosi, Daniele and Papavassiliou, Joannis",
    title = "{Pinch Technique: Theory and Applications}",
    eprint = "0909.2536",
    archivePrefix = "arXiv",
    primaryClass = "hep-ph",
    reportNumber = "ECT*-09-05",
    doi = "10.1016/j.physrep.2009.05.001",
    journal = "Phys. Rept.",
    volume = "479",
    pages = "1--152",
    year = "2009"
}

@article{Degrassi:1992ue,
    author = "Degrassi, Giuseppe and Sirlin, Alberto",
    title = "{Gauge invariant selfenergies and vertex parts of the Standard Model in the pinch technique framework}",
    reportNumber = "NYU-TR-92-05-02",
    doi = "10.1103/PhysRevD.46.3104",
    journal = "Phys. Rev. D",
    volume = "46",
    pages = "3104--3116",
    year = "1992"
}

@article{Cornwall:1989gv,
    author = "Cornwall, John M. and Papavassiliou, Joannis",
    title = "{Gauge Invariant Three Gluon Vertex in QCD}",
    reportNumber = "UCLA/89/TEP/24",
    doi = "10.1103/PhysRevD.40.3474",
    journal = "Phys. Rev. D",
    volume = "40",
    pages = "3474",
    year = "1989"
}

@article{Denner:2018opp,
    author = "Denner, Ansgar and Dittmaier, Stefan and Lang, Jean-Nicolas",
    title = "{Renormalization of mixing angles}",
    eprint = "1808.03466",
    archivePrefix = "arXiv",
    primaryClass = "hep-ph",
    reportNumber = "FR-PHENO-2018-10, ZU-TH 31/18",
    doi = "10.1007/JHEP11(2018)104",
    journal = "JHEP",
    volume = "11",
    pages = "104",
    year = "2018"
}

@article{Dittmaier:2022ivi,
    author = "Dittmaier, Stefan and Rzehak, Heidi",
    title = "{Electroweak renormalization based on gauge-invariant vacuum expectation values of non-linear Higgs representations. Part II. Extended Higgs sectors}",
    eprint = "2206.01479",
    archivePrefix = "arXiv",
    primaryClass = "hep-ph",
    reportNumber = "FR-PHENO-2022-05",
    doi = "10.1007/JHEP08(2022)245",
    journal = "JHEP",
    volume = "08",
    pages = "245",
    year = "2022"
}

@article{Dey:2008jm,
    author = "Dey, Paramita and Kundu, Anirban and Mukhopadhyaya, Biswarup",
    title = "{Some consequences of a Higgs triplet}",
    eprint = "0802.2510",
    archivePrefix = "arXiv",
    primaryClass = "hep-ph",
    reportNumber = "HRI-P08-02-003, HRI-RECAPP-08-02, CU-PHYSICS-02-2008",
    doi = "10.1088/0954-3899/36/2/025002",
    journal = "J. Phys. G",
    volume = "36",
    pages = "025002",
    year = "2009"
}

@article{ParticleDataGroup:2024cfk,
    author = "Navas, S. and others",
    collaboration = "Particle Data Group",
    title = "{Review of particle physics}",
    doi = "10.1103/PhysRevD.110.030001",
    journal = "Phys. Rev. D",
    volume = "110",
    number = "3",
    pages = "030001",
    year = "2024"
}

@article{Bohm:1986rj,
    author = "Bohm, M. and Spiesberger, H. and Hollik, W.",
    title = "{On the One Loop Renormalization of the Electroweak Standard Model and Its Application to Leptonic Processes}",
    doi = "10.1002/prop.19860341102",
    journal = "Fortsch. Phys.",
    volume = "34",
    pages = "687--751",
    year = "1986"
}

@article{Hollik:1988ii,
    author = "Hollik, W. F. L.",
    title = "{Radiative Corrections in the Standard Model and their Role for Precision Tests of the Electroweak Theory}",
    reportNumber = "DESY-88-188",
    doi = "10.1002/prop.2190380302",
    journal = "Fortsch. Phys.",
    volume = "38",
    pages = "165--260",
    year = "1990"
}

@article{Chankowski:2006hs,
    author = "Chankowski, Piotr H. and Pokorski, Stefan and Wagner, Jakub",
    title = "{(Non)decoupling of the Higgs triplet effects}",
    eprint = "hep-ph/0605302",
    archivePrefix = "arXiv",
    reportNumber = "IFT-9-2006",
    doi = "10.1140/epjc/s10052-007-0259-x",
    journal = "Eur. Phys. J. C",
    volume = "50",
    pages = "919--933",
    year = "2007"
}

@article{Denner:1991kt,
    author = "Denner, Ansgar",
    title = "{Techniques for calculation of electroweak radiative corrections at the one loop level and results for W physics at LEP-200}",
    eprint = "0709.1075",
    archivePrefix = "arXiv",
    primaryClass = "hep-ph",
    reportNumber = "PRINT-91-0349 (WURZBURG)",
    doi = "10.1002/prop.2190410402",
    journal = "Fortsch. Phys.",
    volume = "41",
    pages = "307--420",
    year = "1993"
}

@article{Fleischer:1980ub,
    author = "Fleischer, J. and Jegerlehner, F.",
    title = "{Radiative Corrections to Higgs Decays in the Extended Weinberg-Salam Model}",
    reportNumber = "BI-TP-80-18",
    doi = "10.1103/PhysRevD.23.2001",
    journal = "Phys. Rev. D",
    volume = "23",
    pages = "2001--2026",
    year = "1981"
}

@article{Kanemura:2017wtm,
    author = "Kanemura, Shinya and Kikuchi, Mariko and Sakurai, Kodai and Yagyu, Kei",
    title = "{Gauge invariant one-loop corrections to Higgs boson couplings in non-minimal Higgs models}",
    eprint = "1705.05399",
    archivePrefix = "arXiv",
    primaryClass = "hep-ph",
    reportNumber = "OU-HET-930, UT-HET-120",
    doi = "10.1103/PhysRevD.96.035014",
    journal = "Phys. Rev. D",
    volume = "96",
    number = "3",
    pages = "035014",
    year = "2017"
}

@article{Goodsell:2017pdq,
    author = "Goodsell, Mark D. and Liebler, Stefan and Staub, Florian",
    title = "{Generic calculation of two-body partial decay widths at the full one-loop level}",
    eprint = "1703.09237",
    archivePrefix = "arXiv",
    primaryClass = "hep-ph",
    reportNumber = "DESY-17-042, KA-TP-11-2017",
    doi = "10.1140/epjc/s10052-017-5259-x",
    journal = "Eur. Phys. J. C",
    volume = "77",
    number = "11",
    pages = "758",
    year = "2017"
}

@article{Passarino:1978jh,
    author = "Passarino, G. and Veltman, M. J. G.",
    title = "{One Loop Corrections for e+ e- Annihilation Into mu+ mu- in the Weinberg Model}",
    reportNumber = "Print-79-0284 (UTRECHT)",
    doi = "10.1016/0550-3213(79)90234-7",
    journal = "Nucl. Phys. B",
    volume = "160",
    pages = "151--207",
    year = "1979"
}

@article{Braaten:1980yq,
    author = "Braaten, E. and Leveille, J. P.",
    title = "{Higgs Boson Decay and the Running Mass}",
    reportNumber = "COO-881-127",
    doi = "10.1103/PhysRevD.22.715",
    journal = "Phys. Rev. D",
    volume = "22",
    pages = "715",
    year = "1980"
}

@article{Gorishnii:1991zr,
    author = "Gorishnii, S. G. and Kataev, A. L. and Larin, S. A. and Surguladze, L. R.",
    title = "{Scheme dependence of the next to next-to-leading QCD corrections to Gamma(tot) (H0 ---{\ensuremath{>}} hadrons) and the spurious QCD infrared fixed point}",
    doi = "10.1103/PhysRevD.43.1633",
    journal = "Phys. Rev. D",
    volume = "43",
    pages = "1633--1640",
    year = "1991"
}

@article{Chetyrkin:1995pd,
    author = "Chetyrkin, K. G. and Kwiatkowski, A.",
    title = "{Second order QCD corrections to scalar and pseudoscalar Higgs decays into massive bottom quarks}",
    eprint = "hep-ph/9505358",
    archivePrefix = "arXiv",
    reportNumber = "LBL-37269, TTP-95-23",
    doi = "10.1016/0550-3213(95)00616-8",
    journal = "Nucl. Phys. B",
    volume = "461",
    pages = "3--18",
    year = "1996"
}

@article{Larin:1995sq,
    author = "Larin, S. A. and van Ritbergen, T. and Vermaseren, J. A. M.",
    title = "{The Large top quark mass expansion for Higgs boson decays into bottom quarks and into gluons}",
    eprint = "hep-ph/9506465",
    archivePrefix = "arXiv",
    reportNumber = "NIKHEF-H-95-027",
    doi = "10.1016/0370-2693(95)01192-S",
    journal = "Phys. Lett. B",
    volume = "362",
    pages = "134--140",
    year = "1995"
}

@article{Aiko:2020ksl,
    author = "Aiko, Masashi and Kanemura, Shinya and Kikuchi, Mariko and Mawatari, Kentarou and Sakurai, Kodai and Yagyu, Kei",
    title = "{Probing extended Higgs sectors by the synergy between direct searches at the LHC and precision tests at future lepton colliders}",
    eprint = "2010.15057",
    archivePrefix = "arXiv",
    primaryClass = "hep-ph",
    reportNumber = "OU-HET 1075, KA-TP-15-2020",
    doi = "10.1016/j.nuclphysb.2021.115375",
    journal = "Nucl. Phys. B",
    volume = "966",
    pages = "115375",
    year = "2021"
}

@article{Dawson:1993qf,
    author = "Dawson, S. and Kauffman, R.",
    title = "{QCD corrections to Higgs boson production: nonleading terms in the heavy quark limit}",
    eprint = "hep-ph/9310281",
    archivePrefix = "arXiv",
    reportNumber = "BNL-DK-1, BNL-49642",
    doi = "10.1103/PhysRevD.49.2298",
    journal = "Phys. Rev. D",
    volume = "49",
    pages = "2298--2309",
    year = "1994"
}

@article{Spira:1995rr,
    author = "Spira, M. and Djouadi, A. and Graudenz, D. and Zerwas, P. M.",
    title = "{Higgs boson production at the LHC}",
    eprint = "hep-ph/9504378",
    archivePrefix = "arXiv",
    reportNumber = "DESY-94-123, UDEM-GPP-TH-95-16, CERN-TH-95-30, CERN-TH-95-030",
    doi = "10.1016/0550-3213(95)00379-7",
    journal = "Nucl. Phys. B",
    volume = "453",
    pages = "17--82",
    year = "1995"
}

@article{Chetyrkin:1997iv,
    author = "Chetyrkin, K. G. and Kniehl, Bernd A. and Steinhauser, M.",
    title = "{Hadronic Higgs decay to order alpha-s**4}",
    eprint = "hep-ph/9705240",
    archivePrefix = "arXiv",
    reportNumber = "MPI-PHT-97-006",
    doi = "10.1103/PhysRevLett.79.353",
    journal = "Phys. Rev. Lett.",
    volume = "79",
    pages = "353--356",
    year = "1997"
}

@article{Chen:2013dh,
    author = "Chen, Chian-Shu and Geng, Chao-Qiang and Huang, Da and Tsai, Lu-Hsing",
    title = "{$h\rightarrow Z\gamma$ in Type-II seesaw neutrino model}",
    eprint = "1302.0502",
    archivePrefix = "arXiv",
    primaryClass = "hep-ph",
    doi = "10.1016/j.physletb.2013.05.007",
    journal = "Phys. Lett. B",
    volume = "723",
    pages = "156--160",
    year = "2013"
}

@article{Ashanujjaman:2021txz,
    author = "Ashanujjaman, Saiyad and Ghosh, Kirtiman",
    title = "{Revisiting type-II see-saw: present limits and future prospects at LHC}",
    eprint = "2108.10952",
    archivePrefix = "arXiv",
    primaryClass = "hep-ph",
    doi = "10.1007/JHEP03(2022)195",
    journal = "JHEP",
    volume = "03",
    pages = "195",
    year = "2022"
}

@article{ATLAS:2022tnm,
    author = "Aad, Georges and others",
    collaboration = "ATLAS",
    title = "{Measurement of the properties of Higgs boson production at $\sqrt{s} = 13$ TeV in the $H\to\gamma\gamma$ channel using $139$ fb$^{-1}$ of $pp$ collision data with the ATLAS experiment}",
    eprint = "2207.00348",
    archivePrefix = "arXiv",
    primaryClass = "hep-ex",
    reportNumber = "CERN-EP-2022-094",
    doi = "10.1007/JHEP07(2023)088",
    journal = "JHEP",
    volume = "07",
    pages = "088",
    year = "2023"
}

@article{CMS:2021kom,
    author = "Sirunyan, Albert M and others",
    collaboration = "CMS",
    title = "{Measurements of Higgs boson production cross sections and couplings in the diphoton decay channel at $ \sqrt{\mathrm{s}} $ = 13 TeV}",
    eprint = "2103.06956",
    archivePrefix = "arXiv",
    primaryClass = "hep-ex",
    reportNumber = "CMS-HIG-19-015, CERN-EP-2021-038",
    doi = "10.1007/JHEP07(2021)027",
    journal = "JHEP",
    volume = "07",
    pages = "027",
    year = "2021"
}

@article{Kanemura:2013mc,
    author = "Kanemura, Shinya and Kikuchi, Mariko and Yagyu, Kei",
    title = "{Probing exotic Higgs sectors from the precise measurement of Higgs boson couplings}",
    eprint = "1301.7303",
    archivePrefix = "arXiv",
    primaryClass = "hep-ph",
    reportNumber = "UT-HET-077",
    doi = "10.1103/PhysRevD.88.015020",
    journal = "Phys. Rev. D",
    volume = "88",
    pages = "015020",
    year = "2013"
}

@article{Kanemura:2014bqa,
    author = "Kanemura, Shinya and Tsumura, Koji and Yagyu, Kei and Yokoya, Hiroshi",
    title = "{Fingerprinting nonminimal Higgs sectors}",
    eprint = "1406.3294",
    archivePrefix = "arXiv",
    primaryClass = "hep-ph",
    reportNumber = "KUNS-2500, UT-HET-095",
    doi = "10.1103/PhysRevD.90.075001",
    journal = "Phys. Rev. D",
    volume = "90",
    pages = "075001",
    year = "2014"
}

@article{Kanemura:2017gbi,
    author = "Kanemura, Shinya and Kikuchi, Mariko and Sakurai, Kodai and Yagyu, Kei",
    title = "{H-COUP: a program for one-loop corrected Higgs boson couplings in non-minimal Higgs sectors}",
    eprint = "1710.04603",
    archivePrefix = "arXiv",
    primaryClass = "hep-ph",
    reportNumber = "OU-HET-947, UT-HET-122, OU-HET 947, UT-HET 122",
    doi = "10.1016/j.cpc.2018.06.012",
    journal = "Comput. Phys. Commun.",
    volume = "233",
    pages = "134--144",
    year = "2018"
}

@article{Kanemura:2019slf,
    author = "Kanemura, Shinya and Kikuchi, Mariko and Mawatari, Kentarou and Sakurai, Kodai and Yagyu, Kei",
    title = "{H-COUP Version 2: a program for one-loop corrected Higgs boson decays in non-minimal Higgs sectors}",
    eprint = "1910.12769",
    archivePrefix = "arXiv",
    primaryClass = "hep-ph",
    reportNumber = "OU-HET 1024, KA-TP-17-2019",
    doi = "10.1016/j.cpc.2020.107512",
    journal = "Comput. Phys. Commun.",
    volume = "257",
    pages = "107512",
    year = "2020"
}

@article{Aiko:2023xui,
    author = "Aiko, Masashi and Kanemura, Shinya and Kikuchi, Mariko and Sakurai, Kodai and Yagyu, Kei",
    title = "{H-COUP Version 3: A program for one-loop corrected decays of any Higgs bosons in non-minimal Higgs models}",
    eprint = "2311.15892",
    archivePrefix = "arXiv",
    primaryClass = "hep-ph",
    reportNumber = "KEK-TH-2578, OU-HET 1201, NU-EHET 002, TU-1214",
    doi = "10.1016/j.cpc.2024.109231",
    journal = "Comput. Phys. Commun.",
    volume = "301",
    pages = "109231",
    year = "2024"
}

@article{Aiko:2023nqj,
    author = "Aiko, Masashi and Braathen, Johannes and Kanemura, Shinya",
    title = "{Leading two-loop corrections to the Higgs di-photon decay in the inert doublet model}",
    eprint = "2307.14976",
    archivePrefix = "arXiv",
    primaryClass = "hep-ph",
    reportNumber = "DESY-23-104, KEK-TH-2539, OU-HET-1194",
    doi = "10.1140/epjc/s10052-025-14184-z",
    journal = "Eur. Phys. J. C",
    volume = "85",
    number = "5",
    pages = "489",
    year = "2025"
}

@article{Braathen:2019pxr,
    author = "Braathen, Johannes and Kanemura, Shinya",
    title = "{On two-loop corrections to the Higgs trilinear coupling in models with extended scalar sectors}",
    eprint = "1903.05417",
    archivePrefix = "arXiv",
    primaryClass = "hep-ph",
    reportNumber = "OU-HET-1001",
    doi = "10.1016/j.physletb.2019.07.021",
    journal = "Phys. Lett. B",
    volume = "796",
    pages = "38--46",
    year = "2019"
}

@article{Braathen:2020vwo,
    author = "Braathen, Johannes and Kanemura, Shinya and Shimoda, Makoto",
    title = "{Two-loop analysis of classically scale-invariant models with extended Higgs sectors}",
    eprint = "2011.07580",
    archivePrefix = "arXiv",
    primaryClass = "hep-ph",
    reportNumber = "OU-HET-1077, DESY-20-192, DESY 20-192",
    doi = "10.1007/JHEP03(2021)297",
    journal = "JHEP",
    volume = "03",
    pages = "297",
    year = "2021"
}

@article{Braathen:2019zoh,
    author = "Braathen, Johannes and Kanemura, Shinya",
    title = "{Leading two-loop corrections to the Higgs boson self-couplings in models with extended scalar sectors}",
    eprint = "1911.11507",
    archivePrefix = "arXiv",
    primaryClass = "hep-ph",
    reportNumber = "OU-HET-1030",
    doi = "10.1140/epjc/s10052-020-7723-2",
    journal = "Eur. Phys. J. C",
    volume = "80",
    number = "3",
    pages = "227",
    year = "2020"
}

@article{Degrassi:2023eii,
    author = "Degrassi, Giuseppe and Slavich, Pietro",
    title = "{On the two-loop BSM corrections to $h\longrightarrow \gamma \gamma $ in the aligned THDM}",
    eprint = "2307.02476",
    archivePrefix = "arXiv",
    primaryClass = "hep-ph",
    doi = "10.1140/epjc/s10052-023-12097-3",
    journal = "Eur. Phys. J. C",
    volume = "83",
    number = "10",
    pages = "941",
    year = "2023"
}

@article{Degrassi:2024qsf,
    author = "Degrassi, Giuseppe and Slavich, Pietro",
    title = "{On the two-loop BSM corrections to $h\longrightarrow \gamma \gamma $ in a triplet extension of the SM}",
    eprint = "2407.18185",
    archivePrefix = "arXiv",
    primaryClass = "hep-ph",
    doi = "10.1140/epjc/s10052-025-13758-1",
    journal = "Eur. Phys. J. C",
    volume = "85",
    number = "1",
    pages = "49",
    year = "2025"
}

@article{Sturm:2022hak,
    author = "Sturm, Christian and Summ, Benjamin and Uccirati, Sandro",
    title = "{Electroweak corrections to g + g {\textrightarrow} H$_{l,h}$ and H$_{l,h}${\textrightarrow} {\ensuremath{\gamma}} + {\ensuremath{\gamma}} in the Higgs-singlet extension of the Standard model}",
    eprint = "2212.11835",
    archivePrefix = "arXiv",
    primaryClass = "hep-ph",
    doi = "10.1007/JHEP11(2023)113",
    journal = "JHEP",
    volume = "11",
    pages = "113",
    year = "2023"
}

@article{Degrassi:2025pqt,
  author        = {Degrassi, Giuseppe and Gr{\"o}ber, Ramona and Slavich, Pietro},
  title         = {Two-loop BSM contributions to Higgs pair production in the aligned THDM},
  eprint        = {2508.11539},
  archivePrefix = {arXiv},
  primaryClass  = {hep-ph},
  journal       = "arXiv",
  year          = {2025},
  month         = {Aug},
}

@article{Nielsen:1975fs,
    author = "Nielsen, N. K.",
    title = "{On the Gauge Dependence of Spontaneous Symmetry Breaking in Gauge Theories}",
    reportNumber = "Print-75-0792 (AARHUS)",
    doi = "10.1016/0550-3213(75)90301-6",
    journal = "Nucl. Phys. B",
    volume = "101",
    pages = "173--188",
    year = "1975"
}

@article{deBlas:2019rxi,
    author = "de Blas, J. and others",
    title = "{Higgs Boson Studies at Future Particle Colliders}",
    eprint = "1905.03764",
    archivePrefix = "arXiv",
    primaryClass = "hep-ph",
    reportNumber = "DESY-19-079",
    doi = "10.1007/JHEP01(2020)139",
    journal = "JHEP",
    volume = "01",
    pages = "139",
    year = "2020"
}

@article{Misiak:2017bgg,
    author = "Misiak, Mikolaj and Steinhauser, Matthias",
    title = "{Weak radiative decays of the B meson and bounds on $M_{H^\pm }$ in the Two-Higgs-Doublet Model}",
    eprint = "1702.04571",
    archivePrefix = "arXiv",
    primaryClass = "hep-ph",
    reportNumber = "TTP17-004, IFT-1-2017",
    doi = "10.1140/epjc/s10052-017-4776-y",
    journal = "Eur. Phys. J. C",
    volume = "77",
    number = "3",
    pages = "201",
    year = "2017"
}

@article{Misiak:2020vlo,
    author = "Misiak, M. and Rehman, Abdur and Steinhauser, Matthias",
    title = "{Towards $ \overline{B}\to {X}_s\gamma $ at the NNLO in QCD without interpolation in m$_{c}$}",
    eprint = "2002.01548",
    archivePrefix = "arXiv",
    primaryClass = "hep-ph",
    reportNumber = "TTP20-001, P3H-20-005, IFT-01/2020",
    doi = "10.1007/JHEP06(2020)175",
    journal = "JHEP",
    volume = "06",
    pages = "175",
    year = "2020"
}

@article{FileviezPerez:2008jbu,
    author = "Fileviez Perez, Pavel and Han, Tao and Huang, Gui-yu and Li, Tong and Wang, Kai",
    title = "{Neutrino Masses and the CERN LHC: Testing Type II Seesaw}",
    eprint = "0805.3536",
    archivePrefix = "arXiv",
    primaryClass = "hep-ph",
    reportNumber = "MADPH-08-1510, NSF-KITP-08-65",
    doi = "10.1103/PhysRevD.78.015018",
    journal = "Phys. Rev. D",
    volume = "78",
    pages = "015018",
    year = "2008"
}

@article{Asakawa:2005gv,
    author = "Asakawa, Eri and Kanemura, Shinya",
    title = "{The H+- W-+ Z0 vertex and single charged Higgs boson production via W Z fusion at the large hadron collider}",
    eprint = "hep-ph/0506310",
    archivePrefix = "arXiv",
    reportNumber = "YITP-05-31, OU-HET-533",
    doi = "10.1016/j.physletb.2005.08.091",
    journal = "Phys. Lett. B",
    volume = "626",
    pages = "111--119",
    year = "2005"
}

@article{Asakawa:2006gm,
    author = "Asakawa, Eri and Kanemura, Shinya and Kanzaki, Junichi",
    title = "{Potential for measuring the H+- W-+ Z0 vertex from WZ fusion at the Large Hadron Collider}",
    eprint = "hep-ph/0612271",
    archivePrefix = "arXiv",
    reportNumber = "KEK-TH-1122, UT-HET-005",
    doi = "10.1103/PhysRevD.75.075022",
    journal = "Phys. Rev. D",
    volume = "75",
    pages = "075022",
    year = "2007"
}

@article{Godfrey:2010qb,
    author = "Godfrey, Stephen and Moats, Ken",
    title = "{Exploring Higgs Triplet Models via Vector Boson Scattering at the LHC}",
    eprint = "1003.3033",
    archivePrefix = "arXiv",
    primaryClass = "hep-ph",
    doi = "10.1103/PhysRevD.81.075026",
    journal = "Phys. Rev. D",
    volume = "81",
    pages = "075026",
    year = "2010"
}

@article{Huitu:1996su,
    author = "Huitu, K. and Maalampi, J. and Pietila, A. and Raidal, M.",
    title = "{Doubly charged Higgs at LHC}",
    eprint = "hep-ph/9606311",
    archivePrefix = "arXiv",
    reportNumber = "HU-SEFT-R-1996-16, FTUV-96-34A, IFIC-96-40, TURKU-SFL-R-14",
    doi = "10.1016/S0550-3213(97)87466-4",
    journal = "Nucl. Phys. B",
    volume = "487",
    pages = "27--42",
    year = "1997"
}

@article{Melfo:2011nx,
    author = "Melfo, Alejandra and Nemevsek, Miha and Nesti, Fabrizio and Senjanovic, Goran and Zhang, Yue",
    title = "{Type II Seesaw at LHC: The Roadmap}",
    eprint = "1108.4416",
    archivePrefix = "arXiv",
    primaryClass = "hep-ph",
    doi = "10.1103/PhysRevD.85.055018",
    journal = "Phys. Rev. D",
    volume = "85",
    pages = "055018",
    year = "2012"
}

@article{CMS:2025jwz,
    author = "{CMS Collaboration}", 
    title = "{Combined measurements and interpretations of Higgs boson production and decay at $\sqrt s$=13 TeV}",
    reportNumber = "{CMS-PAS-HIG-21-018}",
    journal = "CMS-PAS-HIG-21-018",
    month = "April",
    year = "2025"
}

@article{AidanSean:Feynman,
    author = "{Aidan Randle-Conde}",
    title        = "{Feynman Diagram Tutorial}",
    reportNumber = "{\url{https://www.aidansean.com/feynman/}}",
    journal = "https://www.aidansean.com/feynman/",
    year = "2010"
}

@article{Jurciukonis:2024bzx,
    author = "Jur{\v{c}}iukonis, Darius and Lavoura, Lu{\'{\i}}s",
    title = "{On the extension of the SM through a scalar quadruplet}",
    eprint = "2406.01628",
    archivePrefix = "arXiv",
    primaryClass = "hep-ph",
    journal = " ",
}

@article{Grifols:1980uq,
    author = "Grifols, J. A. and Mendez, A.",
    title = "{The $WZH^\pm$ Coupling in $SU(2) \times U(1)$ Gauge Models}",
    reportNumber = "UAB-FT-58",
    doi = "10.1103/PhysRevD.22.1725",
    journal = "Phys. Rev. D",
    volume = "22",
    pages = "1725",
    year = "1980"
}

@article{Papavassiliou:1997pb,
    author = "Papavassiliou, Joannis and Pilaftsis, Apostolos",
    title = "{Gauge and renormalization group invariant formulation of the Higgs boson resonance}",
    eprint = "hep-ph/9710426",
    archivePrefix = "arXiv",
    reportNumber = "CERN-TH-97-293, MPI-PHT-97-67",
    doi = "10.1103/PhysRevD.58.053002",
    journal = "Phys. Rev. D",
    volume = "58",
    pages = "053002",
    year = "1998"
}

\end{document}